\documentclass[twocolumn]{aastex631}

\usepackage{graphicx} 
\usepackage{color}
\usepackage{float}
\usepackage{tabularx}
\usepackage{comment}
\usepackage{natbib}
\usepackage{amsmath}
\usepackage{booktabs}
\usepackage{array}
\usepackage{hyperref}
\graphicspath{{./}{figures/}}

\begin{document}


\title{SCUBADive I: JWST$+$ALMA Analysis of 289 sub-millimeter galaxies in COSMOS-Web}

\author[0000-0002-6149-8178]{Jed McKinney}
\altaffiliation{NASA Hubble Fellow}
\affiliation{Department of Astronomy, The University of Texas at Austin, Austin, TX, USA}

\author[0000-0002-0930-6466]{Caitlin M. Casey}
\affiliation{Department of Astronomy, The University of Texas at Austin, Austin, TX, USA}

\author[0000-0002-7530-8857]{Arianna S. Long}
\altaffiliation{NASA Hubble Fellow}
\affiliation{Department of Astronomy, The University of Texas at Austin, Austin, TX, USA}

\author[0000-0003-3881-1397]{Olivia R. Cooper}\altaffiliation{NSF Graduate Research Fellow}
\affiliation{Department of Astronomy, The University of Texas at Austin, Austin, TX, USA}

\author[0000-0003-0415-0121]{Sinclaire M. Manning}
\altaffiliation{NASA Hubble Fellow}
\affiliation{Department of Astronomy, University of Massachusetts Amherst, 710 N Pleasant Street, Amherst, MA 01003, USA}

\author[0000-0002-3560-8599]{Maximilien Franco}
\affiliation{Department of Astronomy, The University of Texas at Austin, Austin, TX, USA}

\author[0000-0003-3596-8794]{Hollis Akins}\altaffiliation{NSF Graduate Research Fellow}
\affiliation{Department of Astronomy, The University of Texas at Austin, Austin, TX, USA}

\author[0000-0003-3216-7190]{Erini Lambrides}\altaffiliation{NPP Fellow}
\affiliation{NASA-Goddard Space Flight Center, Code 662, Greenbelt, MD, 20771, USA}

\author[0009-0007-0553-9610]{Elaine Gammon}\altaffiliation{Goldwater Scholar}
\affiliation{Department of Physics \& Astronomy, University of Georgia, Sanford Drive, Athens, GA 30602, USA}

\author[0000-0002-0113-7784]{Camila Silva}
\affiliation{Department of Astronomy and Astrophysics, The University of Chicago, Chicago, IL, USA}
\affiliation{Department of Astronomy, The University of Texas at Austin, Austin, TX, USA}

\author[0000-0002-8008-9871]{Fabrizio Gentile}
\affiliation{University of Bologna - Department of Physics and Astronomy “Augusto Righi” (DIFA), Via Gobetti 93/2, I-40129 Bologna, Italy}
\affiliation{INAF- Osservatorio di Astrofisica e Scienza dello Spazio, Via Gobetti 93/3, I-40129, Bologna, Italy}

\author[0000-0002-7051-1100]{Jorge A. Zavala}
\affiliation{National Astronomical Observatory of Japan, 2-21-1 Osawa, Mitaka, Tokyo 181-8588, Japan}

\author[0000-0002-4465-1564]{Aristeidis Amvrosiadis}
\affiliation{Centre for Extragalactic Astronomy, Durham University, South Road, Durham DH1 3LE, UK}
\affiliation{Institute for Computational Cosmology, Durham University, South Road, Durham DH1 3LE, UK}

\author[0000-0001-6102-9526]{Irham Andika}
\affiliation{Technical University of Munich, TUM School of Natural Sciences, Department of Physics, James-Franck-Str. 1, D-85748 Garching, Germany}
\affiliation{Max-Planck-Institut fur Astrophysik, Karl-Schwarzschild-Str. 1, D-85748 Garching, Germany}

\author[0000-0002-0245-6365]{Malte Brinch}
\affiliation{Cosmic Dawn Center (DAWN), Copenhagen, Denmark}
\affiliation{DTU-Space, Technical University of Denmark, Elektrovej 327, 2800, Kgs. Lyngby, Denmark}

\author[0000-0002-6184-9097]{Jaclyn B. Champagne}
\affiliation{Steward Observatory, University of Arizona, 933 N Cherry Ave, Tucson, AZ 85721, USA}

\author[0000-0003-3691-937X]{Nima Chartab}
\affiliation{Caltech/IPAC, 1200 E. California Blvd., Pasadena, CA 91125, USA}

\author[0000-0003-4761-2197]{Nicole E. Drakos}
\affiliation{Department of Physics and Astronomy, University of Hawaii, Hilo, 200 W Kawili St, Hilo, HI 96720, USA}

\author[0000-0002-9382-9832]{Andreas L. Faisst}
\affiliation{Caltech/IPAC, 1200 E. California Blvd., Pasadena, CA 91125, USA}

\author[0000-0001-7201-5066]{Seiji Fujimoto}
\altaffiliation{NASA Hubble Fellow}
\affiliation{Department of Astronomy, The University of Texas at Austin, Austin, TX, USA}

\author[0000-0001-9885-4589]{Steven Gillman}
\affiliation{Cosmic Dawn Center (DAWN), Copenhagen, Denmark}
\affiliation{DTU-Space, Technical University of Denmark, Elektrovej 327, 2800, Kgs. Lyngby, Denmark}

\author[0000-0002-0236-919X]{Ghassem Gozaliasl}
\affiliation{Department of Computer Science, Aalto University, P.O. Box 15400, FI-00076 Espoo, Finland}
\affiliation{Department of Physics, University of, P.O. Box 64, FI-00014 Helsinki, Finland}

\author[0000-0002-2554-1837]{Thomas R. Greve}
\affiliation{Cosmic Dawn Center (DAWN), Copenhagen, Denmark}
\affiliation{DTU-Space, Technical University of Denmark, Elektrovej 327, 2800, Kgs. Lyngby, Denmark}
\affiliation{Dept. of Physics and Astronomy, University College London, Gower Street, London WC1E 6BT, United Kingdom}

\author[0000-0003-0129-2079]{Santosh Harish}
\affiliation{Laboratory for Multiwavelength Astrophysics, School of Physics and Astronomy, Rochester Institute of Technology, 84 Lomb Memorial Drive, Rochester, NY 14623, USA}

\author[0000-0003-4073-3236]{Christopher C. Hayward}
\affiliation{Center for Computational Astrophysics, Flatiron Institute, 162 Fifth Avenue, New York, NY 10010, USA}

\author[0000-0002-3301-3321]{Michaela Hirschmann}
\affiliation{Institute of Physics, GalSpec, Ecole Polytechnique Federale de Lausanne, Observatoire de Sauverny, Chemin Pegasi 51, 1290 Versoix, Switzerland}
\affiliation{INAF- Osservatorio di Astrofisica e Scienza dello Spazio, Via Gobetti 93/3, I-40129, Bologna, Italy}

\author[0000-0002-7303-4397]{Olivier Ilbert}
\affiliation{Aix Marseille Univ, CNRS, CNES, LAM, Marseille, France  }

\author[0000-0001-9215-7053]{Boris S. Kalita}
\altaffiliation{Kavli Astrophysics Fellow}
\affiliation{Kavli Institute for the Physics and Mathematics of the Universe, The University of Tokyo, Kashiwa, 277-8583, Japan }
\affiliation{Kavli Institute for Astronomy and Astrophysics, Peking University, Beijing 100871, People{\textquotesingle}s Republic of China}
\affiliation{Center for Data-Driven Discovery, Kavli IPMU (WPI), UTIAS, The University of Tokyo, Kashiwa, Chiba 277-8583, Japan} 

\author[0000-0001-9187-3605]{Jeyhan S. Kartaltepe}
\affiliation{Laboratory for Multiwavelength Astrophysics, School of Physics and Astronomy, Rochester Institute of Technology, 84 Lomb Memorial Drive, Rochester, NY 14623, USA}

\author[0000-0002-6610-2048]{Anton M. Koekemoer}
\affiliation{Space Telescope Science Institute, 3700 San Martin Dr., Baltimore, MD 21218, USA} 

\author[0000-0002-5588-9156]{Vasily Kokorev}
\affiliation{Department of Astronomy, The University of Texas at Austin, Austin, TX, USA}

\author[0000-0001-9773-7479]{Daizhong Liu}
\affiliation{Purple Mountain Observatory, Chinese Academy of Sciences, 10 Yuanhua Road, Nanjing 210023, China}

\author[0000-0002-4872-2294]{Georgios Magdis}
\affil{Cosmic Dawn Center (DAWN), Copenhagen, Denmark}
\affiliation{DTU-Space, Technical University of Denmark, Elektrovej 327, 2800, Kgs. Lyngby, Denmark}
\affiliation{Niels Bohr Institute, University of Copenhagen, Jagtvej 128, DK-2200, Copenhagen, Denmark}

\author[0000-0002-9489-7765]{Henry Joy McCracken}
\affiliation{Institut d’Astrophysique de Paris, UMR 7095, CNRS, and Sorbonne Université, 98 bis boulevard Arago, F-75014 Paris, France}

\author[0000-0002-4485-8549]{Jason Rhodes}
\affiliation{Jet Propulsion Laboratory, California Institute of Technology, 4800 Oak Grove Drive, Pasadena, CA 91001, USA}

\author[0000-0002-4271-0364]{Brant E. Robertson}
\affiliation{Department of Astronomy and Astrophysics, University of California, Santa Cruz, 1156 High Street, Santa Cruz, CA 95064, USA}

\author[0000-0003-4352-2063]{Margherita Talia}
\affiliation{University of Bologna - Department of Physics and Astronomy “Augusto Righi” (DIFA), Via Gobetti 93/2, I-40129 Bologna, Italy}
\affiliation{INAF- Osservatorio di Astrofisica e Scienza dello Spazio, Via Gobetti 93/3, I-40129, Bologna, Italy}

\author[0000-0001-6477-4011]{Francesco Valentino}
\affiliation{European Southern Observatory, Karl-Schwarzschild-Str. 2, D-85748, Garching bei Munchen, Germany}
\affil{Cosmic Dawn Center (DAWN), Copenhagen, Denmark}

\author[0000-0002-1905-4194]{Aswin P. Vijayan}
\affiliation{Astronomy Centre, University of Sussex, Falmer, Brighton BN1 9QH,  UK} 

\begin{abstract}
   JWST has enabled detecting and spatially resolving the heavily dust-attenuated stellar populations of sub-millimeter galaxies, revealing detail that was previously inaccessible. In this work we construct a sample of 289 sub-millimeter galaxies with joint ALMA and JWST constraints in the COSMOS field. Sources are originally selected using the SCUBA-2 instrument and have archival ALMA observations from various programs. Their JWST NIRCam imaging is from COSMOS-Web and PRIMER. We extract multi-wavelength photometry in a manner that leverages the unprecedented near-infrared spatial resolution of JWST, and fit the data with spectral energy distribution models to derive photometric redshifts, stellar masses, star-formation rates and optical attenuation. The sample has an average $\langle z\rangle=2.6^{+1.0}_{-0.8}$, $\langle A_V \rangle = 2.5^{+1.5}_{-1.0}$, $\langle{\rm SFR}\rangle=300^{+400}_{-200}\,M_\odot\,{\rm yr^{-1}}$ and $\langle\log(M_*/M_\odot)\rangle=11.1^{+0.3}_{-0.5}$. There are 81 ($30\%$) galaxies that have no previous optical/near-infrared detections, including $75\%$ of the $z>4$ sub-sample ($n=28$). The faintest observed near-infrared sources have the highest redshifts and largest $A_V=4\pm1$. In a preliminary morphology analysis we find that $\sim10\%$ of our sample exhibit spiral arms and $~5\%$ host stellar bars, with one candidate bar found at $z>3$. Finally, we find that the clustering of JWST sources within $10^{\prime\prime}$ of a sub-mm galaxy is a factor of 2 greater than what is expected based on either random clustering or the distribution of sources around any red galaxy irrespective of a sub-mm detection. 
\end{abstract}


\section{Introduction} \label{sec:intro}

Extragalactic sources selected from deep and wide sub-millimeter (mm) fields capture galaxies in an evolutionary stage of rapid mass assembly. The aptly-named ``Sub-mm Galaxies'' (SMGs, $S_{850\,\mu m}\gtrsim 1$mJy) are very luminous in the infrared (IR) with (${\rm L_{IR}/L_\odot}\,>10^{12}$), large stellar masses ($M_*/M_\odot\sim10^{11}$), and typical redshifts of $z \sim 2-3$ (see e.g., \citealt{Casey2014,Hodge2020} for a review, and the initial discovery papers of \citealt{Smail1997,Barger1998,Hughes1998,Eales1999}). SMGs can form up to $100-1000\,\mathrm{M_\odot}$ worth of new stars each year, and are responsible for most of the volume-averaged star formation over the last 10 billion years (i.e., from $0<z<3$, \citealt{Murphy2011,MadauDickinson2014,Zavala2021}). It is the energy from this vigorous star-formation, and in some cases accreting supermassive black holes \citep{McKinney2021agn}, that heats significant reservoirs of cold dust and drives bright sub-mm emission \citep{Sanders1996}. This dust simultaneously attenuates rest-frame UV/optical light leading to typical $A_V\sim 1-5$ in SMGs \citep{daCunha2015,Dudzeviciute2020}. That, combined with $\langle z \rangle = 2.5$ \citep{daCunha2015,Dudzeviciute2020}, causes SMGs to be very faint in the rest-frame UV/optical. As a result, SMGs have been historically difficult to detect with ground- and space-based facilities operating in the optical/near-infrared regime. This is especially true for SMGs at $z>3$, which has lead to a panoply of monikers like ``optically faint radio galaxies'', ``optically-dark'', ``near-infrared-dark'' and ``\textit{HST}-dark'', all used to describe such dust-obscured sources that elude detection by the most sensitive optical/near-infrared telescopes \citep[e.g.,][]{Chapman2004,Casey2009,Wang2016,Franco2018,Williams2019,Smail2021,Manning2022,McKinney2023,Kokorev2023,Price2023}. 

Thanks to JWST, the direct stellar light from SMGs is readily detectable with the NIRCam Long Wavelength filters ($2.4-5\,\mu$m)  up to $z\sim6$ and $A_V\sim6$ \citep{Chen2022,Cheng2022,Gillman2023,McKinney2023,Rujopakarn2023,Wu2023,Zavala2022}. This enables more robust photometric redshifts \citep{cosmos-web}, especially so for the subset of sources with no prior optical/near-IR counterpart. The inclusion of JWST bands can also dramatically impact derived stellar masses \citep{McKinney2023}, thus yielding more precise constraints on the mass assembly in the Universe's most prolific star factories. 

Furthermore, the angular resolution achieved by JWST enables detailed stellar morphology measurements in the SMGs, a historically impossible task given their marginal detections in e.g., \textit{HST} ACS/WFC3. 
Recently, \cite{Gillman2023} studied a sample of SMGs matched to optical counterparts based on their NIRCam colors, recovering the standard SMG redshift posterior and finding compact and clumpy stellar morphologies. 
\cite{LeBail2023} find similarly clumpy NIRCam morphologies and identify sub-regions of quiescence within $z\sim2$ galaxies still undergoing active star-formation. 
\cite{Hodge2024} report a strong correlation between the dust continuum and reddening of NIRCam colors on kpc scales in 13 SMGs favoring dust obscuration as the primary source of their red rest-frame optical colors. 
\cite{McKinney2023} use ALMA and JWST to identify a SMG at $z\sim6$, adding to the growing sample of $z>5$ SMGs that is coming into focus \citep{Williams2019,Chen2021}. Thus JWST is providing a new and powerful perspective on canonical $z\sim1-3$ SMGs while also revealing stellar light in the most distant SMGs for the first time. This is profoundly changing our understanding of their physical nature, as for example it is difficult to test for morphological signatures of mergers without being able to detect, let alone resolve, the stellar light.  

These recent works stand atop decades of thorough analysis of SMGs investigating their nature and distribution across cosmic time and environment. Indeed, multi-wavelength follow-up of SMGs has been an active field, first using radio interferometers like the Very Large Array (VLA) to identify optical counterparts \citep{Barger2000} and then later the Atacama Large Millimetre Array (ALMA) at sub-mm wavelengths \citep{Karim2013,Hodge2013}. 
Near- and mid-infrared follow-up campaigns have also been critical for establishing the physical nature of SMGs \citep{Swinbank2004,Chapman2005,Swinbank2008}. 
In recent years samples have grown beyond the $<100$ sources typical of early pioneering studies to more than $1000$ SMGs with counterparts across the multi-wavelength spectrum \citep{Brisbin2017,Simpson2020,Dudzeviciute2020}. 

In this paper, we make a statistical effort to match a large sample of SMGs identified with the SCUBA-2 sub-mm instrument to their JWST counterparts. We analyze sources in the COSMOS field \citep{Scoville2007,Capak2007,Sanders2007} using data from the largest contiguous JWST mosaic, COSMOS-Web \citep{cosmos-web}. We use archival ALMA data to resolve the low resolution SCUBA-2 data on sub-arcsecond scales in order to identify the optical counterparts for SMGs. By construction every source in our sample has both JWST and ALMA detections. This is the first paper in a series, and precedes a thorough analysis of the JWST morphologies of SMGs, their spatially-resolved stellar properties, and the sub-set at the highest redshifts (Manning et al., in prep.). 
In this work we describe our sample and comment on properties of the population as a whole. 

This paper is structured as follows: In Section \ref{sec:data} we review the multi-wavelength data sets incorporated into our work. In Section \ref{sec:sample} we outline our sample selection criterion. Section \ref{sec:alma} describes our reduction of archival ALMA data sets. Section \ref{sec:phot} describes the manner in which we measure photometry for our sample across the electromagnetic spectrum. We describe our modeling of the galaxy-integrated spectral energy distributions (SEDs) in Section \ref{sec:quant} and present the corresponding results in Section \ref{sec:results}. Finally, we discuss the implications of our findings in Section \ref{sec:discussion}. 
Throughout this work we assume a $\Lambda$CDM cosmology with $H_0=70$\,km\,s$^{-1}$\,Mpc$^{-1}$, $\Omega_m=0.3$, $\Omega_\Lambda=0.7$, and a Chabrier initial mass function (IMF, \citealt{Chabrier2003}). We report AB magnitudes \citep{Oke1974}.

\section{Data\label{sec:data}} 
Our work focuses on sources in the COSMOS extragalactic field which hosts an extensive collection of multi-wavelength data. Most relevant to this work is the JWST Cyle 1 GO program COSMOS-Web, a 0.54 deg$^2$ survey using NIRCam and MIRI (PID \#1727, PIs Kartaltepe \& Casey, \citealt{cosmos-web}) centered in the middle of the 2 deg$^2$ field. The NIRCam filters include F115W, F150W, F277W, and F444W with $5\sigma$ depths between $\sim27-28$ mag, and are accompanied by a 0.19 deg$^2$ non-contiguous MIRI/F770W map with $5\sigma\sim25.8$ mag (Franco et al., in prep., Harish et al., in prep., \citealt{cosmos-web}). We use the extensive compilation of rest-frame optical and near-infrared maps from COSMOS2020 \citep{Weaver2022} including \textit{HST}/ACS imaging from \cite{Koekemoer2007} and \textit{Spitzer}/IRAC from S-COSMOS \citep{Sanders2007,Moneti2022}. 

Our sub-mm sources are selected from the James Clerk Maxwell Telescope SCUBA-2 map of the COSMOS field (S2COSMOS, \citealt{Simpson2019}). We also include \textit{Spitzer}/MIPS $24$\micron\ from \cite{Sanders2007} and \cite{LeFloch2009}, far-IR maps from \textit{Herschel}/PACS+SPIRE \citep{Lutz2011,Oliver2012}, sub-mm imaging from AzTEC on ASTE \citep{Aretxaga2011}, and radio maps with the VLA at 1.4 GHz and 3 GHz \citep{Schinnerer2010,Smolcic2017}. As described in the following section, we supplement these wide-area extragalactic data sets with archival pointed Atacama Large Millimetre Array (ALMA) observations between $850-1250\,\mu$m. 

\section{Sample Selection\label{sec:sample}}
We construct our sample beginning with a $S_{850\mu m}>2$ mJy source catalog from the JCMT/SCUBA-2 map of COSMOS (S2COSMOS, \citealt{Simpson2019}). Given the inhomogeneous S2COSMOS field depth between $\sigma_{850}\sim0.5-1.4$ mJy, a flux-limited selection of 2 mJy corresponds to 1985 sources with ${\rm SNR_{850}>3.5}$. Next, we require overlap with the $0.54$ deg$^2$ COSMOS-Web NIRCam area which contains 706 SCUBA-2 sources with $S_{850\mu m}>2$ mJy. Of these, 60\% have SNR$_{850}\in[3.5,5]$ with the remaining $40\%$ having SNR$_{850}>5$.  

The SCUBA-2 PSF has a FWHM of $14.9^{\prime\prime}$ which allows for many JWST sources per SCUBA-2 beam. Methods using NIRCam colors to identify SCUBA-2 counterparts have had some success \cite[e.g.,][]{Gillman2023}, but can still miss the true source of the dust-emission. The only way to unambiguously associate SMGs with their multi-wavelength counterparts is with high-resolution follow-up using a mm interferometer like ALMA, SMA, NOEMA \citep{Hodge2013,Karim2013,Hezaveh2013,Brisbin2017,Dudzeviciute2020,Simpson2020} or a radio interferometer like the VLA \citep{Chapman2005,Talia2021} although a significant amount of SMGs can go undetected in deep radio maps \citep{Chapman2003,Barger2007}. Therefore, our second selection criterion is the presence of an archival ALMA observation within the FHWM of the SCUBA-2 source in a band well-matched to the SCUBA-2/850$\,\mu$m filter (i.e., Band 6, 7, and/or 8 between $600-1250\,\mu$m). We intentionally omit lower frequency SMG follow-up from this program which are explored by \cite{Zavala2021} and Long et al., in prep. (B4/2mm blank field selection), \cite{Cooper2022} and Cooper et al., in prep. (pointed B4/2mm follow-up), and Gentile et al., in prep. (VLA/3 GHz sources). These lower frequency follow-up techniques can impose different redshift selection functions that may not uniformly capture SCUBA-2 sources (e.g., \citealt{Casey2018}). In any case, $\sim90\%$ of the archival ALMA data over COSMOS is in Bands 6 and 7 so the exclusion of other bands does not exclude a significant fraction of sources. The vast majority of the ALMA coverage comes from Bands 6 ($\lambda_{obs}\sim1250$) and 7 ($\lambda_{obs}\sim850\,\mu$m) and our final sample is drawn exclusively from observations in these bands with 102 pointings in Band 6 and 170 in Band 7.

\section{ALMA Data Reduction and Fluxes\label{sec:alma}}

Using the ALMA archive, we download all observations in Bands 6 and 7 with footprints that overlap a SCUBA-2 source interior to the COSMOS-Web NIRCam area which totals to 272 pointings, 13 of which have pointing centers within $5^{\prime\prime}$ of one-another. We froze our archival analysis in November 2023 and so our data set reflects all observations publicly available on the archive as of September 7th, 2023. Updating the archival search based on public data as of July 2024 would increase the number of ALMA pointings by $7\%$, which we leave to future work. We uniformly re-image all of the observations as individual pointings using $\texttt{tclean}$ and natural weighting in continuum mode with \texttt{CASA v6.6.3} \citep{casa}. The angular resolutions amongst these data range between $0.022^{\prime\prime}-1.76^{\prime\prime}$ with a median of $0.83^{\prime\prime}$ and upper and lower quartiles of $0.36^{\prime\prime}$ and $1.24^{\prime\prime}$ respectively. Only one source was observed in a baseline configuration with resolution below $0.1^{\prime\prime}$. The noise properties across the ALMA data are fairly uniform given the modest RMS used when following-up single-dish SMGs, and given the fact that 41 of the ALMA observations used in this work come from PID: 2013.1.00118.S \citep{Brisbin2017} and 85 from PID: 2016.1.00463.S (AS2COSMOS, \citealt{Simpson2020}). The median $1\sigma$ map noise is $0.14$ mJy with upper and lower quartiles of $0.08$ mJy and $0.19$ mJy respectively. 

For ALMA pointings from the AS2COSMOS program we adopt the source list and associated flux densities reported in \citep{Simpson2020} who perform a thorough search for all sub-mm sources in their Band 7 observations of the brightest ($S_{850}>6.2$ mJy) SCUBA-2 sources in S2COSMOS. This yields 85 ALMA sources. For all other pointings we conduct a blind source extraction using \texttt{PyBDSF} \citep{pybdsf} which was developed and optimized for radio interferometer data. Given that our final sample will include multiple ways to validate counterpart identification, we use the following strategy with \texttt{PyBDSF} to consider all tentative ALMA detections as potentially real sources. We first run \texttt{PyBDSF} blindly on the non primary-beam-corrected maps with a $3\sigma$ integrated flux threshold and a $4\sigma$ peak flux threshold. Then we re-run \texttt{PyBDSF} on the primary beam-corrected ALMA maps using the last iteration for positional priors and to measure primary beam-corrected fluxes. In this manner we identify 255 $>4\sigma$ candidate ALMA sources that may correspond to SCUBA-2/850$\,\mu$m emission and an optical counterpart. From this list we visually check the data comparing the ALMA continuum observations with multi-wavelength maps and remove a further 51 spurious noise peaks all with SNR$\,<5$ that have no obvious counterparts in the JWST data. 
155 (92\%) of SCUBA-2 SNR$\,>5$ sources selected from the original S2COSMOS map are detected with ALMA and matched to a NIRCam counterpart.  There are 47 SCUBA-2 sources that we do not recover a counterpart for which predominantly ($72\%$) have SNR$\,<5$ in SCUBA-2. 

For spatially resolved sources we adopt an integrated flux density that encases the extent of the source, and for unresolved sources we adopt the peak flux. A minority of sources in the sample are resolved ($10\%$), with most being unresolved point sources. Five sources in our sample have ALMA imaging at $850\,\mu$m that likely resolves out extended emission: AzTEC8\_0, z23\_89\_0, z23\_74\_0, z23\_62\_0, and z23\_62\_2. These sources have SNR$\,=6-13$ in SCUBA-2 from \cite{Simpson2019}, and the ALMA detections recover $<30\%$ of the deblended SCUBA-2 flux. For these five sources we do not include the ALMA data in any model fits but the sources are included in the sample as the ALMA detections identify robust JWST counterparts. 

Our final sample includes 290 sources corresponding to 289 galaxies accounting for one doubly imaged lensed source. These galaxies correspond to 219 SCUBA-2/850$\,\mu$m sources. Of these SCUBA-2 sources, 54 host multiples where $2-4$ ALMA sources contribute to the observed  $850\,\mu$m emission. Of these, nine SCUBA-2 sources (4\% of the total) host triplet sub-mm systems, and 44 (20\%) host doublets. One source is resolved into four components all at $z=2.5$ originally discussed in \cite{Wang2016}. Therefore, we measure a multiplicity fraction for the SCUBA-2 sources of $25\%$. This is fairly common for follow-up programs of SMGs \citep{Chen2013,Hezaveh2013,Hodge2013,Karim2013,Brisbin2017,Simpson2020}. 

\begin{figure}
    \centering
    \includegraphics[width=0.47\textwidth]{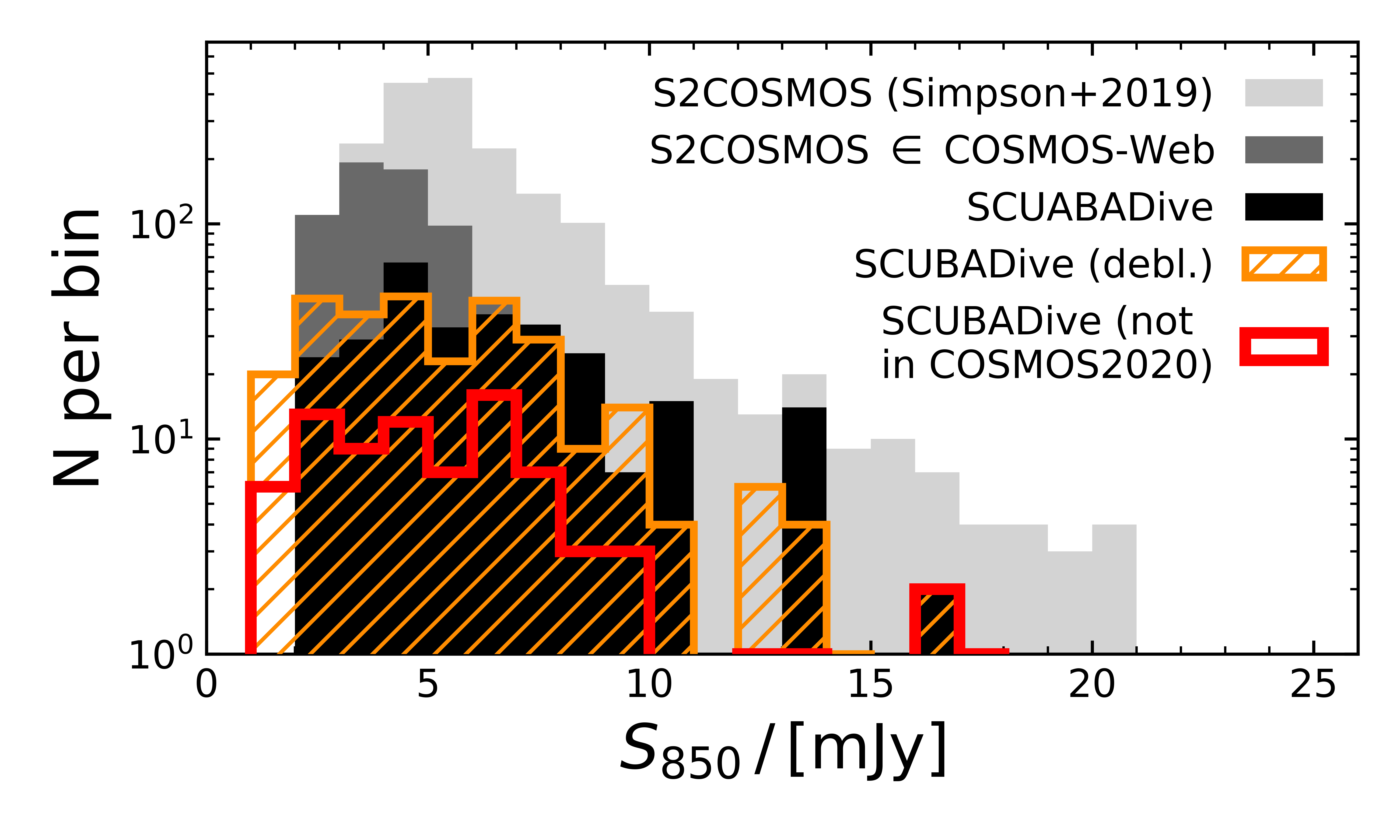}
    \caption{JCMT/SCUBA-2 flux distribution for the parent sample S2COSMOS from \citealt{Simpson2019}, and its overlap with the COSMOS-Web footprint. We find ALMA matches for 219 SCUBA-2 sources (black), which corresponds to 289 individual galaxies that comprise the SCUBADive sample (orange). Of these, 80 had no prior rest-frame optical/near-infrared counterpart (i.e., a detection in the COSMOS2020 catalog, \citealt{Weaver2022}) as shown in red.}
    \label{fig:s850}
\end{figure}

A significant fraction of the galaxies in our sample ($81/289=28\%$) have no counterpart in the COSMOS2020 catalog within 1 arcsec, which is expected given that $20\%-30\%$ of SMGs in pre-JWST studies had no optical/near-IR counterpart in the deep field OIR data ($5\sigma\sim25-26$ mag) prior to JWST \citep{Wardlow2011,Simpson2014,Franco2018,Dudzeviciute2020}.
Figure \ref{fig:s850} shows the final catalog's SCUBA-2 flux distribution which has average $\langle S_{850} \rangle$ from SCUBA-2 of $5.9^{+1.9}_{-2.2}$ mJy. A two-sample Kolmogorov–Smirnov test between the SCUBA-2 source flux distribution of SCUBADive vs.~all S2COSMOS sources in COSMOS-Web returns $(Ks,\log p)=(0.3,-14.8)$ which does not reject the null hypothesis that the samples are drawn from different distributions. 

\subsection{Naming Scheme}
There are many options for naming sources in SCUBADive that reflect the data sets used in this work. We choose to preserve the ALMA source name listed in the ALMA science archive to make it easy for readers to acquire that data. For sources in AS2COSMOS \citep{Simpson2020} we adopt their naming scheme directly. All other sources have their ALMA source names appended by an underscore and a number which follows from our source extraction using \texttt{pyBDSF} and internal naming schemes therein. For example, the ALMA observation targeting ``DSFGS2.60'' from PID \#2015.1.00568.S (PI Casey) reveals three separate sources which in this work are labeled as DSFGS2.60\_0, DSFGS2.60\_1, and DSFGS2.60\_2. 

\subsection{Notable sources}
A handful of our sources in our catalog are worth mentioning individually given their unique nature and discussion in prior works. 
AzTECC71 from \cite{McKinney2023} is an SMG at $z_{phot}=5.7\pm0.6$ and in this work is labeled AzTECC71\_1.  
MAMBO-9 (S2COSMOS.850.50\_1 in this work) is comprised of a pair of unlensed dusty, star-forming galaxies at $z_{spec}=5.850$ (\citealt{Casey2019,Jin2019}, Akins et al., in prep.) residing in an overdensity plausibly representing a protocluster environment \citep{Brinch2024}.
AzTECC3a is a well known lensed galaxy at $z=4.6237$ \citep{Brisbin2017,Miettinen2017,AlvarezCrespo2021} with the two counterpart images denoted AzTECC3a\_0 and AzTECC3a\_1 in this work. 
J1000+0234\_1 is a known dusty, star-forming galaxy at the center of an over-dense environment and Ly$\alpha$ blob \citep{Smolcic2017overdense,Solimano2024}. AzTEC-2 (this work AzTECC2b\_0) is a pair of massive galaxies at $z_{spec}=4.633$ comprising one of the brightest SCUBA-2 sources in COSMOS \citep{JimenezAndrade2020}.
AS2COS$0005.1$ is a lensed SMG at $z=2.625$ \citep{Jin2024}.
Other sources have been reported and studied in prior works but are slightly less-well characterized than those mentioned here \citep[e.g.,][]{Aretxaga2011,Casey2013,Geach2017}.

\begin{figure*}[ht!]
    \centering
    \includegraphics[width=0.97\textwidth]{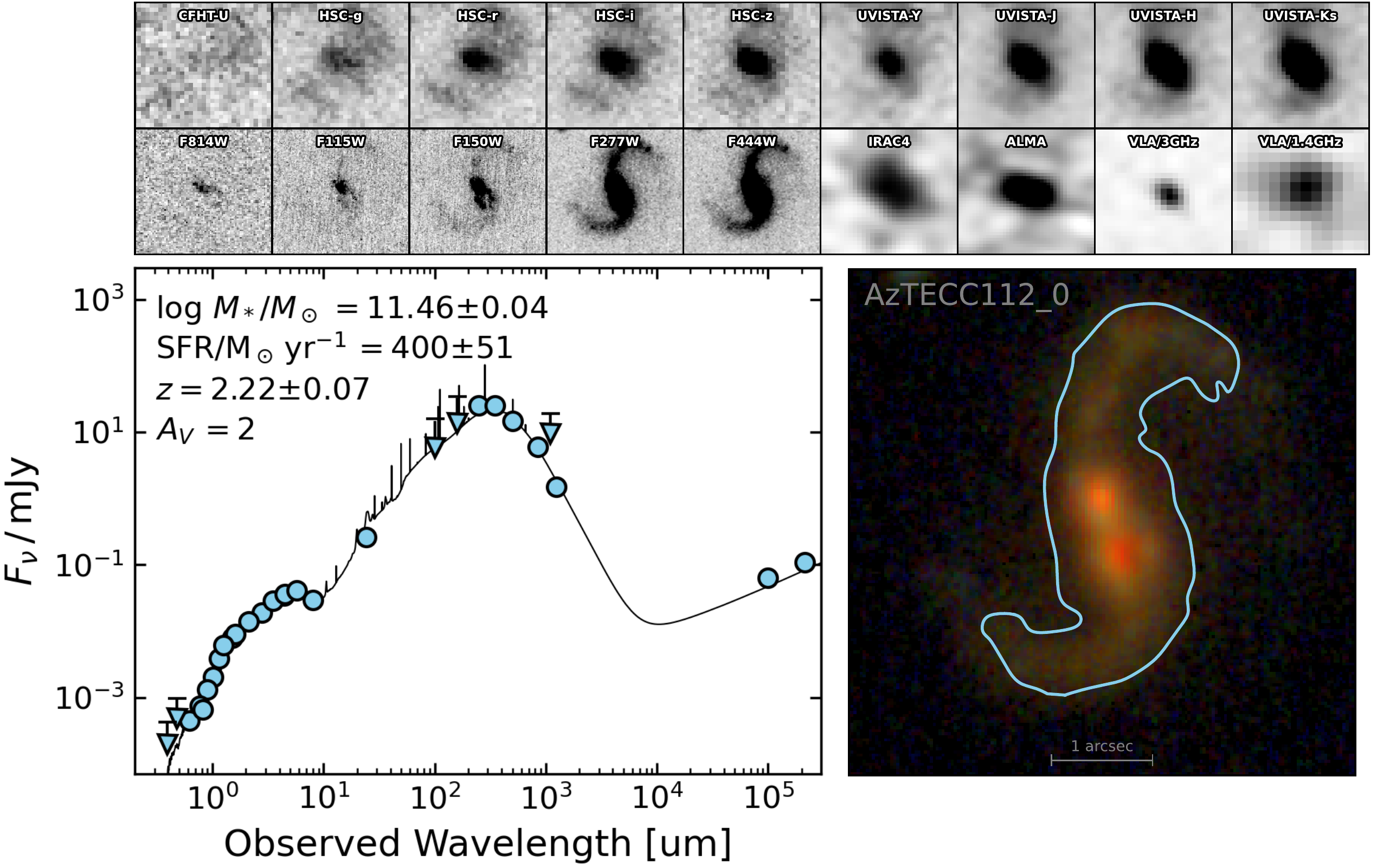}
    \caption{An example of one SCUBADive source in COSMOS-Web and the multi-wavelength data we analyze with \texttt{diver}. (\textit{Top}) $5^{\prime\prime}\times5^{\prime\prime}$ cutouts from COSMOS2020, JWST/COSMOS-Web, ALMA, and the VLA. (\textit{Bottom left}) The optical-through-radio spectral energy distribution with photometry in blue and the best-fit model in black and its corresponding best-fit parameters. (\textit{Bottom right}) B=115+150, G=277, R=444 image of the source AzTECC112\_0. Our custom aperture is shown in blue which we use in the optical through near-infrared photometry.}
    \label{fig:diver}
\end{figure*}

\section{Multi-wavelength Photometry \label{sec:phot}}

\subsection{Custom Optical/Near-IR Aperture Photometry with \texttt{diver}}
The NIRCam counterparts to mm/sub-mm selected galaxies can exhibit complex morphologies that are not optimally captured by classical photometric methods. We tested circular and elliptical apertures from \texttt{SE/SE++} \citep{sourceextractor} and found (i) many examples where the apertures could not encase the extent of the source without adding in substantial noise, (ii) \textit{JWST} fluxes inconsistent with ground-based maps at lower resolution, and (iii) generally under-estimated flux uncertainties relative to what we expect from the map depths \citep{cosmos-web}. In light of these challenges, we developed \texttt{diver} \-- an empirically-motivated photometry tool built to preserve the resolution of JWST when measuring PSF-matched multi-wavelength photometry. We now summarize the methodology of \texttt{diver} below. 

For each of our sources \texttt{diver} proceeds through the following: First, we make $50^{\prime\prime}\times50^{\prime\prime}$ cutouts (to ensure robust noise estimates) around the target in the science image and error/weight image from which a flux will be measured. In this case we use JWST/NIRCam and MIRI data from COSMOS-Web (F115W, F150W, F277W, F444W, F770W) and PRIMER (GO \#1837, PI: Dunlop, \citealt{PRIMER}, imaging in F090W, F200W, F356W, F410M) where available, as well as all multi-wavelength imaging data compiled in the COSMOS field in COSMOS2020 (e.g., CFHT, HSC, \textit{HST}, UVISTA, \textit{Spitzer}, see \citealt{Weaver2022}). In principle any data map can be included in \texttt{diver} if accompanied by a PSF and an error map. Next, we identify the reddest NIRCam band containing the source (typically F444W) and create a segmentation map using \texttt{photutils} \citep{photutils} in some cases with an additional de-blending step included to isolate the source from its neighbors. We save the arbitrarily shaped segmentation boundary down to an SNR-per-pixel of 3, 5, or 10 as the custom aperture through which to measure all multi-wavelength photometry. Due to the larger PSF, the flux density is generally the most extended in F444W, making it a good choice for capturing the source flux with minimal added noise across bands.

When measuring flux through \texttt{diver}'s non-parametric aperture, we make the following calculations to preserve PSF information and properly estimate uncertainties. For maps of poorer spatial resolution than NIRCam/F444W we calculate a PSF correction factor by first convolving the map used for segmentation (F277W or F444W) to the lower resolution using a convolutional PSF kernel derived using \texttt{pypher} \citep{pypher}. We take the ratio of the total convolved flux to the convolved flux interior to the custom aperture as our PSF correction factor, effectively using an empirical model of the source from JWST/NIRCam to estimate flux loss from beam smearing in lower resolution optical/near-IR maps. Then, we measure the flux of the map by summing pixels within the arbitrarily shaped aperture multiplied by this PSF correction factor. To estimate flux errors we follow the uncertainty in CCD-based detector given by 
\begin{equation}
    \begin{split}
            \sigma_{\rm aper}^2 = n_{\rm pix}\sigma_{\rm bg}^2 + (\sigma_{\rm p}^2 + \sigma_{\rm dark}^2 + \sigma_{\rm R}^2) \\ \approx \sigma_{\rm boot,sci}^2 + \sigma_{\rm boot,err}^2
    \end{split}
    \label{eq:noise}
\end{equation}
for an aperture containing $n_{\rm pix}$ pixels, a background noise of $\sigma_{\rm bg}$ (which we estimate locally), a CCD with shot noise $\sigma_p$, and further error terms from dark current ($\sigma_{\rm dark}$) and readout ($\sigma_{\rm R}$). We bootstrap the sky noise and detector level uncertainties by projecting the science map (after source masking) and error/weight maps respectively through the custom aperture randomly placed $1000$ times within the $50^{\prime\prime}\times50^{\prime\prime}$ cutout. We combine these terms in quadrature for a full representation of the flux uncertainty.  

For consistency we test the \texttt{diver} photometry on targets with simple morphologies (compact sources) by comparing against photometric extraction using \texttt{SE/SE++}. For these, 
the fluxes and flux errors measured with \texttt{diver} are generally consistent with the aperture fluxes from \texttt{SE/SE++} within $1\sigma_{\rm diver}$, but not within $1\sigma_{\rm SE/SE++}\ll\sigma_{\rm diver}$. Most notably, the multi-wavelength fluxes from \texttt{diver} are far more consistent between ground-based and JWST data given the expected shape of the rising SEDs into the near-infrared. Our approach to PSF corrections using the NIRCam light profile and convolutional kernels from \texttt{pypher} grounds the multi-wavelength photometry estimator in the most accurate empirical model available. One caveat is that this technique assumes that the F277W or F444W light profile is a good model across the full optical/near-IR spectrum which may or may not be the case; however, it does seem to work well for the purposes of estimating aperture corrections. 

Figure \ref{fig:diver} shows, for one source, the JWST/NIRCam RGB from the COSMOS-Web bands alongside cutouts for most but not all of the optical/near-IR data for which we measure flux densities. We also highlight a system with complex morphology for which \texttt{diver} is exceptionally well-suited for measuring optimal flux densities.

\begin{deluxetable}{lrr}
    \tablecaption{Table Content of SCUBADive Continuum Data \label{tab:oir}}
    \tablehead{\colhead{Instrument/Telescope}&\colhead{Band/Central $\lambda$}&\colhead{Units}}
    \startdata
     ground based \\ 
     MegaCam/CFHT & $u$         & mJy  \\ 
     HSC/Subaru & $g$           & mJy  \\ 
     HSC/Subaru & $r$           & mJy  \\ 
     HSC/Subaru & $i$           & mJy  \\ 
     HSC/Subaru & $z$           & mJy  \\ 
     VIRCAM/VISTA & $Y$         & mJy  \\ 
     VIRCAM/UltraVISTA & $J$    & mJy  \\ 
     VIRCAM/UltraVISTA & $H$    & mJy  \\ 
     VIRCAM/UltraVISTA & $K_s$  & mJy  \\ 
     Band 6/ALMA & $850\,\mu$m & mJy  \\ 
     SCUBA-2/JCMT & $850\,\mu$m & mJy \\ 
     AzTEC/ASTE & 1100$\,\mu$m & mJy \\ 
     Band 7/ALMA & $1250\,\mu$m & mJy  \\ 
     VLA & S/3 GHz & mJy \\ 
     VLA & L/1.4 GHz & mJy \\ 
     \hline 
     space based \\ 
      ACS/\textit{HST} & F814W    &  mJy \\ 
      NIRCam/\textit{JWST}& F090W &  mJy  \\ 
      NIRCam/\textit{JWST}& F115W &  mJy  \\ 
      NIRCam/\textit{JWST}& F150W &  mJy  \\
      NIRCam/\textit{JWST}& F200W &  mJy  \\ 
      NIRCam/\textit{JWST}& F277W &  mJy  \\ 
      NIRCam/\textit{JWST}& F356W &  mJy  \\
      NIRCam/\textit{JWST}& F410M &  mJy  \\ 
      NIRCam/\textit{JWST}& F444W &  mJy  \\ 
      MIRI/\textit{JWST}  & F770W &  mJy  \\ 
     IRAC/\textit{Spitzer} & Ch.~1 & mJy  \\ 
     IRAC/\textit{Spitzer} & Ch.~2 & mJy  \\ 
     IRAC/\textit{Spitzer} & Ch.~3 & mJy  \\ 
     IRAC/\textit{Spitzer} & Ch.~4 & mJy  \\ 
     MIPS/\textit{Spitzer} & 24$\,\mu$m & mJy \\ 
     PACS/\textit{Herschel} & $100\,\mu$m & mJy \\
     PACS/\textit{Herschel} & $160\,\mu$m & mJy \\
     SPIRE/\textit{Herschel} & $250\,\mu$m & mJy \\
     SPIRE/\textit{Herschel} & $350\,\mu$m & mJy \\
     SPIRE/\textit{Herschel} & $500\,\mu$m & mJy \\
    \enddata
    \tablecomments{This table is published in its entirety including measurements from all ground- and space-based photometric bands from the optical through radio, in the machine-readable format. 
    }
\end{deluxetable}

\subsection{X-Rays}
We check for X-ray detections of our sources in the  Chandra COSMOS Legacy Survey (CCLS, \citealt{Civano2016,Marchesi2016}). There are 18 SNR$\,>5$ X-ray counterparts to galaxies in SCUBADive with median ($\pm$16th-84th percentiles) $\log L_{2-10\rm\,keV}/{\rm (erg/s)}=43.9^{+0.3}_{-0.5}$. We check for $3-5\sigma$ detections at the positions of SCUBADive galaxies not found in the X-ray catalogs, and find marginal detections for a further $\sim20\%$ of the sample.

\subsection{Optical/Near-IR Aperture Photometry for NIRCam Point Sources\label{sec:ptsrc}}

A small subset of SCUBADive galaxies are dominated by point-source components in NIRCam and/or MIRI. There are seven such sources, for which we measure aperture photometry using circular apertures of $r=0.5^{\prime\prime}$ and make the appropriate PSF corrections. Through curve-of-growth analysis an $r=0.5^{\prime\prime}$ circular aperture was found to maximize the SNR. 

All seven of the point source-dominated SCUBADive galaxies are detected in X-rays which indicates the presence of a heavily dust-obscured supermassive black hole. In extreme cases a heavily obscured active galactic nucleus (AGN) can contribute significantly to the far-infrared and sub-mm fluxes of IR-luminous galaxies \citep{McKinney2021agn}. We leave a careful analysis of such sources in SCUBADive to future works in part because better empirical constraint on the mid-infrared continuum is needed to disentangle the contribution of of AGN to the SED. 

\subsection{Mid-/far-IR, sub-mm, and radio fluxes}
For mid-IR up through radio fluxes we first check for $>5\sigma$ matches in existing catalogs within $1^{\prime\prime}$ of the ALMA dust continuum centroid. We use S-COSMOS for \textit{Spitzer}/MIPS 24$\,\mu$m (66 matches, \citealt{SCOSMOS,LeFloch2009}), PEP and HERMES for \textit{Herschel} PACS and SPIRE (20 and 39 matches, \citealt{Lutz2011,Oliver2012,Hurley2017}), and the VLA at 1.4 and 3 GHz (55 and 153 matches, \citealt{Schinnerer2010,Smolcic2017}). For objects without matches to catalogs, we set the flux density to that of the pixel containing our target and set the flux uncertainty to that of the total mosaic and flag the data point as an upper limit for the purposes of modeling. As implemented in \cite{McKinney2023}, this can include flux contribution from nearby sources in the \textit{Herschel}/SPIRE bands, a limiting uncertainty that we let propagate into our spectral energy distribution modeling.

As previously mentioned, 88 ALMA sources in our sample belong to doublets where two ALMA sources contribute to the SCUBA-2 flux. A further 27 sources belong to triplets where three sources contribute to the SCUBA-2 flux and one SCUBA-2 system is comprised of four ALMA sources. We take a simple approach to deblending the SCUBA-2 flux by dividing the total amongst the ALMA sources proportional to the ALMA flux ratios. Nearly two thirds (56/88) of the multiples have Band 7 $870\,\mu$m imaging and therefore directly constraining the flux ratios contributing in SCUBA-2. For the other sources we are assuming that the flux ratios at $1100-1200\,\mu$m and $850\,\mu$m are equal. This only falls apart if (a) the targets have very different redshifts or dust temperatures such that $850\,\mu$m probes the cold dust SED peak in one while 1.1mm remains along the RJ tail in its companion, (b) if the sub-mm spectral index $\beta$ varies dramatically source-to-source, and/or (c) if the observations resolve out emission disproportionately. The standard deviation amongst $\beta$ measured for $S_{850}>5$ mJy SMGs is $<20\%$ \citep{daCunha2021,Cooper2022} which makes this an unlikely source of systematic error in deblending SCUBA-2 emission. No SCUBADive source has $z_{phot}>6$ beyond which redshift differences would lead to different sampling on and off the RJ tail. Moreover, the cold dust temperatures of sub-mm detected galaxies appear not to systematically evolve with redshift out to $z\sim2$ \citep{Drew2022}; however, there is some debate as to how this trend extrapolates to higher-redshifts \citep{Schreiber2018,Liang2019,Viero2022,Sommovigo2022,Algera2024}, and of course, individual sources can deviate from these trends because of recent starbursts and/or growing supermassive black holes \citep{Kirkpatrick2015}. Thus the cold dust temperature uncertainties constitute a systematic on deblending that only robust far-infrared SED coverage sampling the peak of the dust continuum can overcome. 

\begin{figure}
    \centering
    \includegraphics[width=0.48\textwidth]{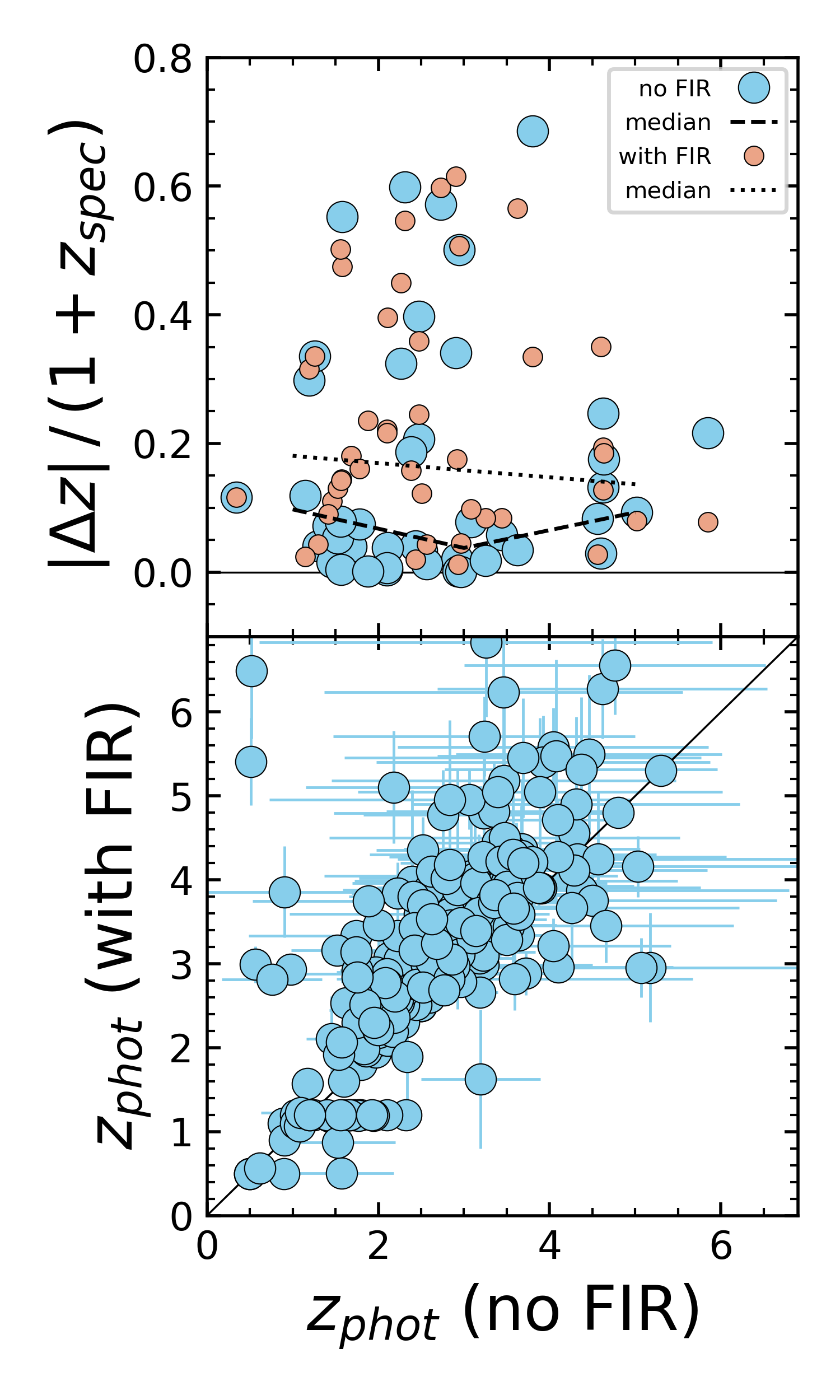}
   \caption{A comparison of photometric redshifts derived using only the optical/near-IR SED up to $8\,\mu$m (no FIR) vs.~all the multi-wavelength photometry out to the radio (with FIR). (\textit{Top}) Recovery of spectroscopic redshifts expressed as the fractional deviation $|z_{phot}-z{spec}|/(1+z_{spec})$ for both $z_{phot}$ (no FIR, blue) and $z_{phot}$ (with FIR, tan), with binned meidans shown as dashed/dotted black lines. Fitting to only the optical/near-IR SED recovers $z_{spec}$ within $\sim10\%$ out to $z=6$. Including the FIR data leads to on-average poorer recovery of $z_{spec}$ to within $\sim30-40\%$.  (\textit{Bottom}) Direct comparison between photometric solutions for $z_{phot}$ (no FIR) and $z_{phot}$ (with FIR).  The inclusion of far-infrared/sub-mm and radio data prefers photometric redshifts that are on-average $\sim30\%$ higher. }
    \label{fig:zcompare}
\end{figure}

\begin{figure}
    \centering
    \includegraphics[width=0.47\textwidth]{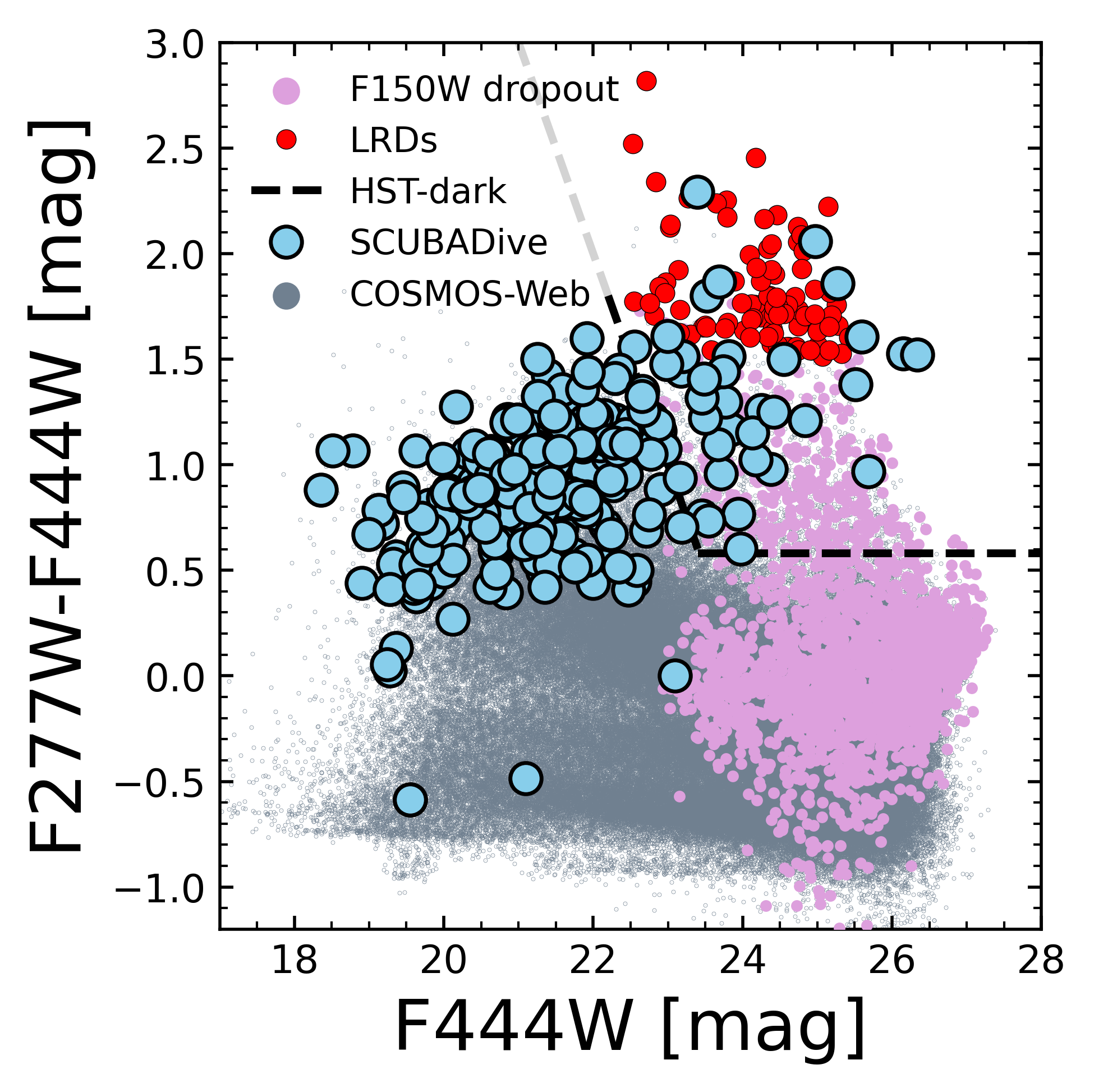}
    \caption{JWST/NIRCam color-magnitude diagram in the F277W and F444W filters for $\sim300,000$ galaxies in the COSMOS-Web survey with SNR$\,>10$ (grey), the subset with no detection in F150W (pink), SCUBADive (blue), and Little Red Dots (LRDs, red, \citealt{Akins2024}). The dashed black line outlines a typical color-magnitude wedge used to select \textit{HST}-dark $z\gtrsim3$ dusty galaxies \citep[e.g.,][]{Barrufet2022}. SMGs exhibit a redder F277W-F444W color than $\sim95\%$ of all sources in COSMOS-Web, and the $z>3$ SCUBADive subset preferentially have F444W$\,>23$ mag and F277W$\,-\,$F444W$\,>0.6$. The reddest SMGs in SCUBADive also overlap with the domain of LRDs, but they do not qualify for the compactness criterion that reflects LRD point-source morphologies (\citealt{Akins2024}). There are $480$ F150W-dropouts in COSMOS-Web that fall within the high$-z$, dusty galaxy selection wedge (black dashed line) and do not belong to either SCUBADive/SMGs nor LRD samples. These F150W-dropout, red galaxies may represent less extreme counterparts to the SMGs and LRDs in terms of their stellar masses, $A_V$ and specific star-formation rates.}
    \label{fig:colormag}
\end{figure}

\section{Derived Quantities\label{sec:quant}}

\subsection{Spectroscopic Redshifts}
We compile spectroscopic redshifts from optical, near-infrared and sub-mm/radio spectra for our sources where possible using matches reported in the following:
AS2COSPEC \citep{Chen2021}, 
\citet{Jin2019,Jin2024},
\cite{Horowitz2022},
\cite{Hasinger2018},
\cite{Shah2020},
\cite{Valentino2020},
\cite{Kartaltepe2015_spec},
\cite{Kashino2019},
\cite{Stott2016},
MAGIC \citep{Epinat2024},
\cite{Casey2019},
\cite{Onodera2015},
\cite{Kriek2015},
QUAIA \citep{StoreyFisher2024},
\cite{Krogager2014},
hCOSMOS \citep{Damjanov2018},
zCOSMOS \citep{Lilly2007}, and 
zFIRE \citep{Nanayakkara2016}.
Most of these matches were compiled as part of the latest update to the spectroscopic redshift compliation of all surveys and published redshifts in the COSMOS field (Khostovan et al., in prep). In total we find spectroscopic redshifts for 59 of the 289 galaxies in our sample with a median $z_{\rm spec}=2.5$ and 16th-84th percentiles of $z_{\rm spec}=2.4$ and $z_{\rm spec}=3.9$ respectively. The highest spectroscopic redshift we include is $z_{\rm spec}=5.850$ for a well-studied system ``MAMBO-9'' \citep{Casey2019,Jin2019}. A majority (60\%) of our $z>3$ spec-z's are from AS2COSPEC \citep{Chen2021} which uses Band 3 ALMA scans to confirm redshifts of dusty galaxies with multiple CO/[C\,I] detections. We also measure and report one new $z_{\rm spec}>5$ from ALMA PID\#2018.1.00874.S (PI Oteo) for AzTECC129\_0 at $z_{\rm spec}=5.0175$,  based on a detection of CO(4-3) in our re-reduction of this archival data set.

\subsection{Spectral Energy Distribution Modeling\label{sec:fits}}
To derive physical properties across our sample and to estimate photometric redshifts we fit the multi-wavelength photometry independently with \texttt{CIGALE} \citep{Boquien2019}. We assume the following constraints on the model SEDs: a delayed star-formation history with a main stellar population of age $0.5-12$ Gyr and an e-folding time $\in[30,6000]$ Myr, combined with a late stage burst $50-300$ Myr old. We assume a Chabrier IMF \citep{Chabrier2003}, solar metallicity, and a modified \cite{Charlot2000} attenuation curve $\propto\lambda^{-0.7}$ with $A_V\in[0,9]$. We model the far-IR SED using templates from \citet{DraineLi2007} (as updated in \citealt{Draine2014}) with $q_{PAH}\in[0.47,4]$, a minimum radiation field density $U_{min}\in[5,50]$ and $dU/dM\propto U^{1-2}$ illuminating $\gamma=1-4\%$ of the dust. We let the redshifts vary between $z=0.5-7$ and we assume a flat prior, unless the target has a spectroscopic redshift in which case we fix the redshift to $z_{\rm spec}$. 
Finally, we include a radio power-law component constrained by our VLA 1.4 GHz and 3 GHz data with a fixed slope of 0.8 and FIR/radio correlation coefficient $q_{\rm IR}\in[1.8,2.6]$ to span the range of values common amongst high$-z$, dusty star-forming galaxies \citep{Delvecchio2021}. For bands with non-detections we let the models fit the photometry measured through the aperture $\pm$ the $1\sigma$ uncertainty. These model and fitting assumptions are successful in reproducing the SEDs of high-redshift dusty, star-forming galaxies as demonstrated by \cite{Gentile2024} who also show they yield results consistent with other modeling codes such as \texttt{magphys} \citep{magphys2008,daCunha2015,magphys-photoz}, specifically developed for dust-obscured galaxies. We also fit our targets with \texttt{magphys} and discuss systematic differences where relevant.  

As mentioned in Section \ref{sec:ptsrc}, seven sources have point-like morphologies and X-ray detections indicating that their optical/near-infrared light might be dominated by an AGN. Out of the remaining sources, $\sim10\%$ have cataloged X-ray detections from \cite{Civano2016} and \cite{Marchesi2016}. This is consistent with X-ray detection statistics amongst SMGs from works using much deeper X-ray data than COSMOS \citep[e.g.,][]{Wang2013}, which also find minimal contribution of AGN to the SED. For these reasons we do not believe AGN are a systematic source of contamination in this sample. Nevertheless, we omit the seven strong X-ray sources with a NIRCam/F444W point source profile from our population analysis on the stellar masses. 

\begin{figure*}
    \centering
    \includegraphics[width=\textwidth]{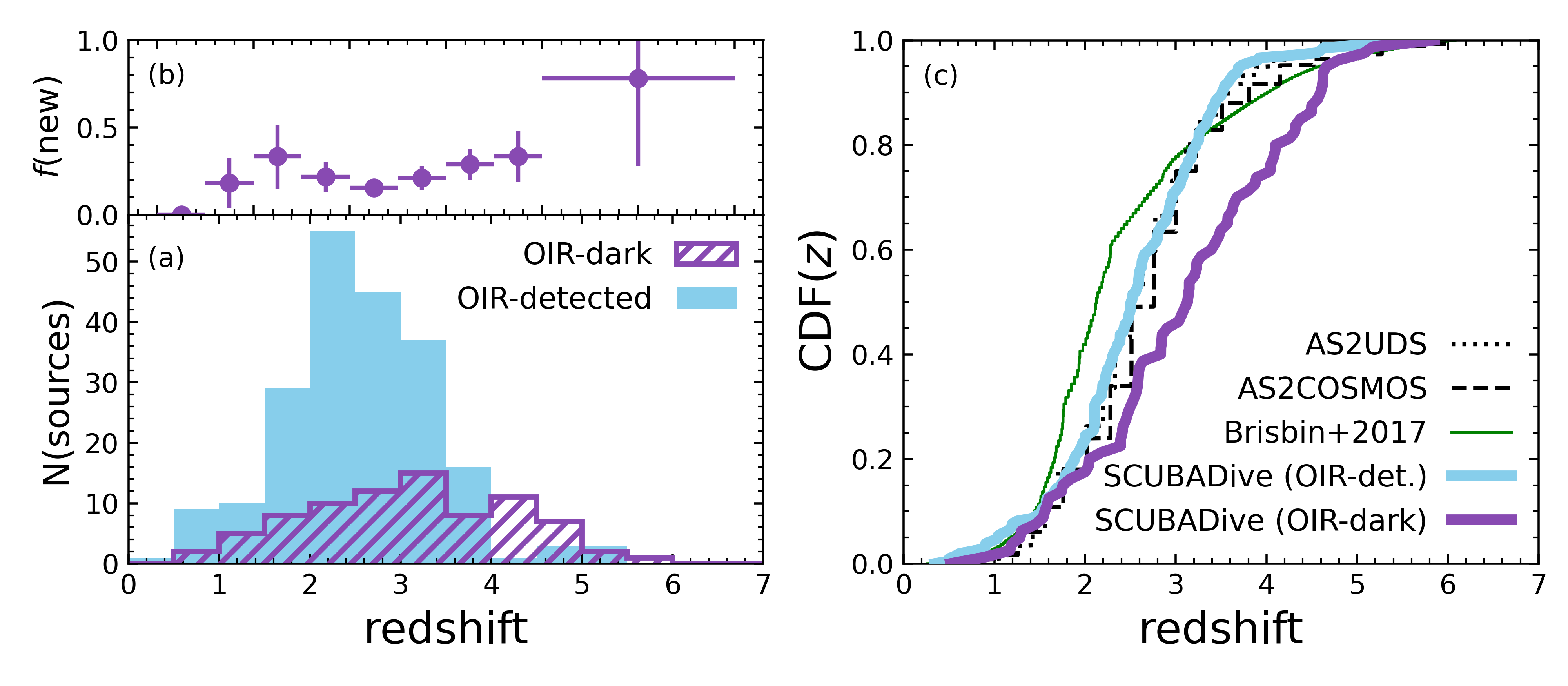}
    \caption{Redshift distributions. \textit{(a)} The redshift histogram for sources in SCUBADive that have counterparts in COSMOS2020 (blue) and the ``new'' OIR-dark subset with no prior counterparts (purple). Above in panel \textit{(b)} we show the fraction of NIR-dark sources in SCUBADive between $z=0-6$, which grows from an average of $20\pm10\%$ at $1<z<4$ to nearly $100\%$ at $z>4$. In panel \textit{(c)} we compare the SCUBADive cumulative redshift distribution function against large analyses of SMGs: AS2COSMOS \citep{Simpson2020}, AS2UDS \citep{Dudzeviciute2020}, and \cite{Brisbin2017}. SCUBADive is consistent with these prior works, and the OIR-dark subset is biased towards higher redshifts.
    }
    \label{fig:zhist}
\end{figure*}

We run the model fits once including all multi-wavelength data and then again fitting only to the optical/near-IR photometry up to $\sim8\,\mu$m in MIRI/F770W and/or IRAC $8\mu$m. The full multi-wavelength fit is used when estimating and reporting derived quantities, but we compare against the optical/near-IR fits to assess systematics imposed by \texttt{CIGALE}'s energy balance with particular attention to derived redshifts and stellar masses that are most susceptible to the inclusion of direct constraint on the dust. As shown in Figure \ref{fig:zcompare}, photometric redshifts derived purely from the optical/near-infrared data are generally lower than those derived from fitting the full multi-wavelenth SED by $\sim10-30\%$. 
We find that $z_{opt}$ does a better job in recovering spectroscopic redshifts as shown in Figure  \ref{fig:zcompare} because of added SED degeneracies in the FIR. Therefore, we adopt $z_{opt}$ as the fiducial photometric redshifts for our sample in the absence of $z_{spec}$. For all other derived quantities we adopt the results from a full multi-wavelength fit (with redshifts fixed to $z_{opt}$) including direct constraint on the dust in the far-IR and sub-mm. Table \ref{tab:phys} lists the redshifts, stellar masses, star-formation rates, and $A_V$ we adopt in this paper from our SED modeling.

As a consistency check on our model assumptions we calculate rest-frame $H-$band and $K-$band mass-to-light ratios from the model SEDs which reflects the stellar ages and star formation history \citep{Hainline2011,Simpson2014,Michalowski2014}. From the \texttt{CIGALE} fits described above we find $\log(M_*/L_{H}\,[M_\odot/L\odot])=-0.33^{+0.37}_{-0.13}$ and $\log(M_*/L_{K}\,[M_\odot/L\odot])=-0.21^{+0.29}_{-0.27}$. The $H-$band mass-to-light ratio we recover is $\sim0.2$ dex less than the average amongst ALESS SMGs derived from \texttt{magphys} in \cite{daCunha2015} as well as for AS2UDS SMGs in \cite{Dudzeviciute2020}. To account for this artificial offset in mass tied to the different modeling assumptions, and for a fair comparison against the literature, we normalize our stellar masses from \texttt{CIGALE} to the average $H-$band mass-to-light ratio of \cite{daCunha2015}. 

\begin{deluxetable}{lrr}
    \tablecaption{Table Contents for Derived Physical Properties in SCUBADive \label{tab:phys}}
    \tablehead{\colhead{Column}&\colhead{Units}&\colhead{Description}}
    \startdata
    \texttt{id} & \nodata & Source ID \\ 
    \texttt{ra} & degrees & RA J2000 \\ 
    \texttt{dec} & degrees & DEC J2000 \\  
    \texttt{z\_opt} & \nodata & OIR photo$-z$  \\ 
    \texttt{z\_full} & \nodata & With FIR photo$-z$ \\ 
    \texttt{z\_spec} & \nodata & Spec$-z$ \\ 
    \texttt{logmstr} & $M_\odot$ & log stellar mass \\ 
    \texttt{sfr} & $M_\odot\,{\rm yr^{-1}}$ & log SFR \\ 
    \texttt{Av} & [mag] & Attenuation \\ 
    \enddata
    \tablecomments{This table is published in its entirety including in the machine-readable format. We include $1\sigma$ uncertainties on derived quantities using the appropriate column names appended by \texttt{\_err}. }
\end{deluxetable}

\section{Results\label{sec:results}}

\subsection{JWST Photometry and NIRCam Colors}

Figure \ref{fig:colormag} shows the NIRCam F277W and F444W color-magnitude space spanned by our sample. In these filters, SCUBADive galaxies are redder than 95\% of all other sources in COSMOS-Web, and the F444W magnitude correlates with the [F277W]$\,-\,$[F444W] color such that fainter sources are redder, consistent with higher redshifts and $A_V$. Out of the total sample containing 289 galaxies, 157 ($54\%$) are not detected in NIRCam/F115W at SNR$\,>5$, and 78 ($27\%$) are not detected in NIRCam/F150W at SNR$\,>5$ which are explored in detail in Manning et al., in prep. Considering a SNR$\,>3$ threshold these detection statistics change to $115$ ($40\%$) and $60$ ($21\%$) for F115W and F150W respectively. The non-detection rate in F150W ($<5\sigma$) is very similar to the fraction of SMGs that historically have not had optical counterparts or are described as ``optically dark'' \citep{Wardlow2011,Simpson2014,Franco2018,Dudzeviciute2020}, even though the NIRCam/F115W depth is much deeper than the UltraVISTA data at the same wavelength. In SCUBADive these galaxies have $\langle \log\,M/M_\odot \rangle\sim11$, $\langle z \rangle = 3.3$ and $\langle A_V \rangle = 3.5$. 
There are thousands of COSMOS-Web sources with F277W$\,-\,$F444W$\,>0.5$ and no F150W detection which may also represent $z>3$ dust-obscured star-forming galaxies (\citealt{Barrufet2022}, see also Gammon et al., in prep.). 
Only nine sources have SNR$\,<5$ in NIRCam/F277W, one of which is MAMBO-9 at $z=5.850$ \citep{Casey2019}, hinting at a rare population of massive and dusty galaxies very faint at $2.7\,\mu$m.

A handful of SMGs in SCUBADive exhibit unusual colors. Most are red in [F277W]$\,-\,$[F444W] between $0.5$ and $1.5$ mag with [F444W]$\,=20-24$ mag, but a small subset ($N=12$) with [F277W]$\,-\,$[F444W]$\,>1.5$, $z=3.5-5$, $\log M_*/M_\odot=11.1-12.3$, and $S_{850}=3-6$ mJy overlap with the NIRCam LW colors of so-called ``Little Red Dots'' (LRDs). However, the SCUBADive galaxies in this parameter space do not have blue SW slopes, nor do they have 40-70\% of their flux within 0.2$^{\prime\prime}$ which is an important compactness criterion for LRD selection (\citealt{Akins2024}, see also Gentile et al., in prep. for discussion of ``not so little red dots'').  

Amongst the bluer sources in the NIRCam LW bands, 12 SCUBADive galaxies have [F277W]$\,-\,$[F444W]$\,<0.3$ mag, two of which have $z\sim1$ and low star-formation rates $\sim50\,M_\odot\,{\rm yr^{-1}}$ relative to the typical range of SMGs \citep{daCunha2015}. One source in our sample, J1000+0234\_1, is unusually blue with [F277W]$\,-\,$[F444W]$\,\sim0$. As previously mentioned this source resides within an over-density at $z=4.54$. \cite{Solimano2024} conduct a NIRSpec/IFU analysis of this source and find very strong [O\,III]5007$\lambda\lambda$ with equivalent width $>1000$\AA\ that must contribute significantly to its F277W photometry hence the very blue [F277W]$\,-\,$[F444W] color \citep[e.g.,][]{McKinney2023agn}. The two bluest sources with [F277W]$\,-\,$[F444W]$\,=-0.59$ mag comprise a foreground galaxy (AS2COS0005.2, $z=0.5$) and lensed galaxy (AS2COS0005.1, $z=2.6$) pair. Given the spatial extent of the foreground source there is likely significant contamination to the background object's NIRCam photometry. 

\begin{figure}
    \centering
    \includegraphics[width=0.46\textwidth]{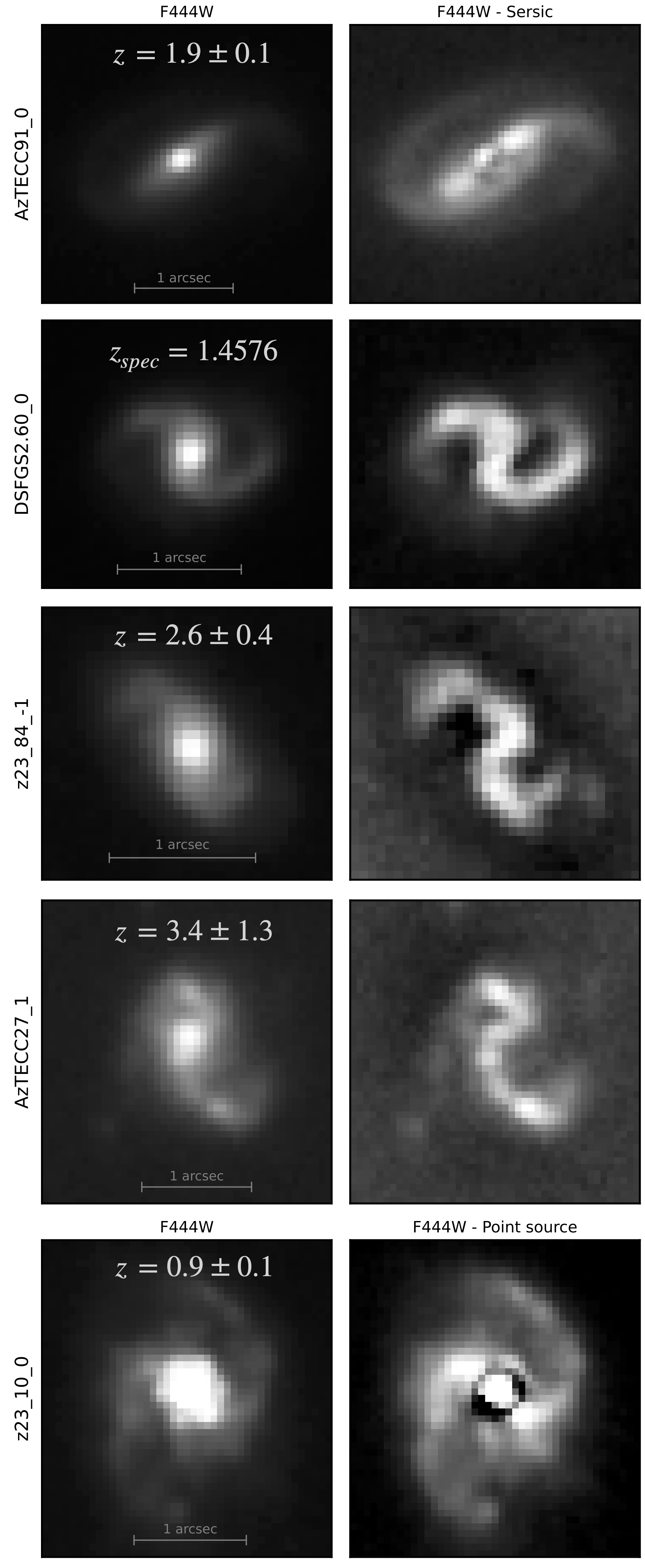}
    \caption{F444W cutouts for SCUBADive sources exhibiting candidate stellar bars, with Sersic-subtracted F444W maps on the right to reveal the bar and spiral arm structures. In the bottom row we show an SMG hosting a nuclear point-source in F444W with a residual also revealing a candidate stellar bar.  
    }
    \label{fig:bars}
\end{figure}

\subsection{NIRCam Morphology}
A major advantage of JWST in addition to its sensitivity is its unprecedented spatial resolution in the near-infrared probing the rest-frame optical emission in galaxies at high$-z$. This has enabled recent studies into the sub-structures of heavily dust-obscured galaxies. \cite{Rujopakarn2023} identify stellar sub-structure in one $z=2.696$ star-forming galaxies with ALMA and JWST. \cite{Wu2023} identify spiral arms in an SMG at $z=3.059$ using NIRCam, and both \cite{Chen2022} and \cite{Cheng2022} identify compact reddened substructures in NIRCam maps of several SMGs at $z\sim2$ with hints of spiral arms and stellar bars. \cite{Hodge2024} find that the surface density of dust continuum and red NIRCam colors correlate on kpc scales, reinforcing the origin of their red colors in dust obscuration rather than older stellar populations. 
In an analysis of 80 SMGs \cite{Gillman2024} find lower S\'{e}rsic indices than less active galaxies of comparable redshift and specific star-formation rate which may indicate lower mass black holes, lower dynamical stability and therefore an increase in secular and/or minor merger triggering of star-formation in the SMGs. 

Appendix 1 contains the RGB postage stamps for our sample. We visually classify galaxies using uniformly scaled RGBs with two classifiers (JM, CS). We identify $50\pm7$ ($17\pm4\%$) SMGs with disturbed morphologies evidence of an on-going or recent major merger. This occurrence of mergers is consistent with the morphology analysis of \cite{Gillman2024}. A further $32\pm5$ ($11\pm3$\%) SMGs exhibit spiral arms and $23\pm6$ SMGs ($8\pm3$\%) show evidence for stellar bars. The remaining sources are either compact and smooth ($\sim40\%$), or clumpy ($\sim20\%$). We investigate the candidate stellar bars further by subtracting out a 2D Sersic profile fit to the central bulge emission in F444W to reveal underlying structure, and also by modeling the radial profile with a series of 2D ellipsoids as described in \cite{Guo2023}. This corroborates the evidence\footnote{(1) Maximum elipticity $e>0.25$ along the bar with constant position angle, and (2) a drop in elipticity of 0.1 or more combined with $>10\deg$ position angle change over the transition from the bar-dominated region to the outer disk. See also \cite{Jogee2004,Marinova2007}.} for 10 bars and leaves the rest as candidates. The barred structures of SCUBADive sources will be explored in further detail in future works (Silva et al., in prep). For now, we show in Figure \ref{fig:bars} examples of stellar bars in SMGs including one source at $z=3.4\pm1.3$ that if verified would be the highest redshift stellar bar detected to-date. One interesting source to note is z23\_10\_0 which has clear spiral arms and a point-source F444W component in the nucleus. After subtracting a point source model there is tentative evidence for a stellar bar, suggesting that gas can still feed a central AGN despite the resonance trapping imposed by bars. 
The fact that (1) the fraction of SMGs exhibiting bars and/or spirals is about equal to those exhibiting disturbed morphologies, and (2) most SMGs have smooth morphologies means that major mergers cannot be invoked to fully explain the sub-mm-bright starburst phase \citep{Hopkins2008}, and that in-situ mechanisms for extreme gas compaction are certainly at work \citep{Hodge2016}. This is consistent with \cite{Gillman2024} and also the results of sub-mm imaging that find a high incidence of disk-like morphologies in SMGs \citep{Cowie2018,Hodge2020,Gullberg2019}.

\subsection{Redshifts}
The redshift distribution of sub-mm galaxies has been the focus of many works dating back to the earliest spectroscopic follow-up campaigns at optical \citep{Chapman2003,Chapman2005} and radio \citep{Weiss2009} wavelengths. Amongst SMGs with $S_{850}\gtrsim1$ mJy the field has mostly converged $\langle z \rangle \sim2.5$ with lower and upper quartiles of $z=1.8$ and $z=3.4$ respectively \citep[e.g.,][]{daCunha2015,Brisbin2017,Simpson2019,Dudzeviciute2020}. As shown in Figure \ref{fig:zhist} we recover this range with a median $z=2.6$ and lower/upper quartiles spanning $z=1.6-3.5$. However, the sub-set of SMGs with no optical counterparts (prior to JWST) have had highly unconstrained redshifts \citep{Casey2018,Franco2018}. ALMA follow-up has revealed they overwhelmingly have $z_{spec}\gtrsim3$ \citep{Chen2021}, but such confirmation usually requires $>1$ hour per source to complete a full Band 3 scan. With JWST we can now derive well-constrained photometric redshifts for sources that have not been detected by prior ground- and space-based maps. For the SCUBADive sources with no counterparts in COSMOS2020 (i.e., having lacked a secure optical source until now) we find a median $z$ and upper/lower quartiles of $\langle z\rangle=3.1^{+1.2}_{-1.3}$, with $26\%$ having $z>4$. Remarkably, the fraction of sources with no prior optical/near-IR counterpart before JWST is $\sim20\%$ between $1<z<3$ (Fig.~\ref{fig:zhist} \textit{b}). These galaxies have $A_V\in[1,4]$ with $\langle A_V\rangle = 2$ and $\langle\log M_\odot/M_*\rangle=11.3$ demonstrating that even massive systems at lower-redshift have eluded optical detection because of their dust attenuation. At $z>4$ the fraction of sources with no prior optical/near-IR counterpart grows to $75\%$. 

\subsection{Stellar Masses\label{sec:mass}}

Owing to their high attenuation and likely stochastic star-formation histories the stellar mass distribution of SMGs has been difficult to assess, and impossible to measure for the $\sim30\%$ of SMGs that have had no optical counterpart prior to JWST. In fact the stellar masses amongst the same samples of highly dust-obscured galaxies can change by up to an order of magnitude depending on assumptions and the available data \citep{Hainline2011,Michalowski2012}. Here we report the stellar mass statistics measured for our sample and compare to previous works in the literature. Note that in this section (and throughout the rest of this work) we consider the stellar masses after applying the mass-to-light ratio correction factor discussed in Section \ref{sec:fits}.

\begin{figure}
    \centering
    \includegraphics[width=0.48\textwidth]{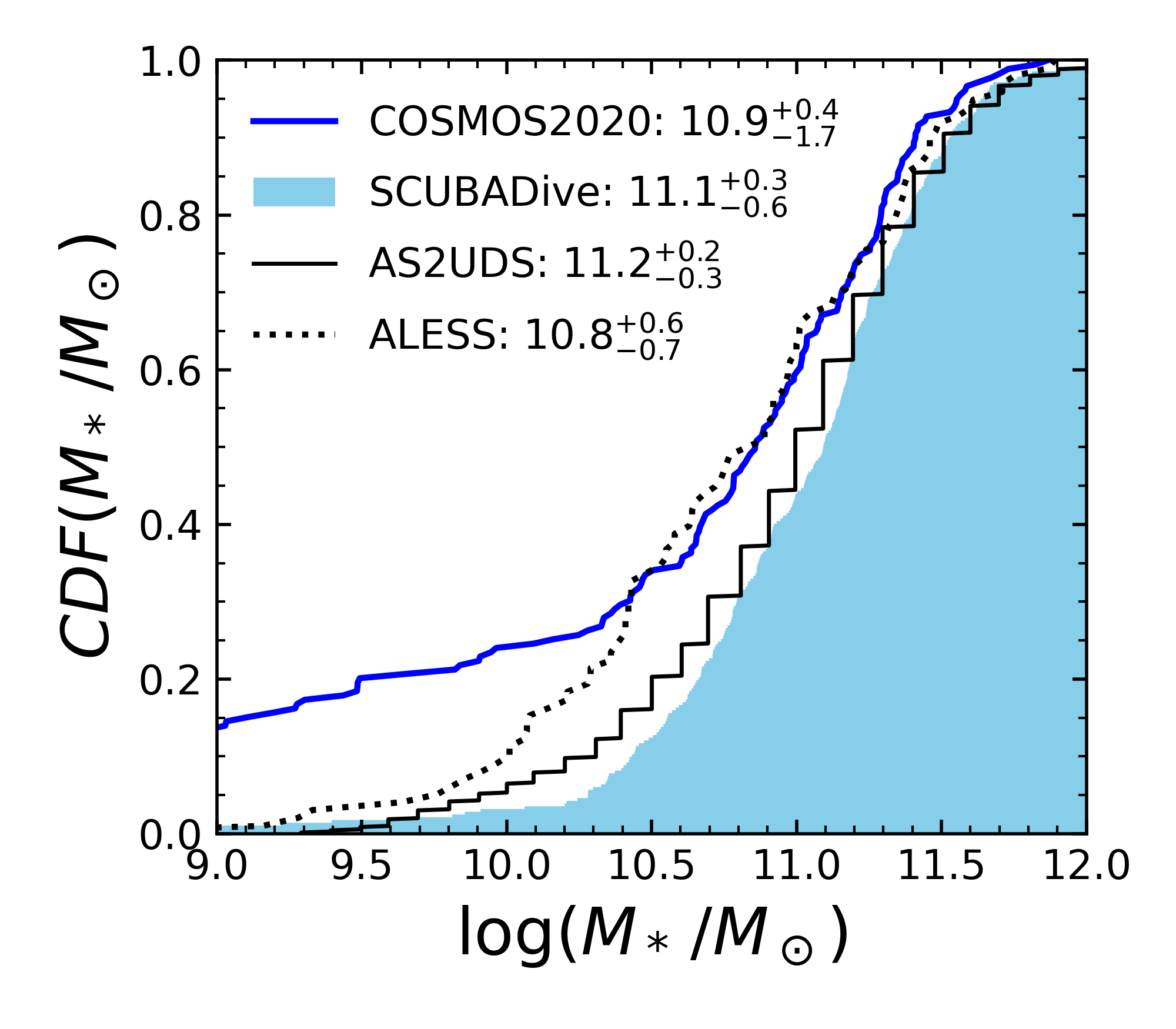}
    \caption{Stellar mass cumulative distribution function for SCUBADive (shaded blue) compared to the AS2UDS  \citep{Dudzeviciute2020} and ALESS \citep{daCunha2015} samples of SMGs. We report the 16th, 50th, and 84th percentiles for each distribution. The mass CDF for SCUBADive prior to JWST (i.e., from COSMOS2020, \citealt{Weaver2022}) is shown in a dashed blue line.}
    \label{fig:pmass}
\end{figure}

The median stellar mass in our sample is $\log M_*/M_\odot=11.1^{+0.3}_{-0.5}$ and spans the range $\log M_*/M_\odot=9-12.2$. The average $1\sigma$ stellar mass uncertainty from our model fits is $0.07^{+0.11}_{-0.04}$ dex, which is less than the $0.1$ dex systematic variance of stellar masses from the choice of star-formation history \citep[e.g.,][]{Carnall2019}. Our median mass is greater than that of canonical analyses of SMGs by $\sim0.2$ dex \citep[e.g.,][]{Borys2005,Hainline2011,Michalowski2012}. More recently, ALMA has been used to identify counterparts to SMGs in deep near-infrared maps and \textit{HST} images which have led to revised stellar mass measures. As shown in Figure \ref{fig:pmass} the stellar mass distribution of SCUBADive is fully consistent with AS2UDS SMGs from \cite{Dudzeviciute2020} which were constrained primarily with K-band and IRAC 3.6$\,\mu$m maps with $3\sigma$ depths of $25.7$ mag and $23.5$ mag respectively. The SCUBADive stellar masses are generally consistent with those found for SMGs in the ALESS sample \citep{daCunha2015}, but the median of SCUBADive is $0.3$ dex greater than that of ALESS. 

\begin{figure}
    \centering
    \includegraphics[width=0.47\textwidth]{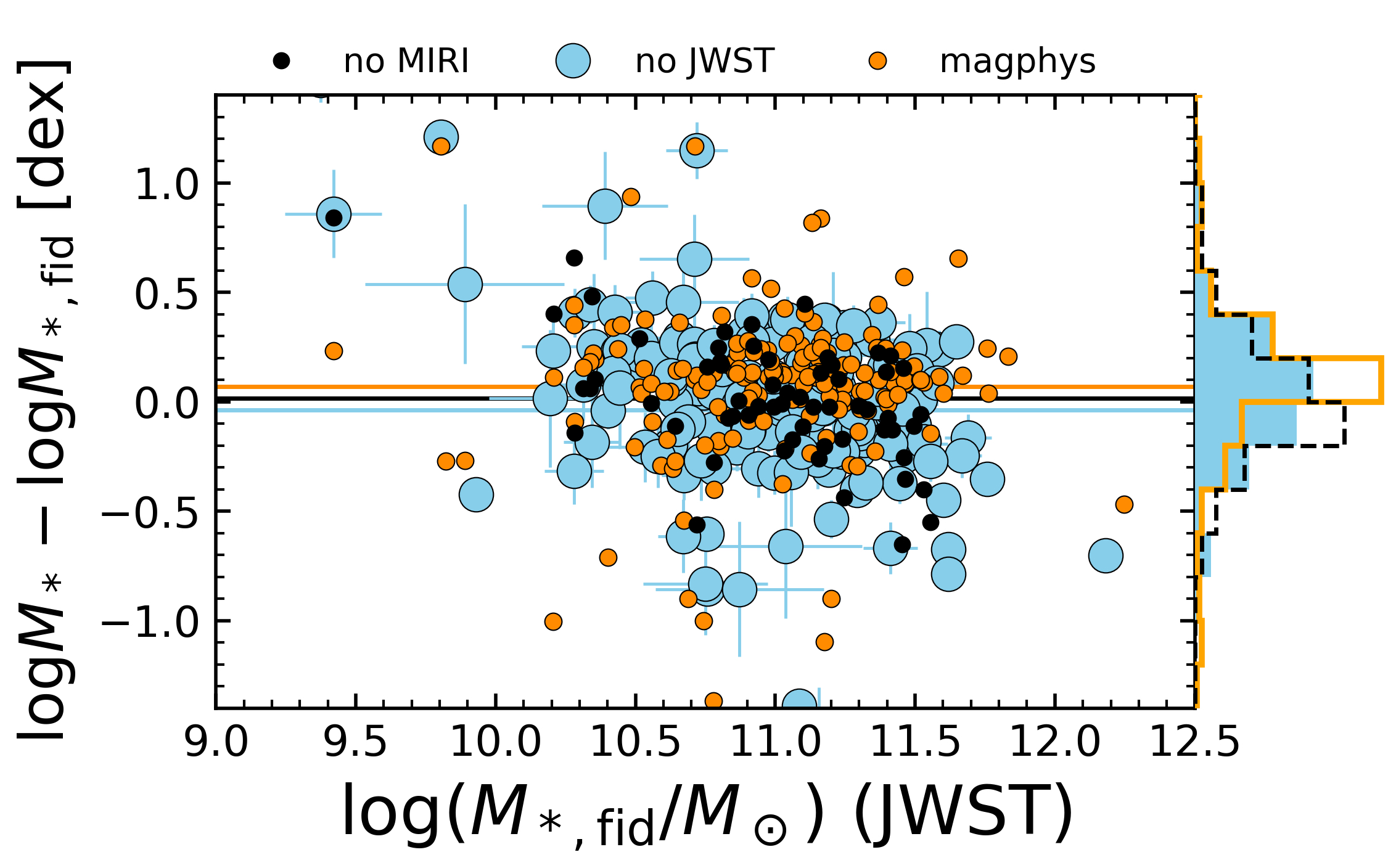}
    \caption{The difference in stellar masses derived for our sample under various modeling assumptions and data inclusion. Our fiducial stellar mass ($M_{\rm *,fid}$) refers to the \texttt{CIGALE} fits outlined in Section \ref{sec:fits} normalized to the $H-$band mass-to-light ratio \cite{daCunha2015}. We show the change in $M_*$ with and without the inclusion of JWST data by comparing against derived stellar masses from the COSMOS2020 \texttt{FARMER} catalog (\citealt{Weaver2022}, blue circles). We also compare against $M_*$ inferred from leaving out the JWST/MIRI data from our fits (black) and when using the \texttt{magphys} SED fitting code (orange). 
    The solid lines shows the average change in stellar mass for each modeling approach which are all within $0.08$ dex of zero. There is no statistically significant correlation between the mass offset when leaving out JWST data and the total stellar mass as measured by JWST ($r_p=-0.21\pm0.04$, $p=0.03\pm0.03$). 
    }
    \label{fig:masscompare}
\end{figure}

Figure \ref{fig:masscompare} shows the change in derived stellar mass for SCUBADive under three different model/data combinations. First, we compare the masses from the COSMOS2020 \texttt{FARMER} catalog \citep{Weaver2022} and our updated masses that include the JWST constraint. There is no systematic offset between these cases; however, sources with masses in COSMOS2020 $>0.5$ dex away from our derivations were previously fit by models that assumed no dust ($A_V\sim0$) and underestimated the redshifts by $1.2$ on-average with $50\%$ having $(z_{C2020}-z_{\rm SCUBADive})<-2$. In total, 66 ($23\%$) of SCUBADive galaxies were significantly misclassified with respect to their stellar mass, $A_V$ and redshift. The most extreme cases correspond to SMGs misidentified as $z=0.6-0.8$ dwarf galaxies ($N=23$, $8\%$) and $z\sim0-5$ post-starburst/quiescent galaxies ($N=14$, $5\%$) with ${\rm SFR}/M_*<10^{-10}\,yr^{-1}$. This is unlikely to drastically impact the number density of either interloper population, but represents a significant fraction of the SMG sample. 
Figure \ref{fig:masscompare} also compares the stellar masses we estimate using \texttt{CIGALE} vs.~\texttt{magphys} which on-average agree with with one-another as expected given the $M_*/L_H$ correction we implement for consistency with \cite{daCunha2015}. The $\sim0.3$ dex scatter in \texttt{CIGALE} vs.~\texttt{magphys} stellar masses most likely reflects lingering effects of the star-formation history assumptions between the two models: although we normalize to the mean $M_*/L_H$ of \cite{daCunha2015} both their work and ours exhibit a $1\sigma$ $M_*/L_H$ of $0.2-0.4$ dex. 
Finally, we also re-run the SED fits leaving out MIRI/F770W where available to test how the reddest SED sampling from JWST systematically changes our results. On average the masses do not change if the MIRI data is removed from the fits, and the mass uncertainties only decrease by $1-7\%$ when fitting includes MIRI/F770W.

\subsection{Optical Attenuation}
\begin{figure*}
    \centering
    \includegraphics[width=\textwidth]{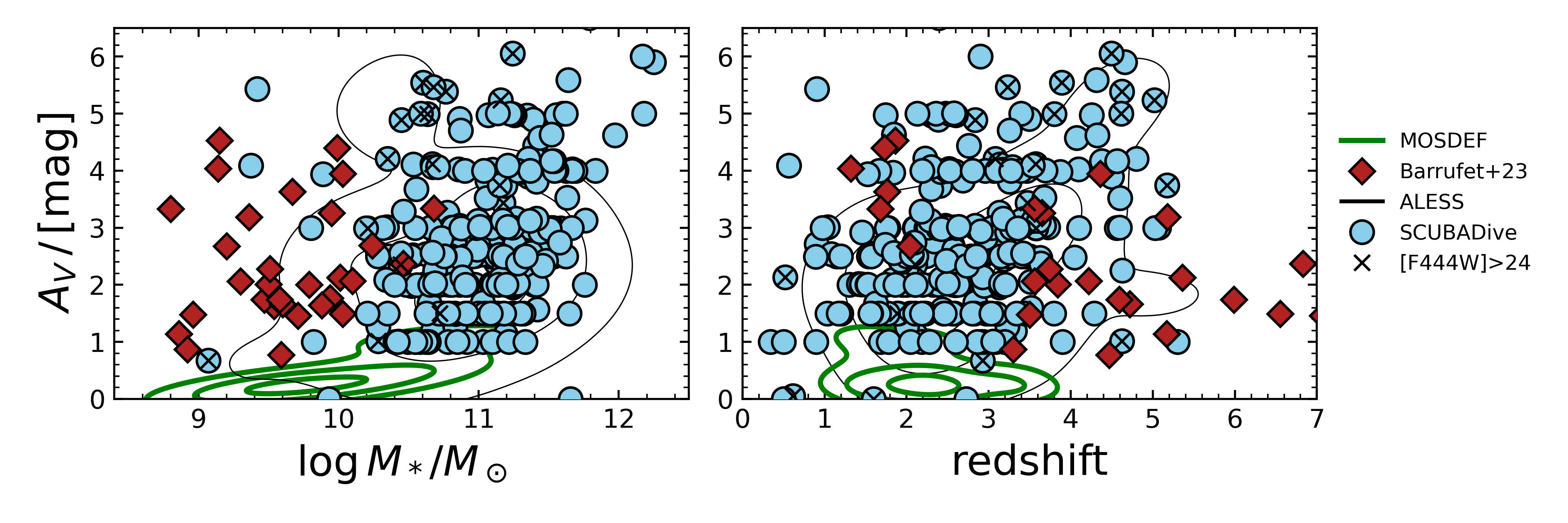}
    \vspace{-30pt}
    \caption{Stellar optical attenuation $A_V$ vs.~total stellar mass (\textit{Left}) and redshift (\textit{right}). SCUBADive is shown in blue compared to normal $z\sim1-3$ galaxies from MOSDEF (green contours, \citealt{Reddy2015,Shivaei2015,Kriek2015}), ALESS SMGs (black contours, \citealt{daCunha2015}), and the \textit{HST}-dark sample of \cite{Barrufet2022,Barrufet2024} as red diamonds. 
    Black $\times$ symbols indicate SCUBADive sources with [F444W]$\,>24$ mag.
    }
    \label{fig:avmass}
\end{figure*}

In Figure \ref{fig:avmass} we show the distribution in $A_V$ as a function of total stellar mass and redshift. SCUBADive has $\langle A_V\rangle=2.6^{+1.5}_{-1.0}$ which overlaps with the attenuation inferred amongst SMGs from ALESS \citep{daCunha2015}, as well as the \textit{HST}-dark NIRCam-selected galaxy sample from \cite{Barrufet2022,Barrufet2024} that exhibit comparable $A_V$ at $\sim2$ dex lower stellar mass. These samples of dust-obscured galaxies all exhibit higher $A_V$ for fixed stellar mass than the more normal sample of $z\sim1-4$ galaxies from MOSDEF \citep{Reddy2015,Shivaei2015,Kriek2015}. Galaxies without a prior counterpart in COSMOS2020 have $\langle A_V \rangle = 3.0^{+2.0}_{-1.5}$, $\sim25\%$ greater than the median of those with a match in COSMOS2020, and the faintest subset of SCUBADive in F444W with [F444W]$\,>24$ mag have $\langle A_V\rangle=4^{+1}_{-2}$ and tend to have $z>3$ as shown on Fig.~\ref{fig:avmass} (\textit{right}). The optically-faint subset ($<4$ detections in optical/near-IR photometric bands) of ALESS has $\langle A_V\rangle=2.9^{+0.3}_{-0.3}$. The added constraint by JWST aids in robustly identifying SMGs with higher $A_V$ and higher redshift, but that still exhibit the high stellar masses and star-formation rates common amongst SMGs in general. 

\section{Discussion\label{sec:discussion}}
\subsection{The volume density of SMGs}

Most SMGs have $z\sim1-3$ with an average of $z\sim2.5$ \citep{Chapman2005,Walter2012,daCunha2015,Brisbin2017,Dudzeviciute2020,Simpson2020} and a high-end tail at $z>3-4$ associated with those lacking optical counterparts \citep[e.g.,][]{Franco2018,Williams2019,Manning2022}. The $z>3$ sub-set of the SMG population has been of growing interest as (a) models of dust formation in the early Universe demonstrate how difficult creating early dust reservoirs might be \citep{Gall2011,Popping2017}, and (b) they are good progenitor candidates for the emergent $z\sim3$ quiescent galaxy population \citep{Toft2014,Long2023}. In both regards, the volume-averaged number density of SMGs is a key measurable that models must reproduce as it captures the formation history, source density, and duty cycle of $z>3$ SMGs. Prior to JWST the volume density of $z>3$ SMGs exhibited over 1 dex in scatter owing to small survey volumes, small number statistics, and highly uncertain photometric redshifts \citep{Zavala2021,Long2023}. Programs like AS2COSPEC \citep{Chen2021} are making significant progress in this regime with 52 spectroscopic redshifts for $z>2$ SMGS in COSMOS; however, spectroscopic follow-up with ALMA is time expensive. Given the significant photometric redshift constraint added by JWST on optically faint sources coupled to the wide area of COSMOS-Web as compared to other JWST extragalactic fields, SCUBADive is well-suited to constrain the $z>3$ volume density of SMGs at high-precision. 

\begin{figure}
    \centering
    \includegraphics[width=0.48\textwidth]{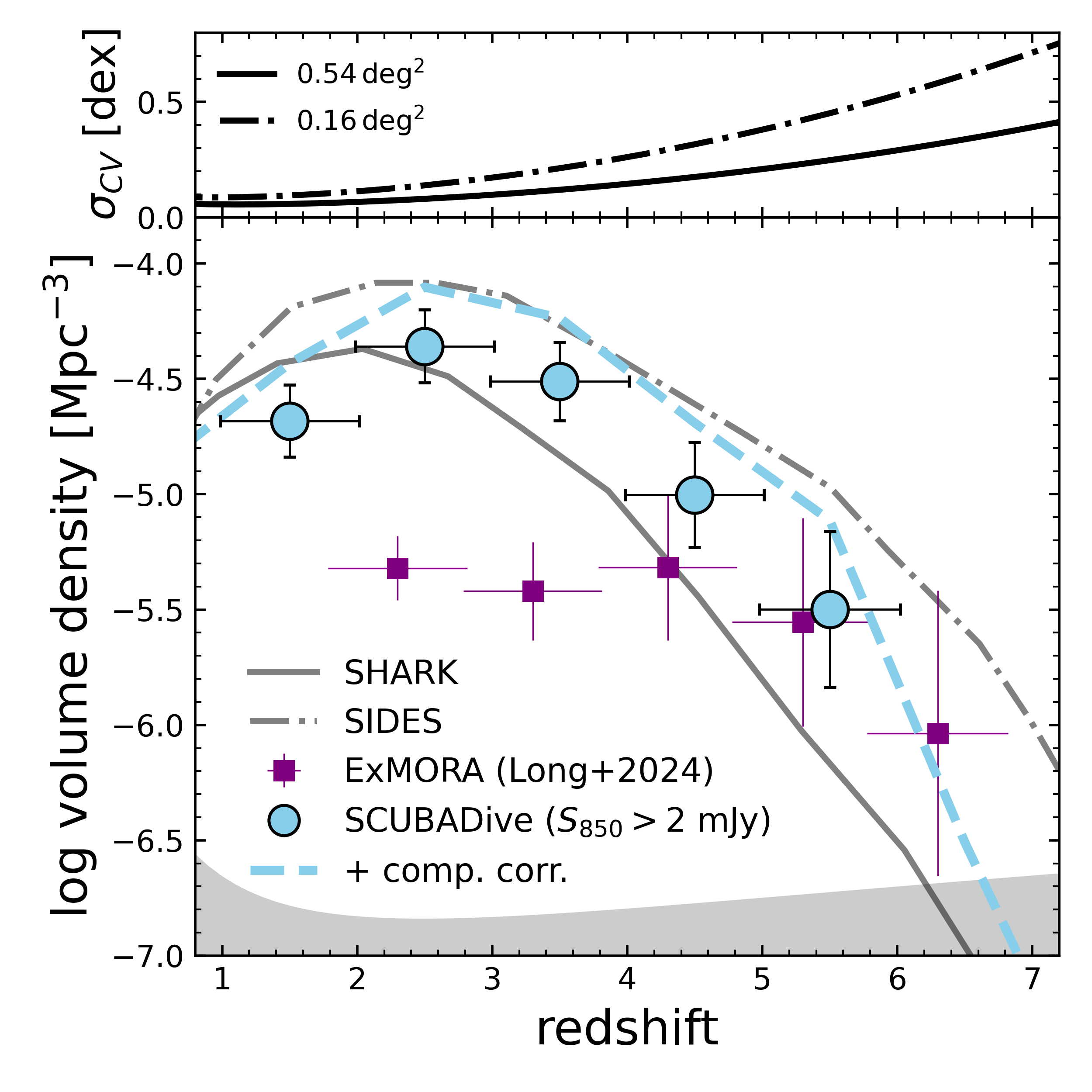}
    \caption{The volume density of dusty star-forming galaxies inferred from SCUBADive (blue circles) out to $z=6$ in bins of redshift with $\Delta z=1$. We compare to the semi-analytic model predictions of SHARK (grey solid, \citealt{Lagos2019,Lagos2020}) and SIDES (grey dashed, \citealt{Bethermin2017}), as well as results from the 0.16 deg$^2$ 2mm blank field ExMORA (purple, Long et al., in prep.). The dashed blue line represents the maximal volume density allowed by the completeness of S2COSMOS \citep{Simpson2019}, and the shaded grey region corresponds to the limit of one source per bin over the COSMOS-Web area. 
    The top panel shows the cosmic variance uncertainty inherent to volume densities by their survey area, assuming bins in redshift of $\Delta z=1$, stellar masses $>10^{10}\,M_\odot$, and following the formalism in \cite{Moster2011}. We show the $1\sigma$ cosmic variance for a 0.54 deg$^2$ field (e.g., COSMOS-Web) and a $0.16$ deg$^2$ field (e.g., CEERS, ExMORA). Uniform selection over large areas like COSMOS-Web are critical for overcoming the limitations of cosmic variance and for discriminating between galaxy formation models at $z\gtrsim3$. 
    }
    \label{fig:nz}
\end{figure}

Figure \ref{fig:nz} shows the volume density of SCUBADive SMGs with $\log(M_*/M_\odot)>10^{10}$ and $S_{850}>2$ mJy between $z=1-6$ assuming redshift bins of $\Delta z=1$. We restrict our volume density measurement to SCUBADive sources with deblended $S_{850}>2$ mJy (87\% of total) for completeness reasons. The error bars capture cosmic variance uncertainty, the $\sqrt{N}$ Poisson uncertainty, and the photometric redshift uncertainty which we fold in by bootstrapping the measurement using 1000 iterations of the sample's redshifts perturbed by their $1\sigma$ uncertainties. 
There are 487 SMGs in the COSMOS-Web area that do not have a secure JWST/NIRCam counterpart for lack of ALMA data. We have corrected for this by scaling the counts in each redshift bin assuming the SCUBADive redshift posterior applies to the unmatched SMGs. We also correct for the completeness of the SCUBA-2 map, which can be as low as 50\% around the SNR limit of the S2COSMOS catalog \citep{Simpson2019}. 44\% of SCUBADive correspond to SCUBA-2 sources with $S_{850}>6.4$ mJy where the average completeness of the S2COSMOS map is $>90\%$ \citep{Simpson2019}. We estimate a completeness-correction factor from the range of SCUBA-2 850$\,\mu$m flux densities in SCUBADive and the completeness function of \cite{Simpson2019}. On-average this increases the volume density by $0.2$ dex at most, and it is worth noting that COSMOS-Web overlaps with the deepest portion of the SCUBA-2 map. Table \ref{tab:vol} lists the SCUBADive volume density. 

Volume density measurements must be accompanied by proper estimates on their uncertainties attributed to cosmic variance: variation in observed source counts attributable to large-scale density fluctuations \citep{Moster2011}. This is especially true when discussing favored models and those ruled out by the observations \citep{Trenti2008}, and certainly factors into the $\sim2$ dex range in reported volume densities on $z>3$ SMGs found in the literature. More often than not the error bars quoted on volume densities only reflect the Poisson statistics of the sample. Here we outline our method for calculating the volume density uncertainty from cosmic variance which follows the formalism developed by \citealt{Moster2011} (see Section 3.4 therein). First we start with the dark matter root cosmic variance ($\sigma_{dm}$) reported by \cite{Moster2011} for the COSMOS field (7045 arcmin$^2$). We next scale this to the the area of COSMOS-Web (1944 arcmin$^2$) by interpolating between $\sigma_{dm}$ reported for the UDF, GOODS, GEMS, the EGS and COSMOS in \cite{Moster2011}. Next we calculate the galaxy bias $b(m_*,\bar z)$ for the stellar mass range in our sample ($>10^{10.5}\,M_\odot/M_*$, see Fig.~\ref{fig:pmass}) and at the center of each redshift bin ($\bar z$) where we calculate a volume density. The bias ranges from $2$ at $z=1$ to $22$ at $z=5$. We next compute the root cosmic variance for galaxies $\sigma_{CV}=b(m_*,\bar z)\sigma_{dm}(\bar z, \Delta z=0.2)\sqrt{0.2/\Delta z}$. The last term is there to match the redshift bin widths we adopt of $\Delta z=1$. For comparison we also calculate the galaxy root cosmic variance assuming a field area of 700 arcmin$^2$ to compare against surveys like CEERS in the EGS and ExMORA in COSMOS. For the COSMOS-Web area and a redshift bin width of $\Delta z=1$ we estimate $\sigma_{CV}$ for SMGs to be $0.12$ dex at $3<z<4$, $0.18$ dex at $4<z<5$, and $0.25$ dex at $5<z<6$. For a 700$^2$ arcmin area the root cosmic variance at $z>3$ is $\sim0.3$ dex greater in each bin, highlighting the importance of wide fields when studying the volume densities of massive sources like SMGs.

For a comparison against semi-analytic model predictions we follow \citet[][See their Section 4.4]{Long2023} in extracting mock samples of actively star-forming galaxies with $\log\,M_*/M_\odot>10.5$ and $\log\,L_{IR}/L_\odot>11$ from SHARK \citep{Lagos2019,Lagos2020} which does a good job matching $850\,\mu$m number counts, and SIDES \citep{Bethermin2017} which successfully reproduces the number counts of \textit{Herschel}/PACS and SPIRE maps. 
The SCUBADive volume densities are in good agreement with SHARK at $z<3$, but accounting for completeness in S2COSMOS could favor SIDES. At $z>3$ the SCUBADive volume density is $0.3-0.4$ dex above SHARK and $\sim0.5$ dex below SIDES. A major conclusion of SHARK is that $S_{850}>2$ mJy SMGs contribute negligibly to the volume-averaged star-formation rate density at $z>3$ which the SCUBADive measurements challenge by allowing for a $\sim3\times$ greater SMG volume density at $z>3$ after accounting for incompleteness. This enhancement is exacerbated by the fact that an $850\,\mu$m selection on the highest redshift dusty, star-forming galaxies might yield a sub-dominant population to those more efficiently selected at 2 mm (e.g., \citealt{Casey2021,Zavala2021,Cooper2022}, Long et al., in prep.). Indeed we compare against the ExMORA 2 mm blind survey over COSMOS from Long et al., in prep which falls between the SHARK and SIDES model at $z>5$. Combined with SCUBADive the data does not rule out $S_{850}>2$ mJy SMGs as potentially important contributors to the star-formation rate density at $z>3$ if they are confirmed to be $\sim1$ dex more numerous than the SHARK semi-analytic model.

\begin{deluxetable}{lllll}
    \tabletypesize{\footnotesize}
    \setlength{\tabcolsep}{5pt}
    \tablecaption{SCUBADive Volume Density over COSMOS-Web for SCUBA-2 sources with $S_{850}>2$ mJy, and the relevant error terms from Poisson noise ($\sigma_p$), cosmic variance ($\sigma_{CV}$) and redshift uncertainties ($\sigma_z$). \label{tab:vol}}
    \tablehead{\colhead{$z$}&\colhead{$\log (n/{\rm Mpc^3})$}&\colhead{$\sigma_{p}$}&\colhead{$\sigma_{CV}$}&\colhead{$\sigma_{z}$}}
    \startdata
    $1<z<2$ & $-4.68\pm0.15$ & $0.06$ & $0.06$ & $0.13$ \\ 
    $2<z<3$ & $-4.35\pm0.16$ & $0.04$ & $0.08$ & $0.13$ \\ 
    $3<z<4$ & $-4.51\pm0.17$ & $0.05$ & $0.12$ & $0.11$ \\ 
    $4<z<5$ & $-5.00\pm0.23$ & $0.09$ & $0.18$ & $0.11$ \\ 
    $5<z<6$ & $-5.50\pm0.34$ & $0.18$ & $0.25$ & $0.15$ \\ 
    \enddata
    \tablecomments{These table values have not been corrected for the incompleteness of the parent SCUBA-2 map, which could increase the volume density by $0.2$ dex on-average in each redshift bin. They have been corrected for the SCUBADive sampling of S2COSMOS SMGs over COSMOS-Web \ref{fig:nz}.}
\end{deluxetable}

\subsection{Selecting high-redshift DSFGs with \textit{JWST}-alone?}

SCUBADive being a verified high-redshift dusty, star-forming galaxy sample matched to NIRCam and MIRI counterparts allows for tests on selection methods using JWST alone. These are of interest for sub-sets of high$-z$ dusty galaxies that may harbor significant dust reservoirs but are not detected in wide-field SCUBA-2 sub-mm maps. 

In SCUBADive there are 82 ($28\%$) sources that satisfy the \textit{HST}-dark selection criterion for massive, high-redshift and obscured galaxies from \cite{Barrufet2022} and \cite{Gottumukkala2024}: [F160W]$\,-\,$[F444W]$\,>2.3$. This sub-set of SCUBADive indeed consists of the highest redshift members; however, it misses $44$ $(75\%)$ of $z>3$ SCUBADive galaxies and 9 ($32\%$) with $z>4$. 

\begin{figure}
    \centering
    \includegraphics[width=0.47\textwidth]{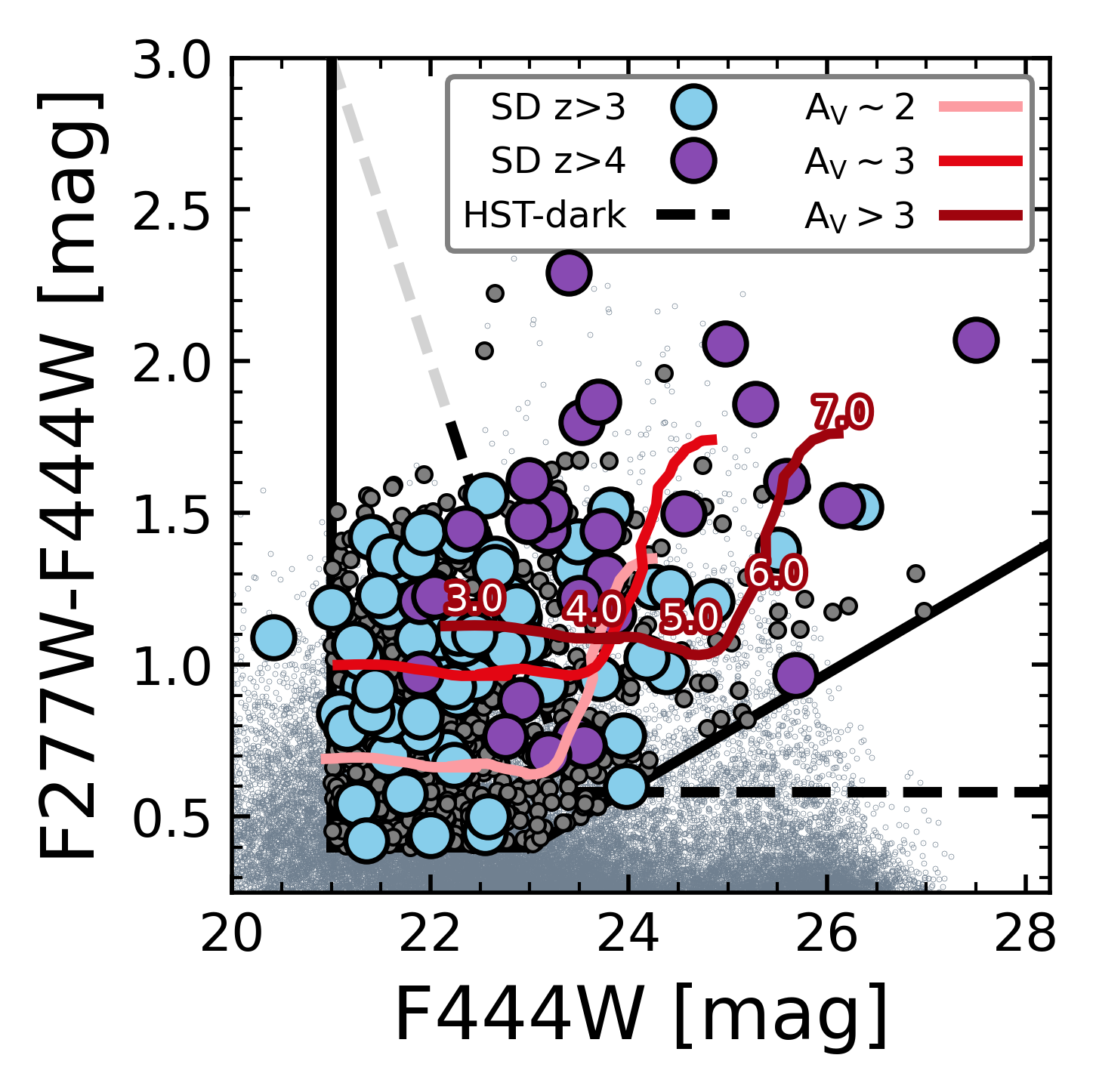}
    \caption{Zoom-in on the faint and red sub-set of COSMOS-Web sources in in JWST/NIRCam's F277W and F444W color-magnitude space. In blue and purple we show SCUBADive sources at $z>3$ and $z>4$ respectively. The black solid line shows our color-magnitude wedge that encases $z>3$ SCUBADive galaxies, some of which are missed by JWST-calibrated selections on massive galaxies that exclude sources brighter than F444W$\,\lesssim23$ mag (black dashed line, \citealt[e.g.,][]{Barrufet2022,Gottumukkala2024}). Grey circles within the solid black wedge represent 762 COSMOS-Web sources within the FWHM of a SCUBA-2 SMG from S2COSMOS \citep{Simpson2019}, which given the domain/range of both SCUBADive, and $z=3-7$ redshifted SED tracks for $A_V\sim2,3$ and $A_V>3$ ALESS SMGs from \citealt{daCunha2015}, are plausibly $z>3$ dusty, star-forming galaxies.}
    \label{fig:colormag_highz}
\end{figure}

Figure \ref{fig:colormag_highz} shows a zoom-in to the F277W and F444W color-magnitude space also shown on Fig.~\ref{fig:colormag}, now highlighting the domain of $z>3$ SMGs. We find that the following color and magnitude cuts in the F277W and F444W filters encases most of the high$-z$ subset of SCUBADive with minimal conamination of lower redshift SMGs:

\begin{flalign*}
    (i)\qquad\qquad\quad\,\,\,\,\, {\rm [F444W]} & > 21   \\ 
    (ii)\,\,\,\, {\rm [F277W]}-{\rm [F444W]} & > 0.4  \\
    (iii)\,\, {\rm [F277W]}-{\rm [F444W]} & > \frac{{\rm [F444W]}}{5.25} - 3.98
\end{flalign*}

We construct these color/magnitude cuts based on the limits of $z>3$ SCUBADive sources, and because 46 of the 63 $z>3$ SMGs in SCUBADive have SNR$\,<5$ in F115W and 29 have SNR$\,<5$ in F150W. NIRCam/LW bands are best-suited for selecting $z>3$ SMGs in part due to their faint fluxes below $\lambda_{obs}=2\,\mu$m. The colors of $z>3$ SCUBADive SMGs are consistent with $z=3-7$ evolution in [F277W]$-$[F444W] and [F444W] for template SEDs from ALESS at $A_V\geq2$ \citep{daCunha2015}.  

While the above selection wedge encases all but one of the $z>3$ SMGs in SCUBADive, it does not exclusively capture this population. Galaxies of similar color can include ``Little Red Dots'', but these are point sources and can therefore be morphologically separated from SMGs \citep{Kokorev2024}. No SCUBADive galaxy is exclusively a point source in NIRCam and none overlap with the sample of LRDs in COSMOS-Web from \cite{Akins2024}. Many host point-source profiles in F444W but always on top of other structure apparent in F444W and the shorter NIRCam bands. Recently different works using a selection function combining red JWST/NIRCam colors with a lack of detection in \textit{HST} bands has uncovered a population of dust-obscured galaxies apparently below the SMG limit on dust masses but with non-negligble dust obscuration \citep{Barrufet2022,Williams2023}. Based on spectroscopic follow-up of 24 objects these sources have $z\sim4-5$, $A_V>2$ mag and $\log\,M_*/M_\odot>9.8$ \citep{Barrufet2024}. Given the masses and attenuation found amongst SCUBADive these NIRCam-selected sources could be a less massive, less extreme tail of the $z>3$ SMG population. Indeed recent models suggest that $z=4-5$ galaxies harbor $\sim5\times$ more dust-obscured star-formation than their $z=0-2$ mass-matched counterparts \citep{Zimmerman2024}, which supports a heterogeneous population of $z>3$ dust-obscured, star-forming galaxies including the most extreme SMGs. Ultimately, differential dust geometry across an SMG can cause its optical/near-infrared colors to appear far less extreme \citep[e.g.,][]{Howell2010,Casey2014b,Cochrane2024}, hence an overlap in colors with samples like those of \cite{Barrufet2022,Barrufet2024} and \cite{Gottumukkala2024} is unsurprising.

In addition to the aforementioned confusion with other galaxy populations, we should approach with caution NIRCam selection of SMG counterparts without high-resolution interferometer follow-up. There are some cases where red NIRCam colors can be a useful prior in predicting likely SMG counterparts \citep[e.g.,][]{Gillman2023}, but the technique is not guaranteed to select the true SMG. Discounting SMGs resolved into 2 or 3 ALMA sources, $50\%$ of SCUBADive galaxies are not the reddest NIRCam source within the footprint of their SCUBA-2 detection, consistent with the findings of \cite{Gillman2024}. Indeed within our selection wedge on Figure \ref{fig:colormag_highz} there are 708 NIRCam sources that are also within the FWHM of the SCUBA-2 beam and therefore plausible counterparts to the sub-mm emission. 
Applying the $z>3$ fraction in SCUBADive to the whole sample, there should be 157 $z>3$ SMGs in S2COSMOS$+$COSMOS-Web yet to have confirmed NIRCam counterparts. For these sources there are on-average $4$ candidate JWST counterparts that occupy the red and faint color space in Fig.~\ref{fig:colormag_highz}, highlighting the need for ALMA, NOEMA, SMA and/or VLA follow-up to isolate the true optical counterparts to SMGs.

\subsection{The clustering of sources around SMGs \label{sec:clustering}}
Many works have proposed SMGs as sign posts of over-dense environments with respect to the large scale structure of the Universe \citep{Clements2014,Dannerbauer2014,Casey2015,Clements2016,Hung2016,WangTao2016,FloresCacho2016,Calvi2023}. However, there is outstanding controversy on this issue \citep[e.g.,][]{Miller2015,Cornish2024} and ultimately high-fidelity redshifts are needed to leverage SMGs and their companions to select (proto-)cluster galaxy populations (\citealt{Miller2015,Hayward2018}, and \citealt{AlbertsNoble2022} for a recent review). Given the improvement on photometric redshifts with the added JWST data now available for COSMOS-Web \citep{Casey2022} we conduct a simple test on the over-density around SCUBADive SMGs. 

\begin{figure}
    \centering
    \includegraphics[width=0.48\textwidth]{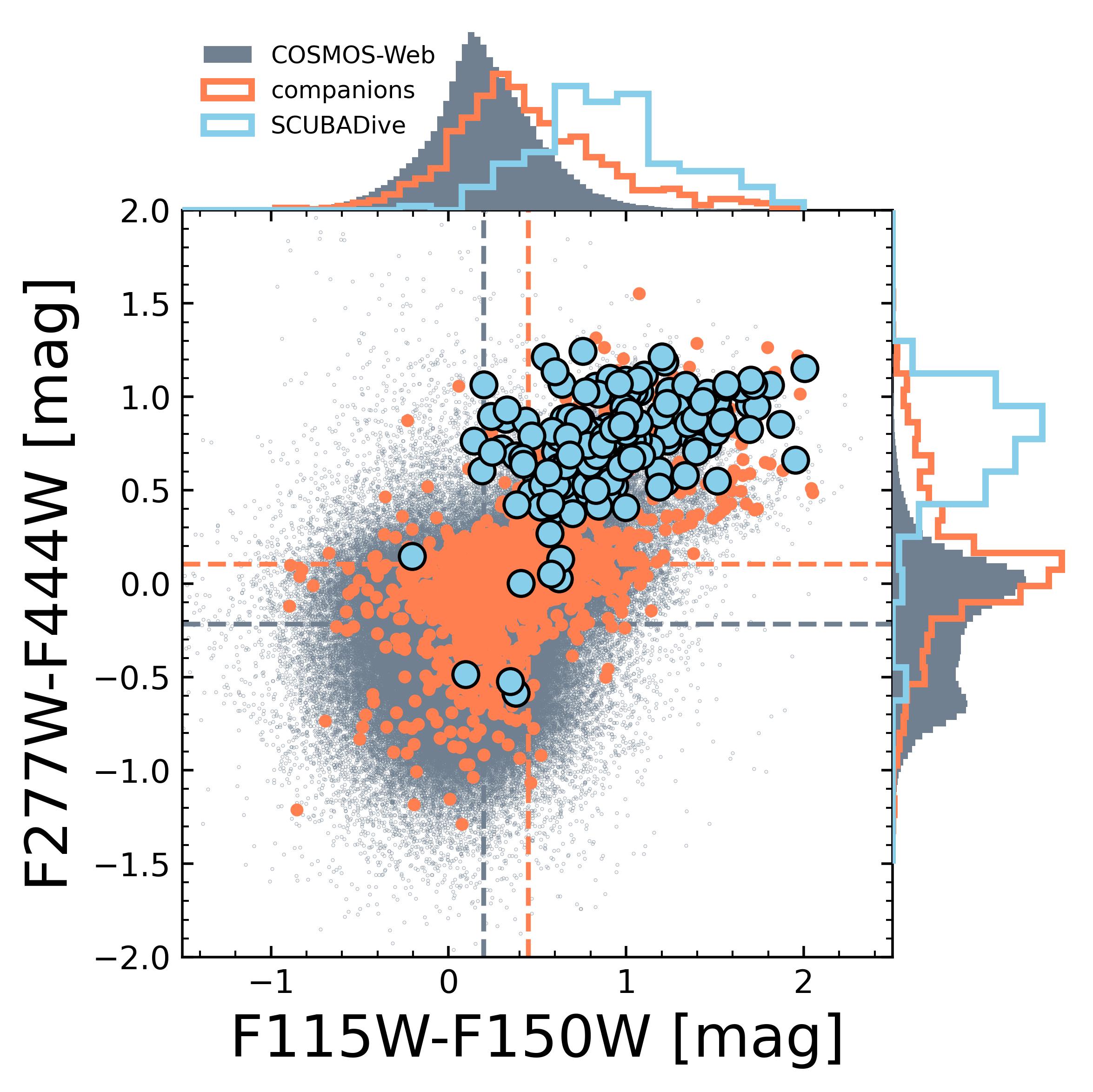}
    \caption{NIRCam SW and LW color-color space for SCUBADive detected in all four COSMOS-Web bands (blue), their candidate companions (orange), and the rest of COSMOS-Web sources (grey). The respectively colored dashed lines correspond to the average colors of the COSMOS-Web and candidate companion data sets. The companions to SCUBADive galaxies exhibit redder NIRCam colors in both SW and LW bands than what would be expected from a randomly drawn sample from all COSMOS-Web sources.}
    \label{fig:compcolor}
\end{figure}

\begin{figure}
    \centering
    \includegraphics[width=.49\textwidth]{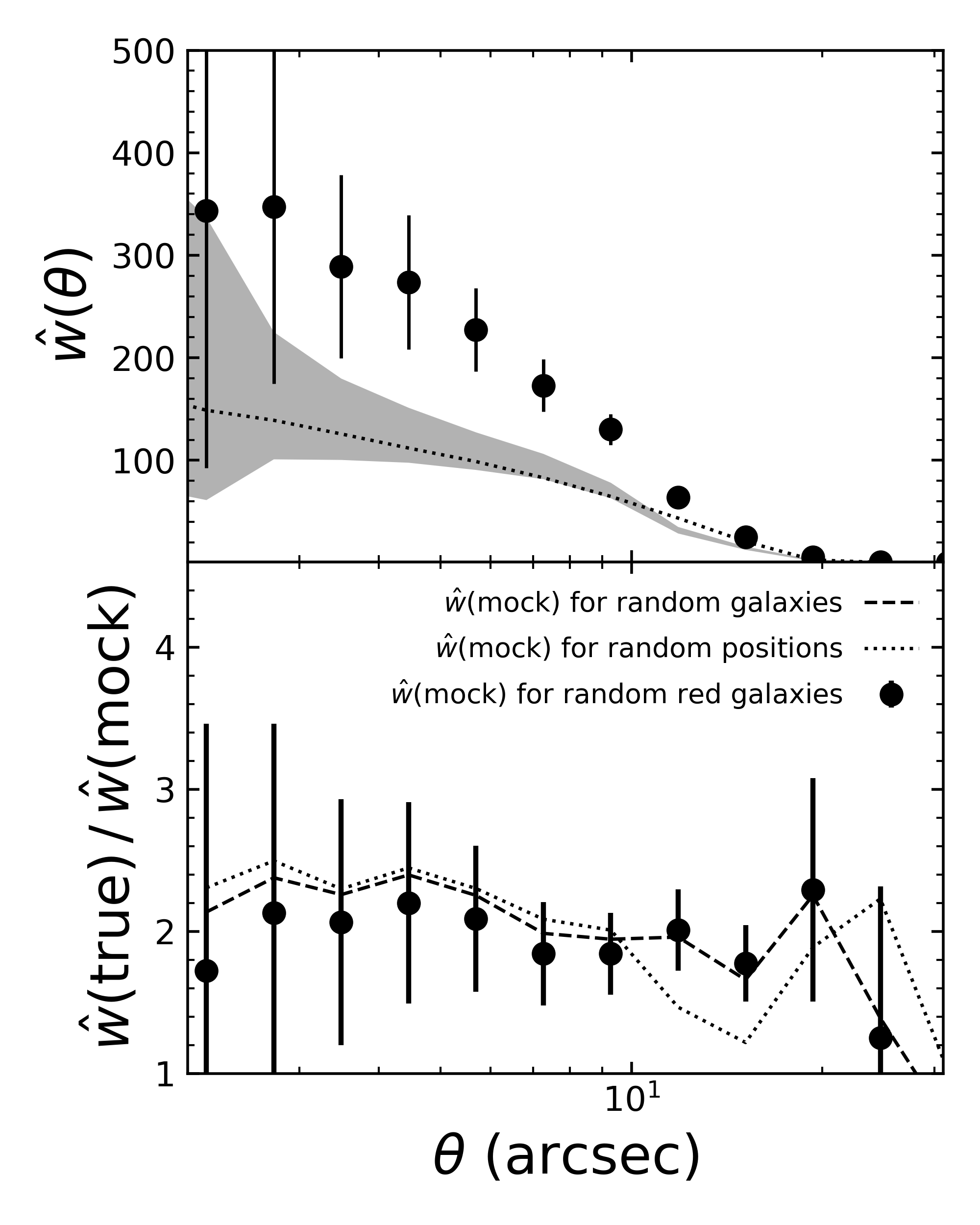}
    \caption{The clustering signal for sources around SCUBADive galaxies with consistent redshifts. (\textit{Top}) Two-point angular correlation function $\hat{w}(\theta)$. With black circles we show $\hat{w}(\theta)$ for $1870$ candidate companion galaxies within 10$^{\prime\prime}$ of a SCUBADive source and with a consistent photometric redshift within $1\sigma$. The shaded grey region encases the bootstrapped range in $\hat{w}(\theta)$ derived for 1000 mock clustering catalogs generated semi-randomly from the NIRCam LW space of SCUBADive galaxies as described in Section \ref{sec:clustering}. The black dotted line shows the two-point correlation function around random coordinates in COSMOS-Web, a baseline test against the significance of the SCUBADive clustering signal. 
    \textit{(Bottom)} The ratio of our measured two-point angular correlation function to that of the average mock clustering catalogs for three different random selections as listed in the legend. 
    Between $\sim5-20^{\prime\prime}$ the clustering of our candidate companion sample is statistically significant at $\sim2\times$ what is expected based on the clustering around random galaxies and random positions in the field.
    }
    \label{fig:2pt}
\end{figure}

We construct a sample of candidate counterparts to SCUBADive galaxies, i.e., possible physical associations, by first searching COSMOS-Web for all NIRCam sources within 10$^{\prime\prime}$ of a SCUBADive coordinate. This corresponds to a physical search radius of approximately $100$ kpc given the redshift range of SCUBADive, covering $\sim30\%-40\%$ the virial radius of a $10^{13}\,M_\odot$ halo within which SCUBADive sources likely fall given their stellar masses \citep{Behroozi2010}. We then only consider the candidate companion galaxies with photometrically-derived redshifts within $1\sigma$ of the SCUBADive source. While this is not the strictest method for removing objects with chance alignment, it does reflect allowable associations given the data at hand. In this manner we find 1870 candidate counterparts to SCUBADive galaxies corresponding to $7\pm3$ associations per source. For random points in the field with the same redshift distribution we expect as many as $10\pm4$ per source based on monte carlo simulations drawn from the catalog. Figure \ref{fig:compcolor} shows the NIRCam SW (F115W-F150W) vs.~LW (F277W-F444W) color-color space occupied by both SCUBADive galaxies and their candidate companions which exhibit colors inconsistent with random draws from the overall COSMOS-Web distribution. Figure \ref{fig:2pt} shows the two-point correlation function for the candidate companions which we compute using \texttt{astroML} \citep{astroML,astroML-book}. 

To assess the statistical significance of clustering around SCUBADive sources we create a series of test samples containing 289 sources/positions drawn from (a) the NIRCam color space occupied by SCUBADive, (b) the entire COSMOS-Web catalog, and (c) random positions in COSMOS-Web. For each of these mock samples we then follow the previously described method for selecting candidate counterparts to build a mock candidate catalog, which we then use to calculate a two-point correlation function. We repeat this process 1000 times to boot-strap the two-point correlation function range. 
As shown in Figure \ref{fig:2pt}, the two-point correlation function around SCUBADive sources is $\sim2\times$ that of all the mock catalogs between $\sim5^{\prime\prime}-20^{\prime\prime}$ and is statistically significant given the propagated uncertainties. In other words, the clustering strength around SMGs in SCUBADive is $2\times$ that of the baseline for arbitrarily selected galaxies in COSMOS-Web, galaxies with similar colors to the SMGs in SCUBADive, and random positions in the field. 
We acknowledge that the significance of the two-point correlation function clustering signal around SMGs would most likely increase with better constraint on redshifts across the board and a more careful analysis of candidate companions; however, this analysis does corroborate the hypothesis that sub-mm-selected galaxies and more generally, dusty star-forming galaxies, reside within massive dark matter halos at the centers of over-dense environments as discussed by many works in the literature \citep{Casey2016,Greenslade2018,Cheng2019,GomezGuijarro2019,AlbertsNoble2022,Popescu2023}. Future clustering work will investigate in more detail. 

\section{Conclusions}
In this work we present the SCUBADive project. We uncover the COSMOS-Web \citep{cosmos-web} \textit{JWST}/NIRCam counterparts to galaxies detected in the SCUBA-2 $850\,\mu$m map of the COSMOS field \citep{Simpson2019}. Using every publicly-available archival ALMA observation in Band 6 and 7 over COSMOS-Web we resolve on sub-arcsecond scales the sub-mm emission from 219 SCUBA-2/$850\,\mu$m sources into 289 galaxies. We then match these sources to the COSMOS-Web data, investigate their near-infrared morphologies, measure their NIRCam photometry and fit the galaxy spectral energy distributions (SEDs) to derive redshifts and other physical properties. Our main conclusions are:
\begin{enumerate}
    \item We resolve 25\% of 219 SCUBA2 sources into multiple galaxies, 44 of which are doublets and 9 of which are triplets. In some cases the member galaxies are obviously interacting with one-another as evidenced by their NIRCam imaging. 
    \item The NIRCam/LW colors of SMGs are redder than 95\% of all sources in COSMOS-Web. The highest redshift SMGs in our catalog have ${\rm [F277W]}-{\rm[F444W]}>1.5$ and ${\rm[F444W]}>23$, overlapping the color-magnitude space of other high-redshift dusty galaxy populations like Little Red Dots and \textit{HST}-dark galaxies. No SMG in SCUBADive is in the COSMOS-Web LRD sample of \cite{Akins2024}.
    \item $\sim50\%$ of the NIRCam counterparts to SMGs in SCUBADive do not have the reddest ${\rm [F277W]}-{\rm [F444W]}$ color amongst nearby galaxies within the extent of the SCUBA-2 detection. Thus NIRCam-alone selection methods may miss the true optical counterpart to an SMG $\sim50\%$ of the time, highlighting the need for interferometer follow-up to confirm counterpart association. 
    \item Fitting the optical-through-radio SEDs we measure stellar masses of $\log\,M_*/M_\odot=11.1^{+0.3}_{-0.5}$, fully consistent with other large samples of SMGs \citep[e.g.,][]{Dudzeviciute2020}. We find $\langle A_V\rangle=2.5^{+1.5}_{-1.0}$ also in good agreement with past estimates of attenuation in SMGs; however, the faint subset of SCUBADive with ${\rm[F444W]}>24$ that also have $z>3$ exhibit $\langle A_V\rangle=4^{+1}_{-2}$ consistent with these sources having both higher redshifts and higher attenuation.
    \item 81 NIRCam sources have no prior optical/near-infrared counterpart in COSMOS2020. These represent the majority of $z>4$ sources in SCUBADive, and surprisingly a constant $\sim25\%$ fraction of SMGs between $1<z<4$. This ``OIR-dark'' sub-set has $\langle A_V\rangle=3^{+2}_{-1.5}$ and $\log\,M_*/M_\odot=11.1^{+0.4}_{-0.5}$.
    \item The majority of SCUBADive exhibit either smooth Sersic-like morphologies or are irregular/clumpy. We visually identify 50 candidate mergers, 32 SMGs with spiral arms and 23 SMGs with candidate stellar bars. Most of the barred SMGs are at $z\sim1-2$, but we also report a candidate stellar bar caught in the act of forming at $z=3.4\pm1.3$. We observe dust emission in the cores of bar hosts and even nuclear NIRCam point sources suggesting that stellar bars might positively affect the growth of a central stellar population as well as the supermassive black hole. 
    \item We measure the volume density of SMGs in SCUBADive out to $z\sim6$ and collapse the volume density uncertainty by a factor of 6 at $z>4$ compared to prior works given (a) our source density at this epoch, (b) the reliability of our photometric redshifts, and most importantly (c) the large survey area of COSMOS-Web. Our volume densities fall between semi-analytic models like SHARK \citep{Lagos2019,Lagos2020} and SIDES \citep{Bethermin2017}. 
    \item Based on a simple search for companion galaxies with similar redshifts to the SMGs in SCUBADive we find the clustering of NIRCam-detected galaxies around SMGs to be greater than the clustering around randomly selected galaxies and random positions by a factor of 2. This is statistically significant but warrants future work to precisely measure the over-density of companions around SMGs.  
\end{enumerate}
SCUBADive is an ongoing statistical project to study the JWST counterparts to SMGs. We have built a large sample of 289 galaxies complete with JWST and ALMA data which are telling new stories about the in-situ mass assembly of stars and supermassive black holes and the physical nature of SMGs. However, the sample remains incomplete. Within the COSMOS-Web area there are 706 SMGs of which we have ALMA Band 6 and 7 detections for just 219 ($\approx 31\%$). Assuming the multiplicity fraction in SCUBADive would imply $\sim600$ high-redshift, dusty, star-forming galaxies yet to be unambiguously associated with their optical counterpart in existing JWST imaging. There are on-going efforts with longer wavelength ALMA imaging (ExMORA, Long et al., in prep.) and radio counterpart matching with the VLA (Talia et al., in prep., Gentile et al., in prep.) that may chip away at this unmatched population but a dedicated follow-up program is needed to achieve a flux-limited SCUBADive data set. The outcome of such a program would be a multi-wavelength census of SMGs out to $z\sim6$ over the COSMOS field including $\sim30-50$ more sources at $z>4$ and $\sim100$ with $1<z<4$ that still elude detection by ground-based facilities and \textit{HST}. 

\vspace{10pt}
{\small
J.~McKinney acknowledges the tremendous support of the NRAO Helpdesk Team in restoring calibrated data products from the ALMA archive and for making their remote computing resources readily accessible: many thanks to S. Wood, T. Ashton, and M. Sanchez in particular for facilitating access to the data. J.~McKinney also acknowledges A.~Pope for her guidance, K.~Guo for her discussion on stellar bars, and I.~Shivaei for providing MOSDEF catalogs. J.~McKinney and ASL acknowledge the invaluable labor of the maintenance and clerical staff at our institutions, whose contributions make our scientific discoveries a reality. JM, ASL, SMM, and SF thank NASA and acknowledge support through the Hubble Fellowship Program, awarded by the Space Telescope Science Institute, which is operated by the Association of Universities for Research in Astronomy, Inc., for NASA, under contract NAS5-26555.
CMC thanks the National Science Foundation for support through grants AST-2009577 and AST- 2307006 and to NASA through grant JWST-GO-0172. 

Authors from UT Austin acknowledge that UT is an institution that sits on indigenous land. The Tonkawa lived in central Texas, and the Comanche and Apache moved through this area. We pay our respects to all the American Indian and Indigenous Peoples and communities who have been or have become a part of these lands and territories in Texas. We are grateful to be able to live, work, collaborate, and learn on this piece of Turtle Island. 

The National Radio Astronomy Observatory is a facility of the National Science Foundation operated under cooperative agreement by Associated Universities, Inc. This paper makes use of the following ALMA data: ADS/JAO.ALMA\#:2012.1.00978.S 2013.1.00034.S 2013.1.00118.S 2013.1.00151.S 2013.1.00208.S 2013.1.00815.S 2013.1.00884.S 2013.1.01292.S 2015.1.00055.S 2015.1.00137.S 2015.1.00260.S 2015.1.00379.S 2015.1.00568.S 2015.1.01074.S 2015.1.01345.S 2015.1.01495.S 2016.1.00279.S 2016.1.00463.S 2016.1.00478.S 
2016.1.00646.S 2016.1.00735.S 2016.1.00804.S 2016.1.01184.S 2016.1.01208.S 2016.1.01604.S 2017.1.00046.S 2017.1.00326.S 2017.1.00428.L 2017.1.01276.S 2018.1.00635.S  2018.1.00874.S 2018.1.01044.S 2018.1.01281.S 2018.1.01871.S 2019.1.01142.S 2019.1.01275.S 2021.1.01133.S 2021.1.01328.S 2022.1.01493.S. ALMA is a partnership of ESO (representing its member states), NSF (USA) and NINS (Japan), together with NRC (Canada), MOST and ASIAA (Taiwan), and KASI (Republic of Korea), in cooperation with the Republic of Chile. The Joint ALMA Observatory is operated by ESO, AUI/NRAO and NAOJ. 

This work is based [in part] on observations made with the NASA/ESA/CSA James Webb Space Telescope. The data were obtained from the Mikulski Archive for Space Telescopes at the Space Telescope Science Institute, which is operated by the Association of Universities for Research in Astronomy, Inc., under NASA contract NAS 5-03127 for JWST. These observations are associated with program \#1727 and \#1837. This work was funded [in part] by support for program \#1727 provided by NASA through a grant from the Space Telescope Science Institute, which is operated by the Association of Universities for Research in Astronomy, Inc., under NASA contract NAS 5-03127. 

}

\appendix\label{sec:rgbs}
This appendix shows {JWST/NIRCam RGBs (B=115+150, G=277, R=444) for 240 SCUBADive galaxies using data from the COSMOS-Web program. ALMA contours are drawn in white at $3\sigma, 4\sigma, 5\sigma$ and $10\sigma$ with the ALMA beam shown in the lower left corner. Each image is $5^{\prime\prime}\times5^{\prime\prime}$. 
 begin figure blocks 
\begin{figure*}
    \centering
\includegraphics[width=0.24\textwidth]{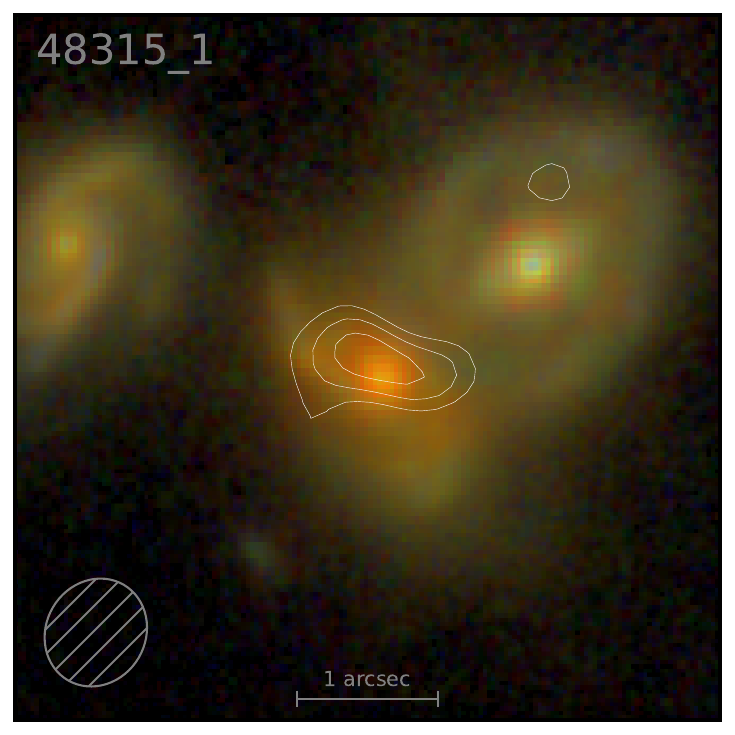}
\includegraphics[width=0.24\textwidth]{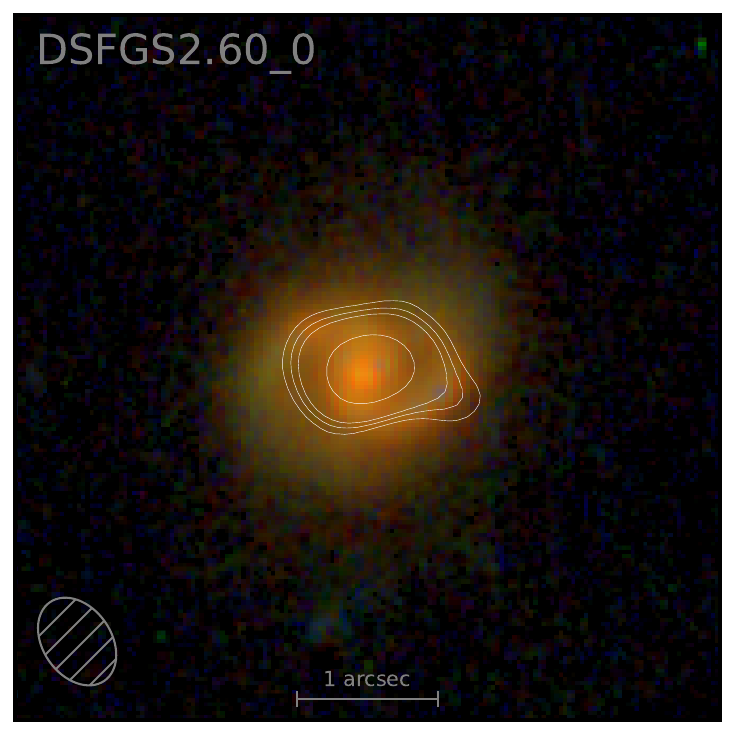}
\includegraphics[width=0.24\textwidth]{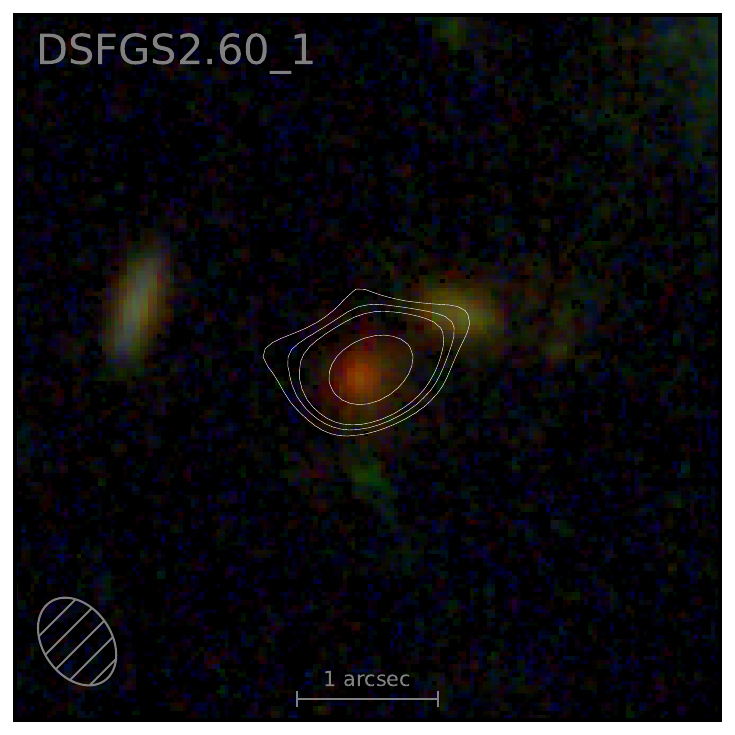}
\includegraphics[width=0.24\textwidth]{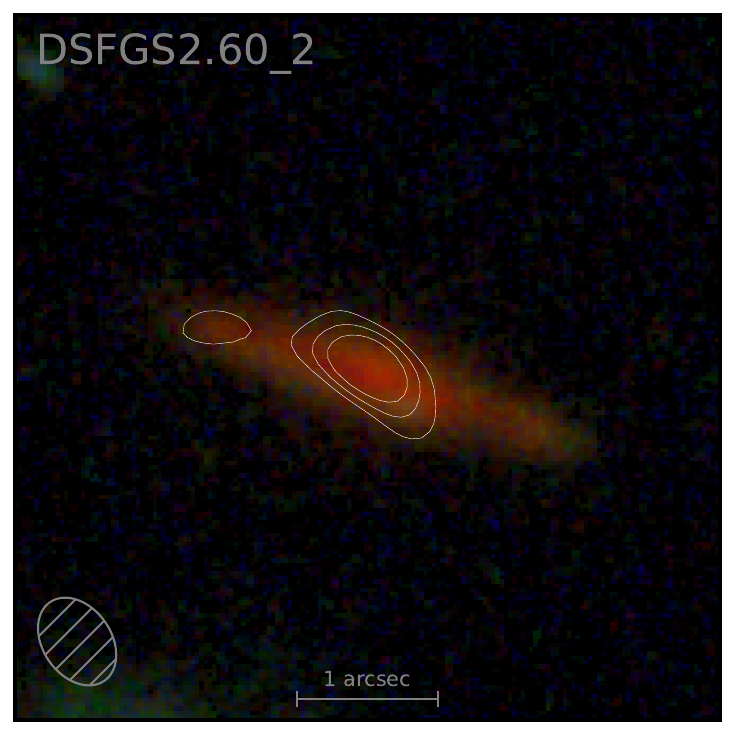}
\includegraphics[width=0.24\textwidth]{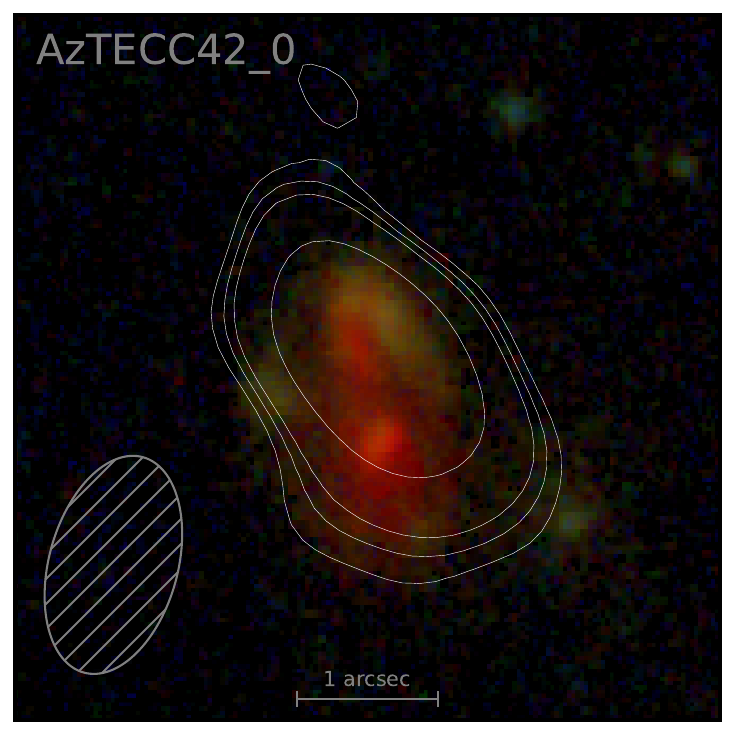}
\includegraphics[width=0.24\textwidth]{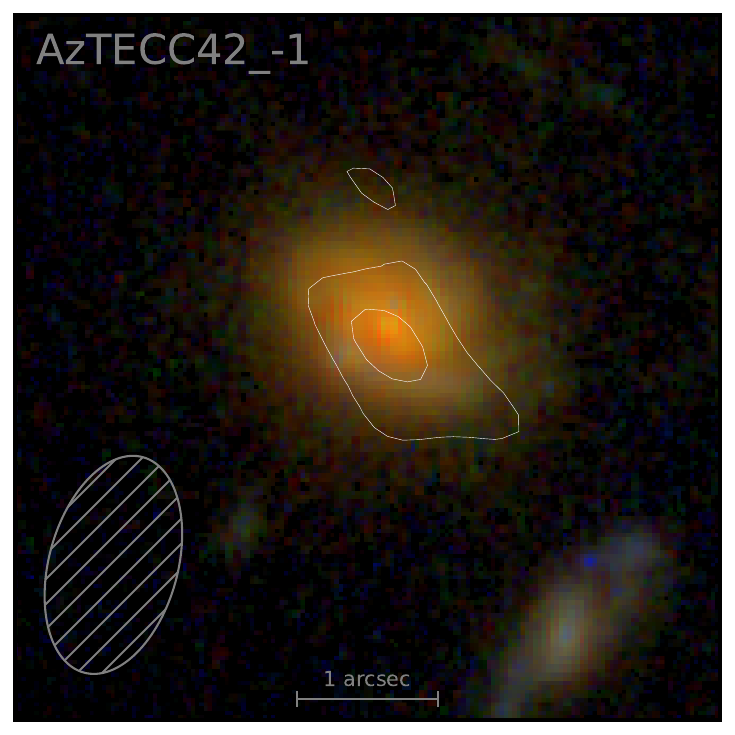}
\includegraphics[width=0.24\textwidth]{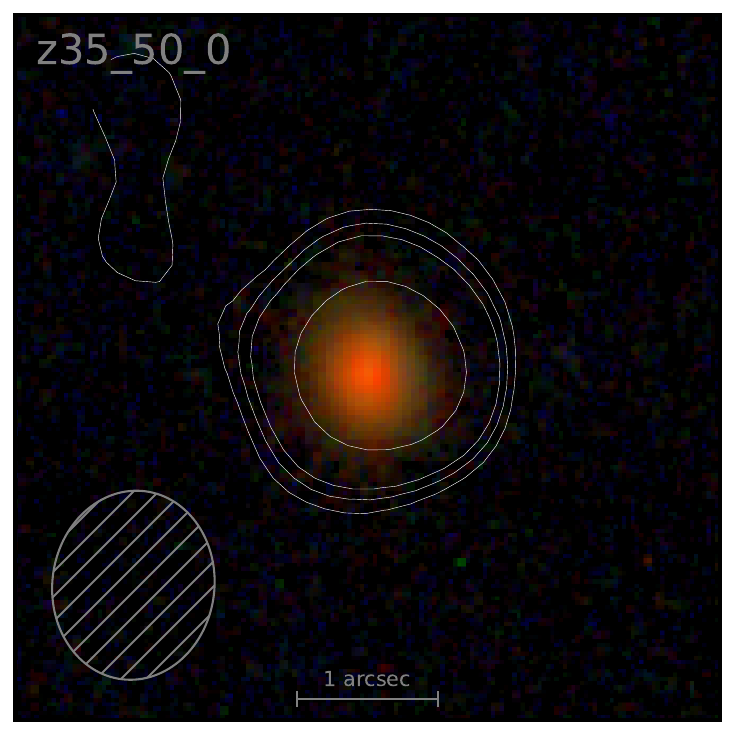}
\includegraphics[width=0.24\textwidth]{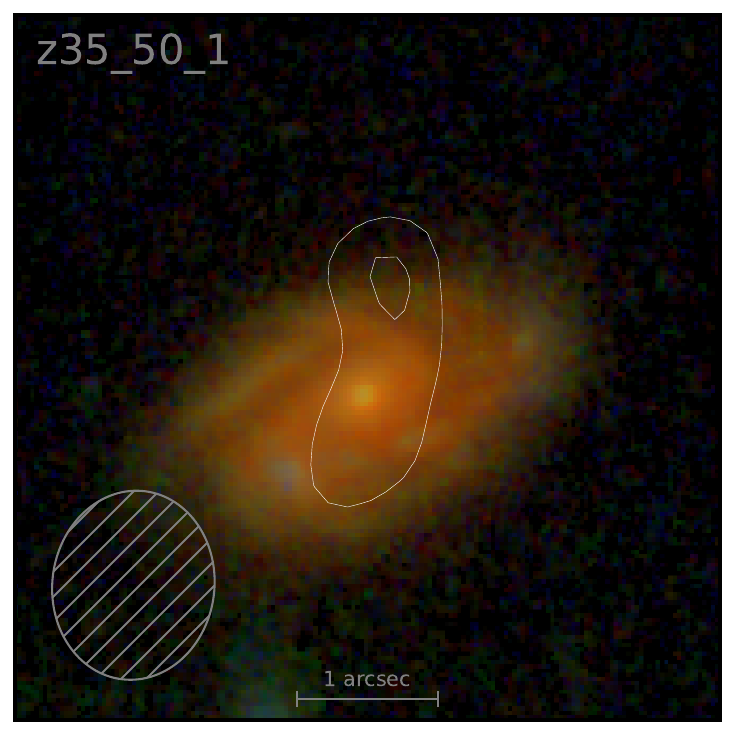}
\includegraphics[width=0.24\textwidth]{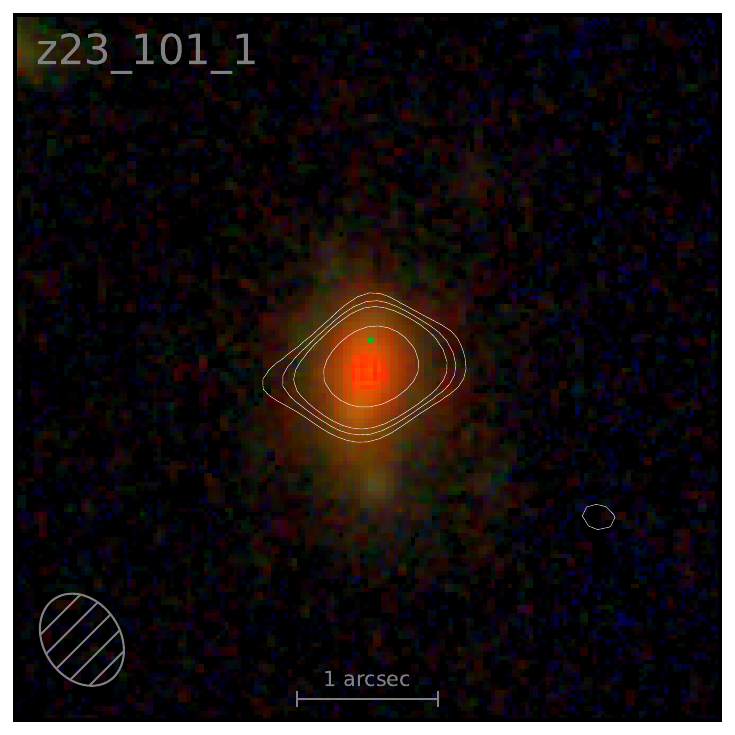}
\includegraphics[width=0.24\textwidth]{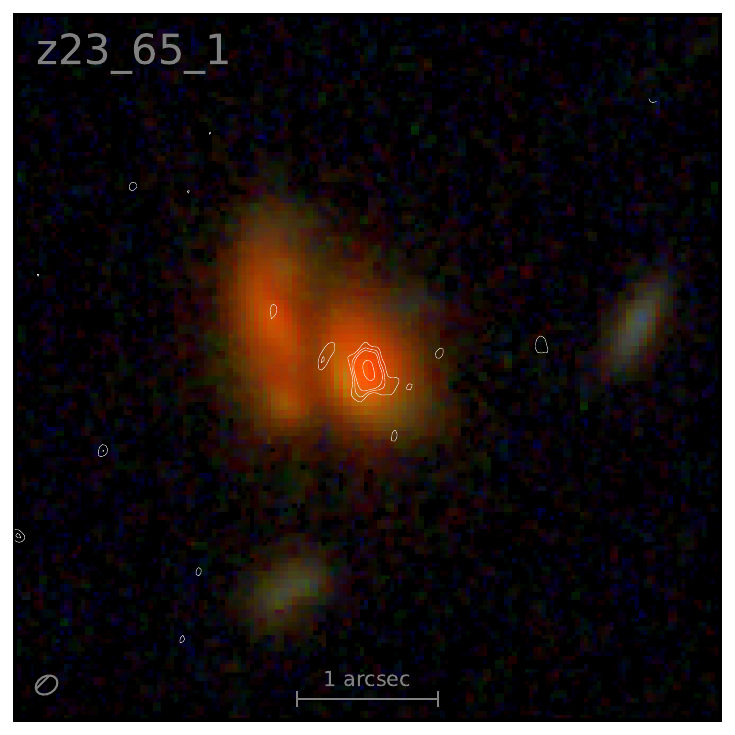}
\includegraphics[width=0.24\textwidth]{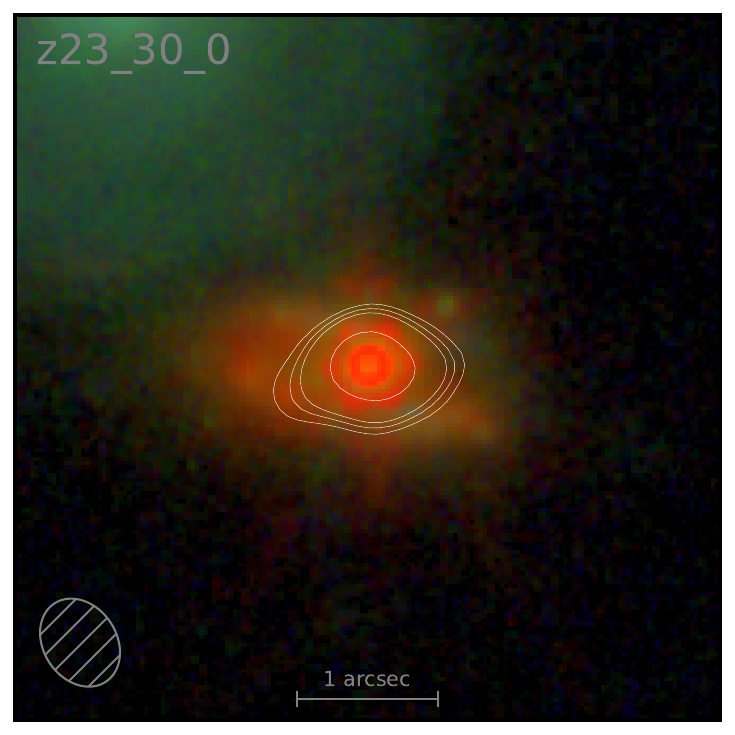}
\includegraphics[width=0.24\textwidth]{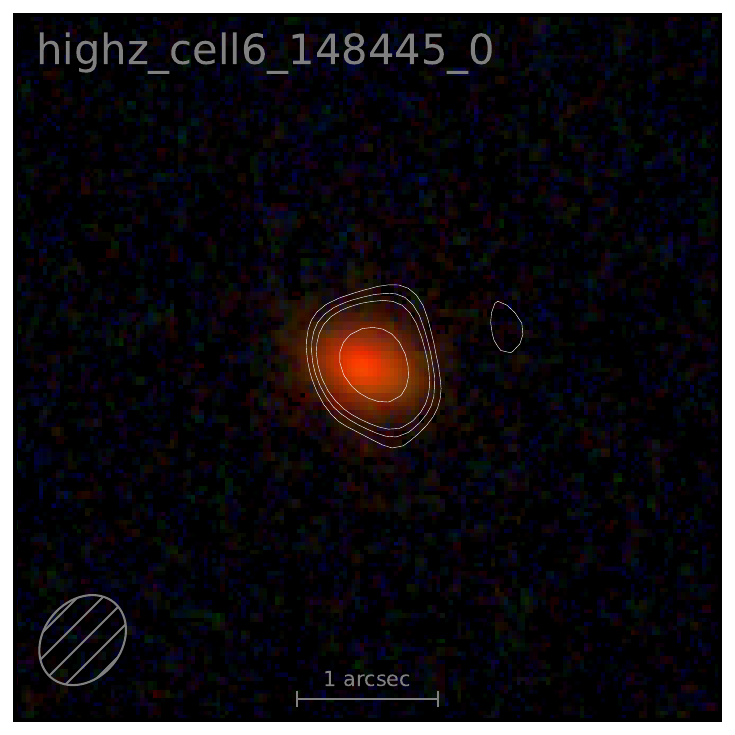}
\includegraphics[width=0.24\textwidth]{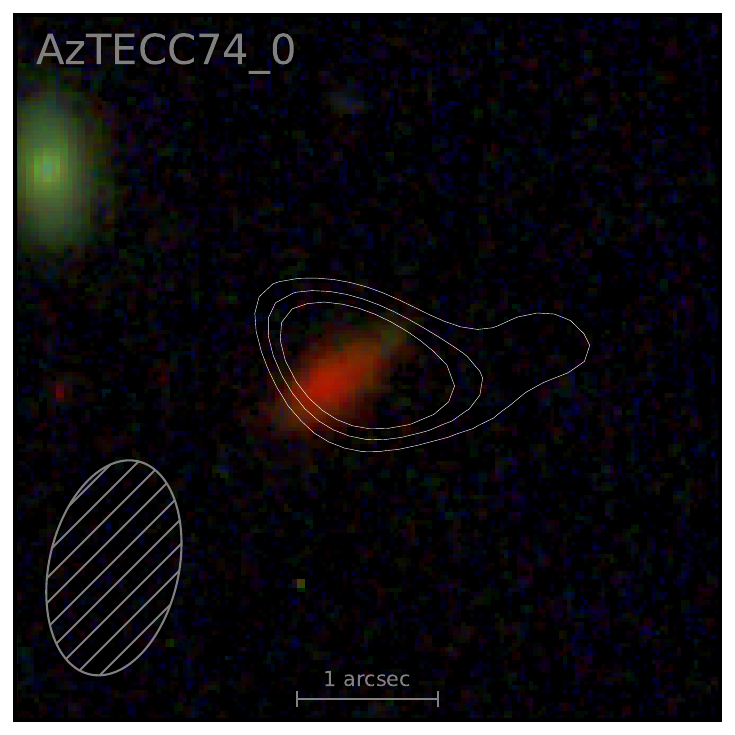}
\includegraphics[width=0.24\textwidth]{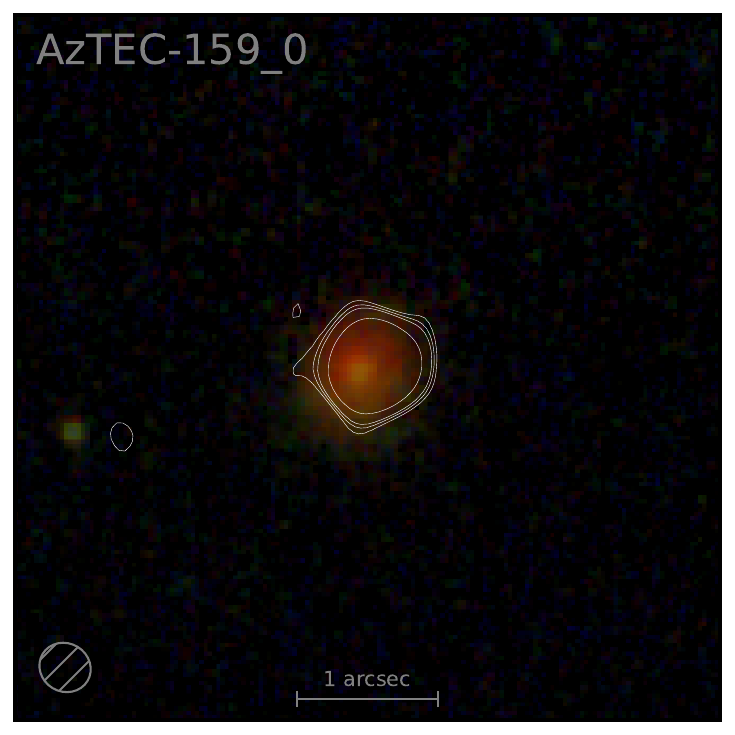}
\includegraphics[width=0.24\textwidth]{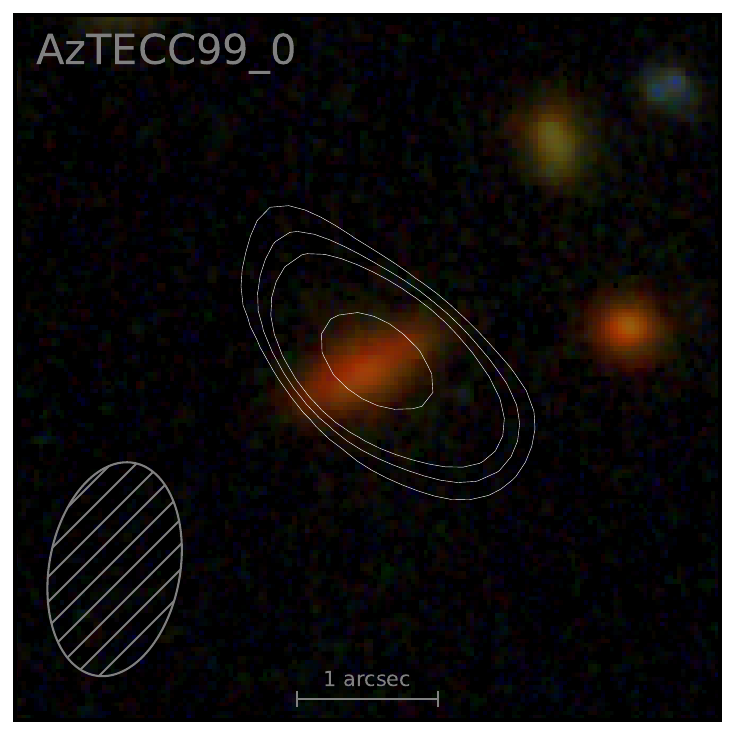}
\includegraphics[width=0.24\textwidth]{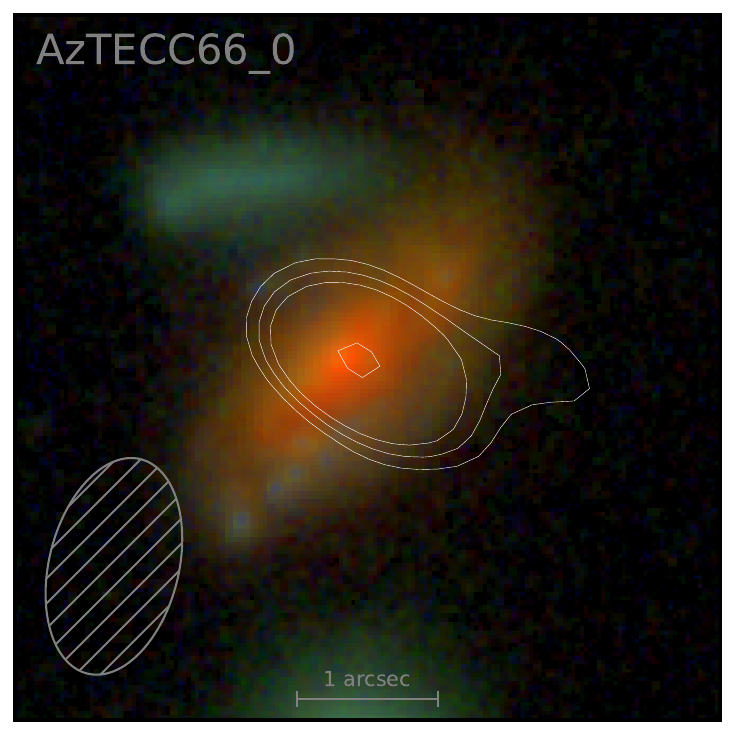}
\includegraphics[width=0.24\textwidth]{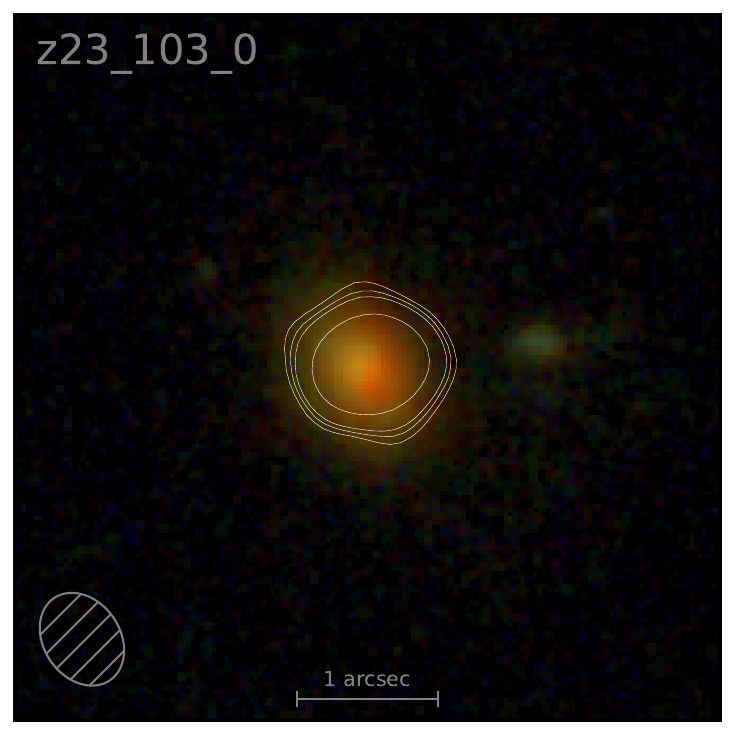}
\includegraphics[width=0.24\textwidth]{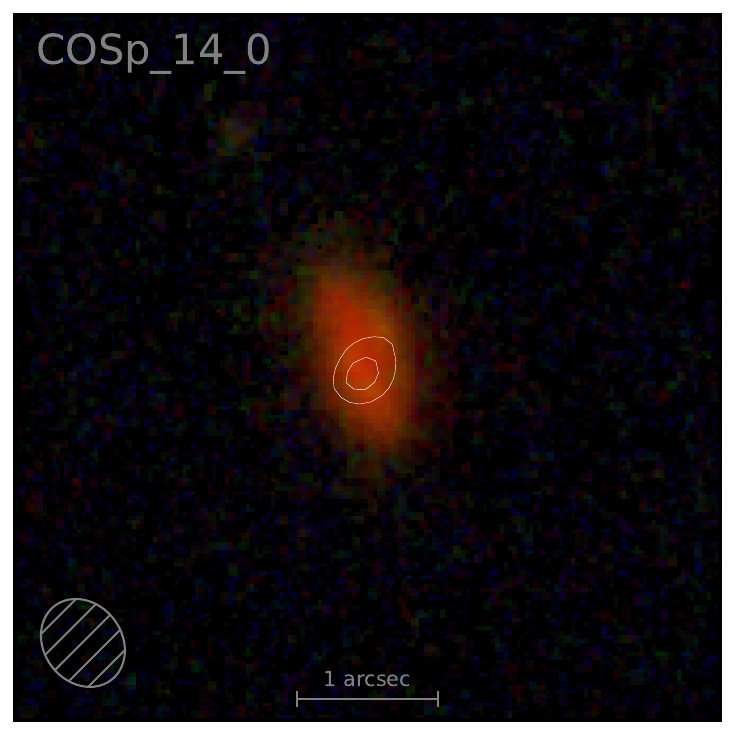}
\includegraphics[width=0.24\textwidth]{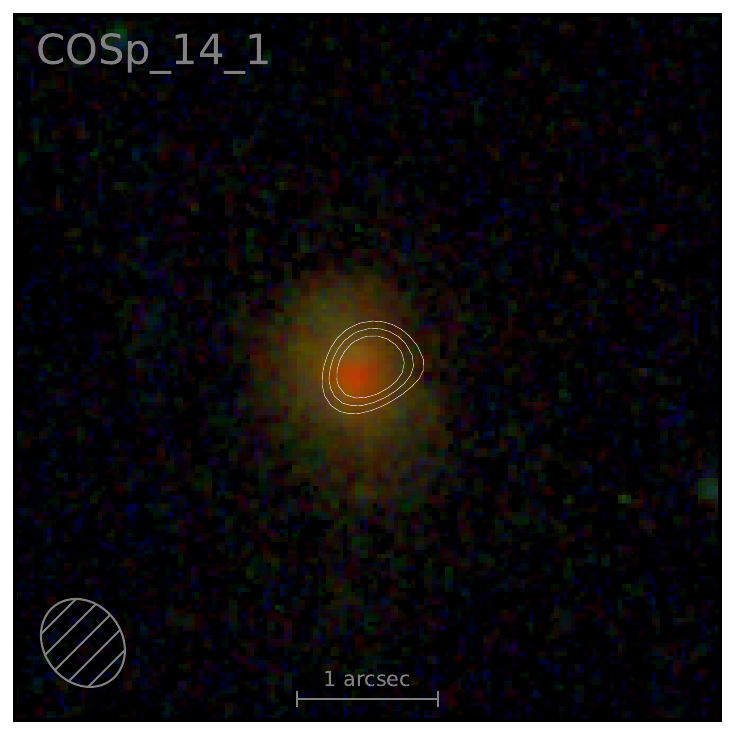}
\includegraphics[width=0.24\textwidth]{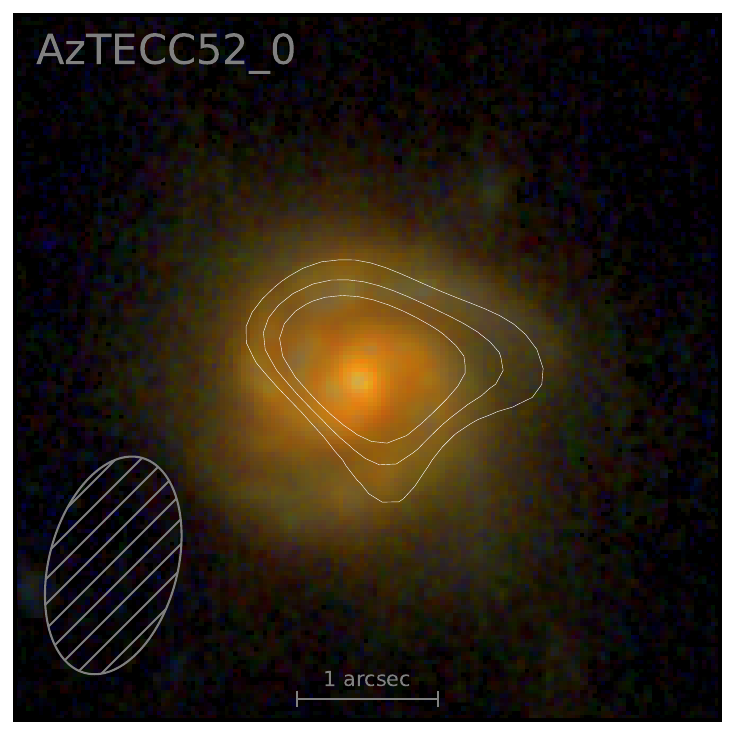}
\end{figure*}
\clearpage

\begin{figure*}
\centering
\includegraphics[width=0.24\textwidth]{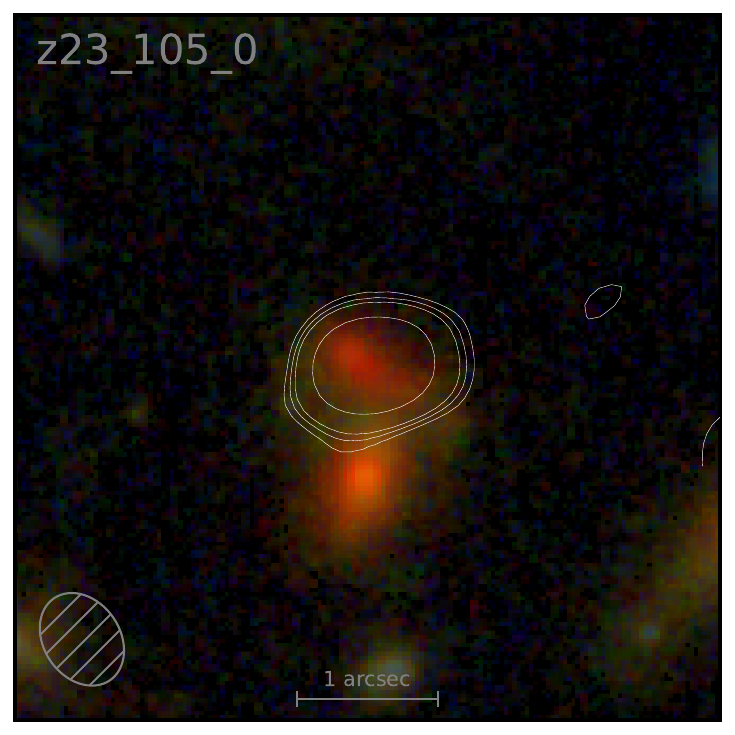}
\includegraphics[width=0.24\textwidth]{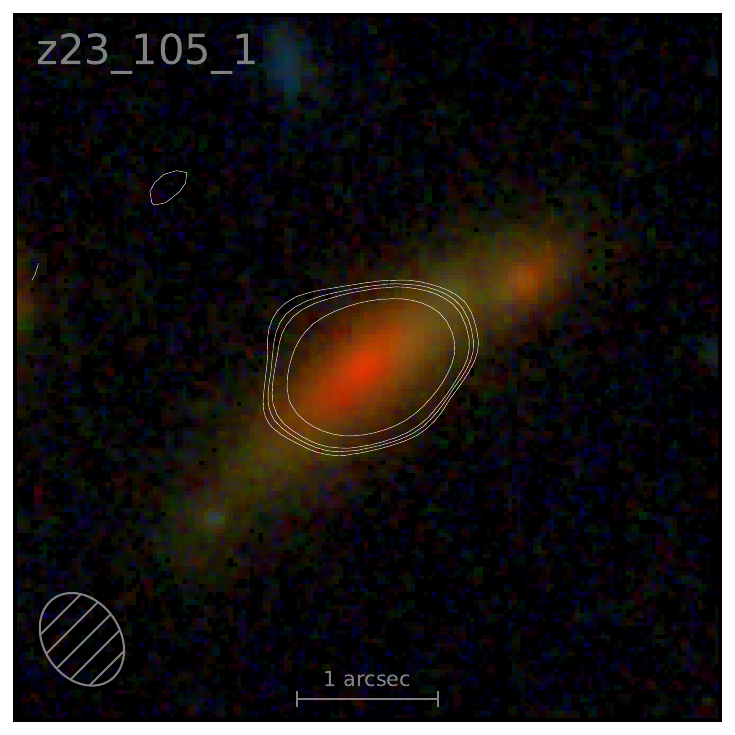}
\includegraphics[width=0.24\textwidth]{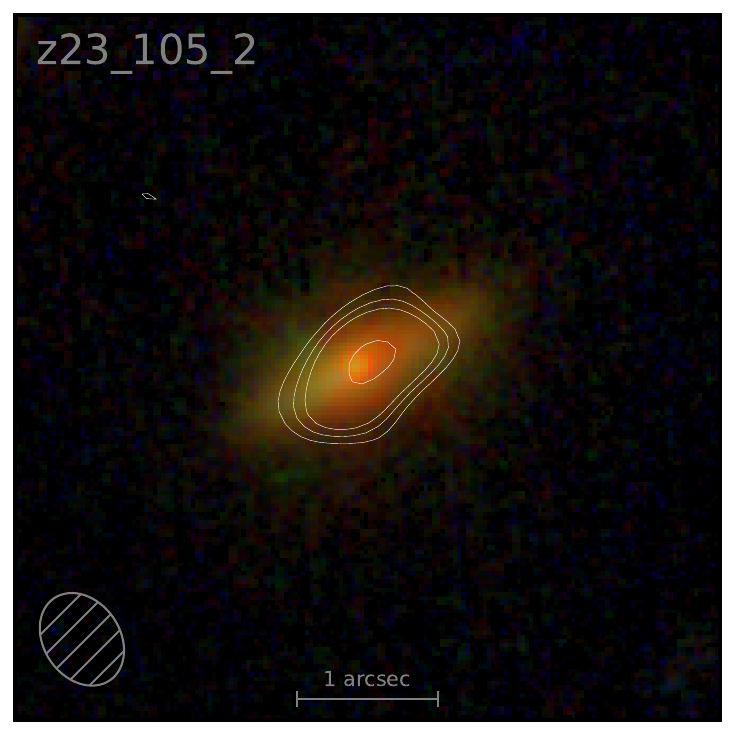}
\includegraphics[width=0.24\textwidth]{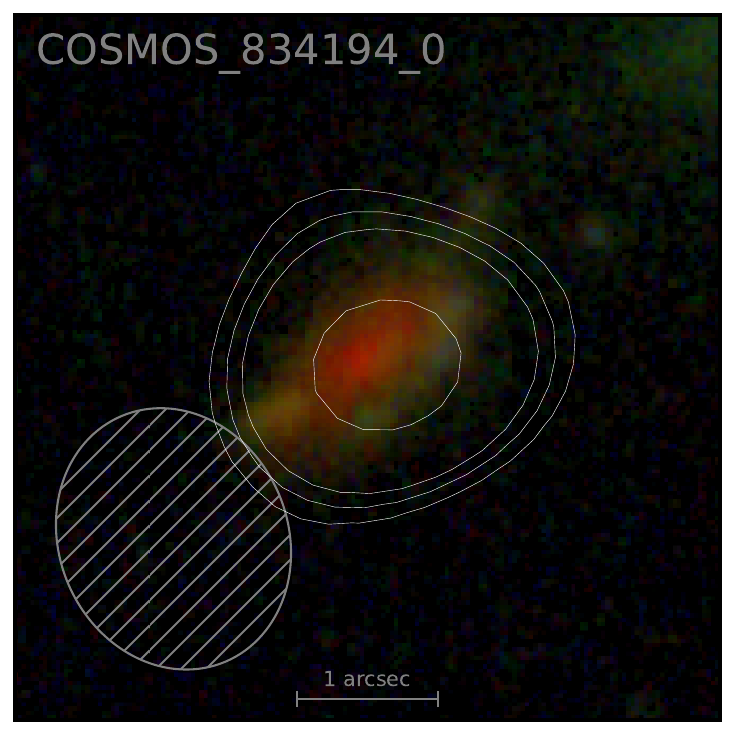}
\includegraphics[width=0.24\textwidth]{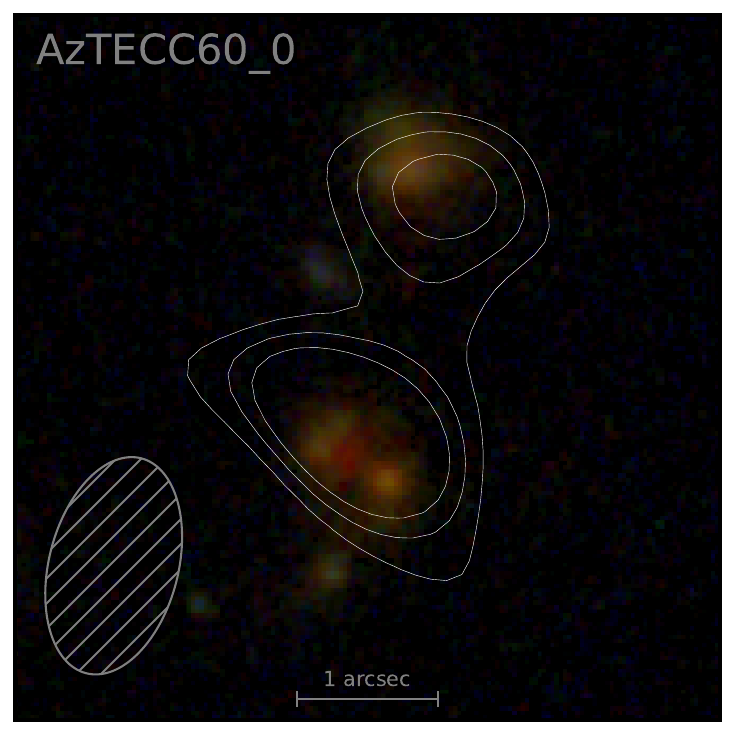}
\includegraphics[width=0.24\textwidth]{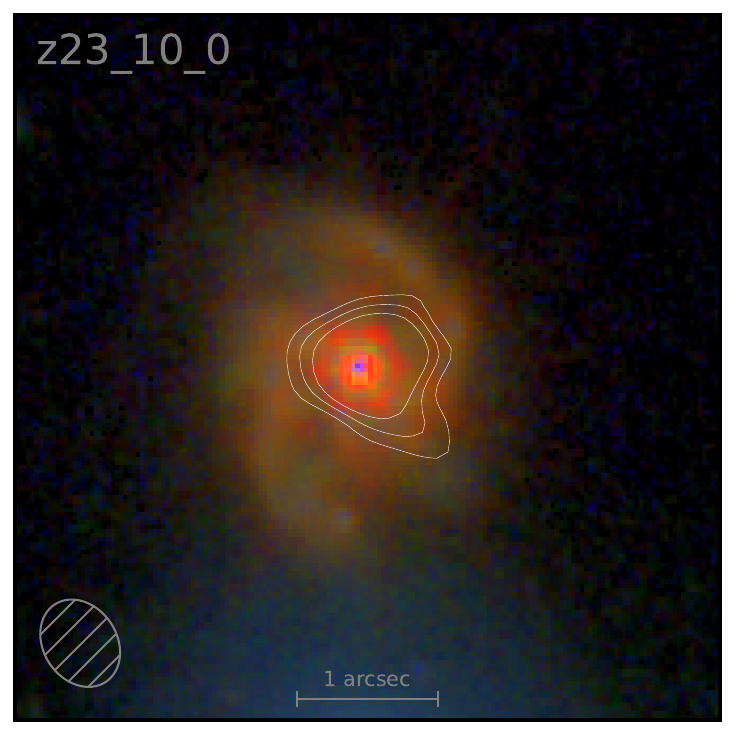}
\includegraphics[width=0.24\textwidth]{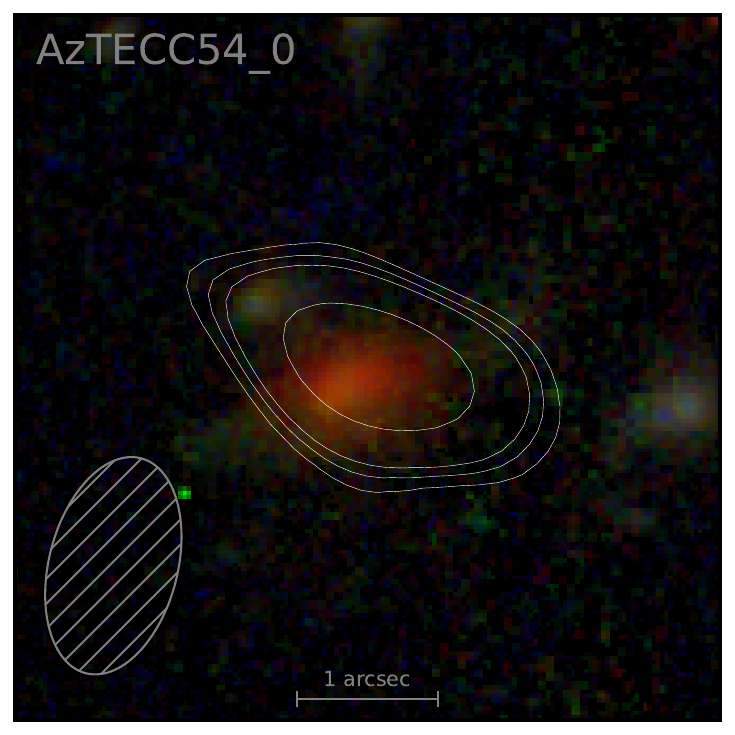}
\includegraphics[width=0.24\textwidth]{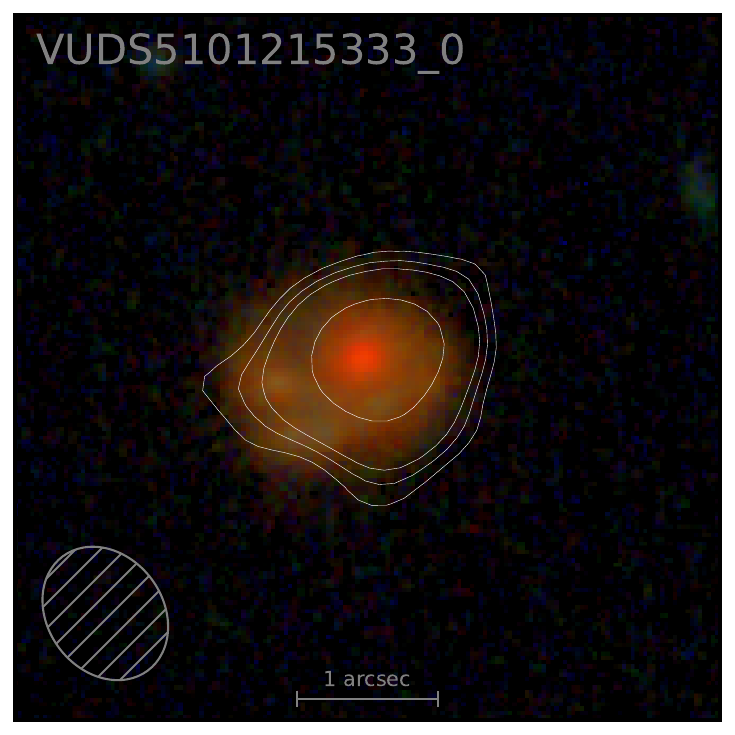}
\includegraphics[width=0.24\textwidth]{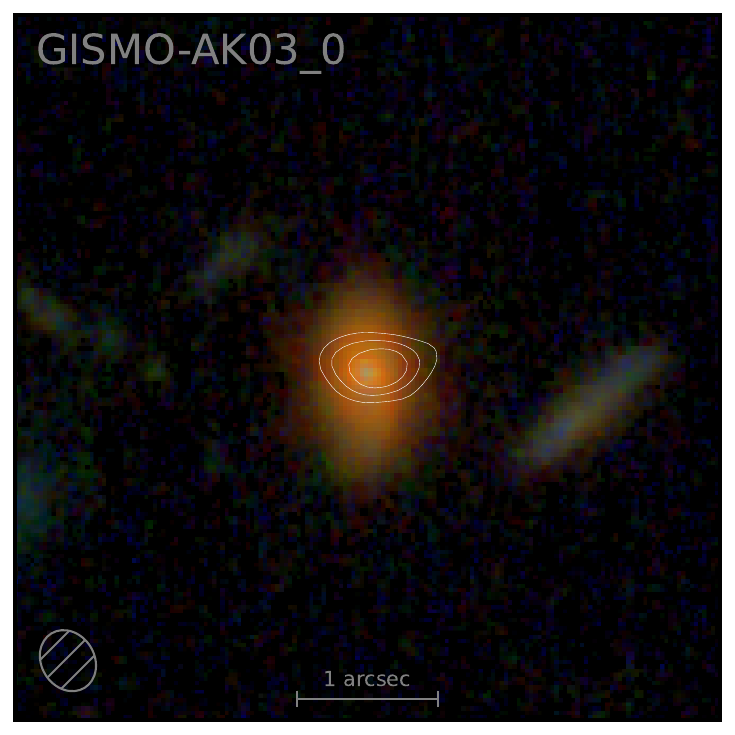}
\includegraphics[width=0.24\textwidth]{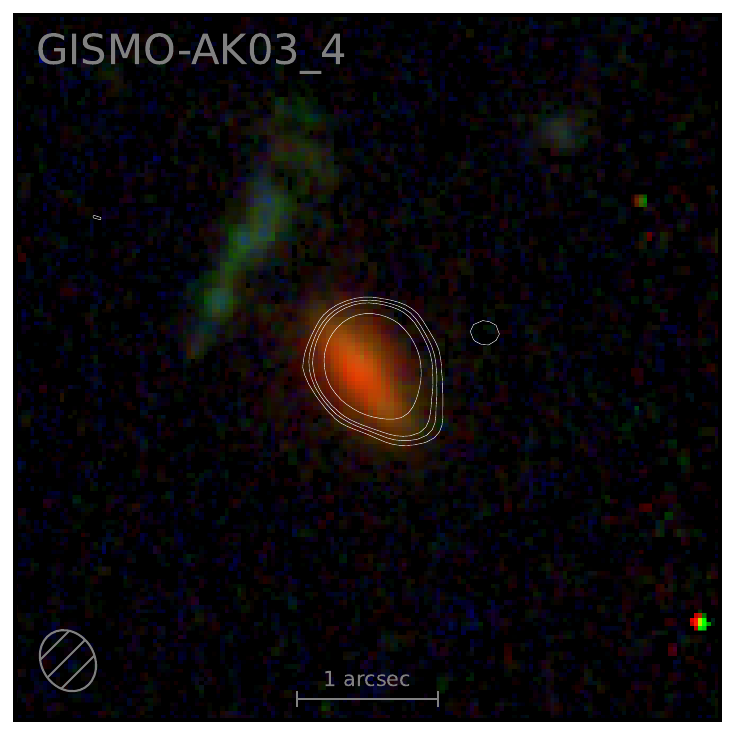}
\includegraphics[width=0.24\textwidth]{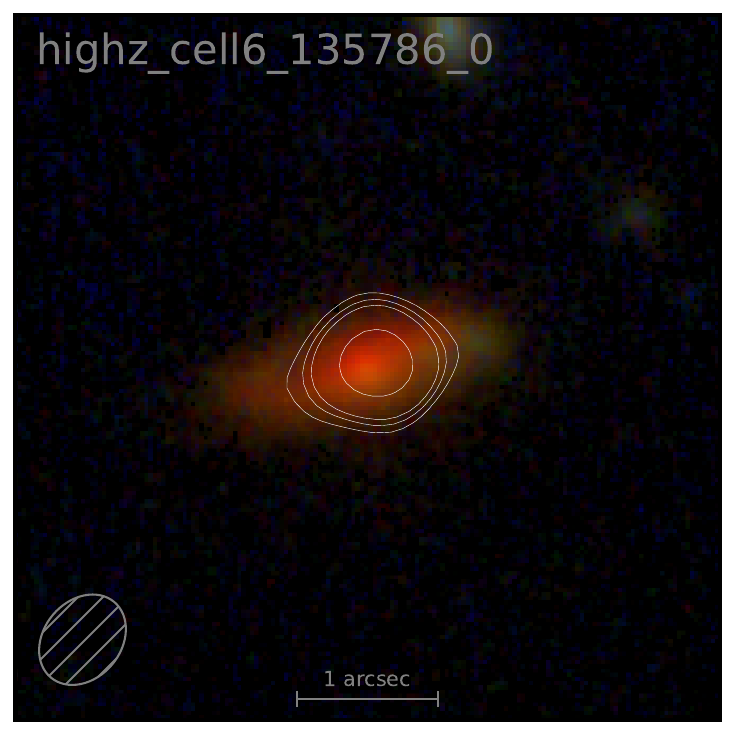}
\includegraphics[width=0.24\textwidth]{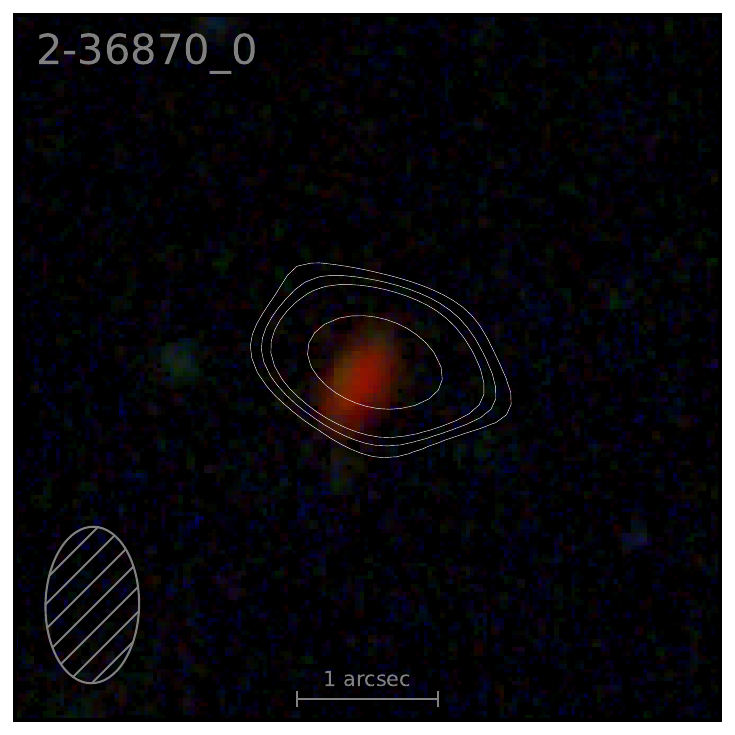}
\includegraphics[width=0.24\textwidth]{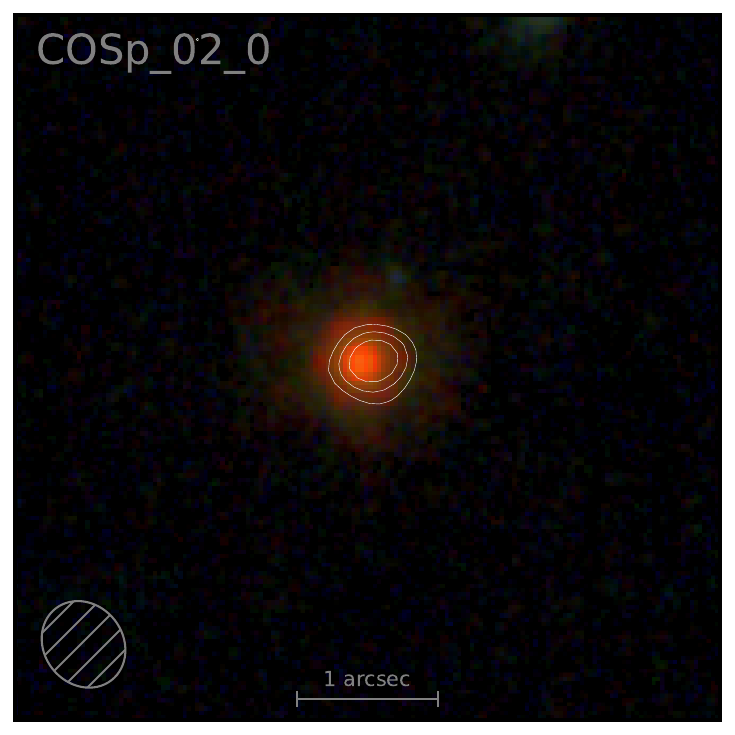}
\includegraphics[width=0.24\textwidth]{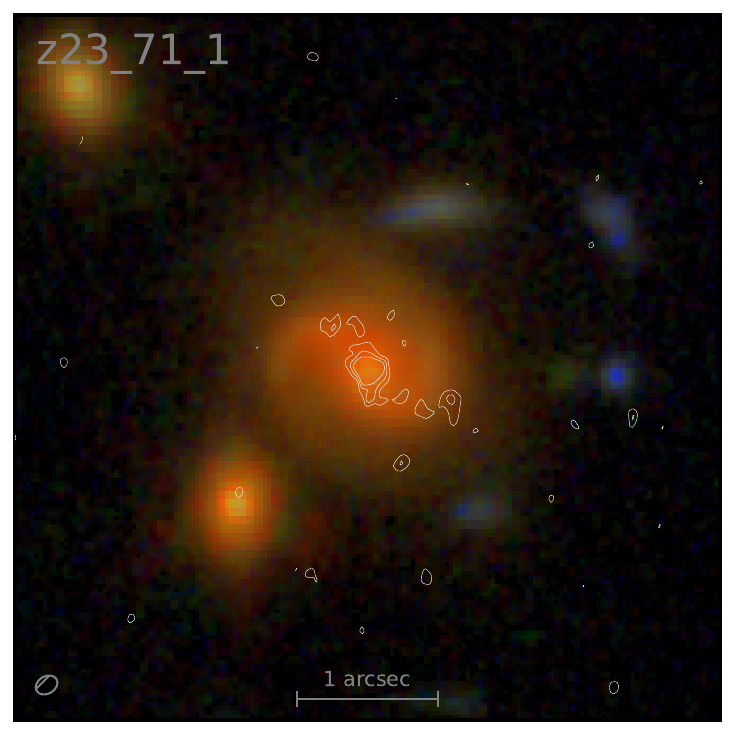}
\includegraphics[width=0.24\textwidth]{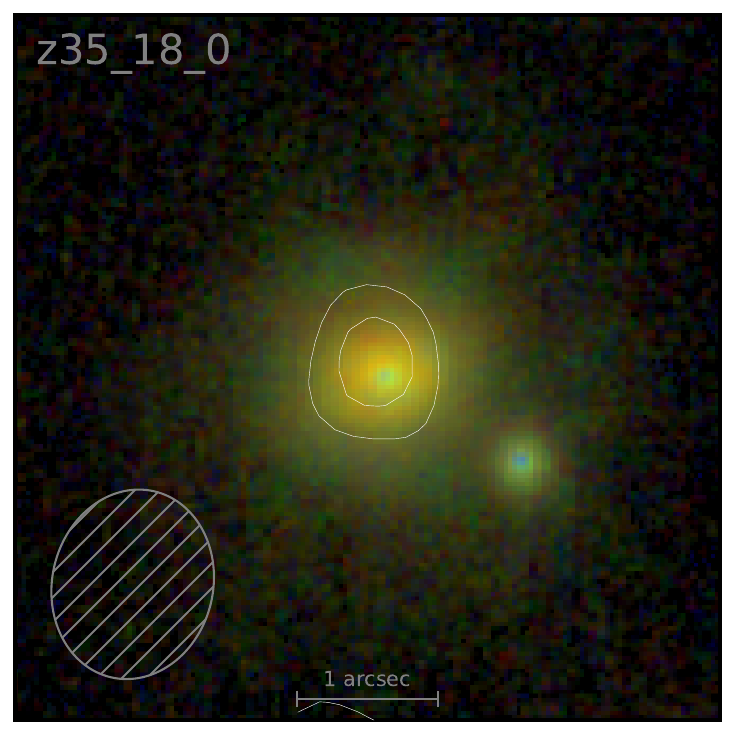}
\includegraphics[width=0.24\textwidth]{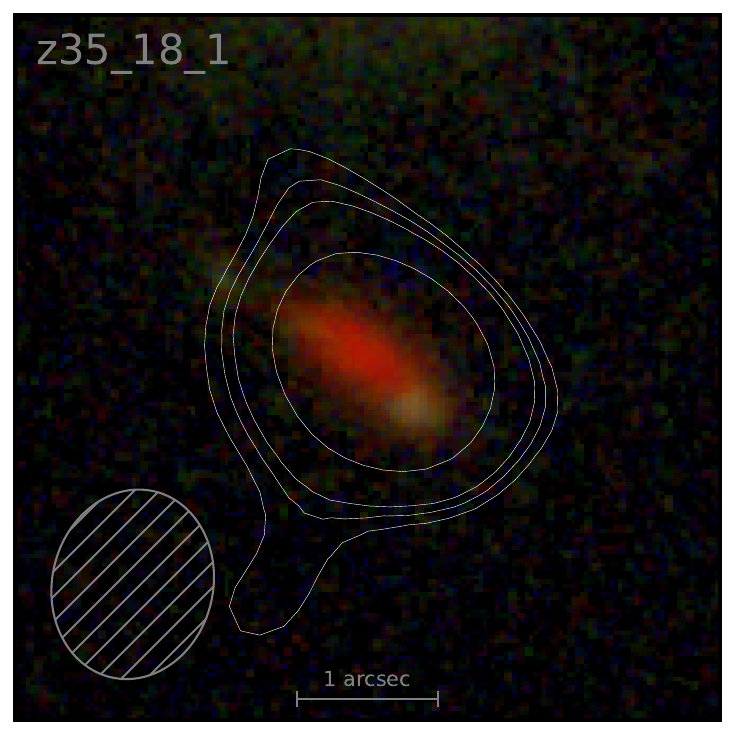}
\includegraphics[width=0.24\textwidth]{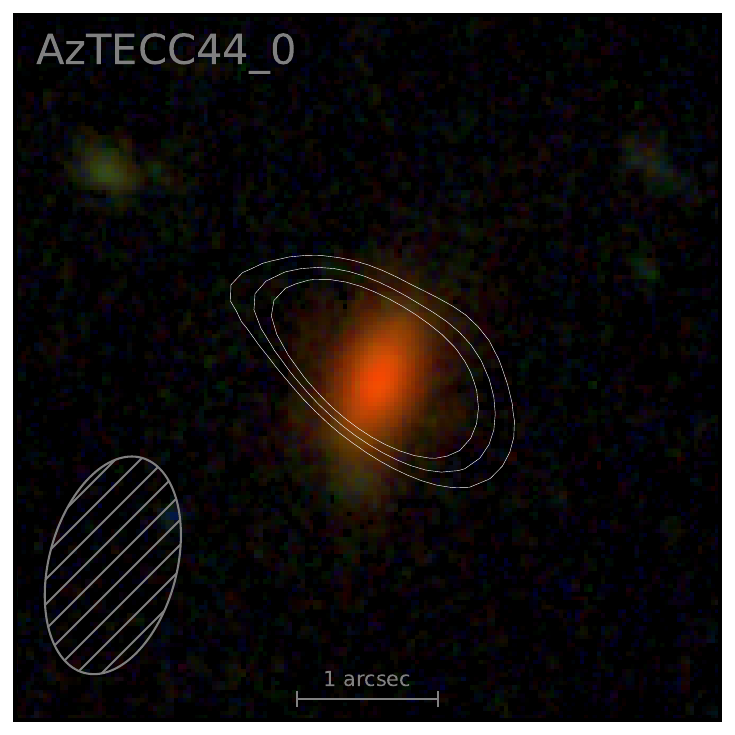}
\includegraphics[width=0.24\textwidth]{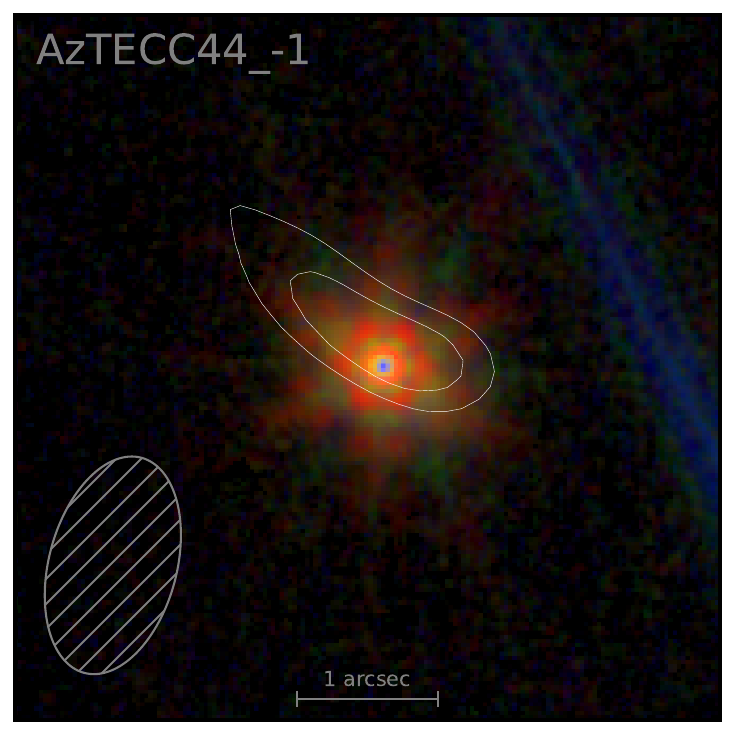}
\includegraphics[width=0.24\textwidth]{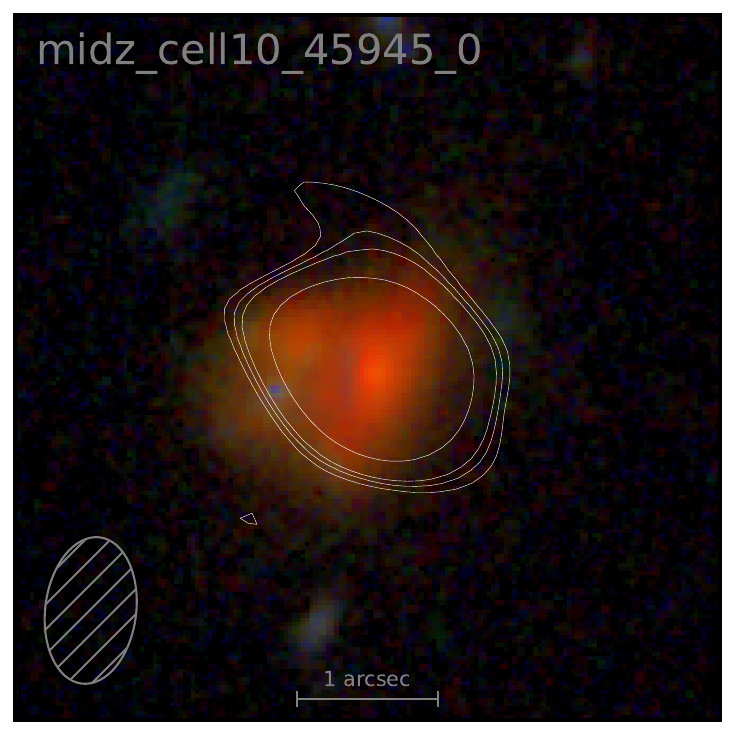}
\includegraphics[width=0.24\textwidth]{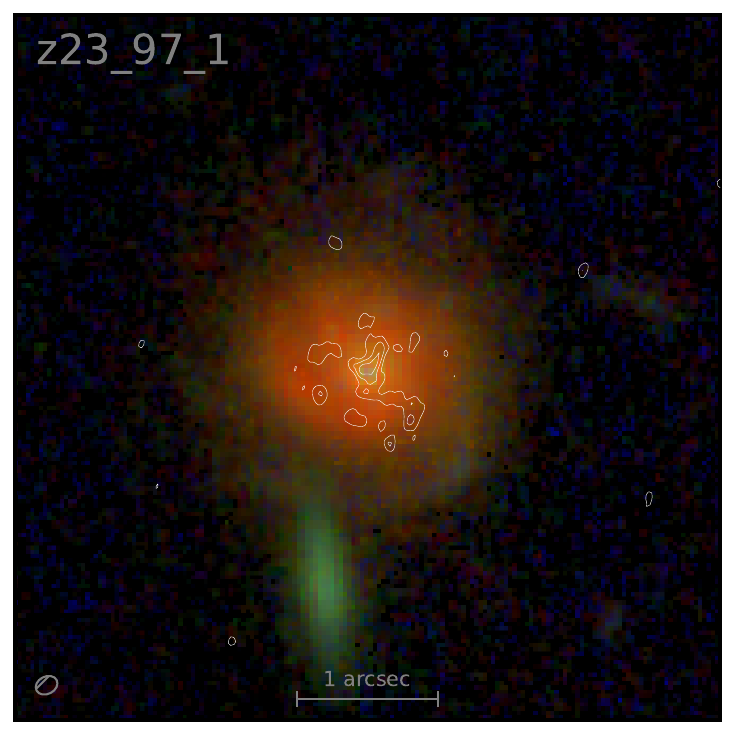}
\end{figure*}
\clearpage
\begin{figure*}
\centering
\includegraphics[width=0.24\textwidth]{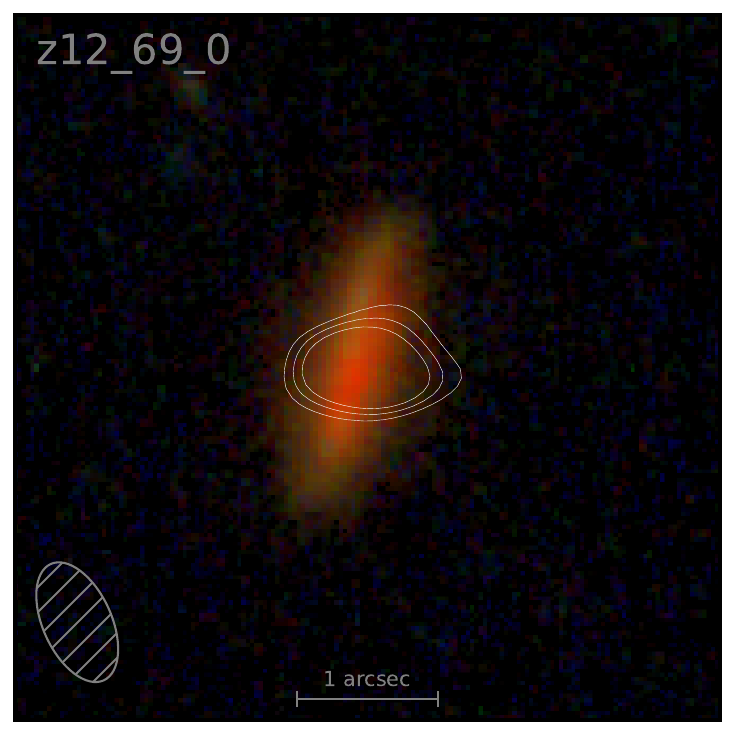}
\includegraphics[width=0.24\textwidth]{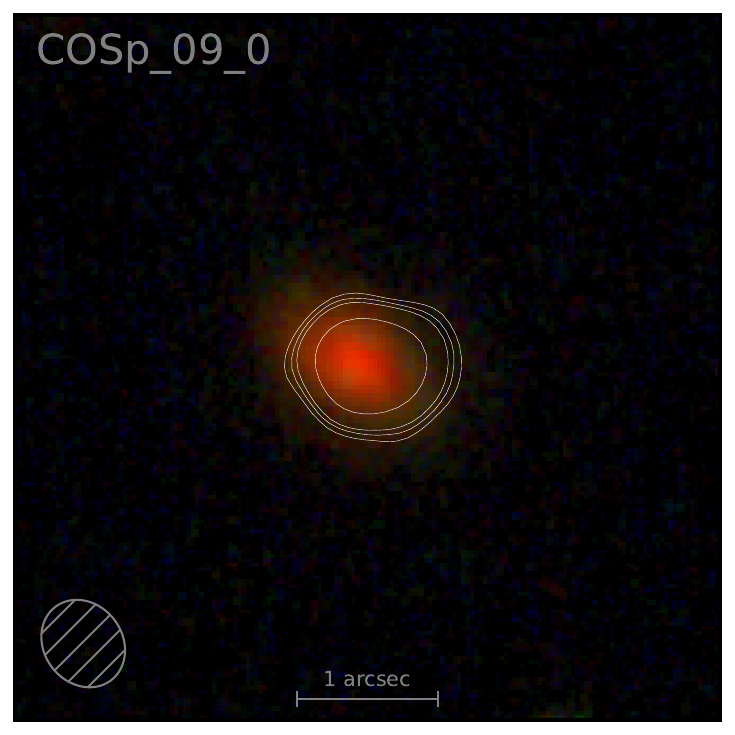}
\includegraphics[width=0.24\textwidth]{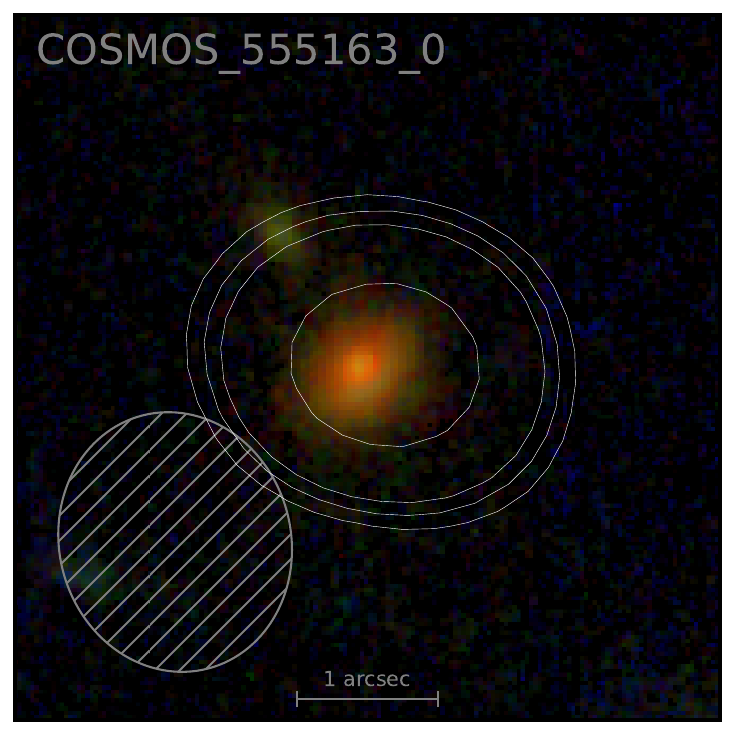}
\includegraphics[width=0.24\textwidth]{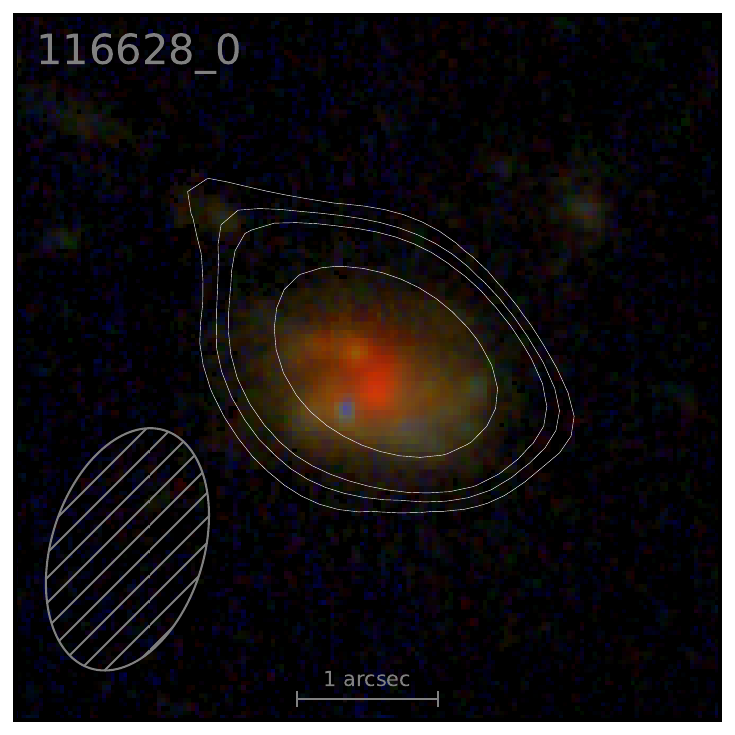}
\includegraphics[width=0.24\textwidth]{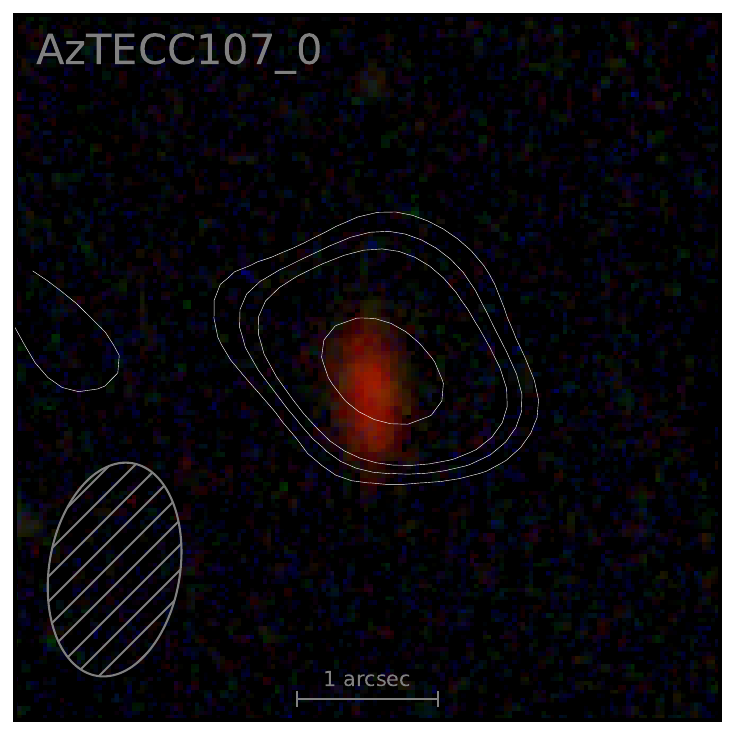}
\includegraphics[width=0.24\textwidth]{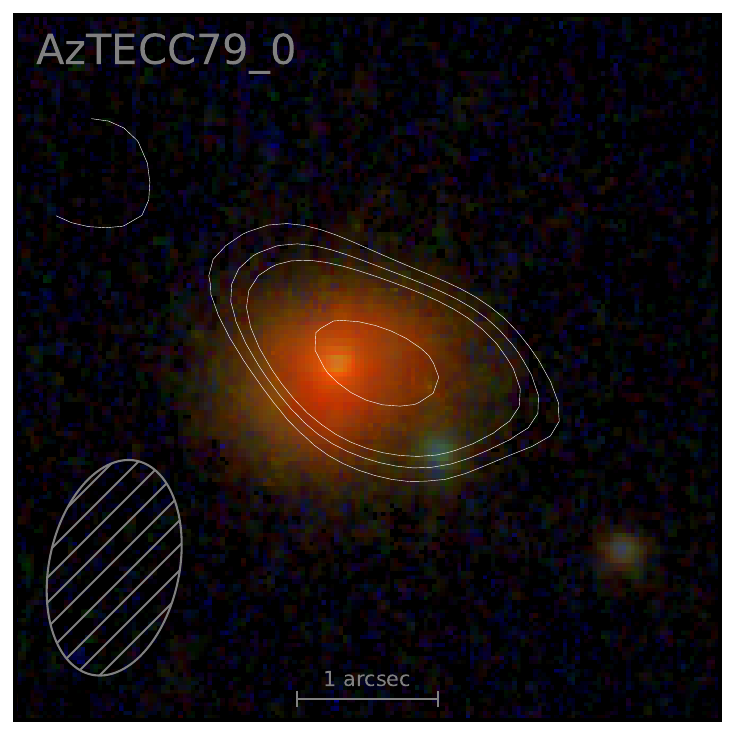}
\includegraphics[width=0.24\textwidth]{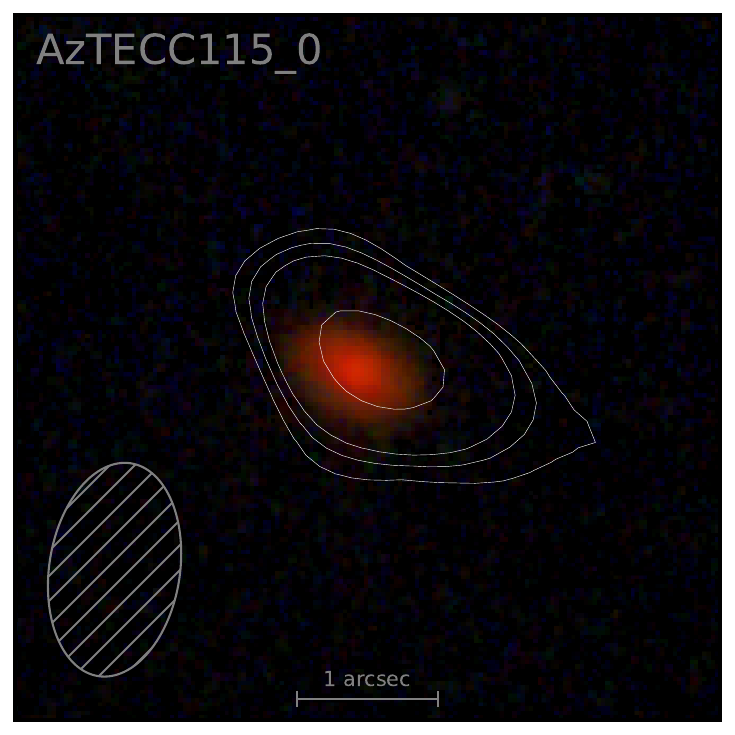}
\includegraphics[width=0.24\textwidth]{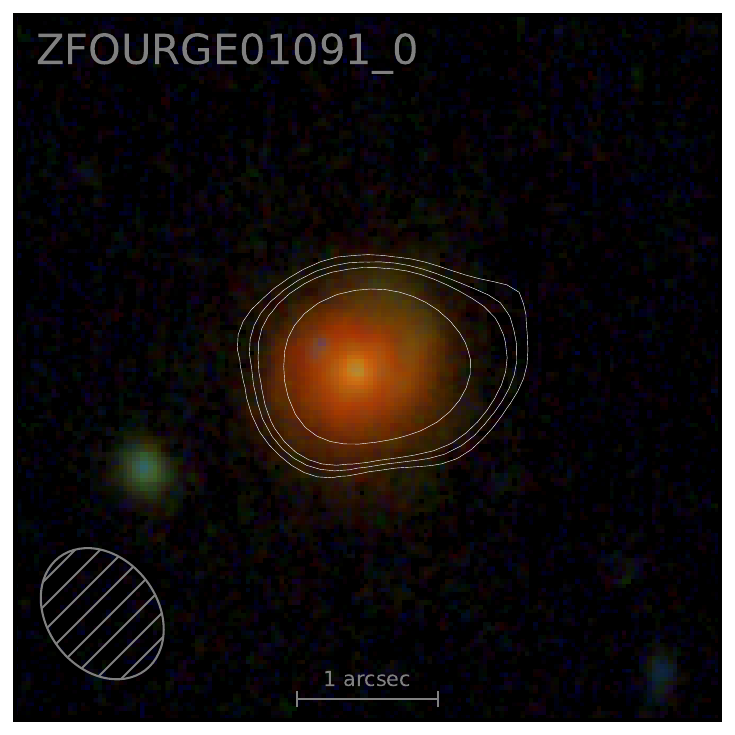}
\includegraphics[width=0.24\textwidth]{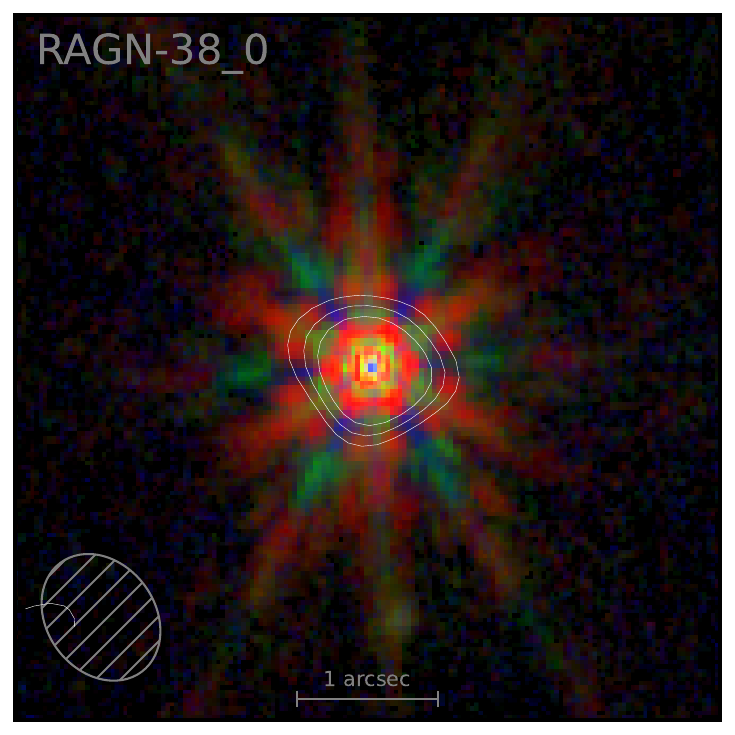}
\includegraphics[width=0.24\textwidth]{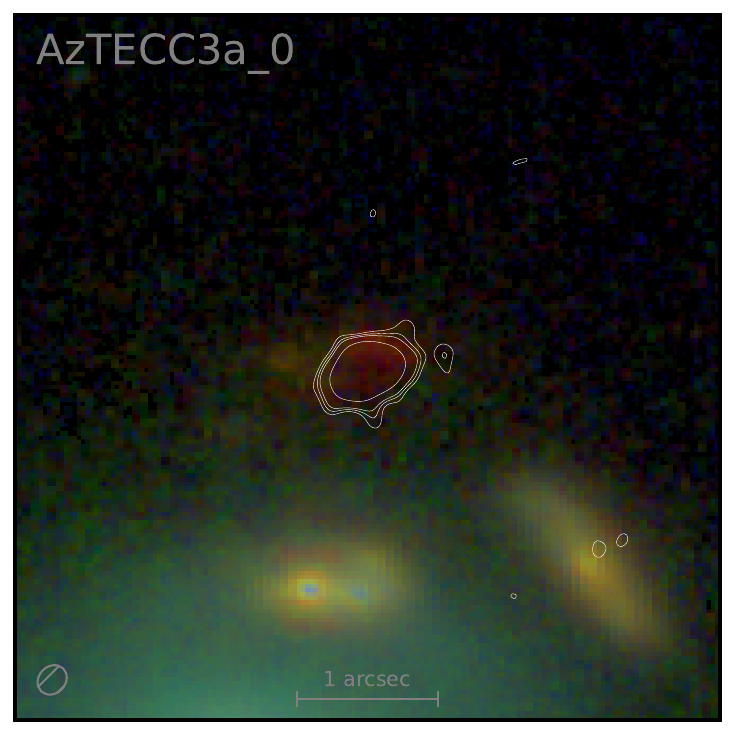}
\includegraphics[width=0.24\textwidth]{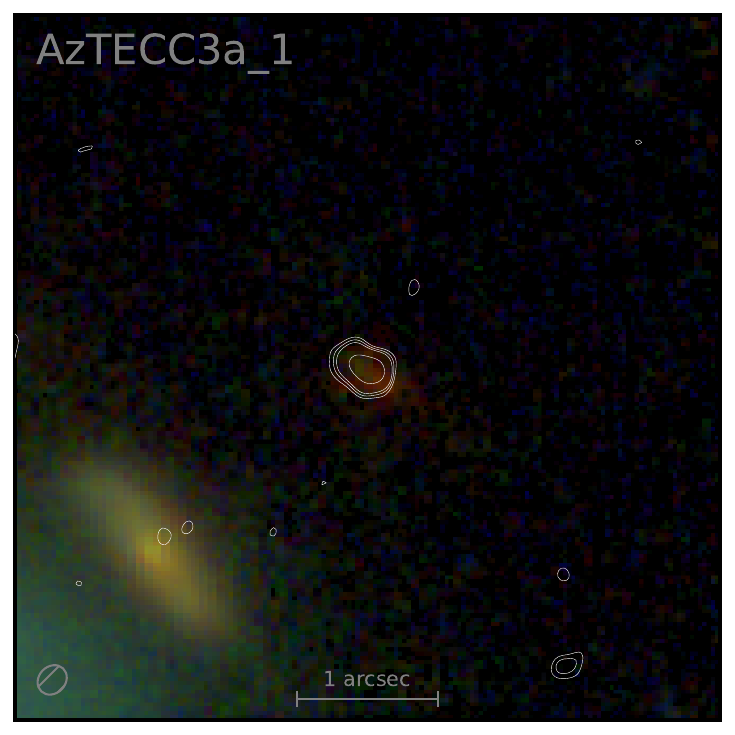}
\includegraphics[width=0.24\textwidth]{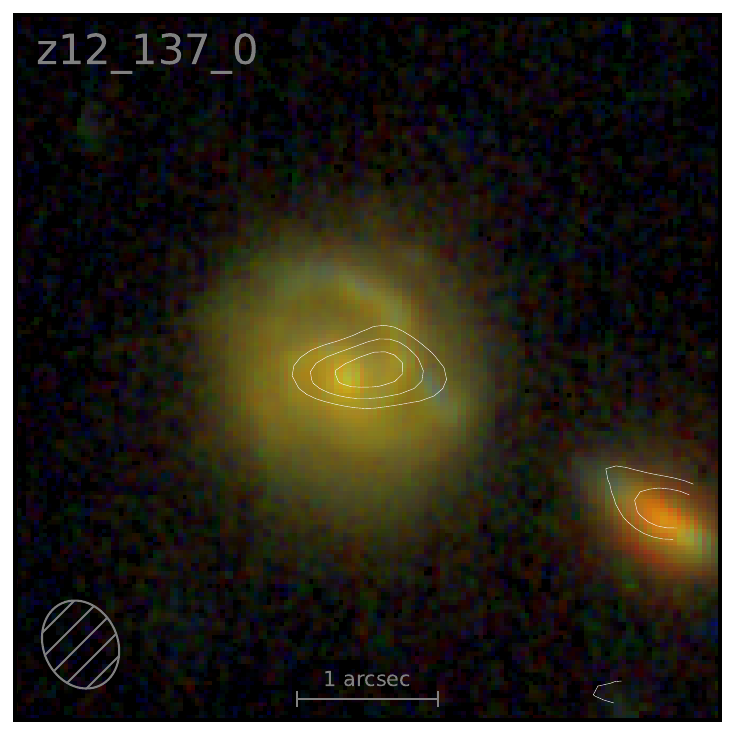}
\includegraphics[width=0.24\textwidth]{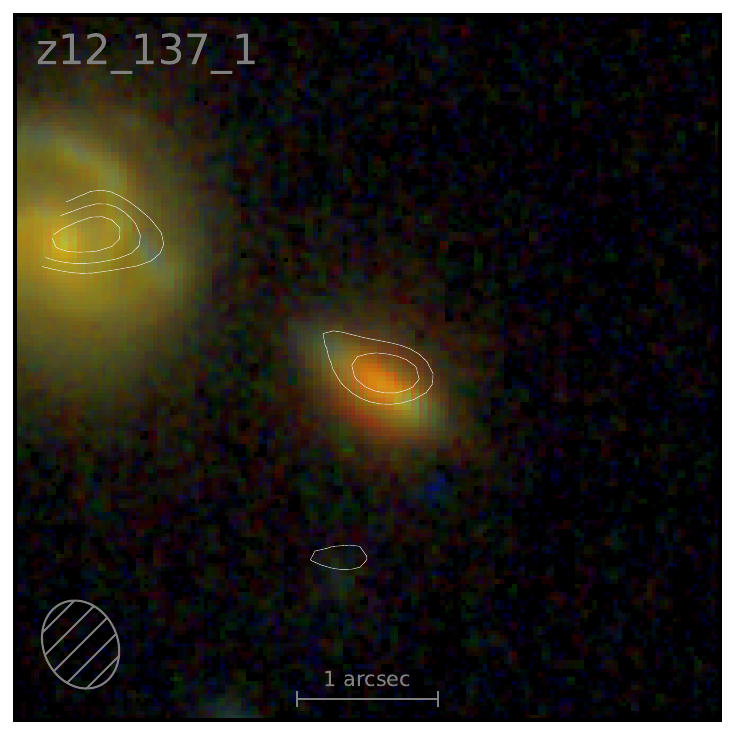}
\includegraphics[width=0.24\textwidth]{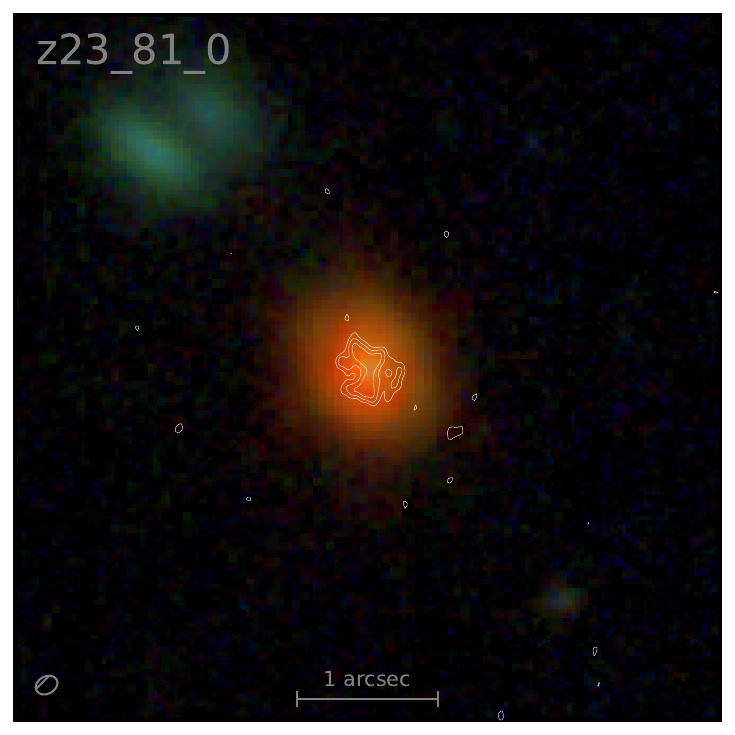}
\includegraphics[width=0.24\textwidth]{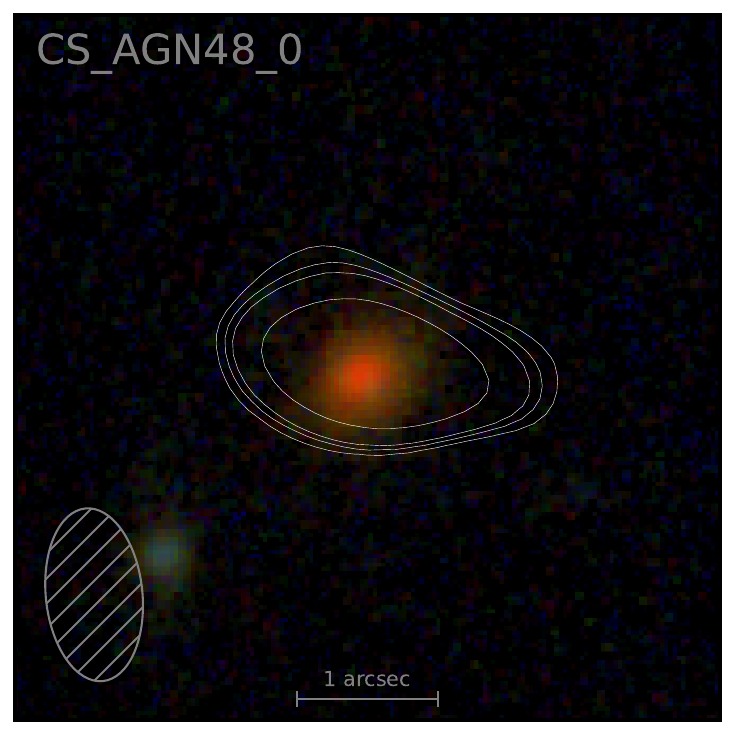}
\includegraphics[width=0.24\textwidth]{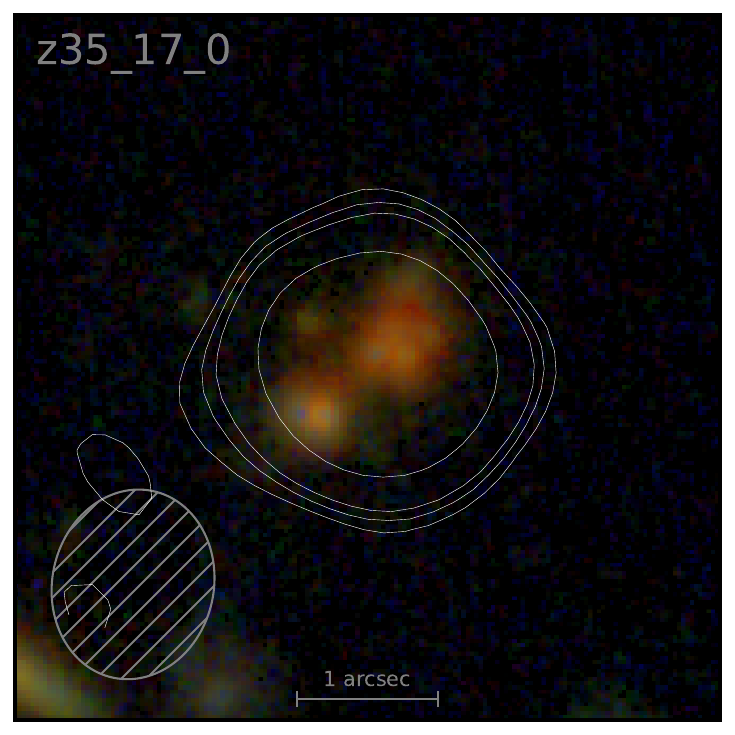}
\includegraphics[width=0.24\textwidth]{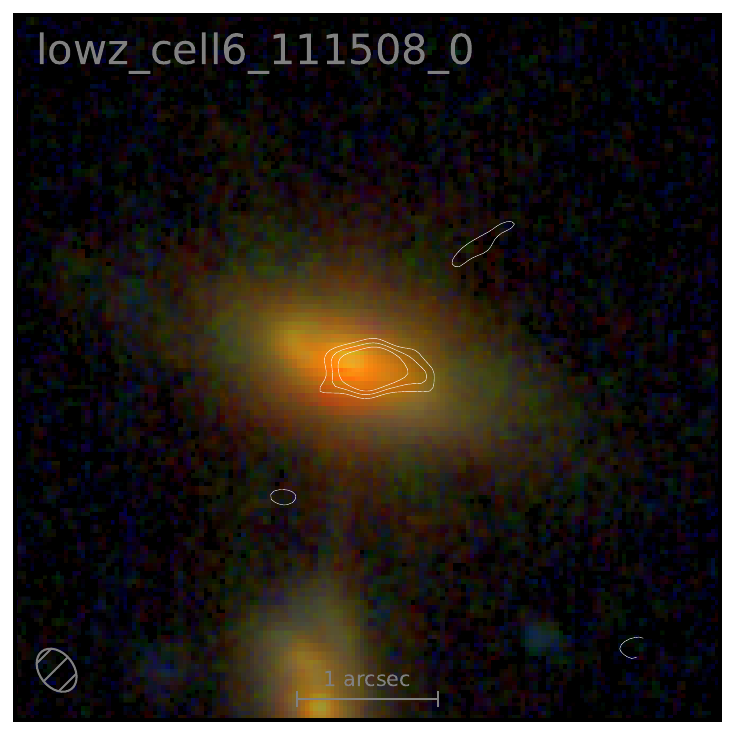}
\includegraphics[width=0.24\textwidth]{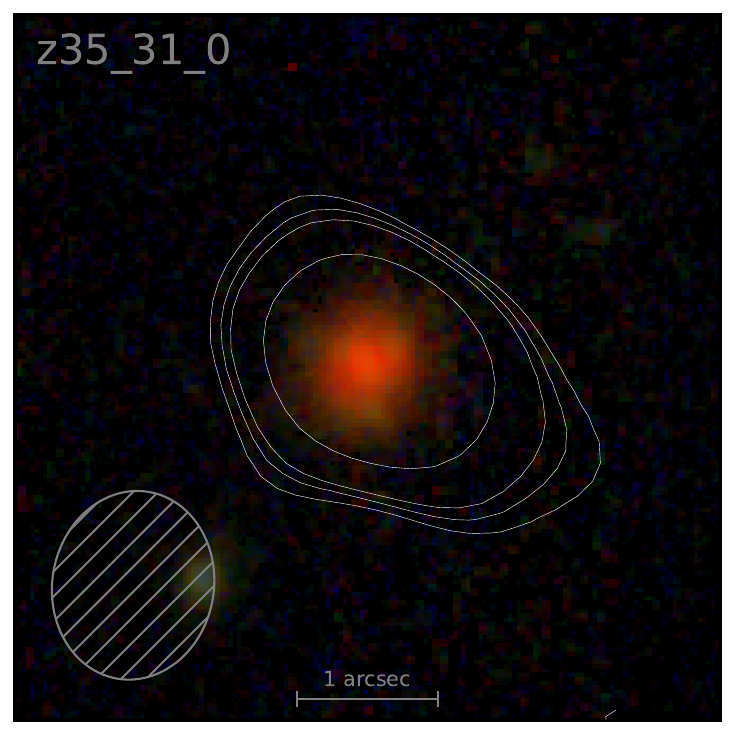}
\includegraphics[width=0.24\textwidth]{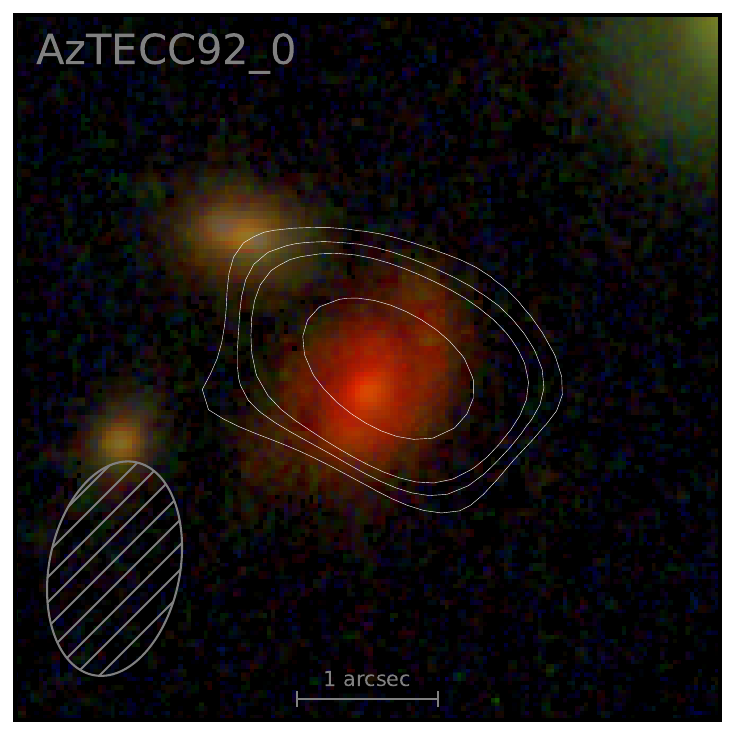}
\includegraphics[width=0.24\textwidth]{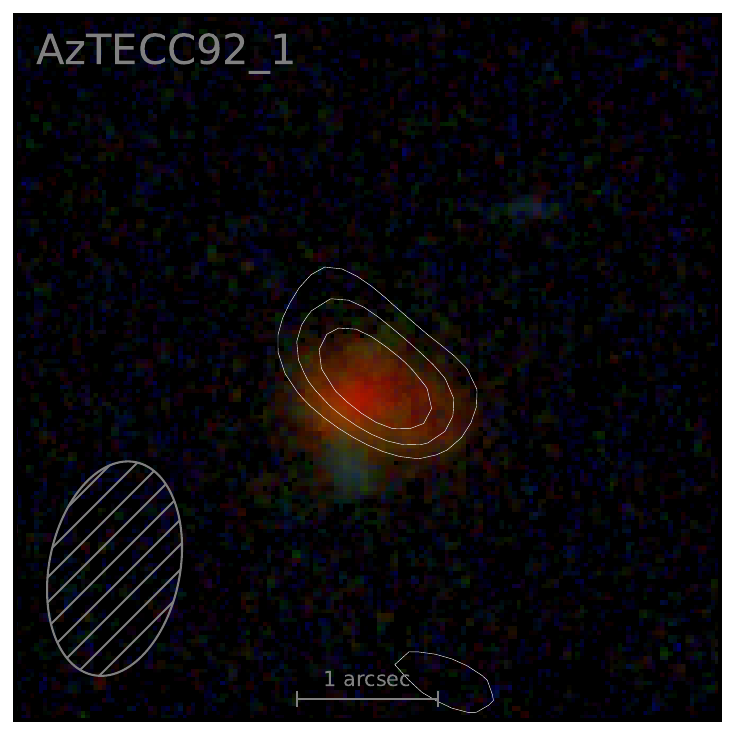}
\end{figure*}
\begin{figure*}
\centering
\includegraphics[width=0.24\textwidth]{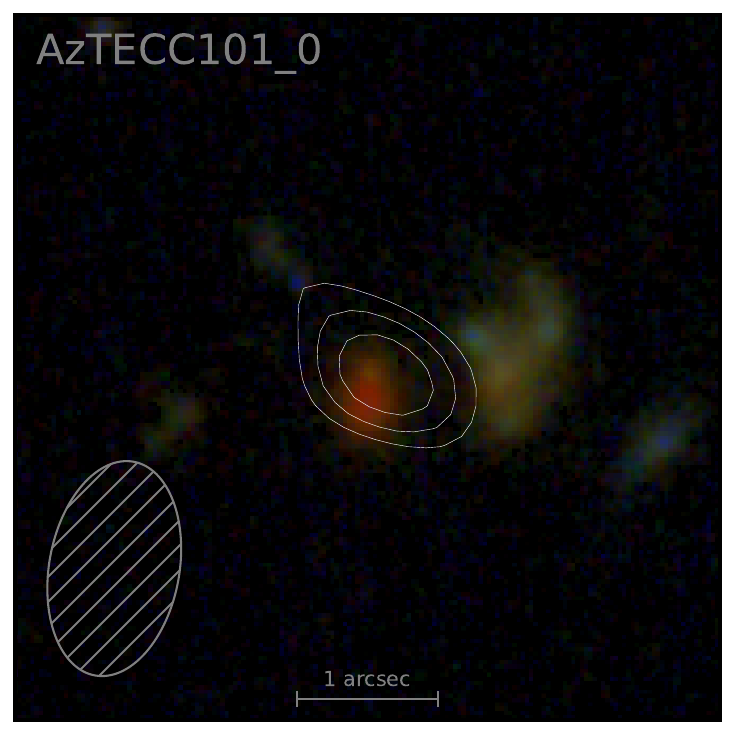}
\includegraphics[width=0.24\textwidth]{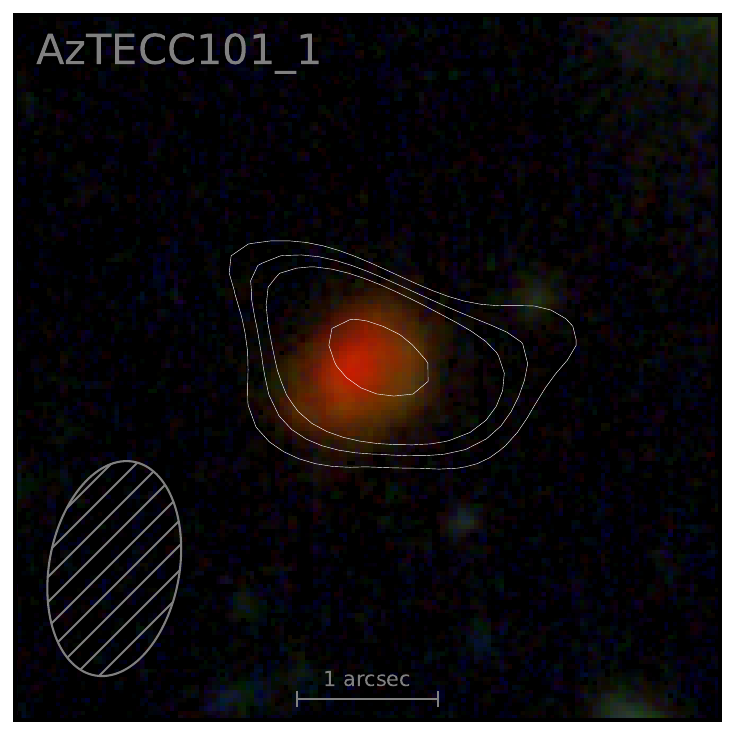}
\includegraphics[width=0.24\textwidth]{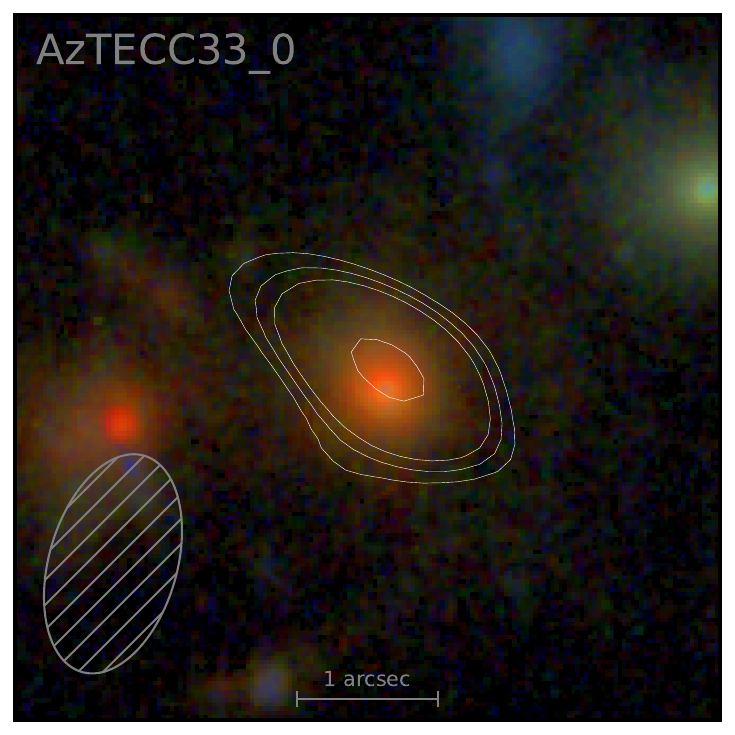}
\includegraphics[width=0.24\textwidth]{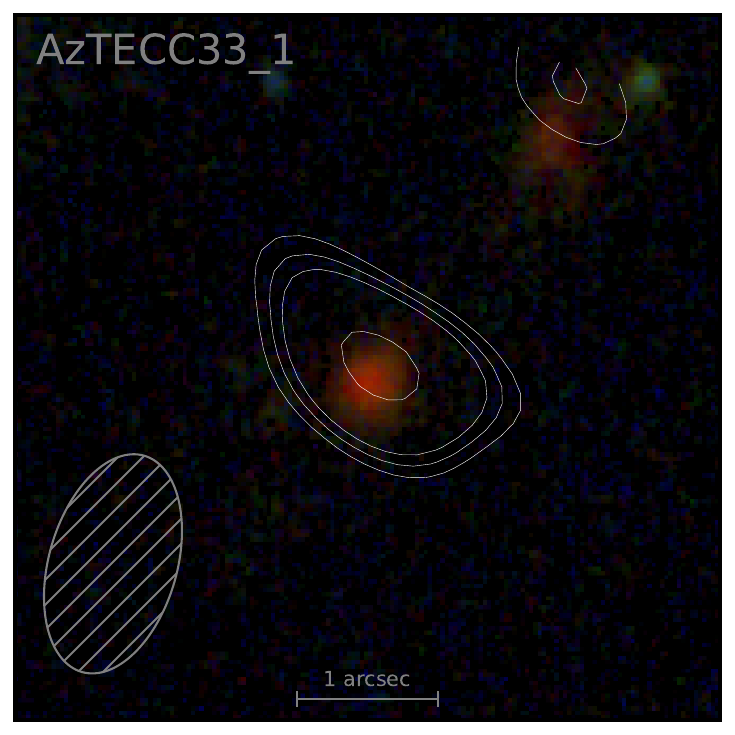}
\includegraphics[width=0.24\textwidth]{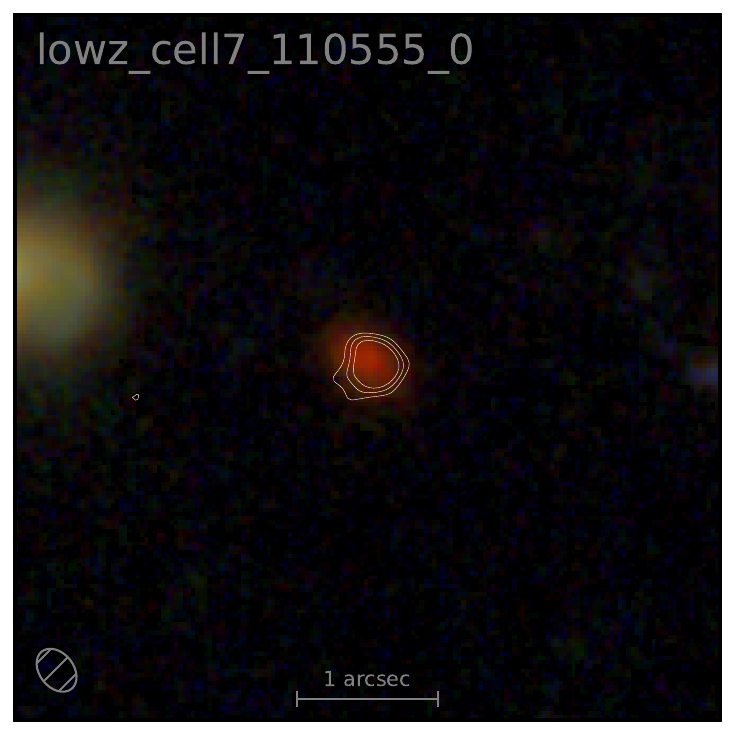}
\includegraphics[width=0.24\textwidth]{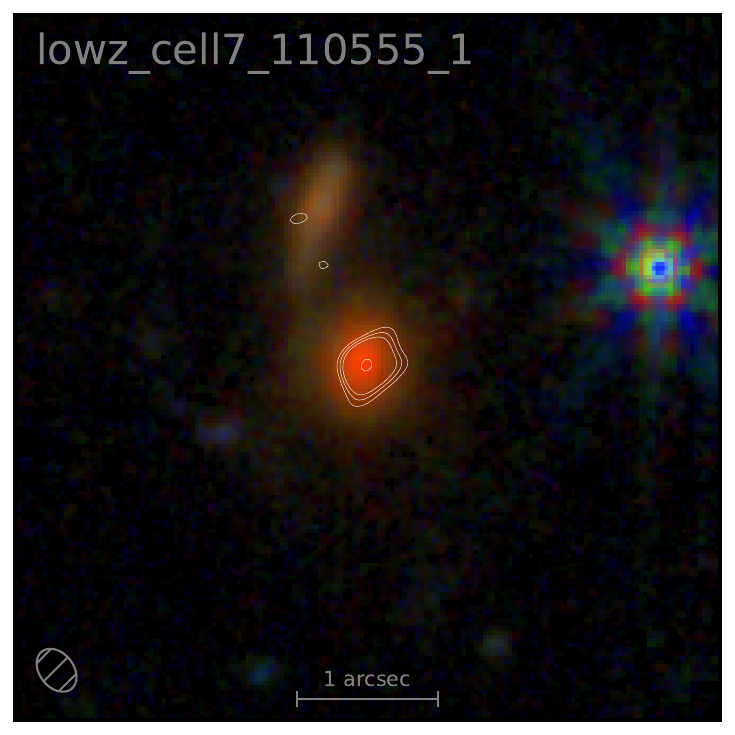}
\includegraphics[width=0.24\textwidth]{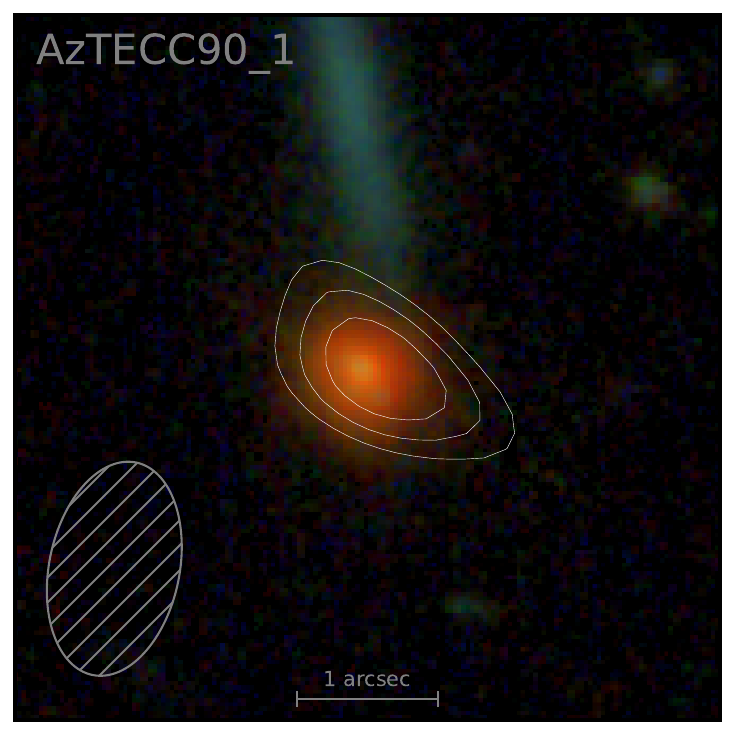}
\includegraphics[width=0.24\textwidth]{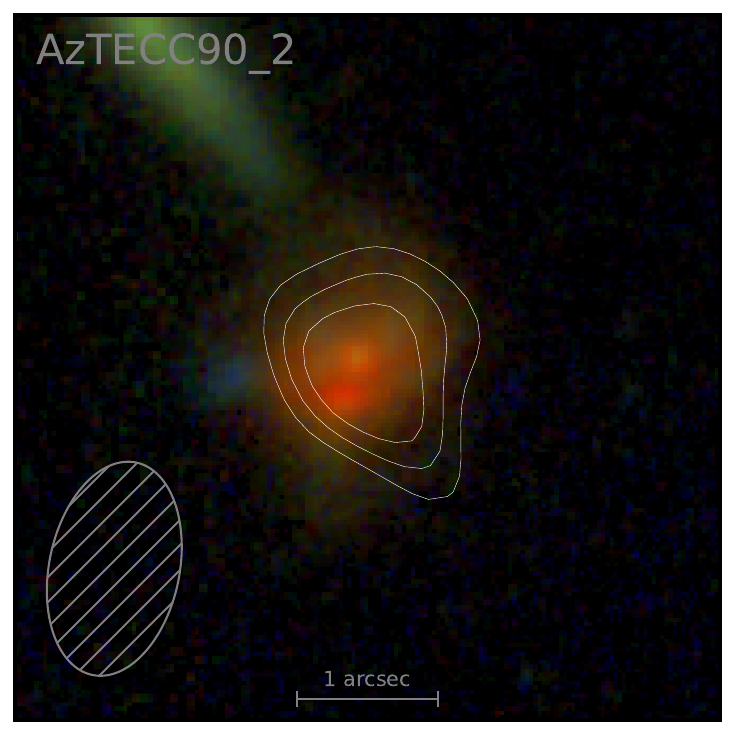}
\includegraphics[width=0.24\textwidth]{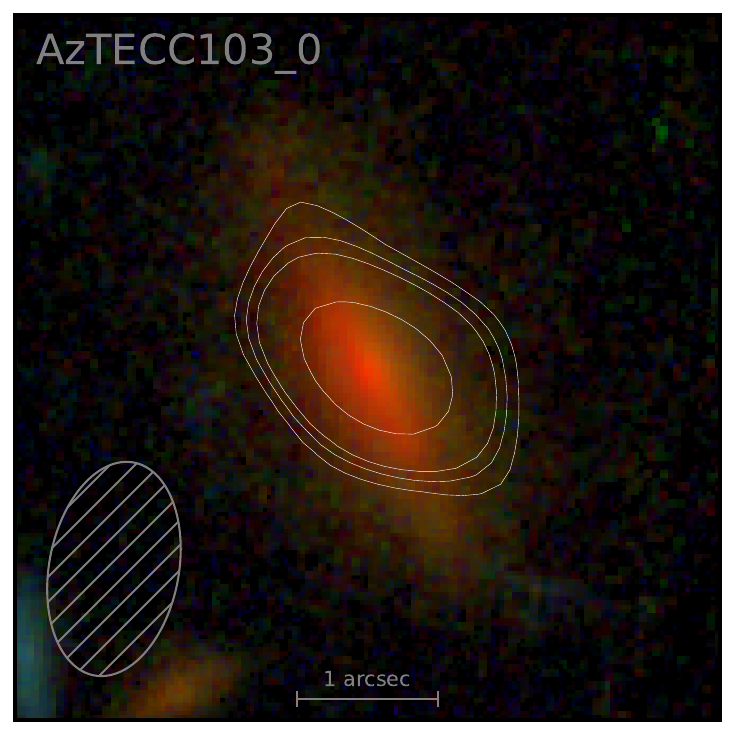}
\includegraphics[width=0.24\textwidth]{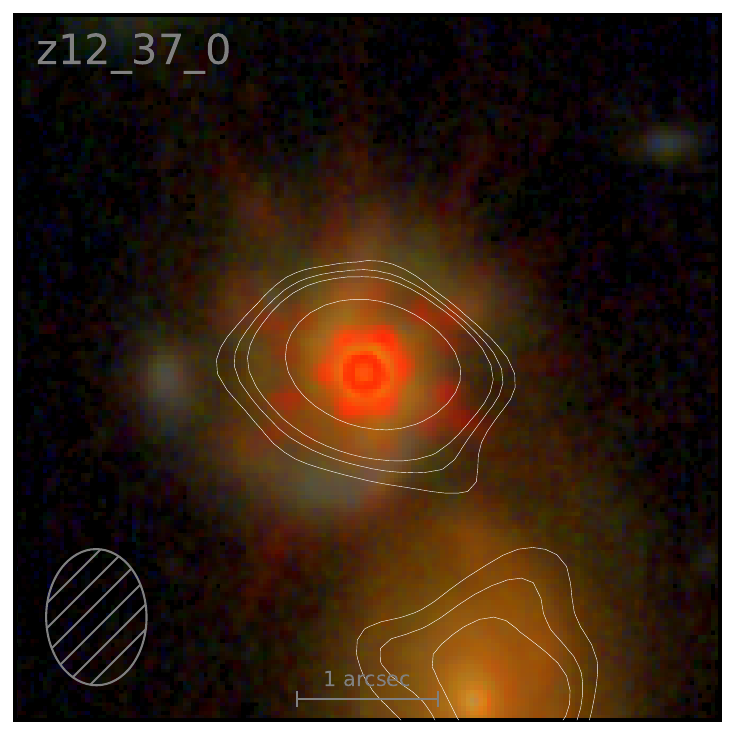}
\includegraphics[width=0.24\textwidth]{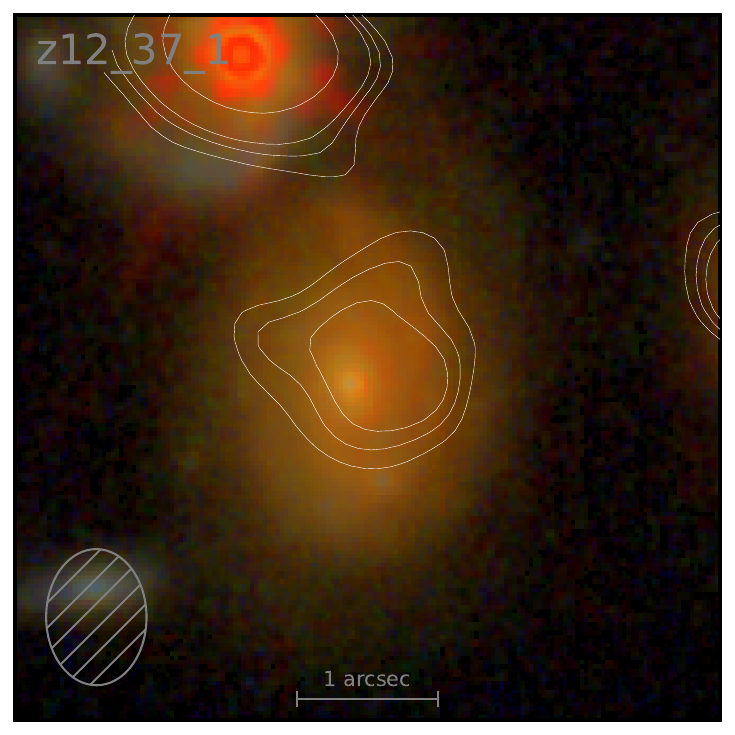}
\includegraphics[width=0.24\textwidth]{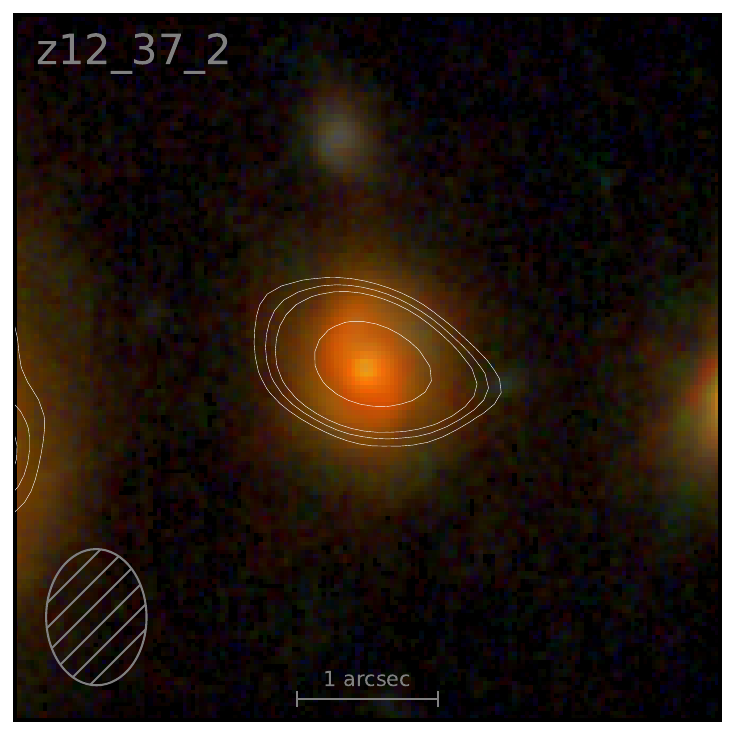}
\includegraphics[width=0.24\textwidth]{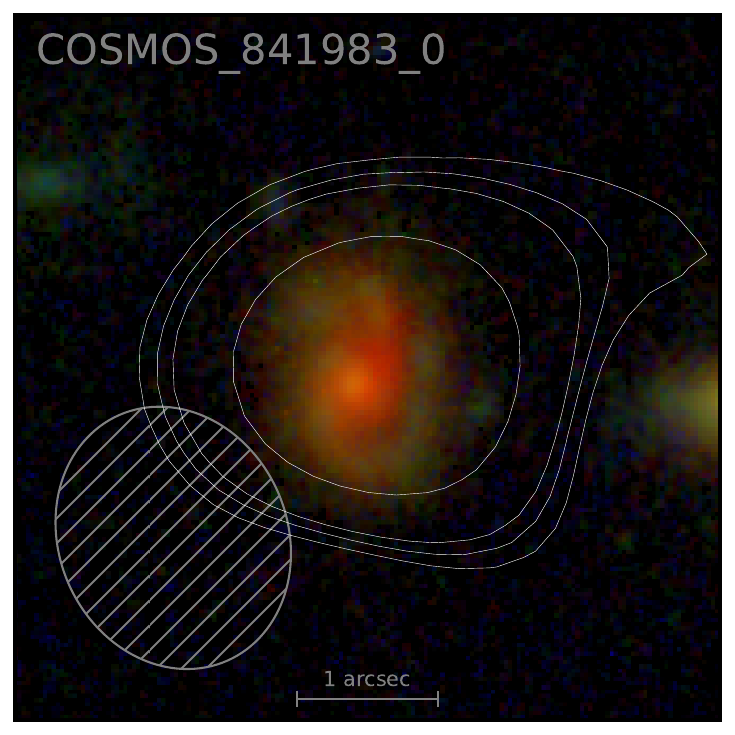}
\includegraphics[width=0.24\textwidth]{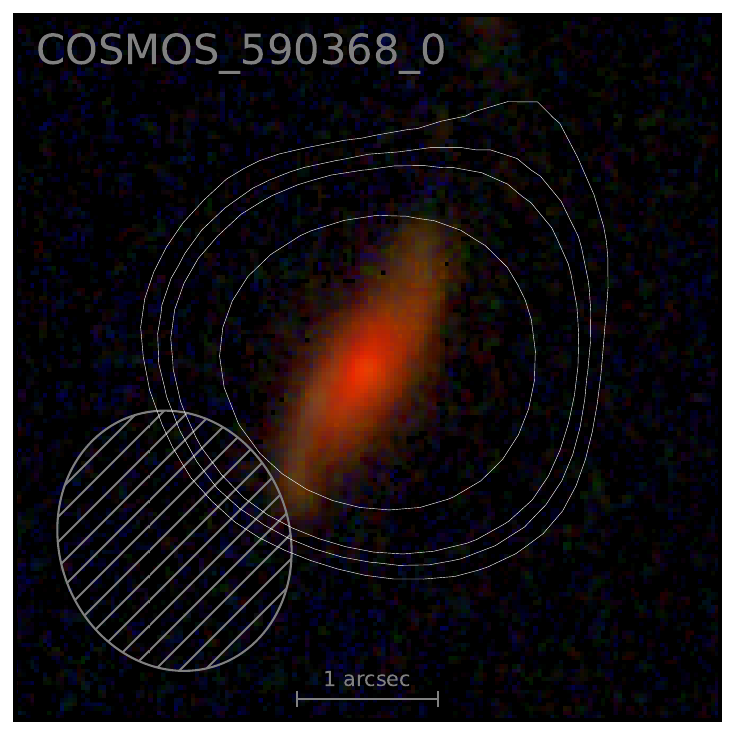}
\includegraphics[width=0.24\textwidth]{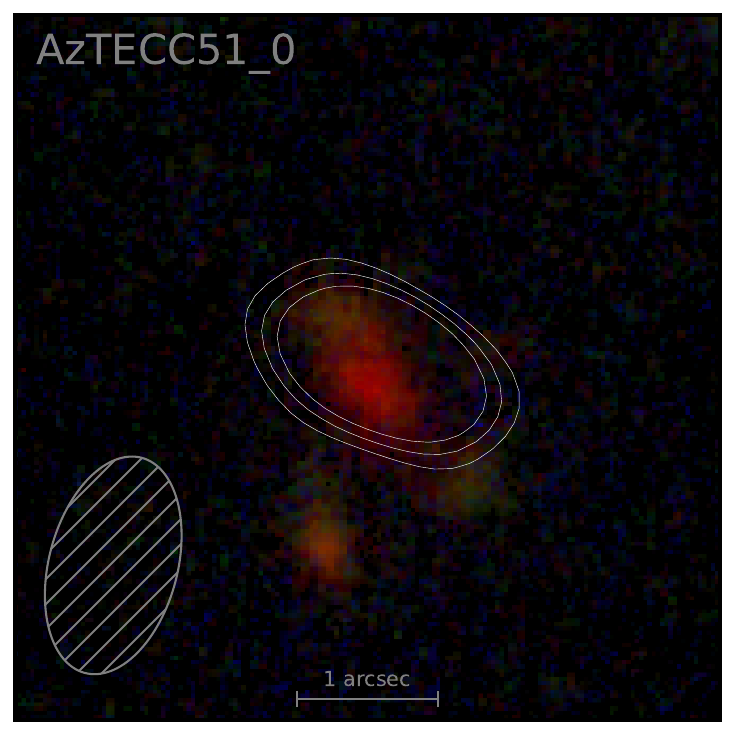}
\includegraphics[width=0.24\textwidth]{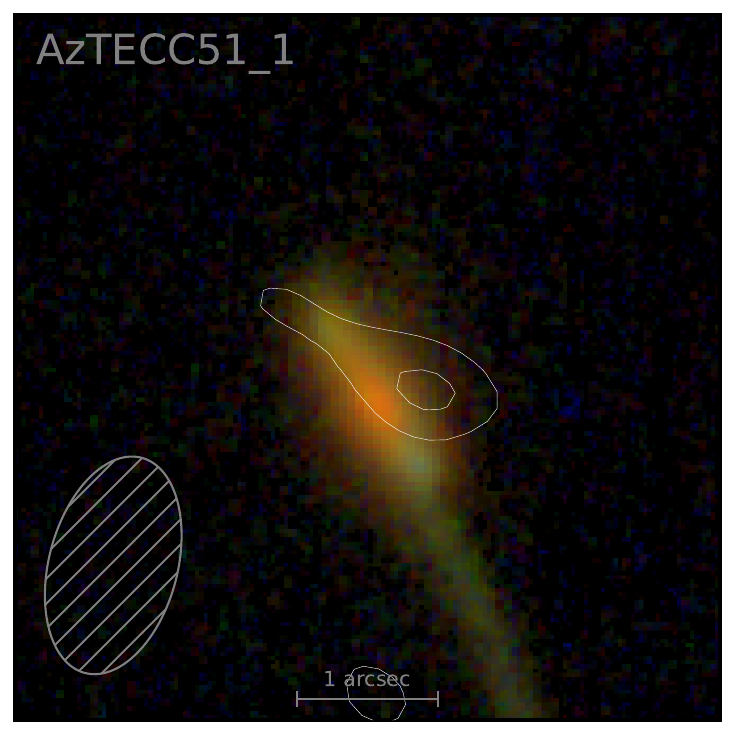}
\includegraphics[width=0.24\textwidth]{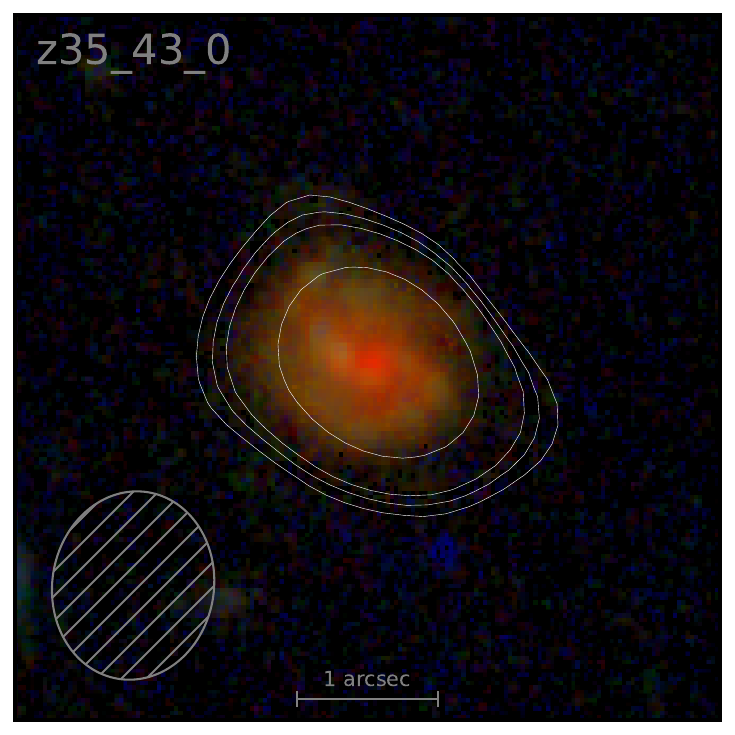}
\includegraphics[width=0.24\textwidth]{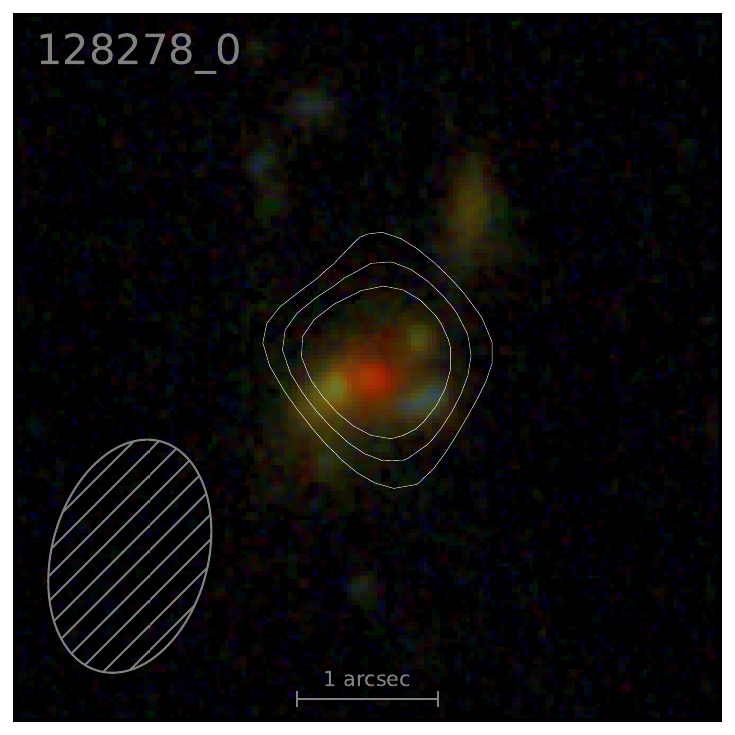}
\includegraphics[width=0.24\textwidth]{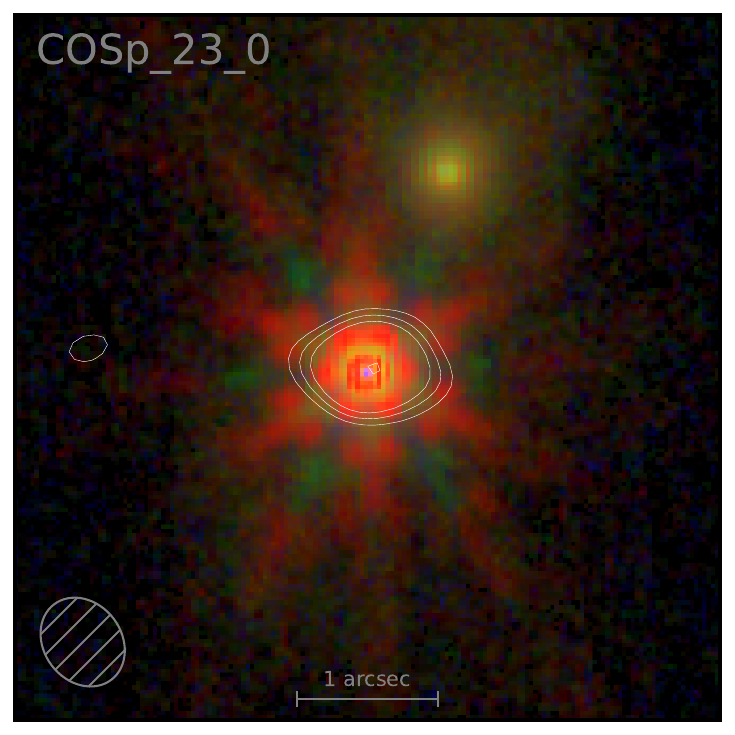}
\includegraphics[width=0.24\textwidth]{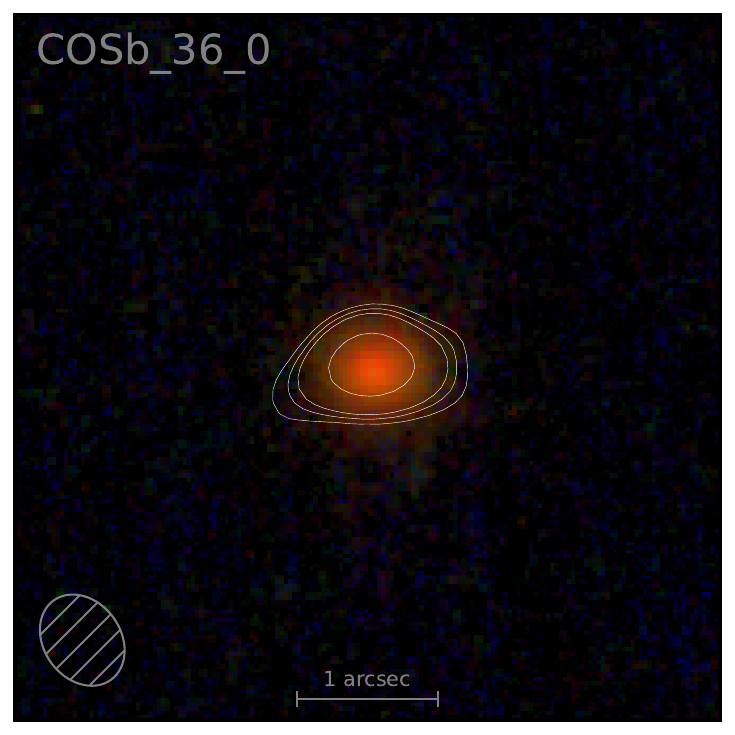}
\end{figure*}
\begin{figure*}
\centering
\includegraphics[width=0.24\textwidth]{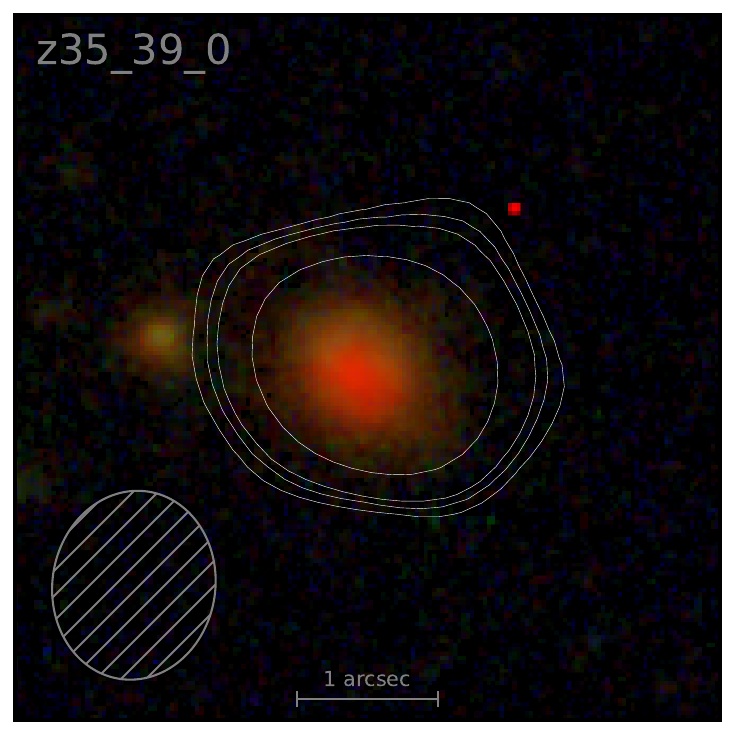}
\includegraphics[width=0.24\textwidth]{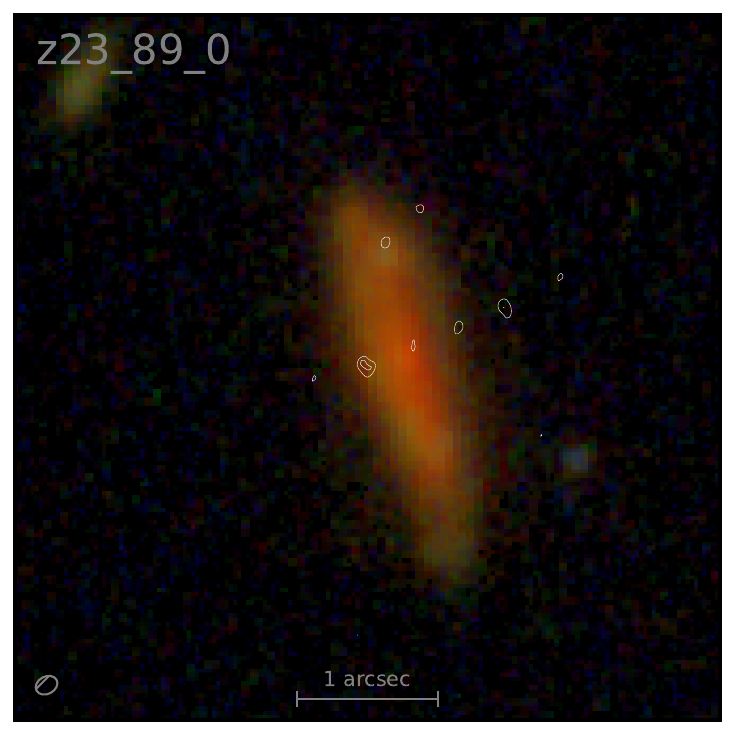}
\includegraphics[width=0.24\textwidth]{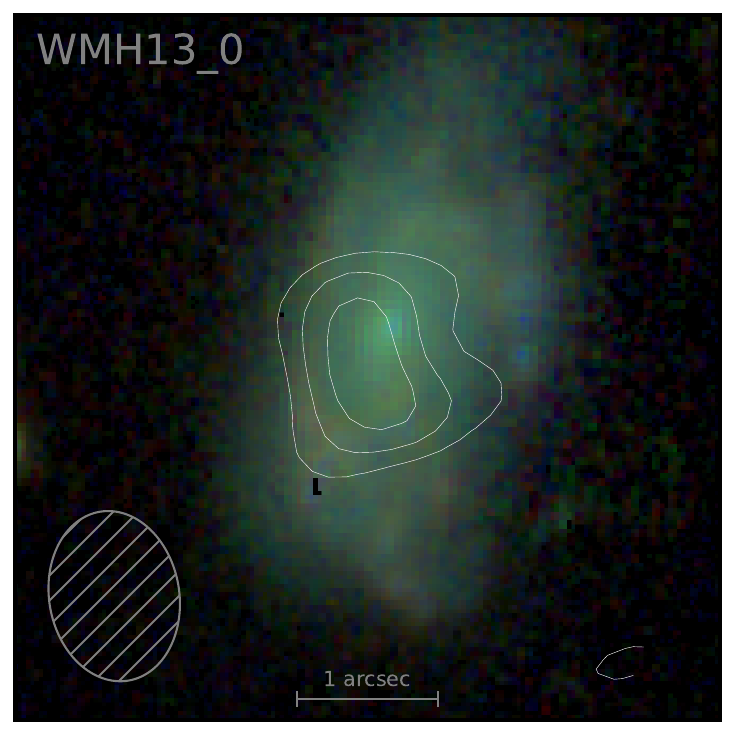}
\includegraphics[width=0.24\textwidth]{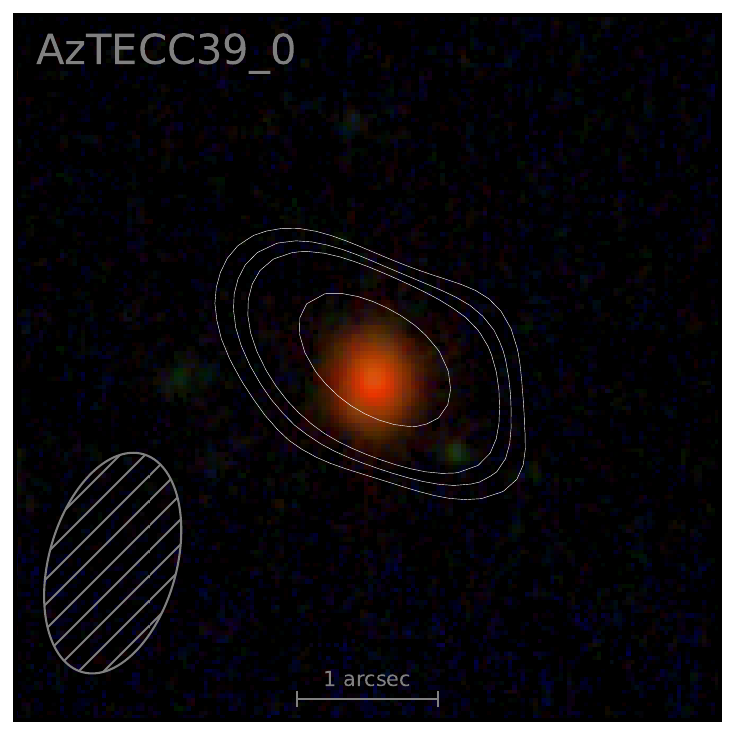}
\includegraphics[width=0.24\textwidth]{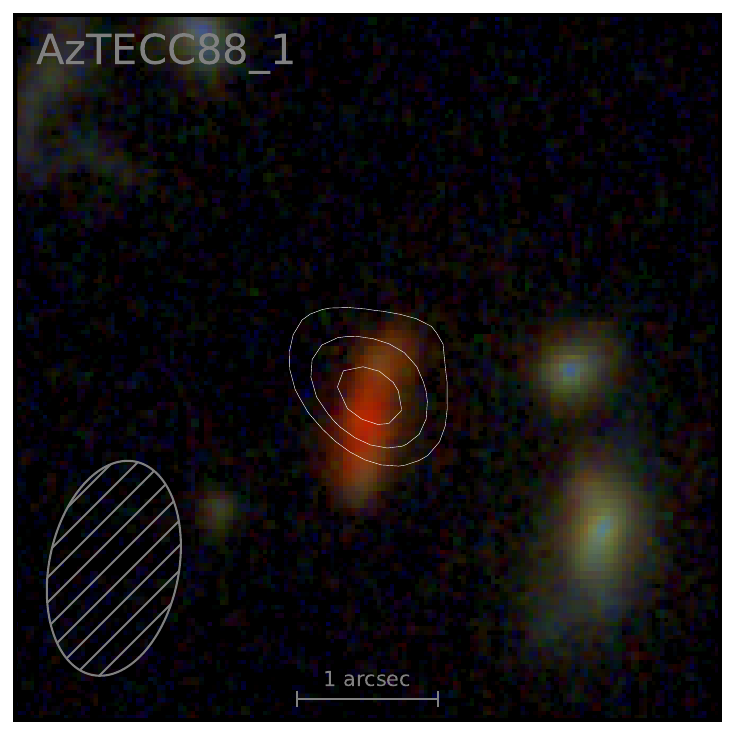}
\includegraphics[width=0.24\textwidth]{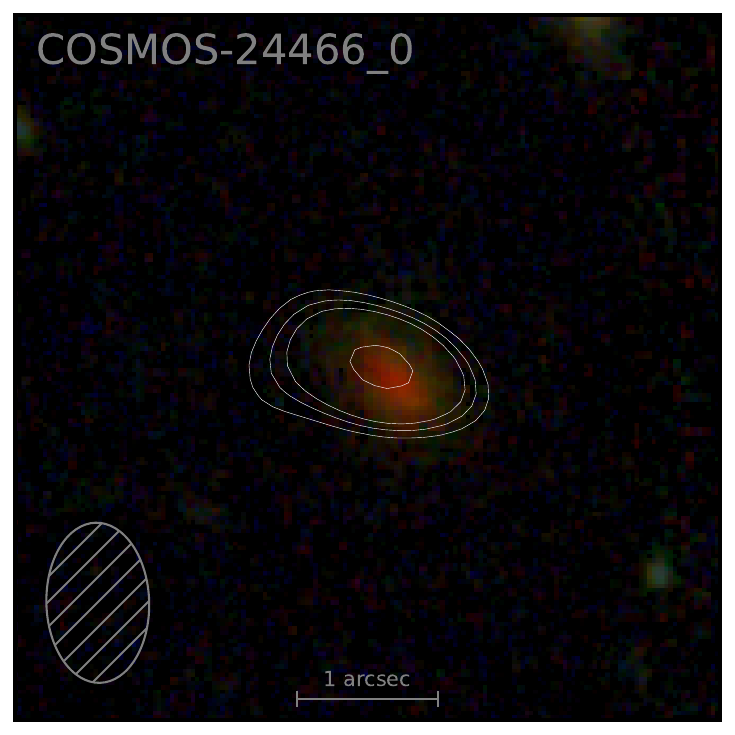}
\includegraphics[width=0.24\textwidth]{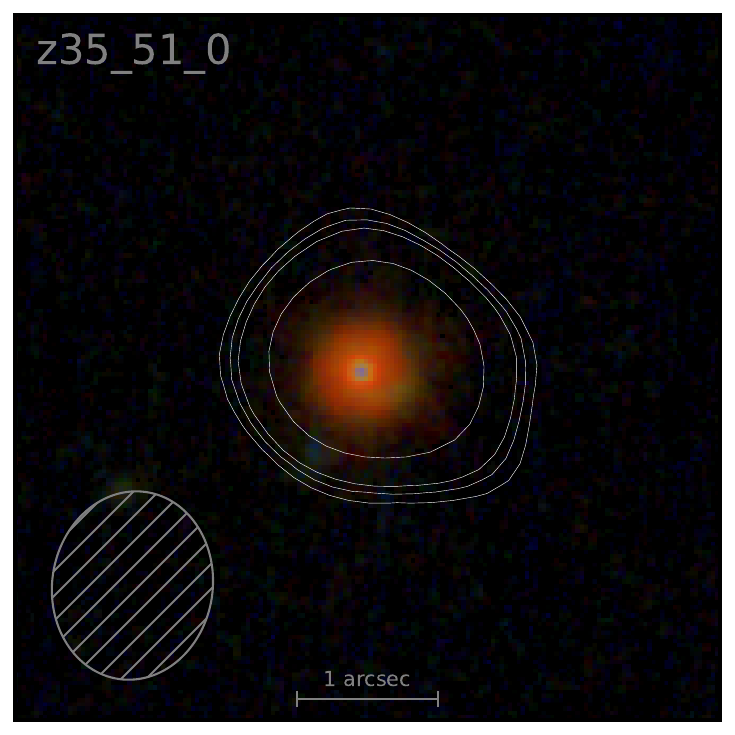}
\includegraphics[width=0.24\textwidth]{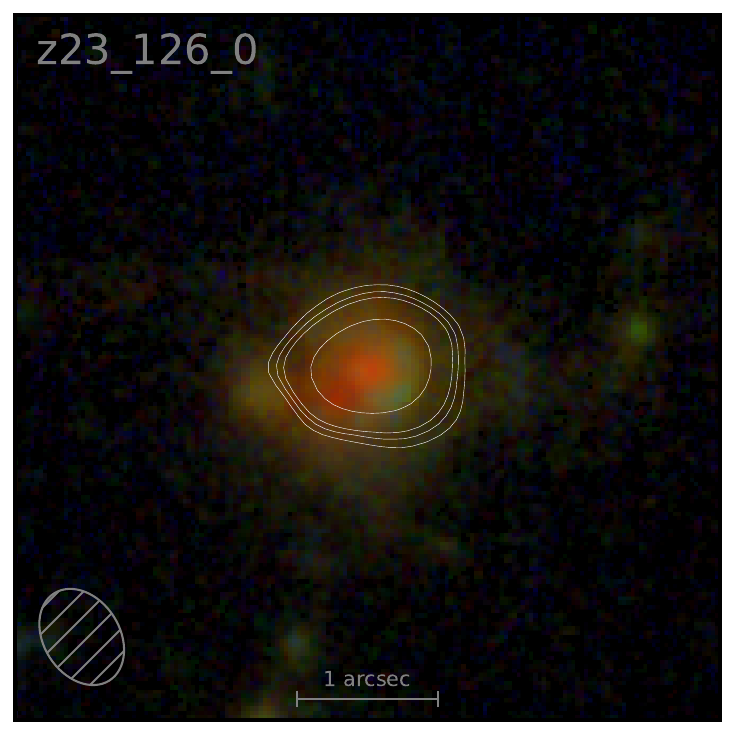}
\includegraphics[width=0.24\textwidth]{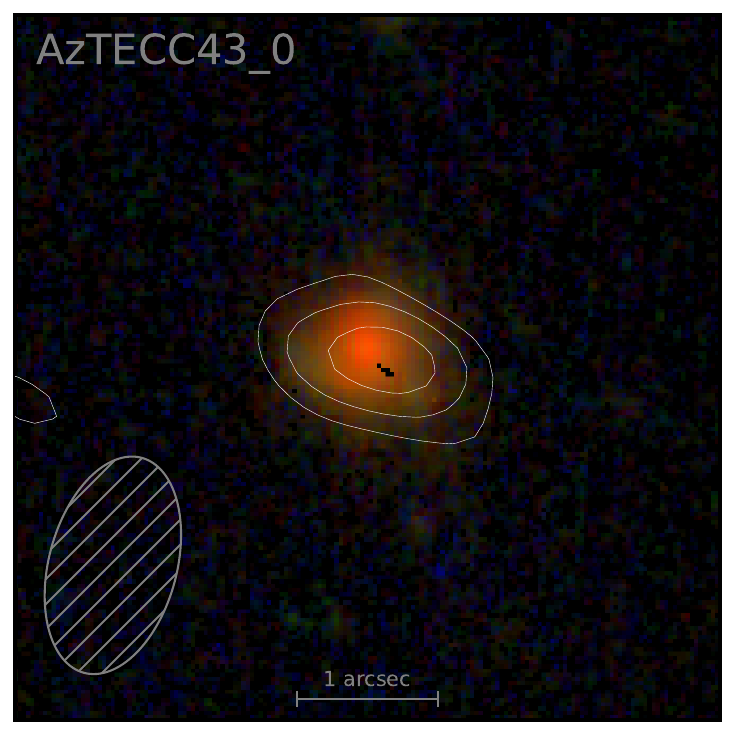}
\includegraphics[width=0.24\textwidth]{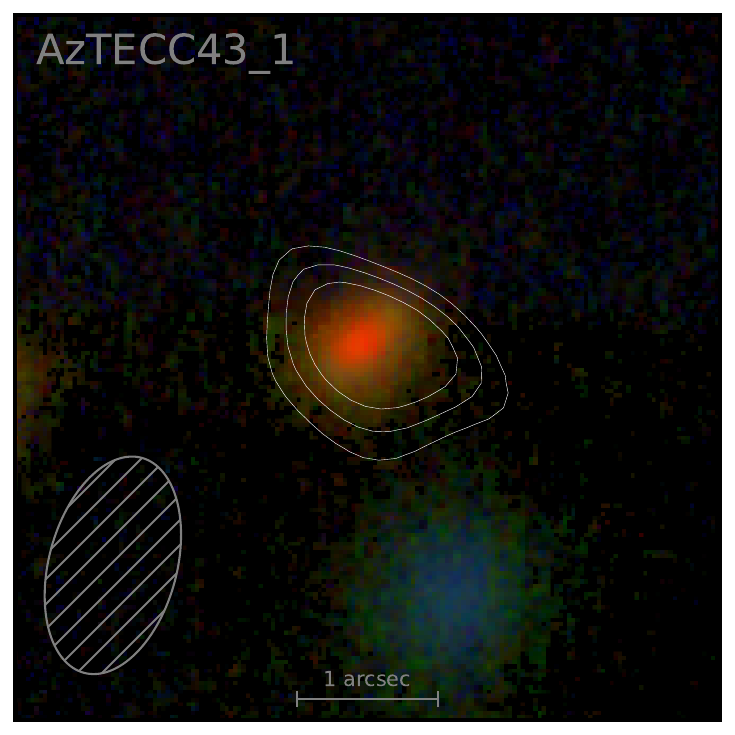}
\includegraphics[width=0.24\textwidth]{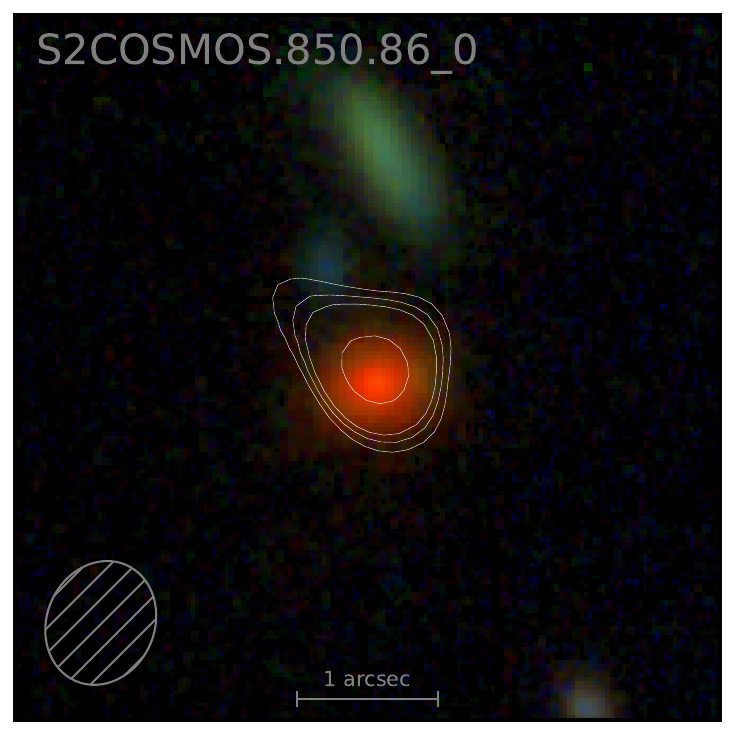}
\includegraphics[width=0.24\textwidth]{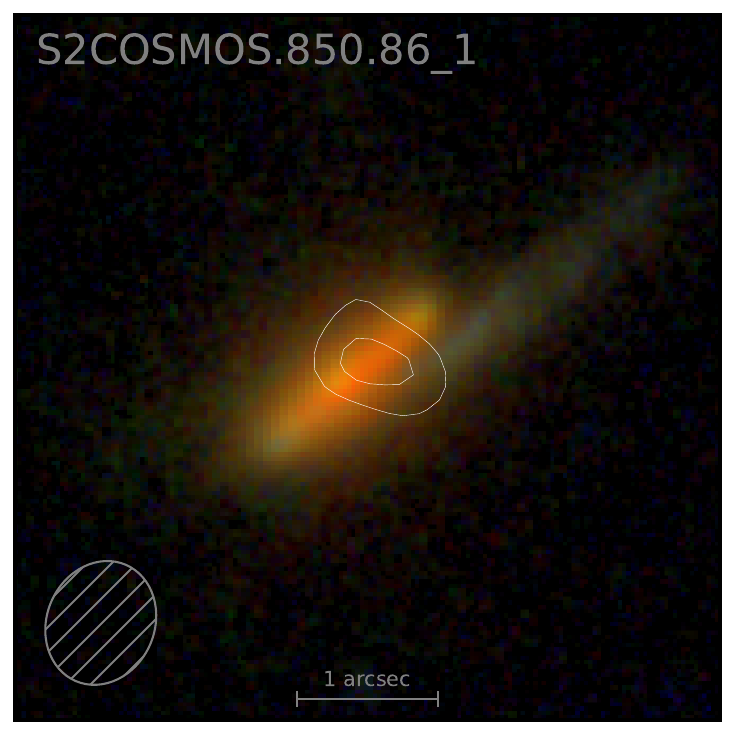}
\includegraphics[width=0.24\textwidth]{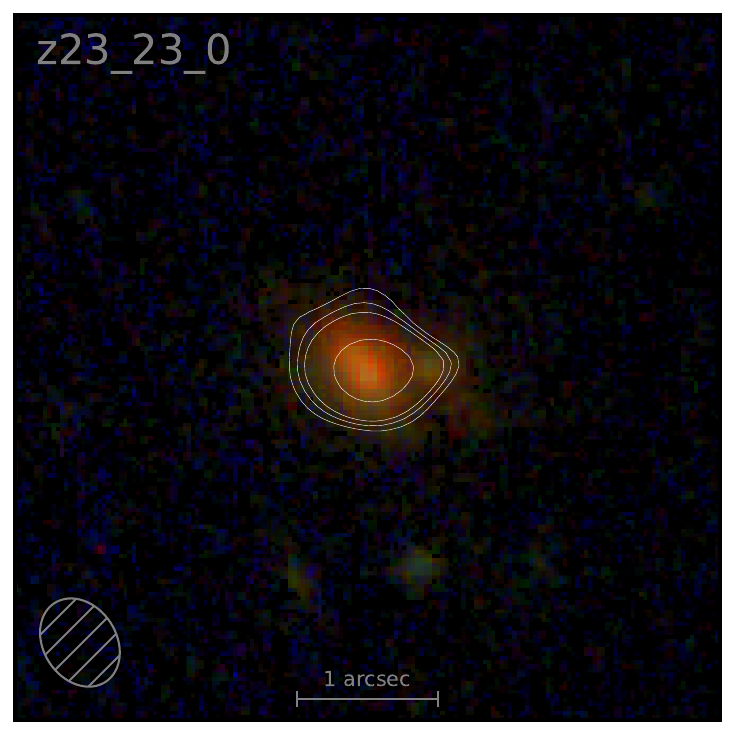}
\includegraphics[width=0.24\textwidth]{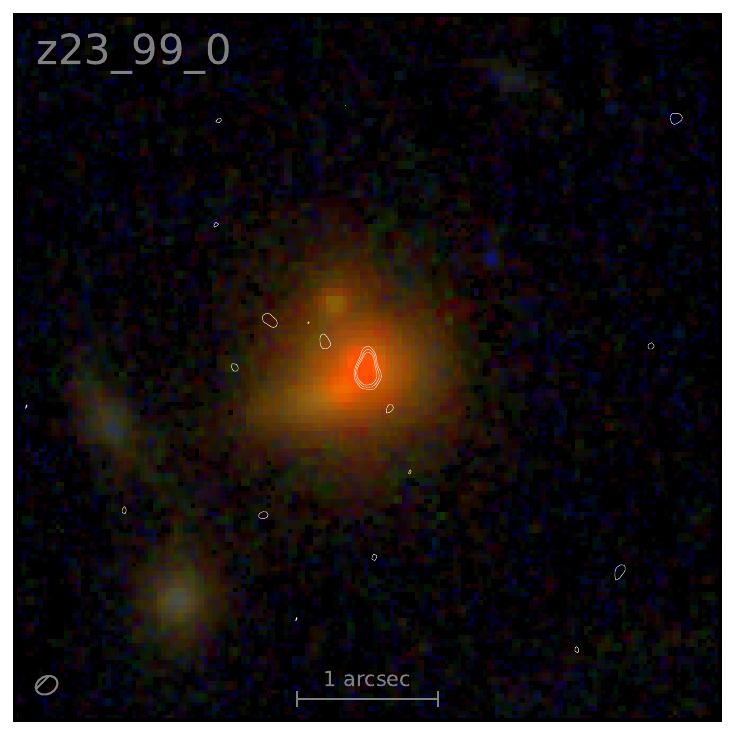}
\includegraphics[width=0.24\textwidth]{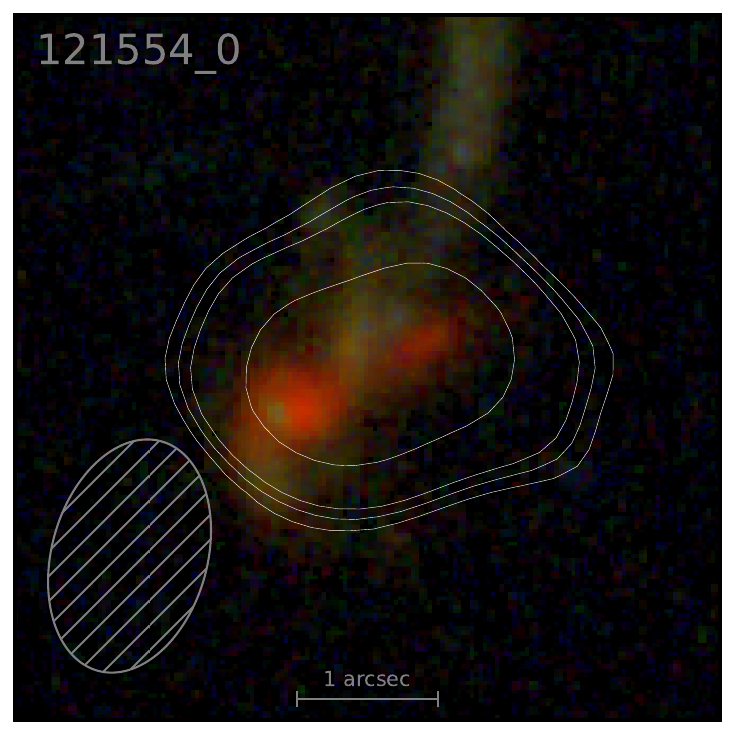}
\includegraphics[width=0.24\textwidth]{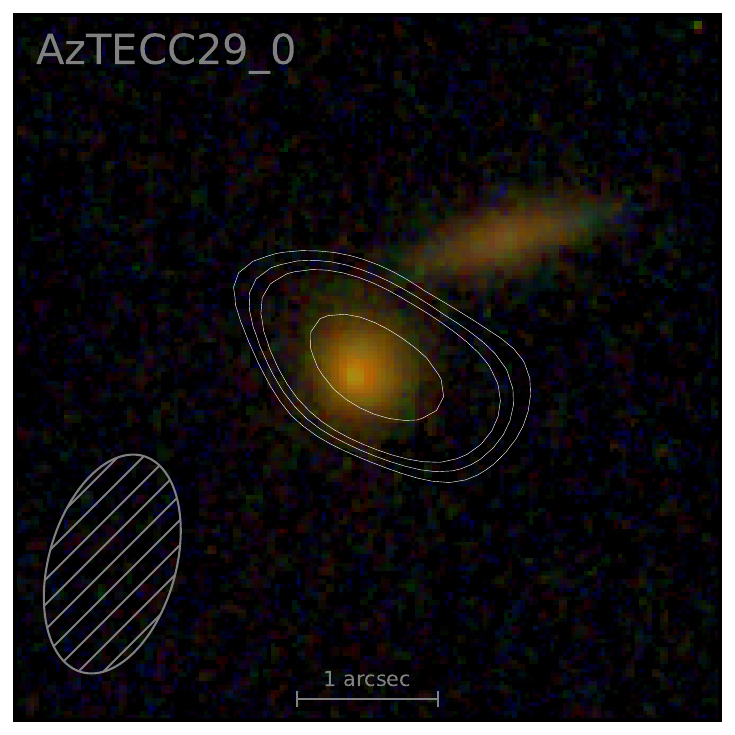}
\includegraphics[width=0.24\textwidth]{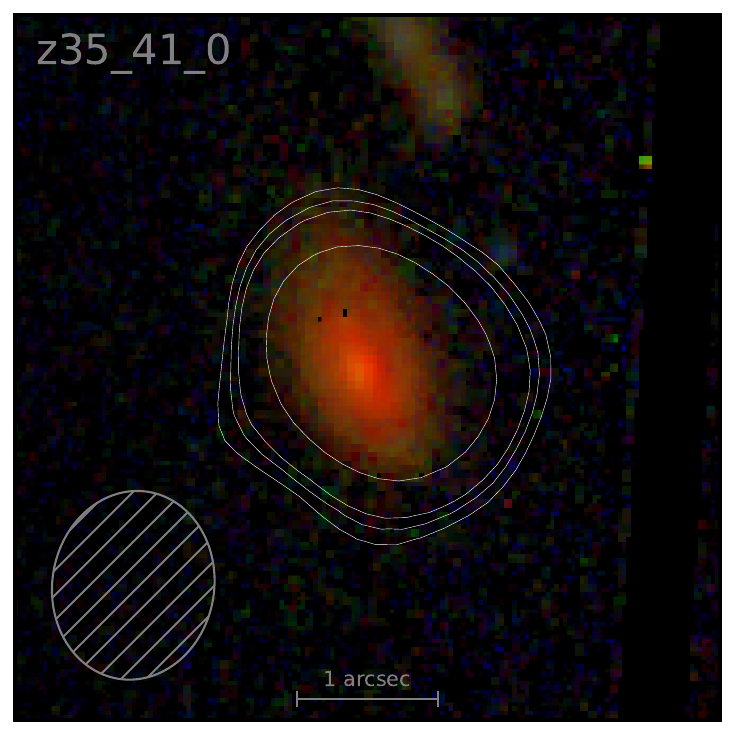}
\includegraphics[width=0.24\textwidth]{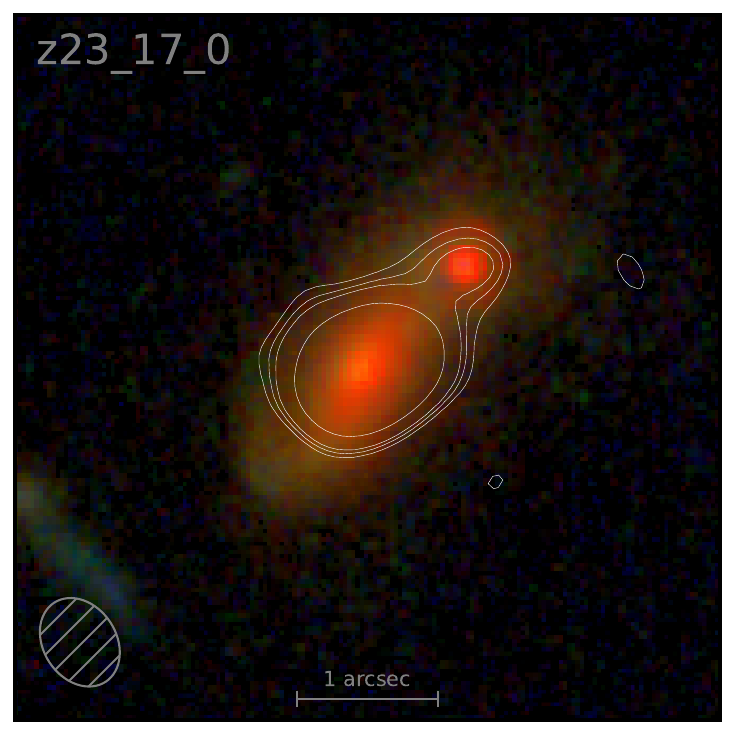}
\includegraphics[width=0.24\textwidth]{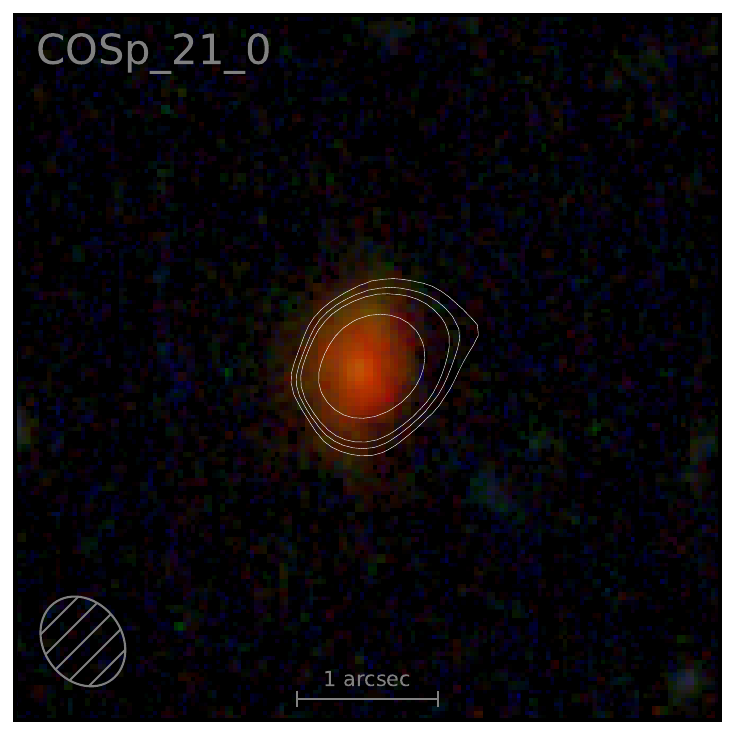}
\includegraphics[width=0.24\textwidth]{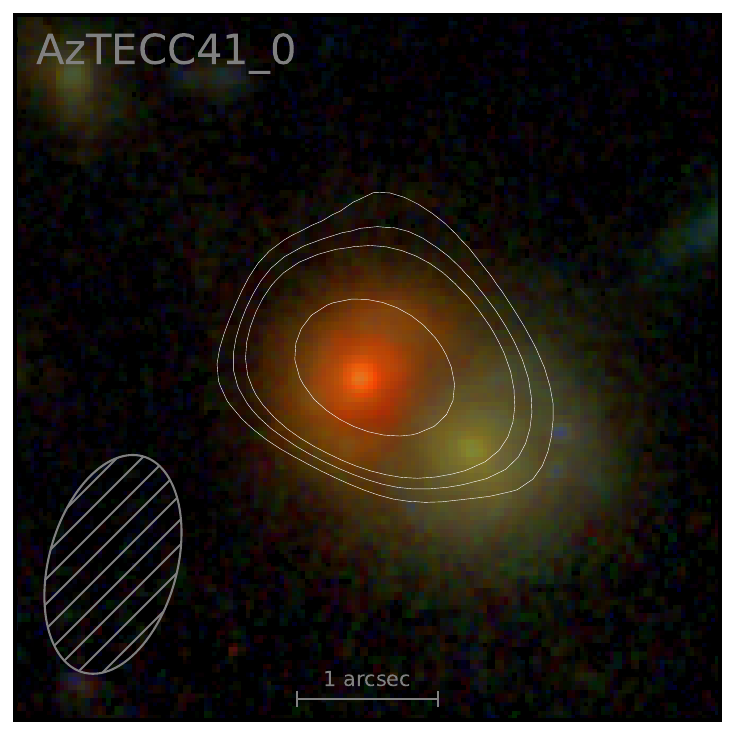}
\end{figure*}
\begin{figure*}
\centering
\includegraphics[width=0.24\textwidth]{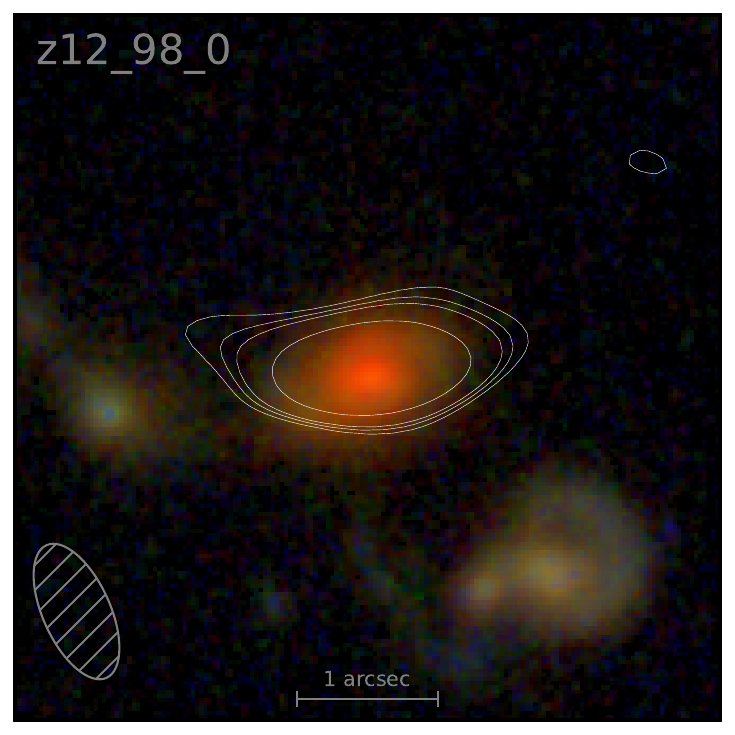}
\includegraphics[width=0.24\textwidth]{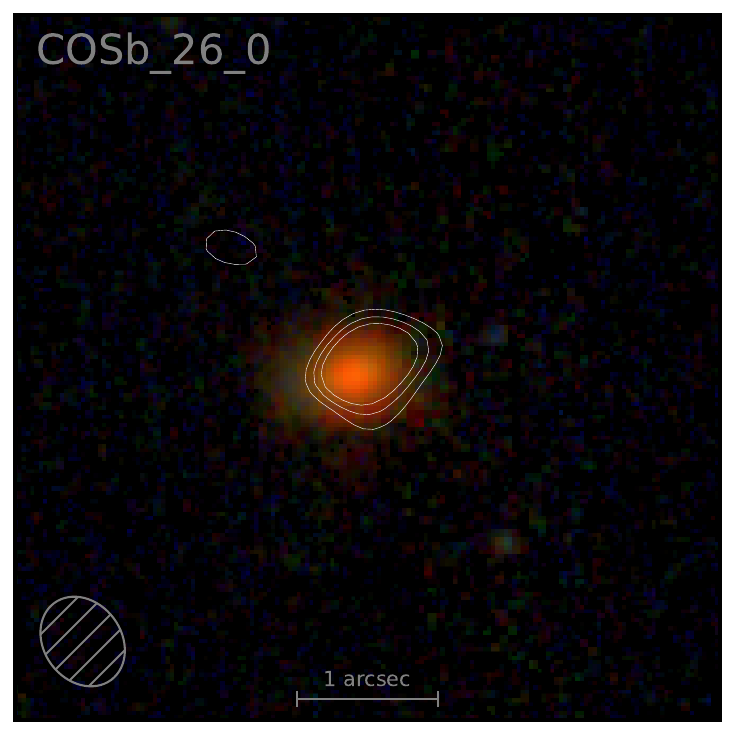}
\includegraphics[width=0.24\textwidth]{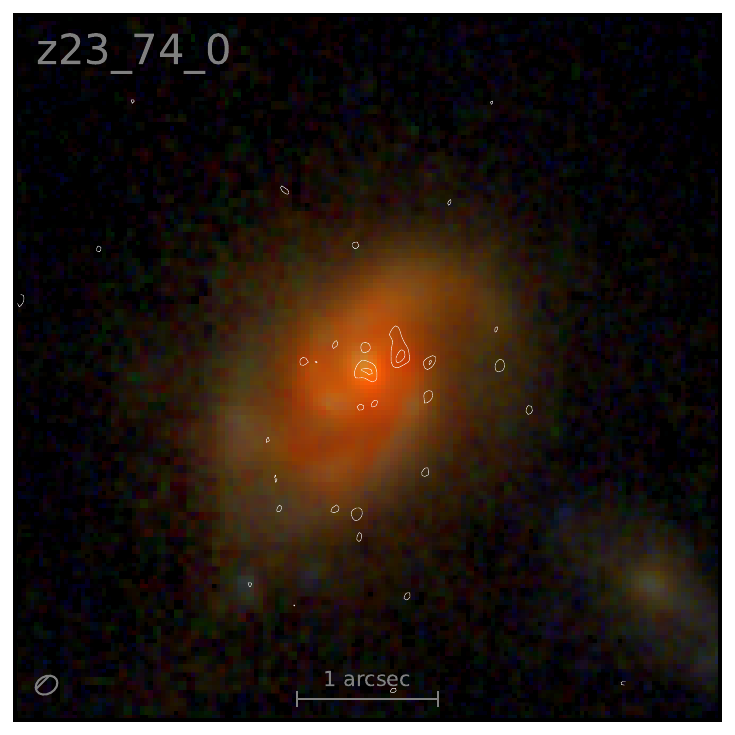}
\includegraphics[width=0.24\textwidth]{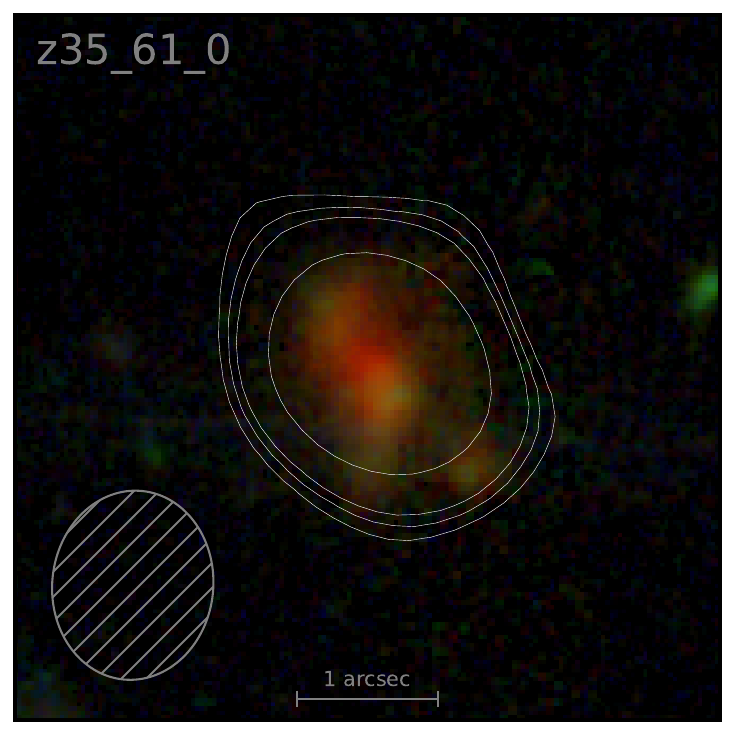}
\includegraphics[width=0.24\textwidth]{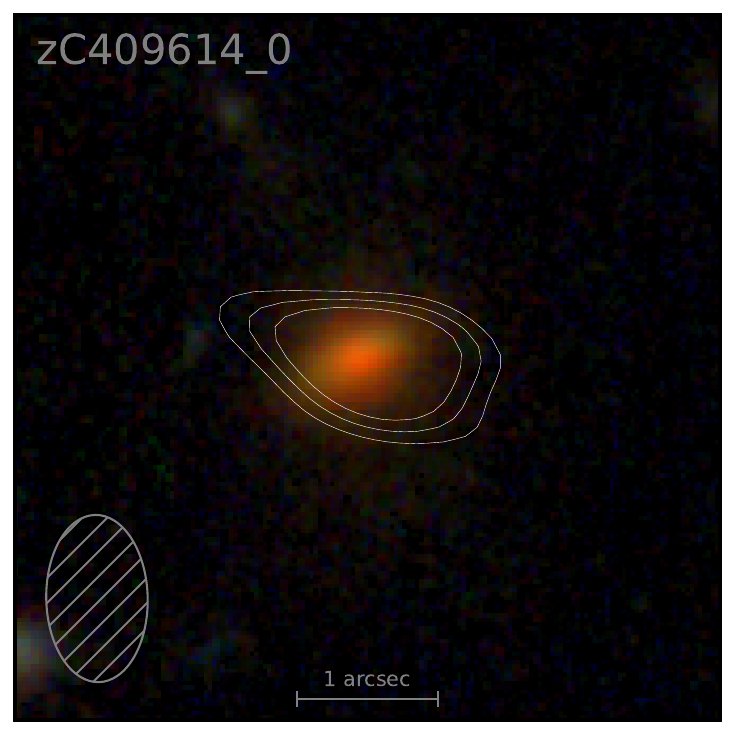}
\includegraphics[width=0.24\textwidth]{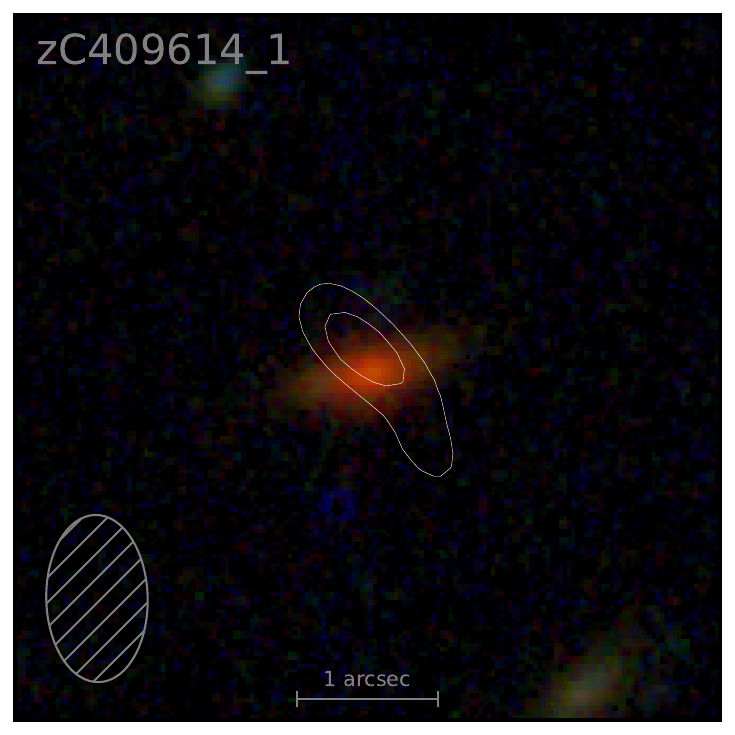}
\includegraphics[width=0.24\textwidth]{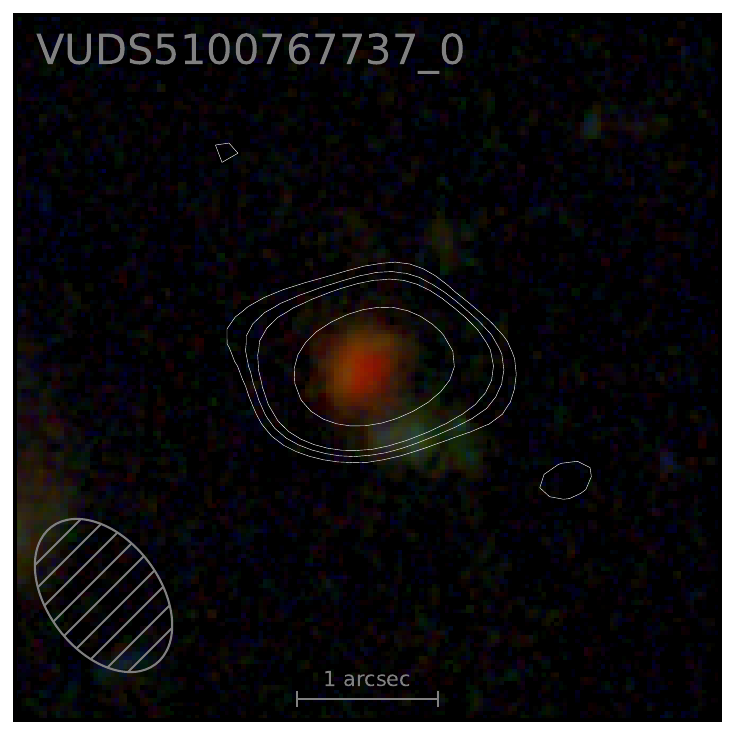}
\includegraphics[width=0.24\textwidth]{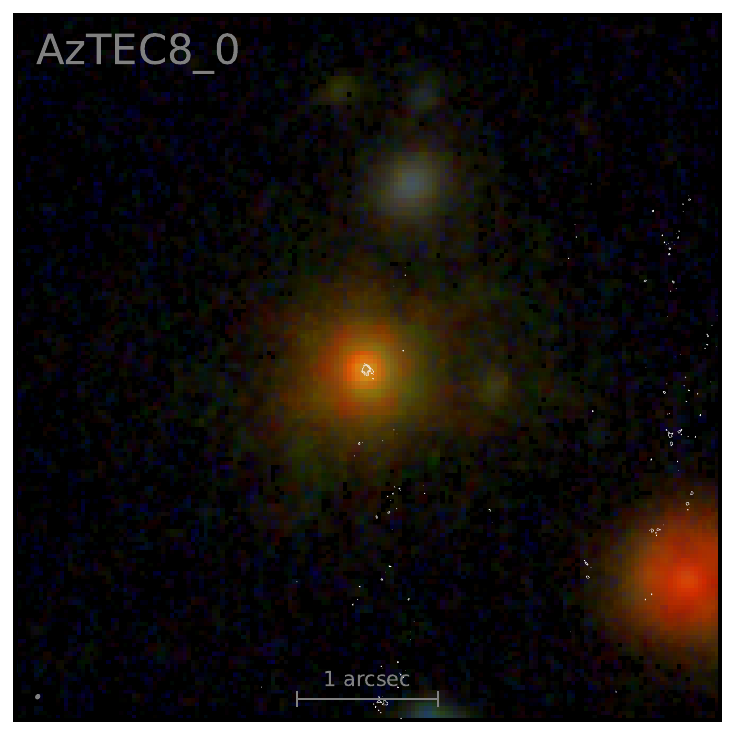}
\includegraphics[width=0.24\textwidth]{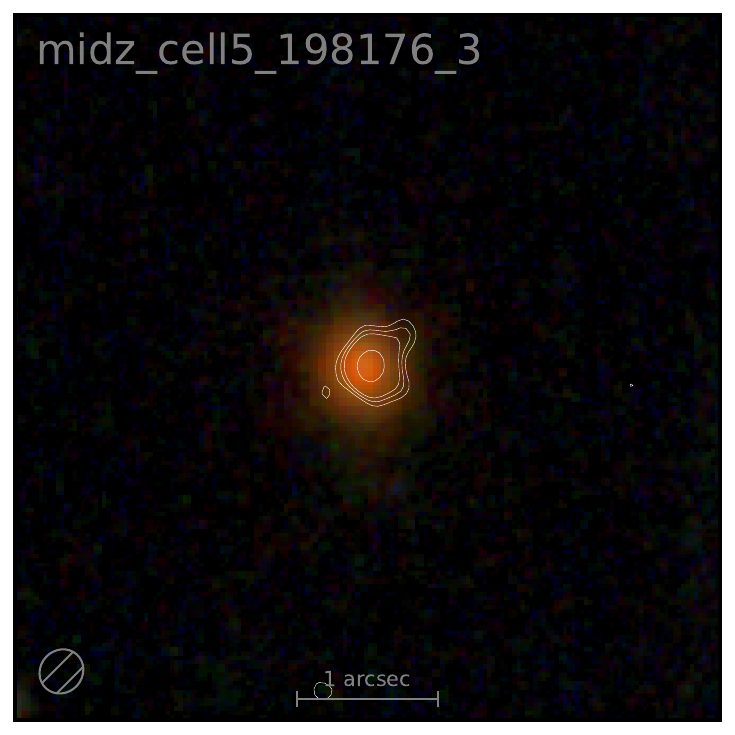}
\includegraphics[width=0.24\textwidth]{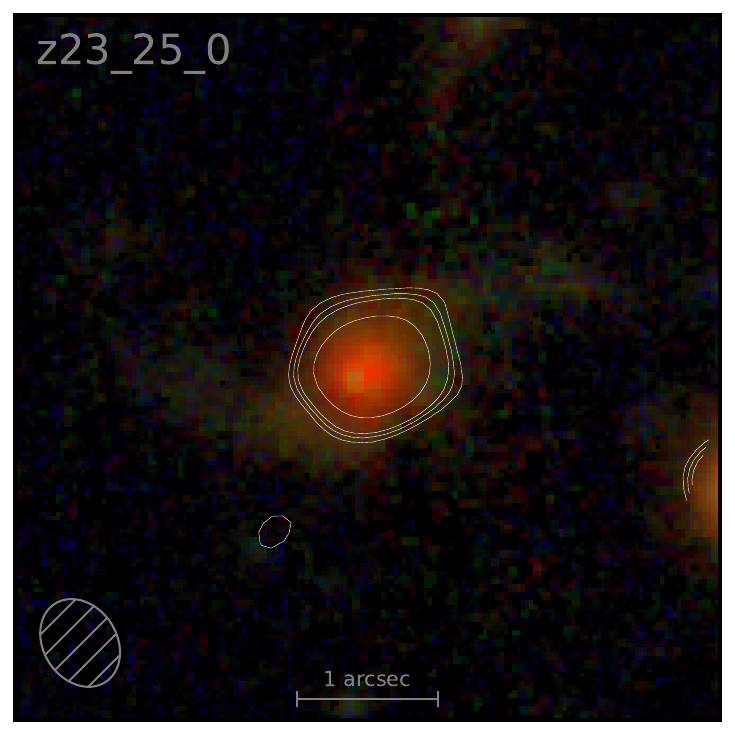}
\includegraphics[width=0.24\textwidth]{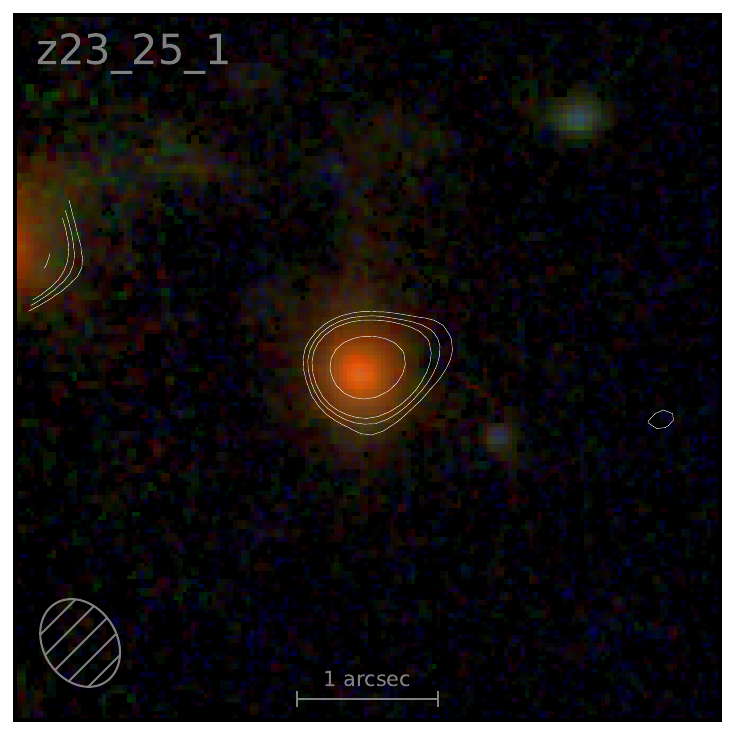}
\includegraphics[width=0.24\textwidth]{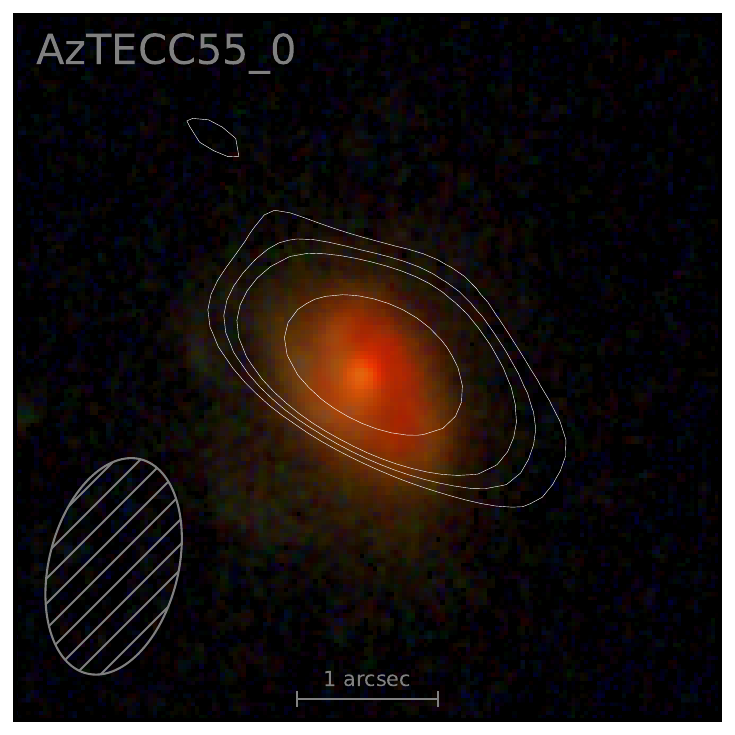}
\includegraphics[width=0.24\textwidth]{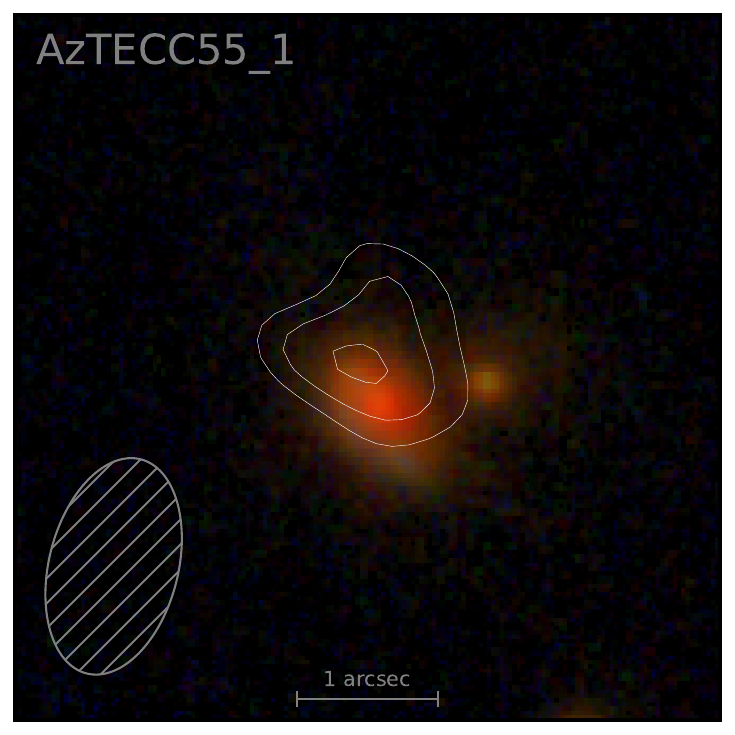}
\includegraphics[width=0.24\textwidth]{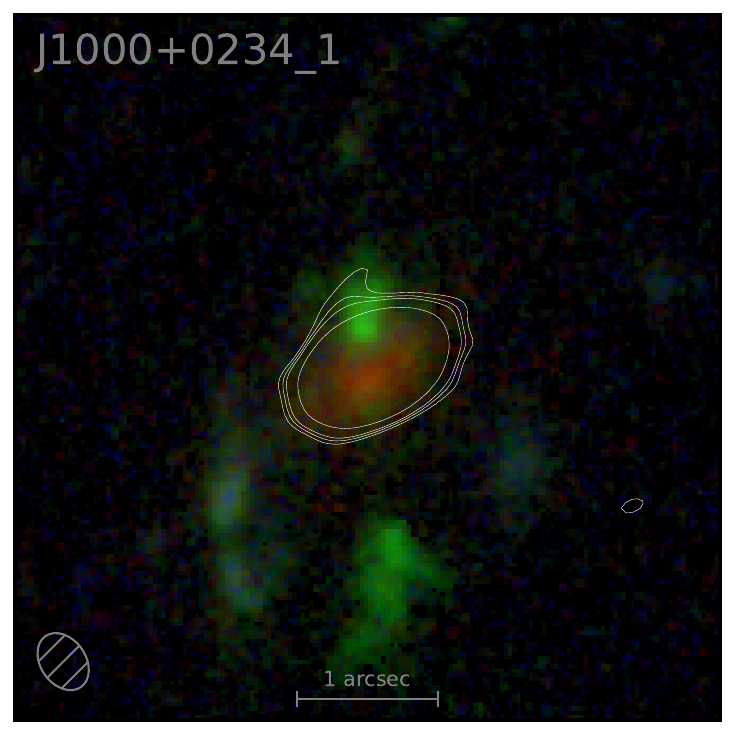}
\includegraphics[width=0.24\textwidth]{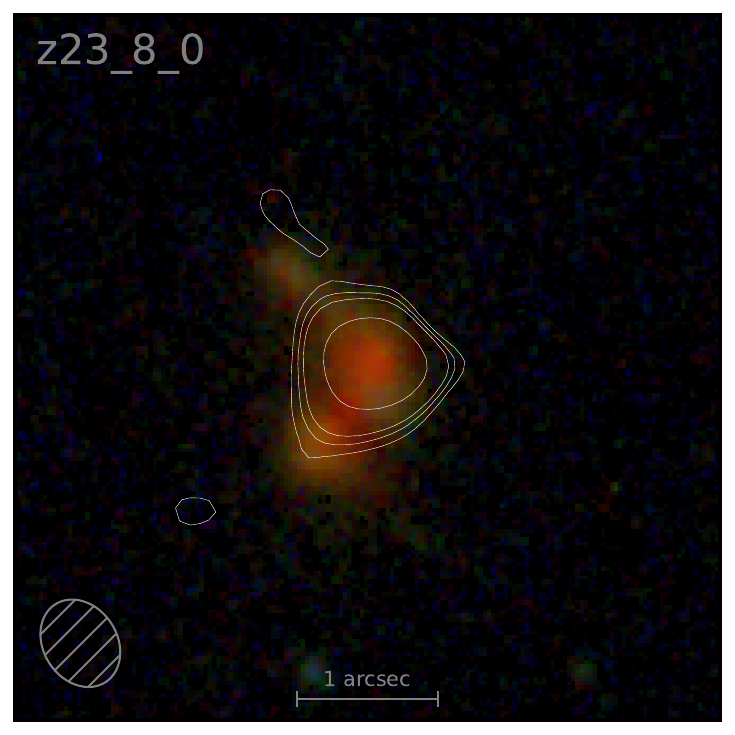}
\includegraphics[width=0.24\textwidth]{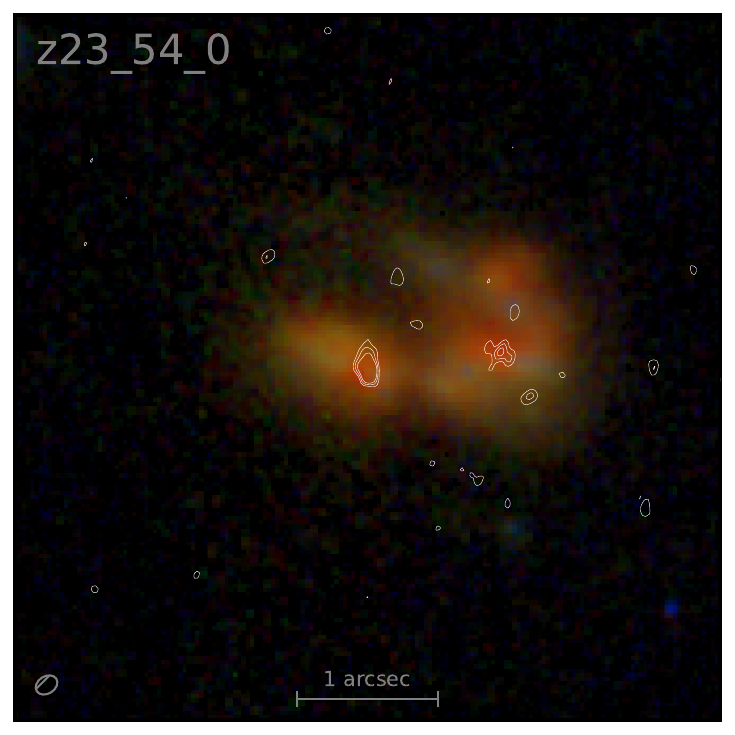}
\includegraphics[width=0.24\textwidth]{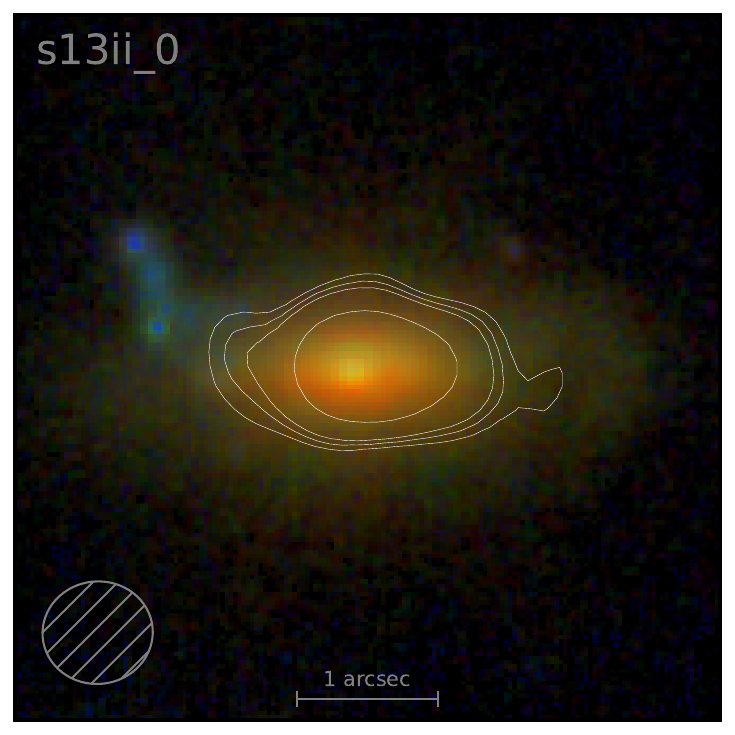}
\includegraphics[width=0.24\textwidth]{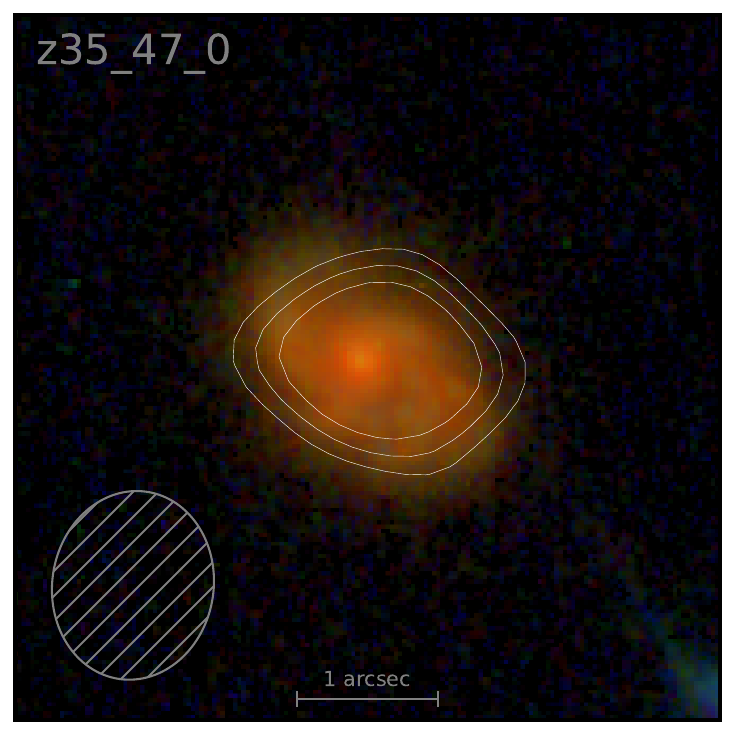}
\includegraphics[width=0.24\textwidth]{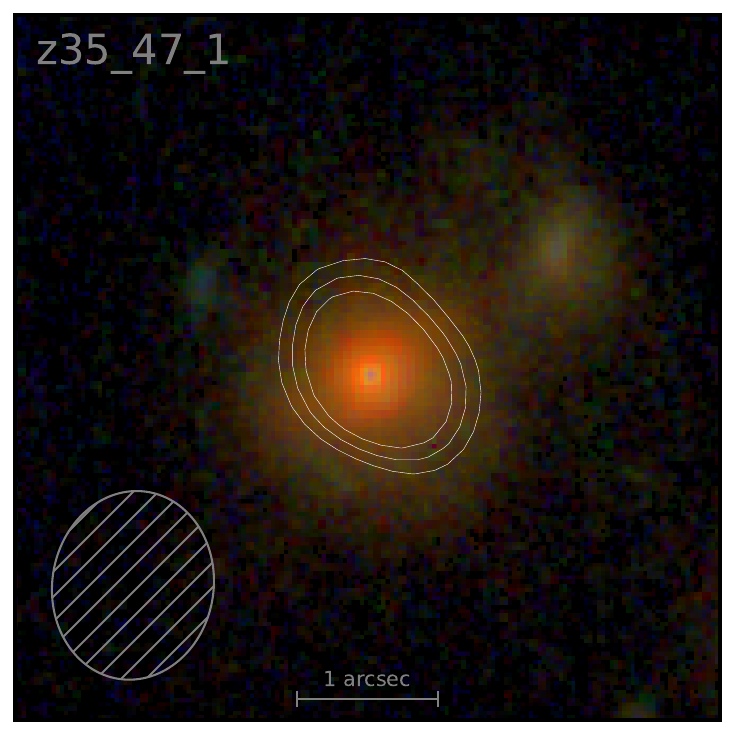}
\includegraphics[width=0.24\textwidth]{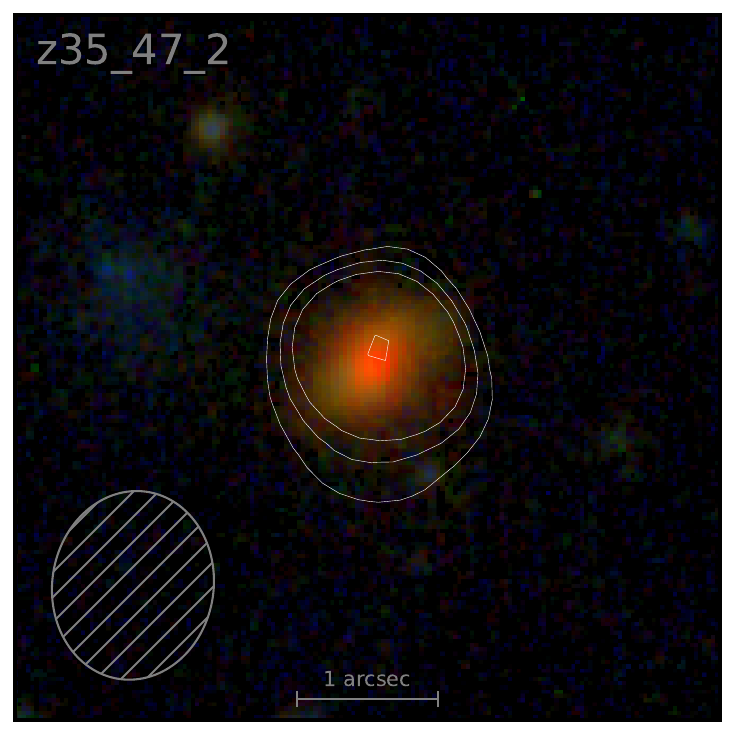}
\end{figure*}
\begin{figure*}
\centering
\includegraphics[width=0.24\textwidth]{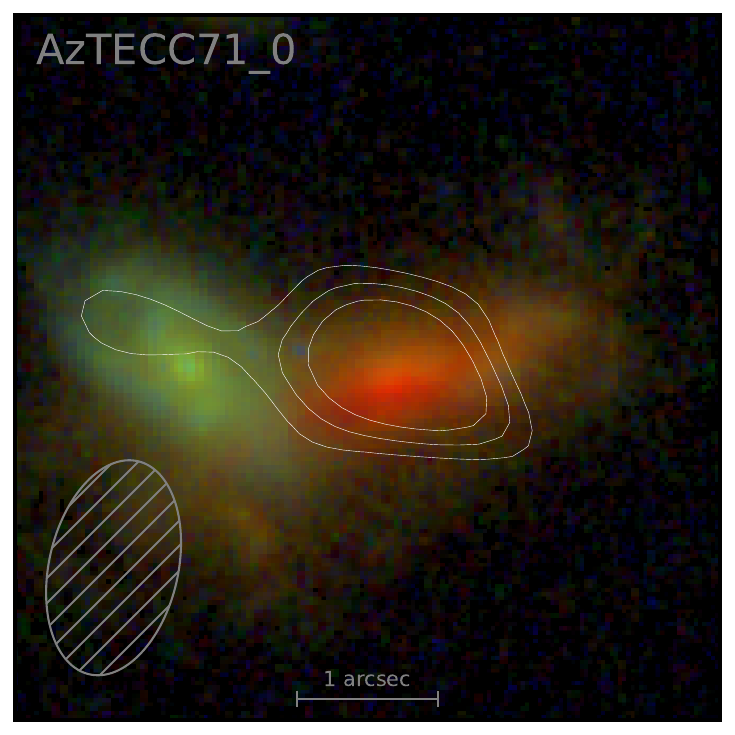}
\includegraphics[width=0.24\textwidth]{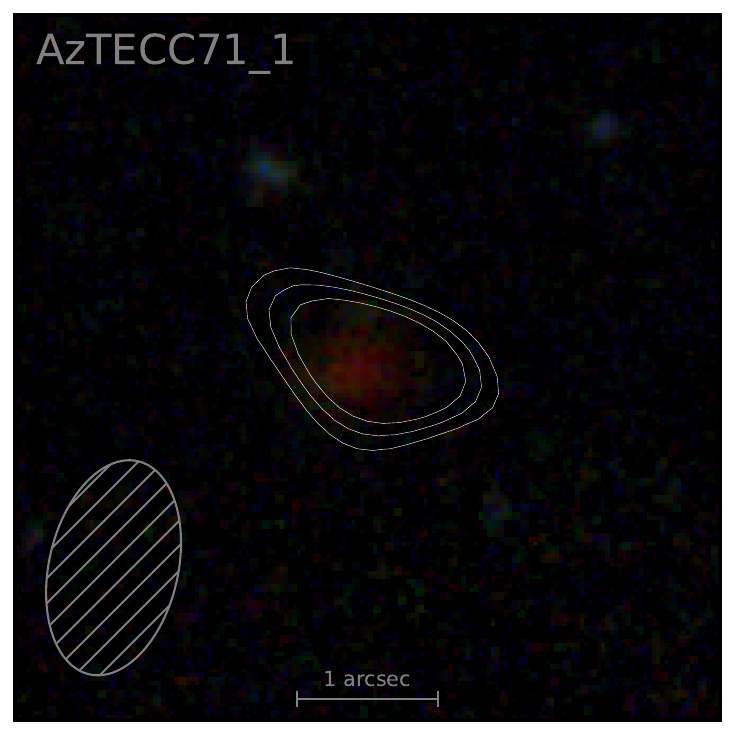}
\includegraphics[width=0.24\textwidth]{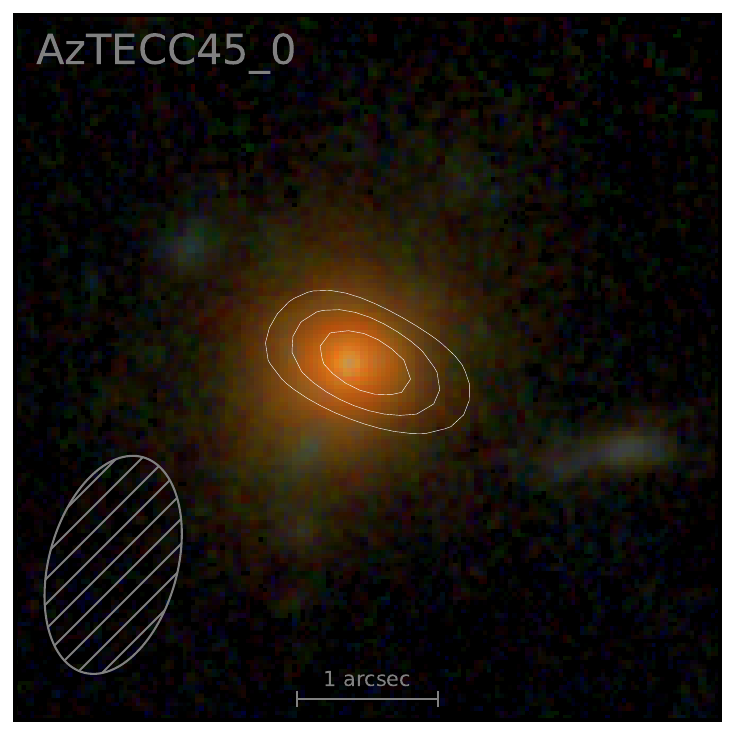}
\includegraphics[width=0.24\textwidth]{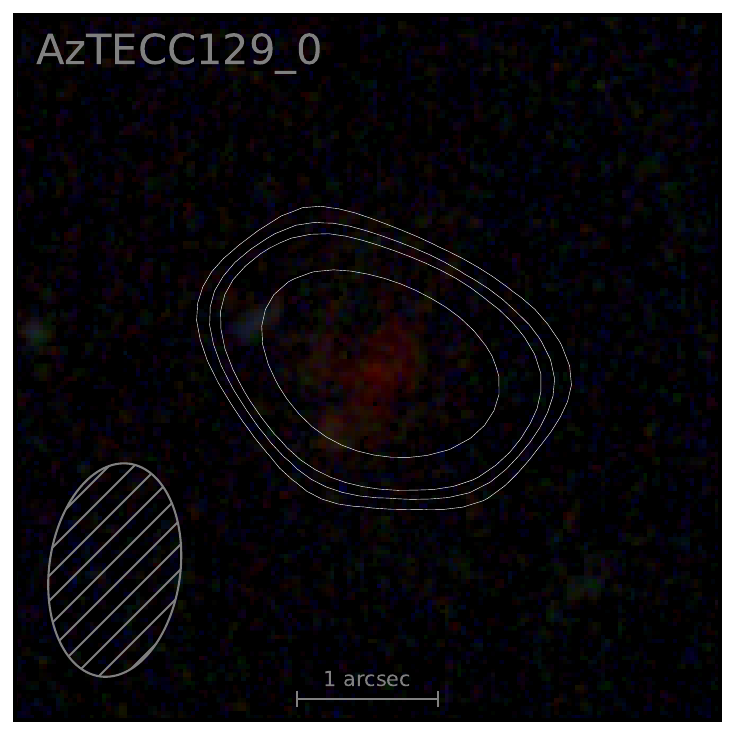}
\includegraphics[width=0.24\textwidth]{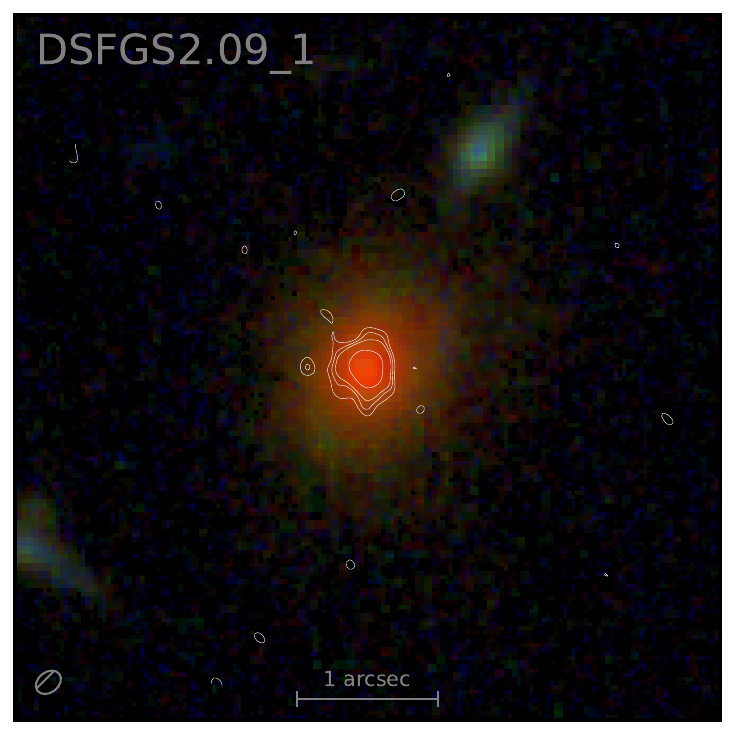}
\includegraphics[width=0.24\textwidth]{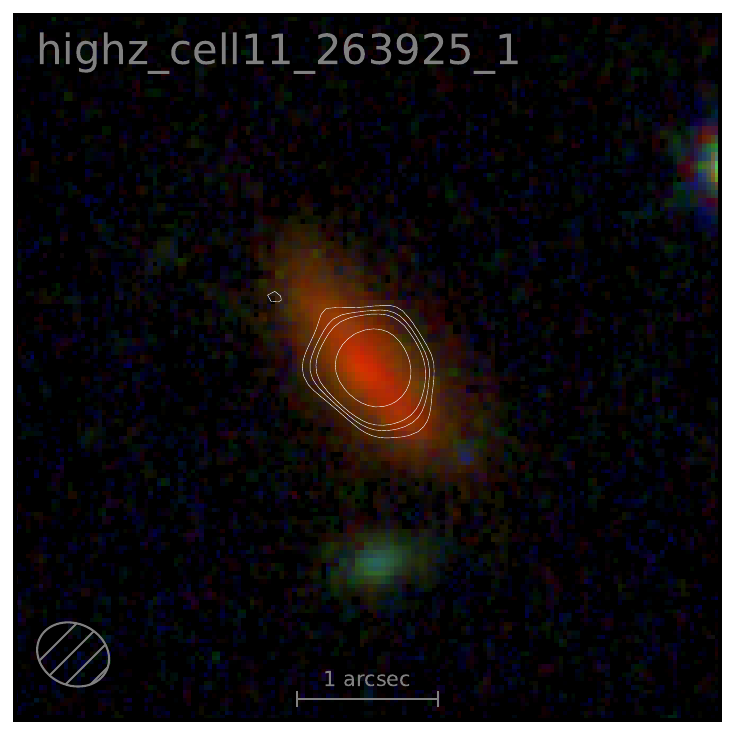}
\includegraphics[width=0.24\textwidth]{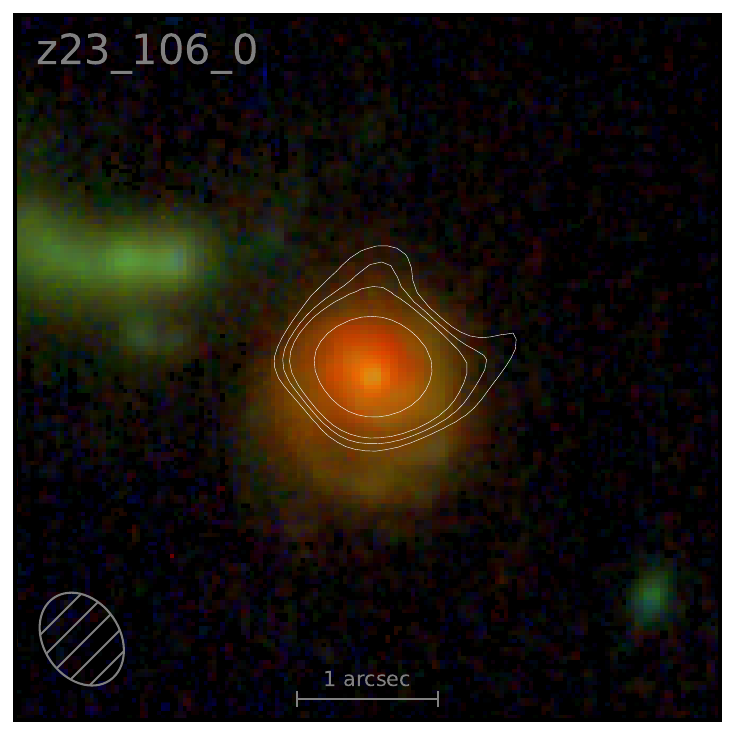}
\includegraphics[width=0.24\textwidth]{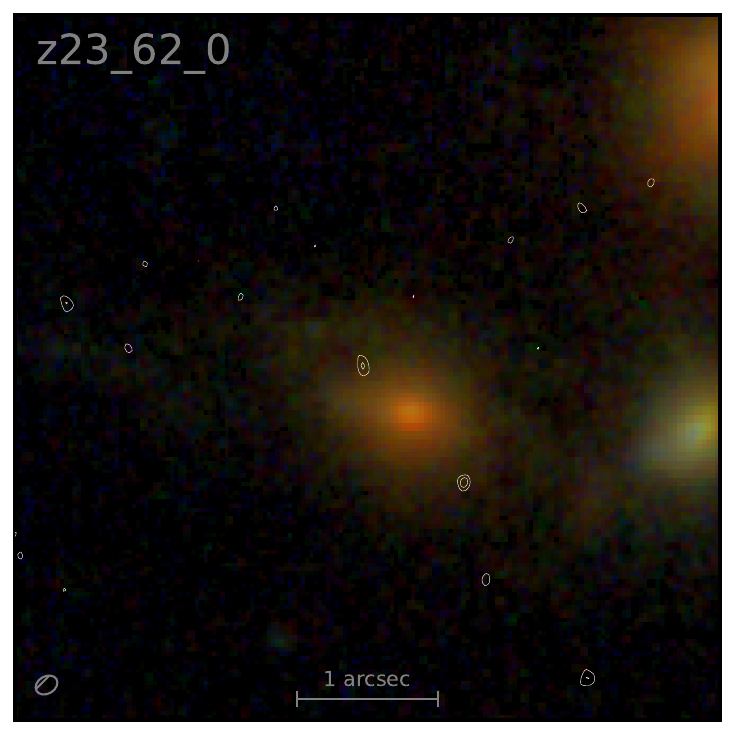}
\includegraphics[width=0.24\textwidth]{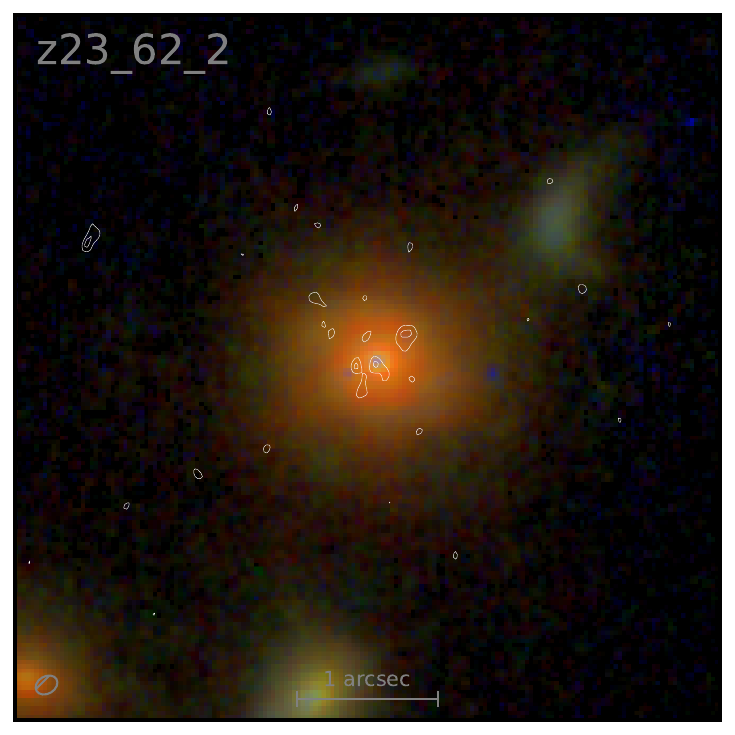}
\includegraphics[width=0.24\textwidth]{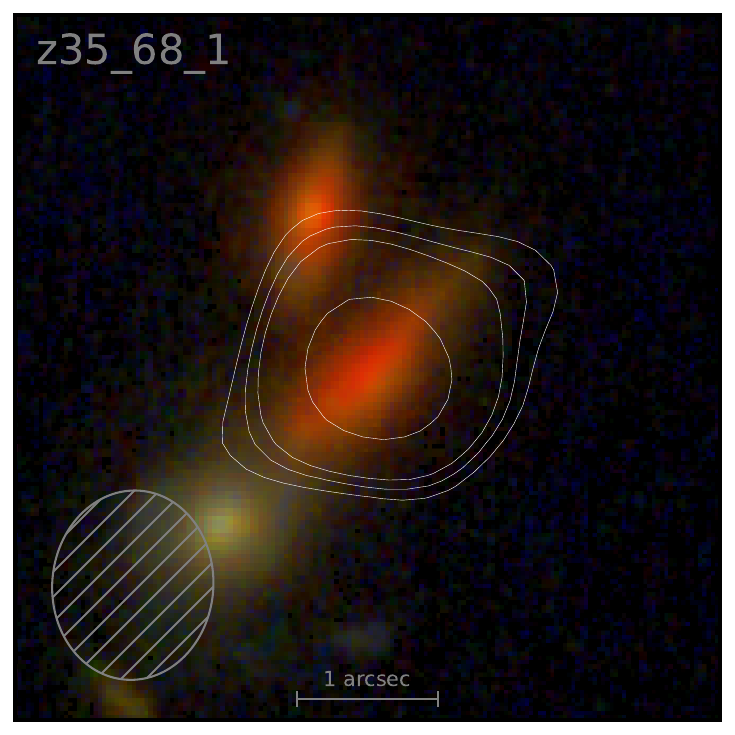}
\includegraphics[width=0.24\textwidth]{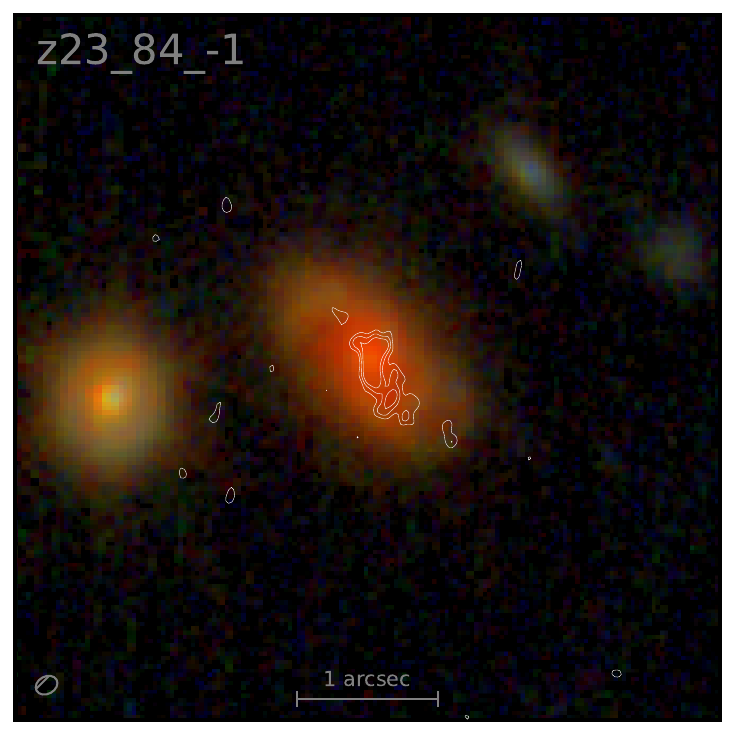}
\includegraphics[width=0.24\textwidth]{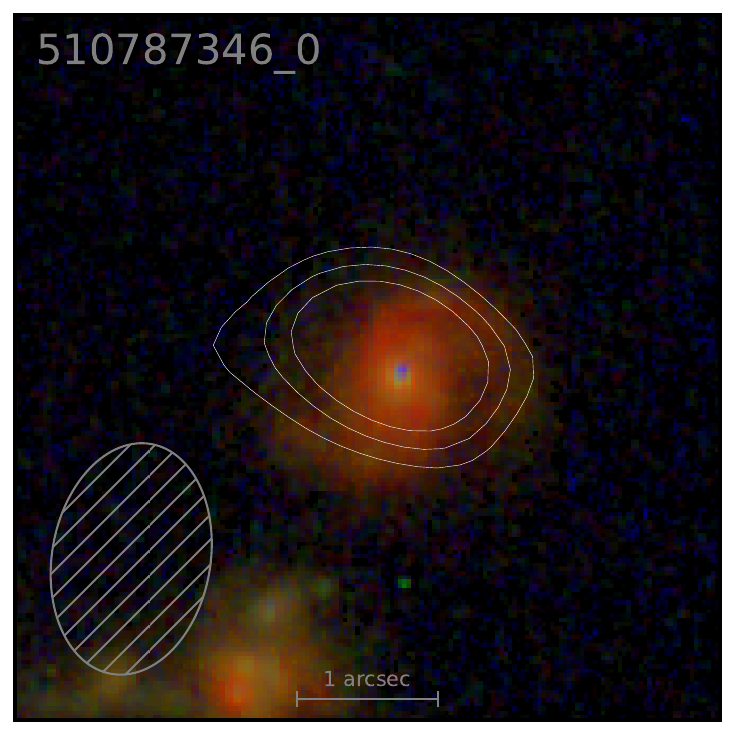}
\includegraphics[width=0.24\textwidth]{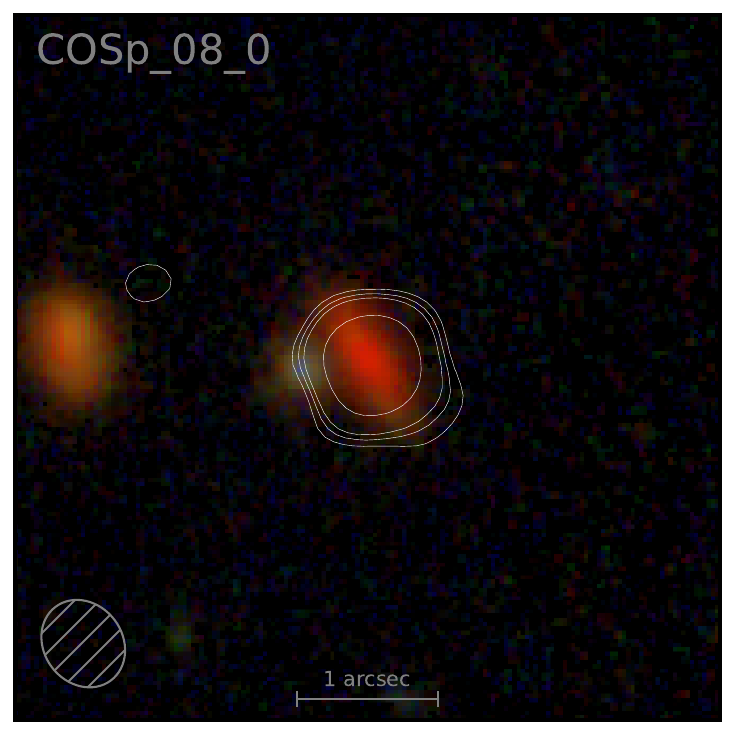}
\includegraphics[width=0.24\textwidth]{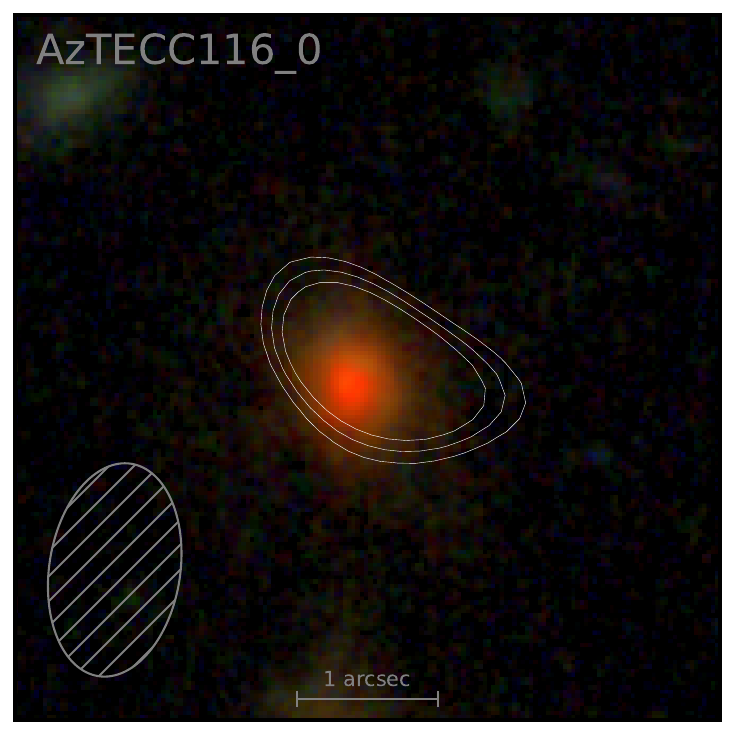}
\includegraphics[width=0.24\textwidth]{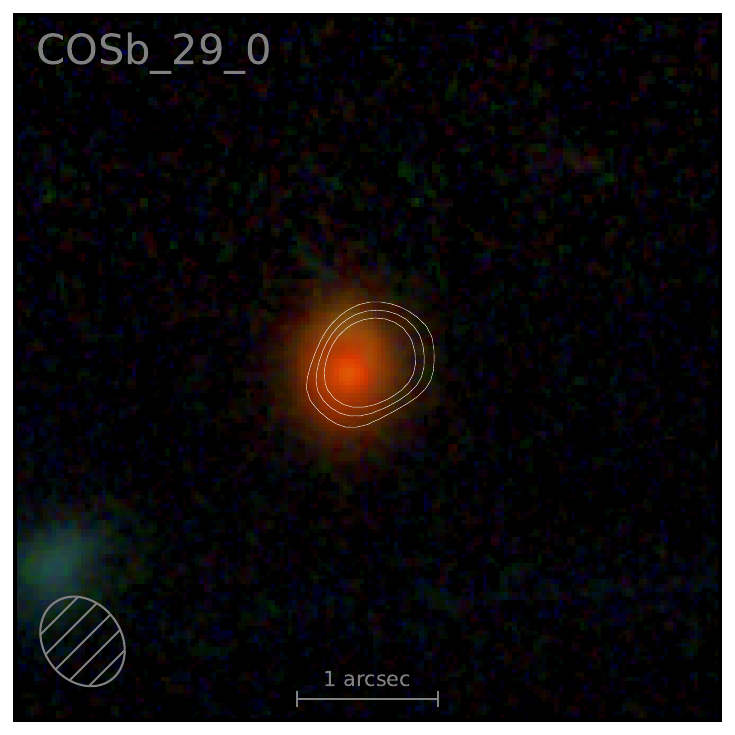}
\includegraphics[width=0.24\textwidth]{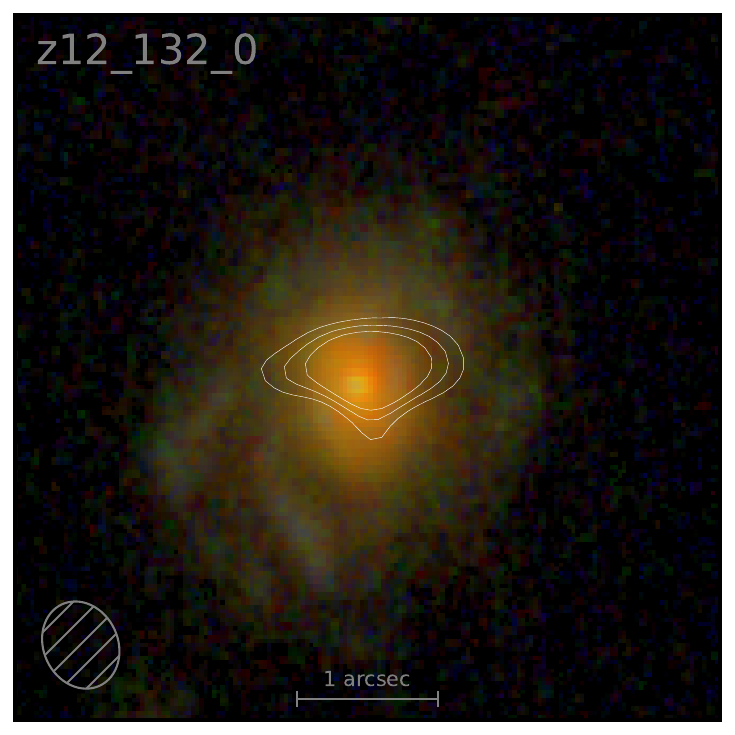}
\includegraphics[width=0.24\textwidth]{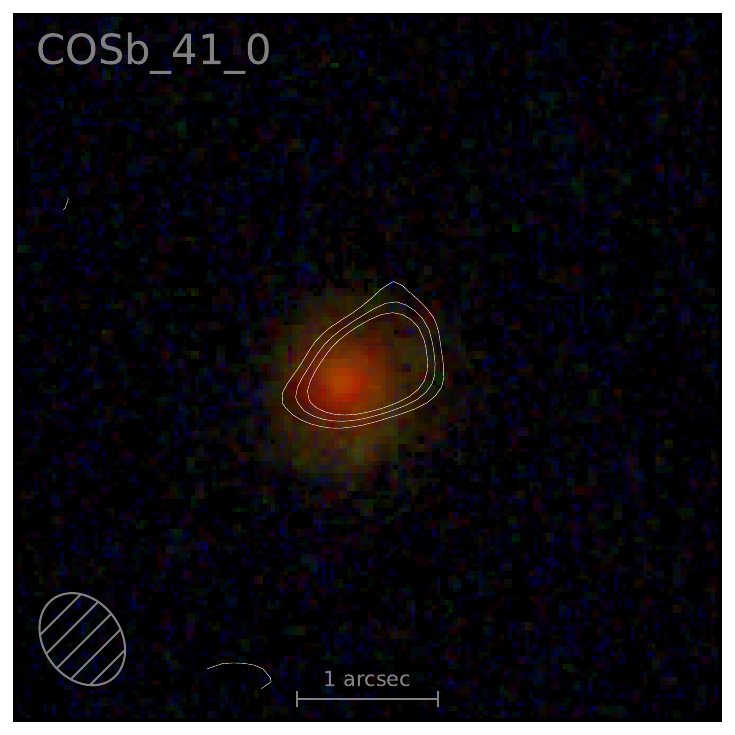}
\includegraphics[width=0.24\textwidth]{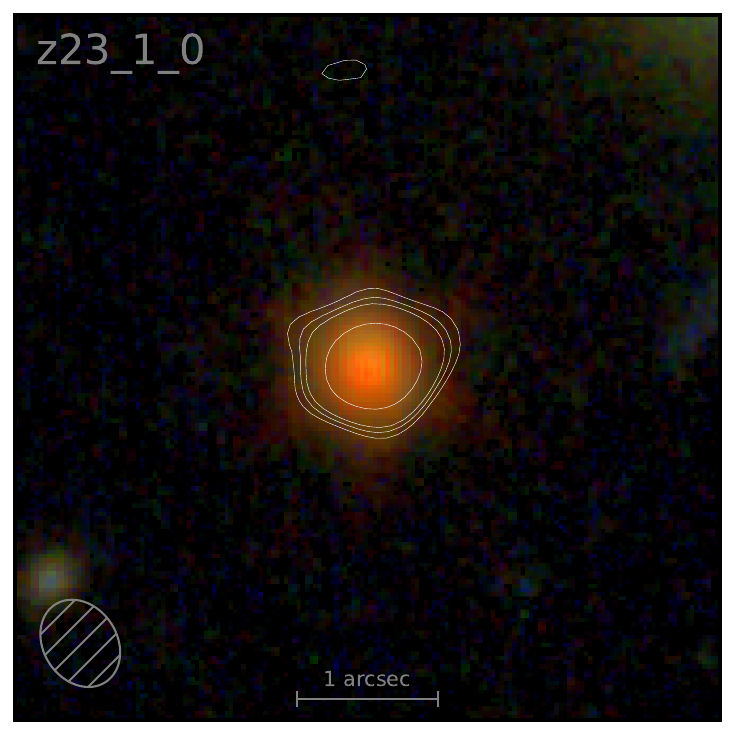}
\includegraphics[width=0.24\textwidth]{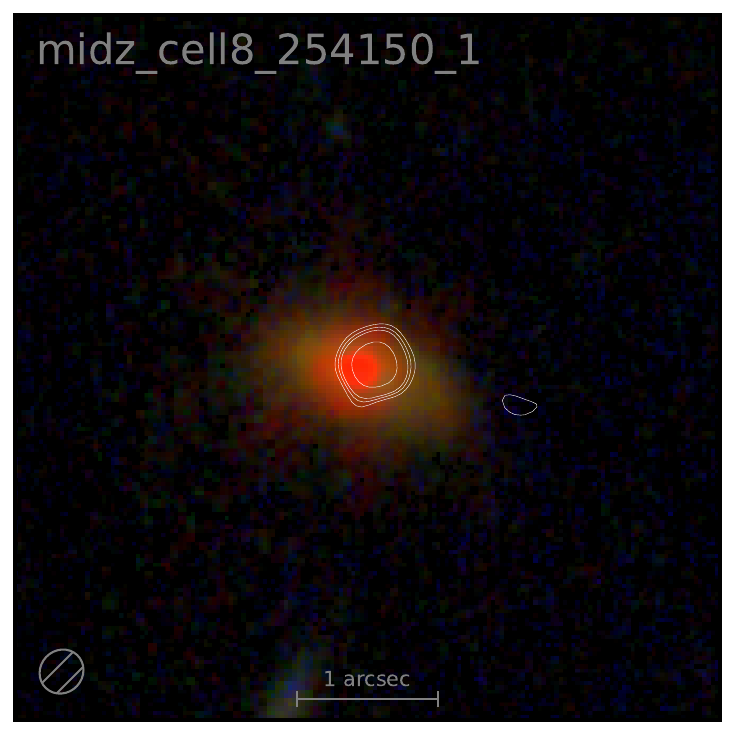}
\includegraphics[width=0.24\textwidth]{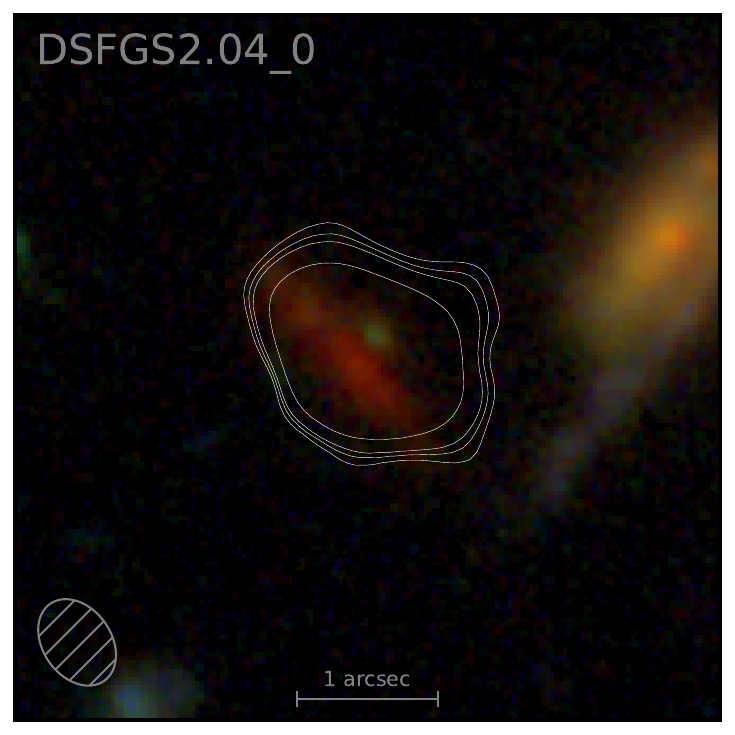}
\end{figure*}
\begin{figure*}
\centering
\includegraphics[width=0.24\textwidth]{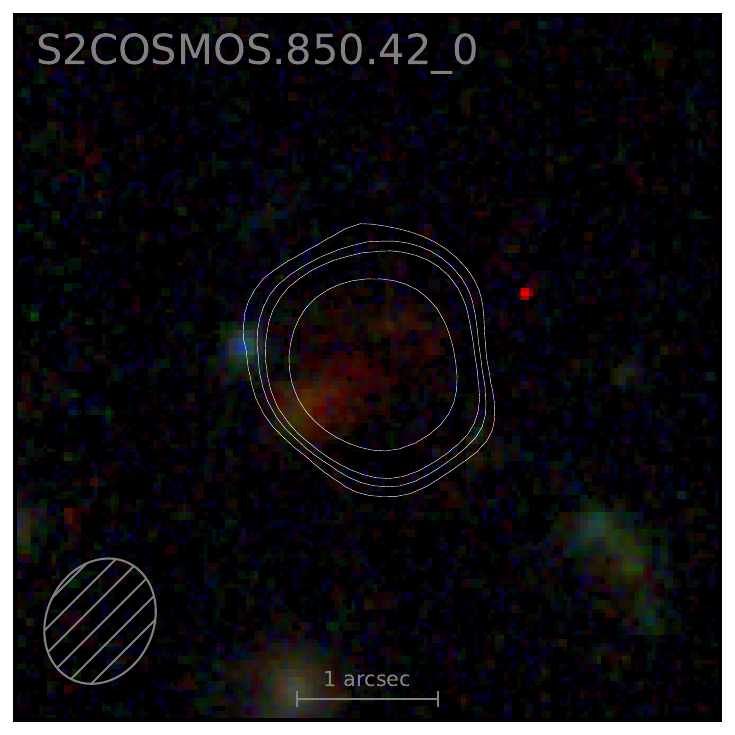}
\includegraphics[width=0.24\textwidth]{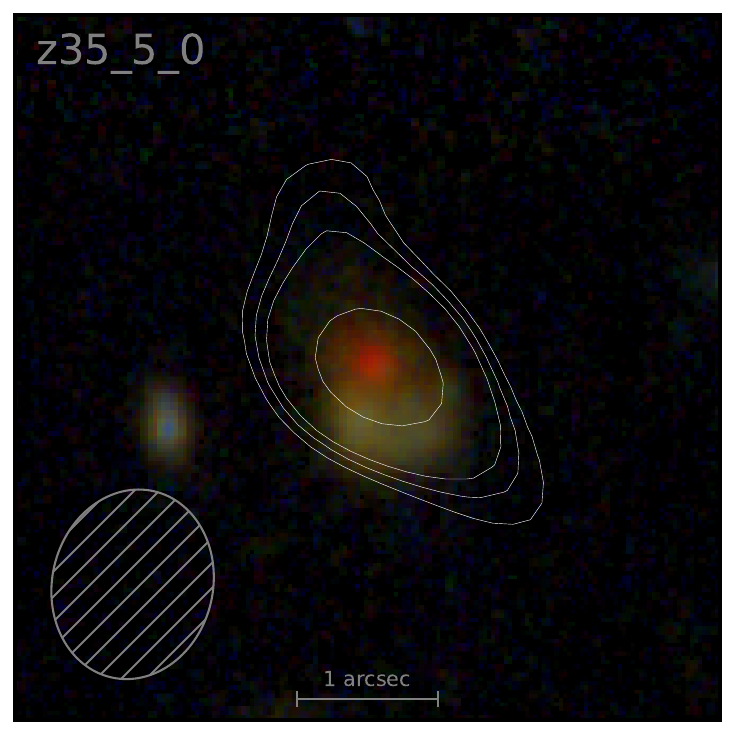}
\includegraphics[width=0.24\textwidth]{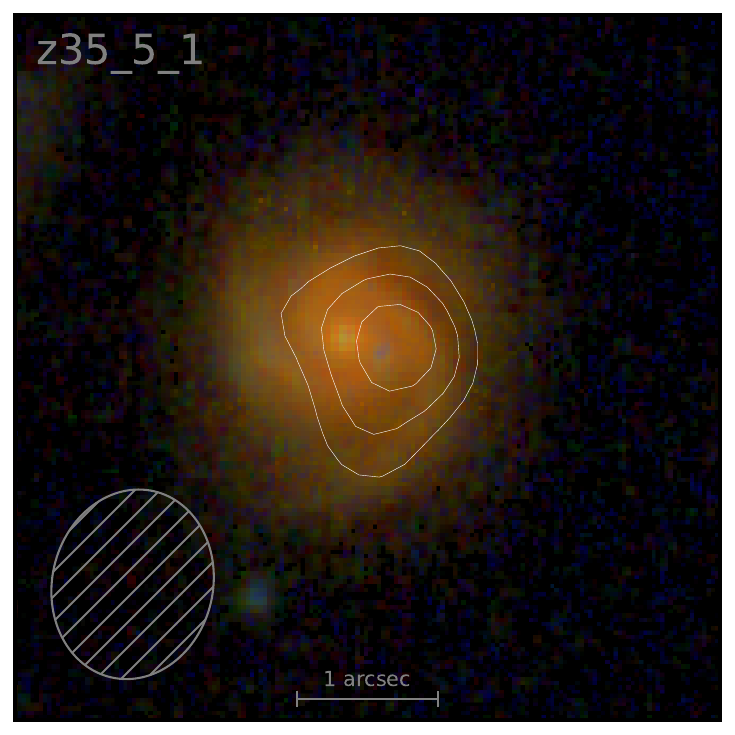}
\includegraphics[width=0.24\textwidth]{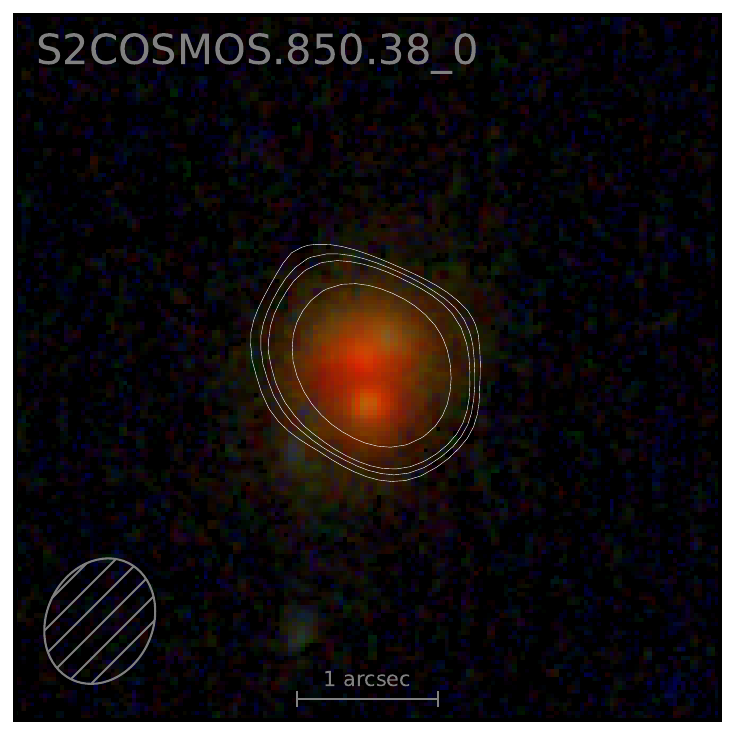}
\includegraphics[width=0.24\textwidth]{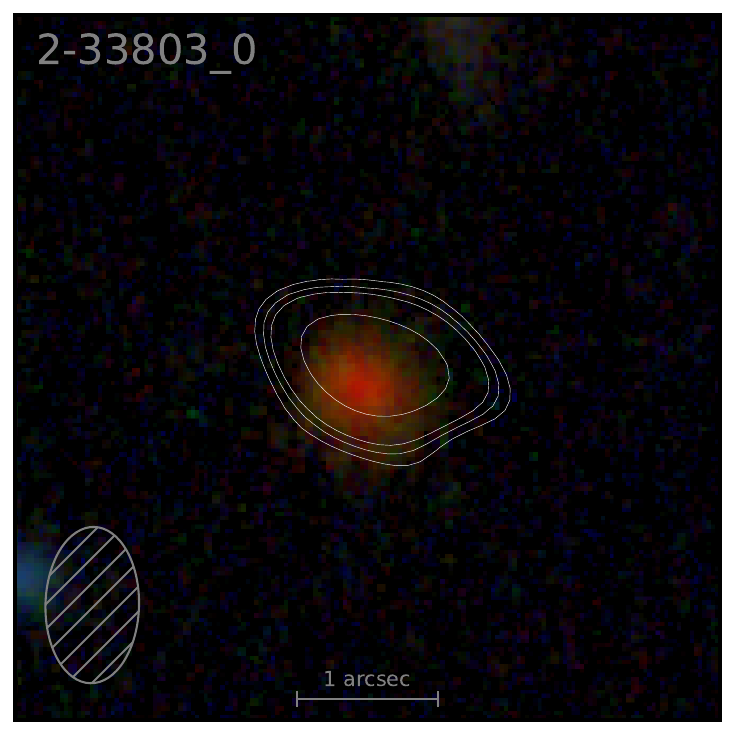}
\includegraphics[width=0.24\textwidth]{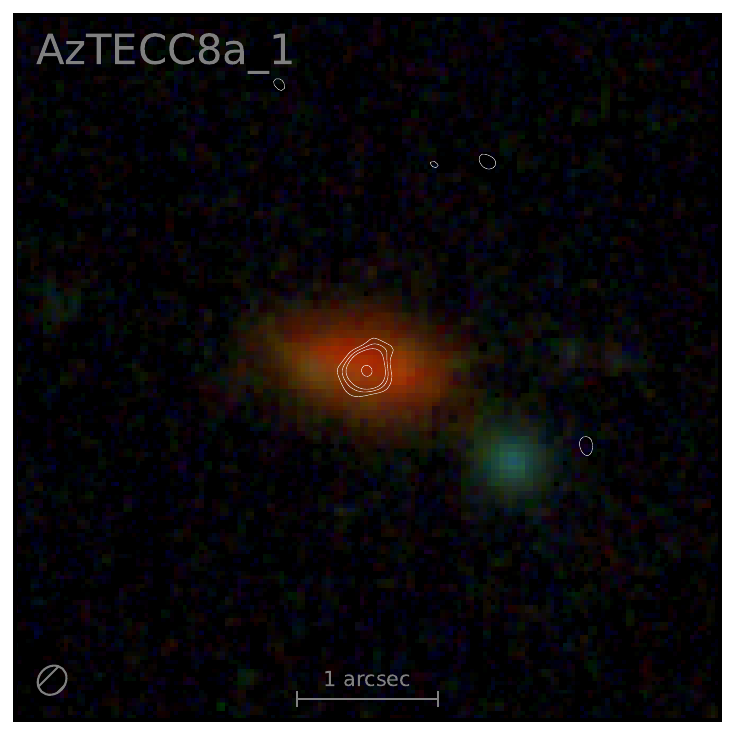}
\includegraphics[width=0.24\textwidth]{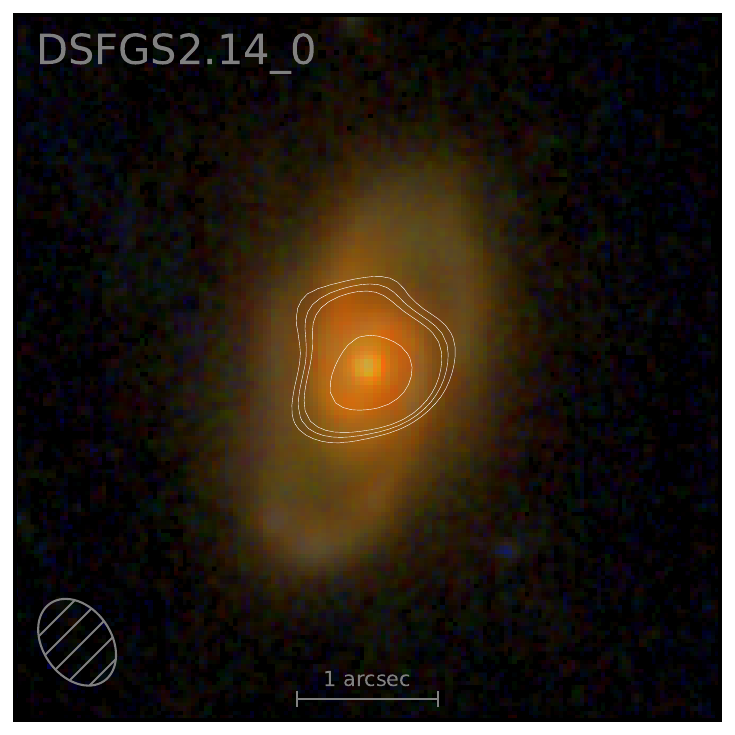}
\includegraphics[width=0.24\textwidth]{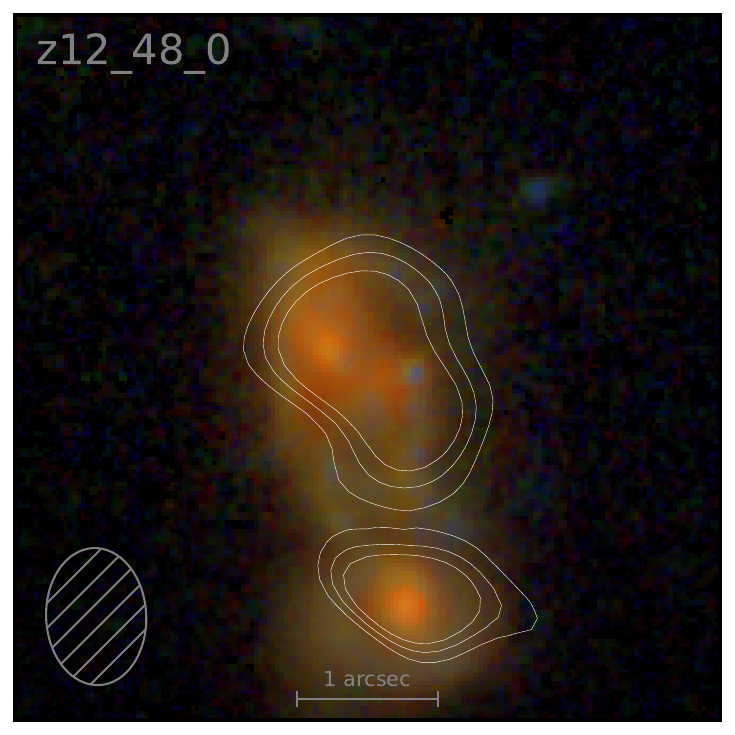}
\includegraphics[width=0.24\textwidth]{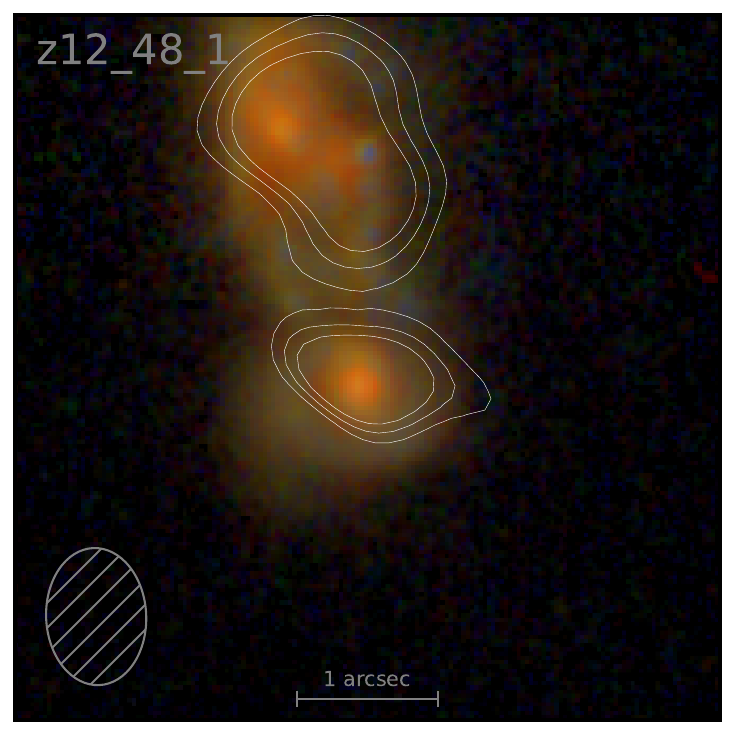}
\includegraphics[width=0.24\textwidth]{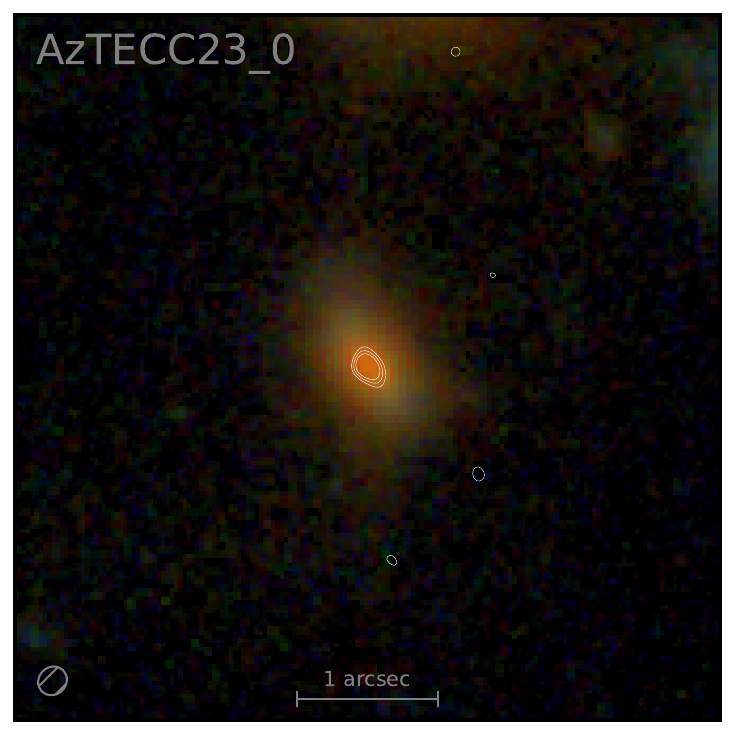}
\includegraphics[width=0.24\textwidth]{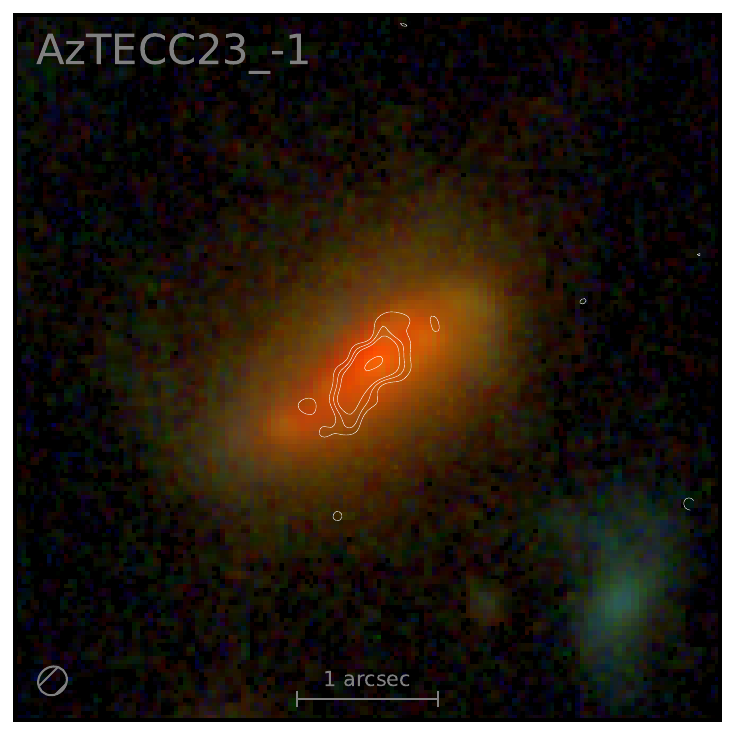}
\includegraphics[width=0.24\textwidth]{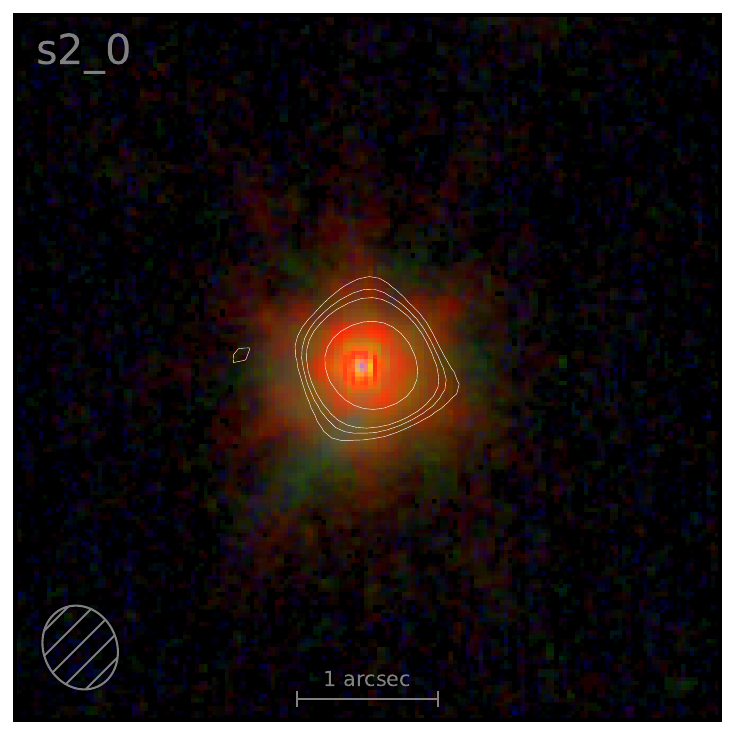}
\includegraphics[width=0.24\textwidth]{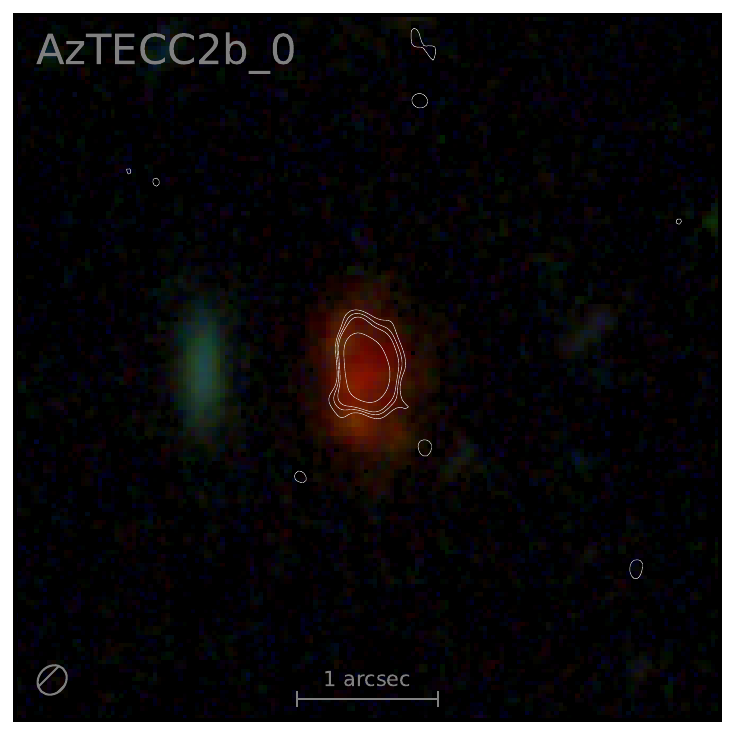}
\includegraphics[width=0.24\textwidth]{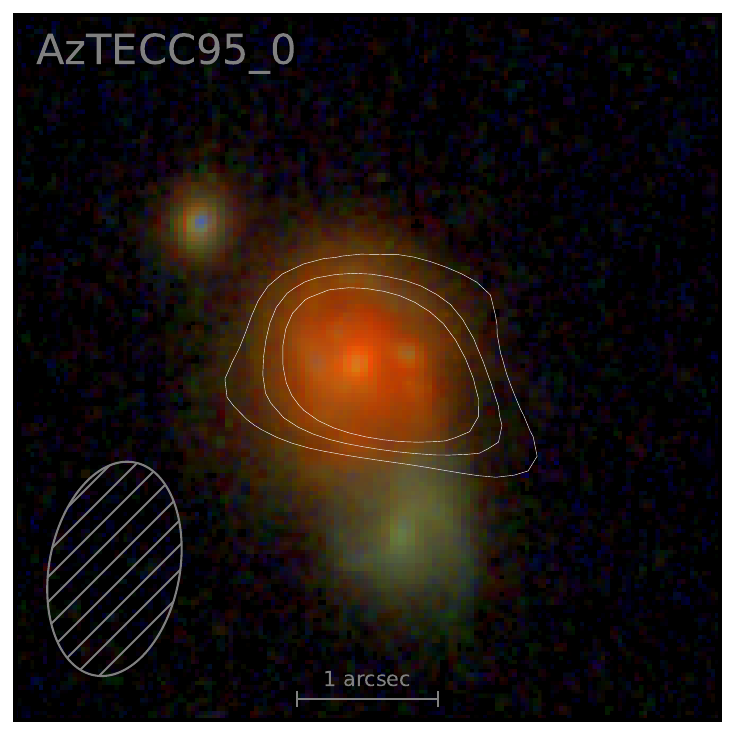}
\includegraphics[width=0.24\textwidth]{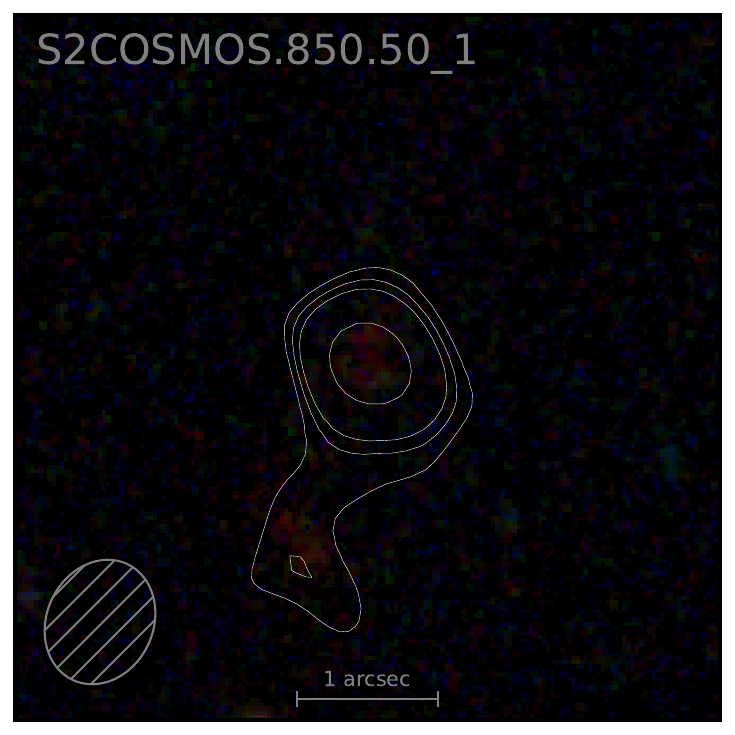}
\includegraphics[width=0.24\textwidth]{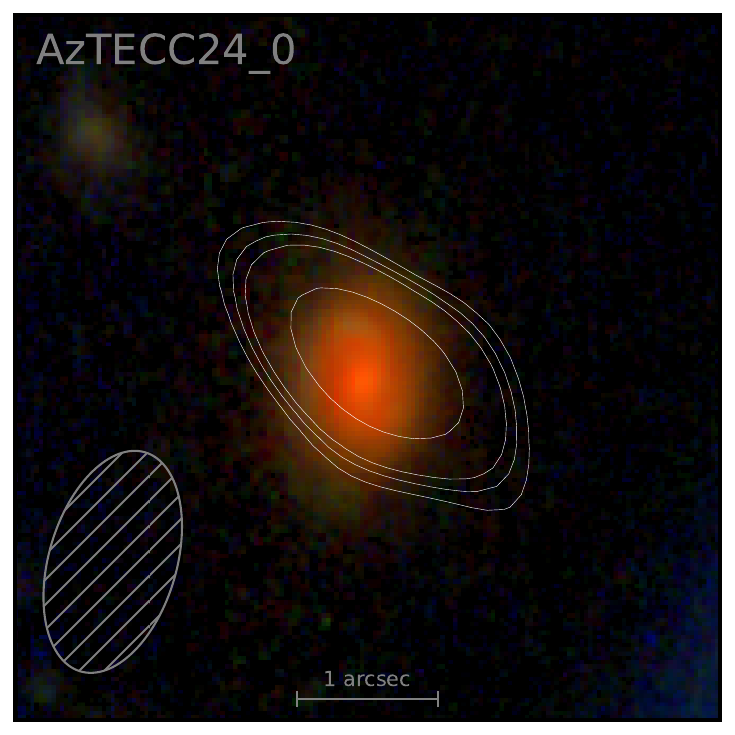}
\includegraphics[width=0.24\textwidth]{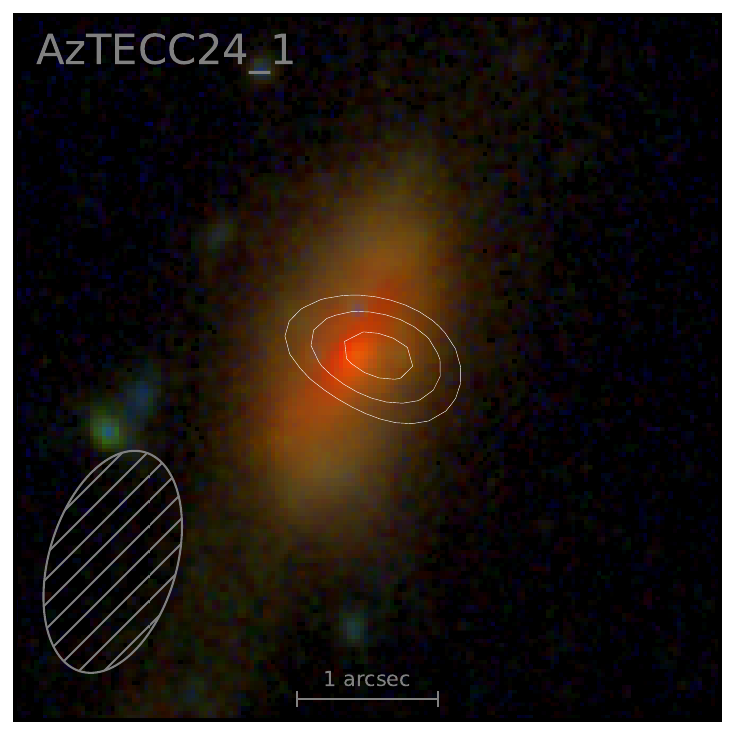}
\includegraphics[width=0.24\textwidth]{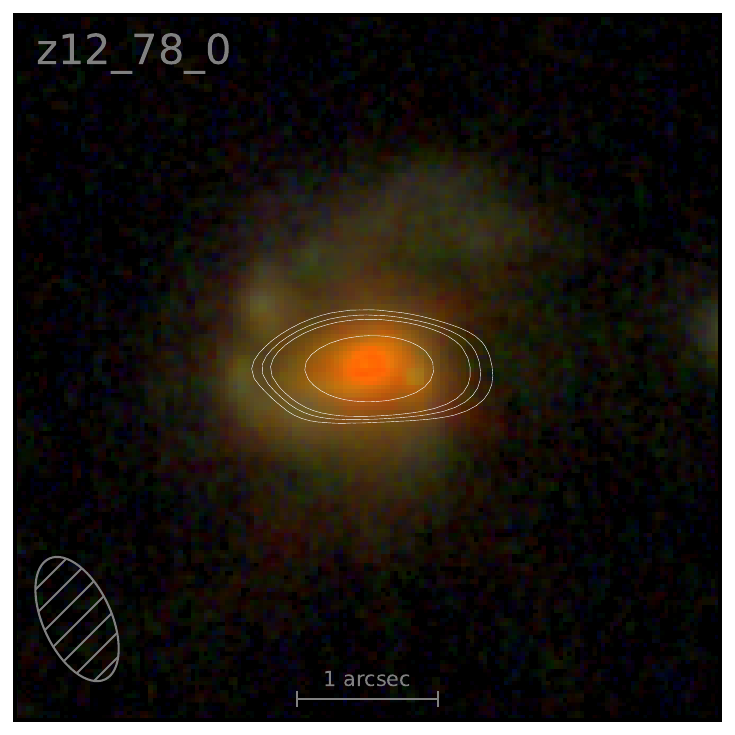}
\includegraphics[width=0.24\textwidth]{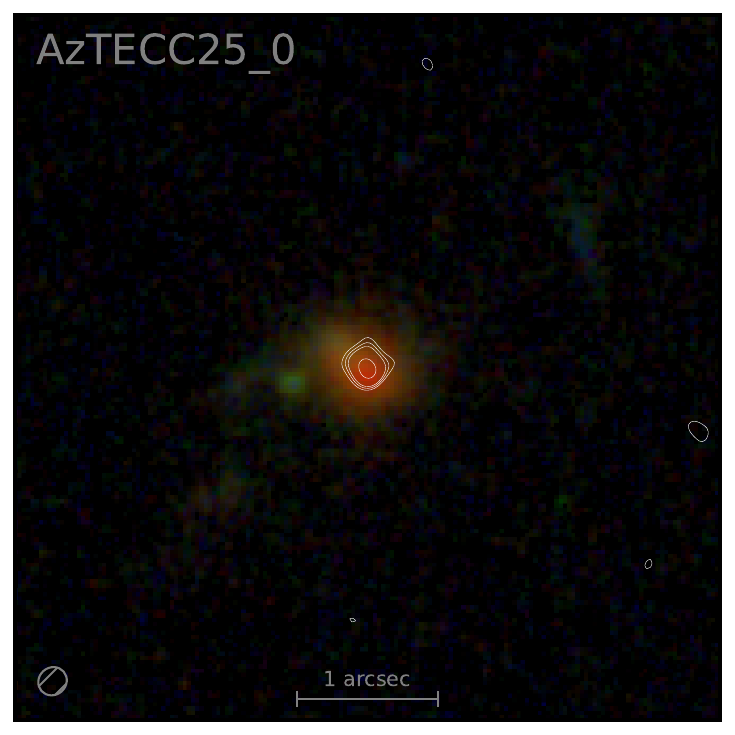}
\includegraphics[width=0.24\textwidth]{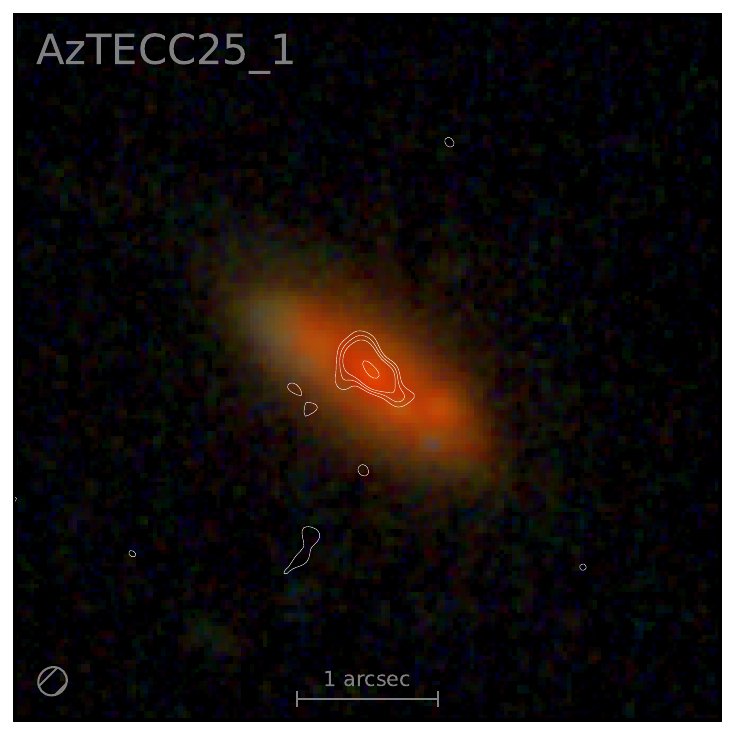}
\end{figure*}
\begin{figure*}
\centering
\includegraphics[width=0.24\textwidth]{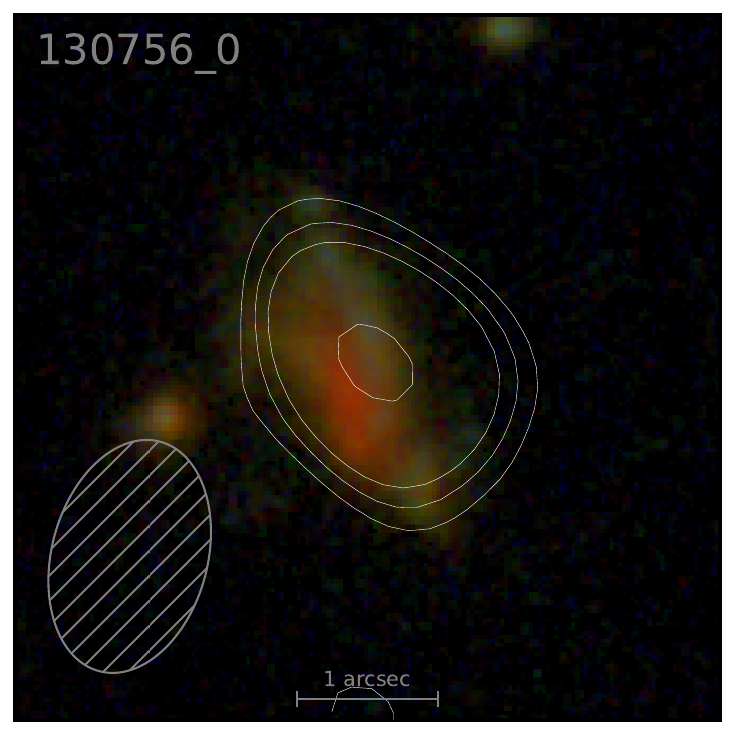}
\includegraphics[width=0.24\textwidth]{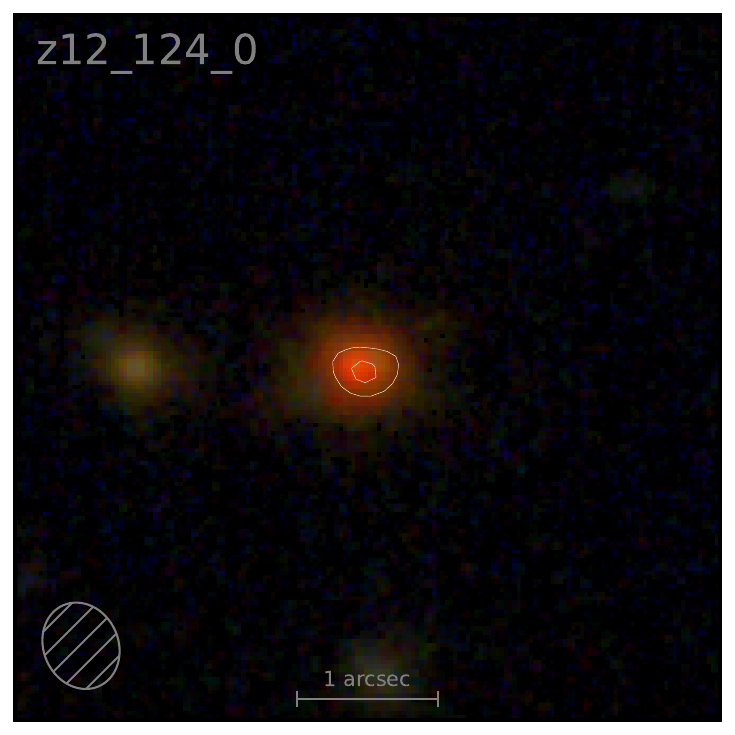}
\includegraphics[width=0.24\textwidth]{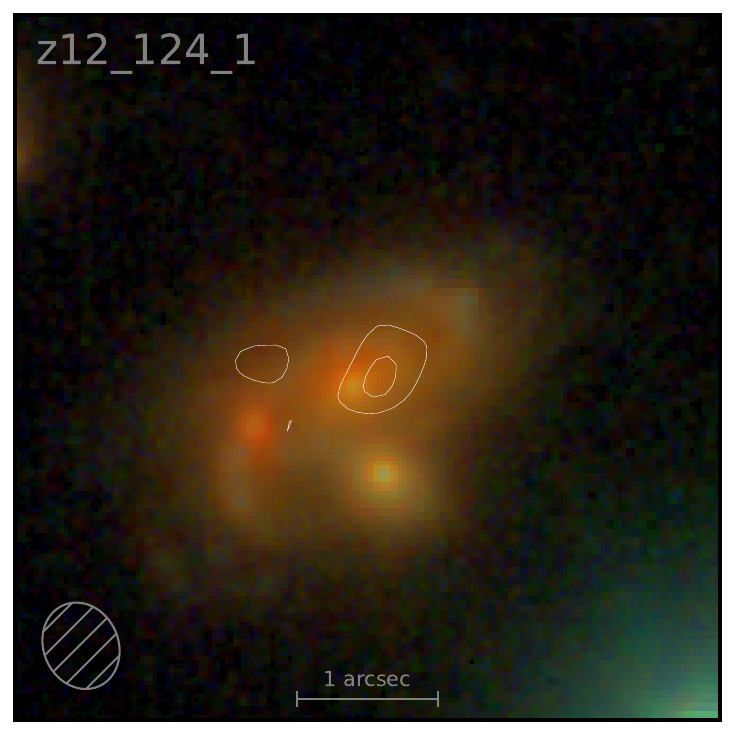}
\includegraphics[width=0.24\textwidth]{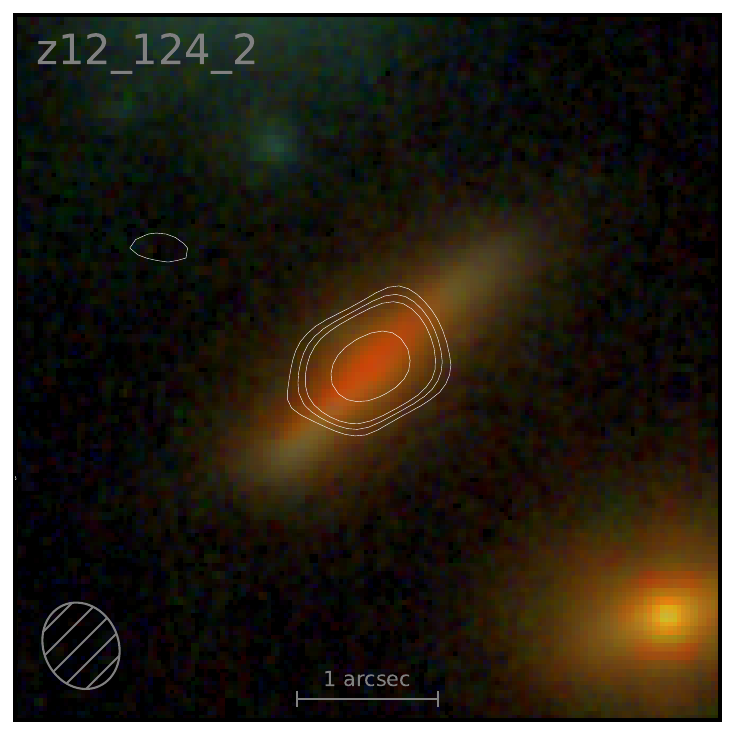}
\includegraphics[width=0.24\textwidth]{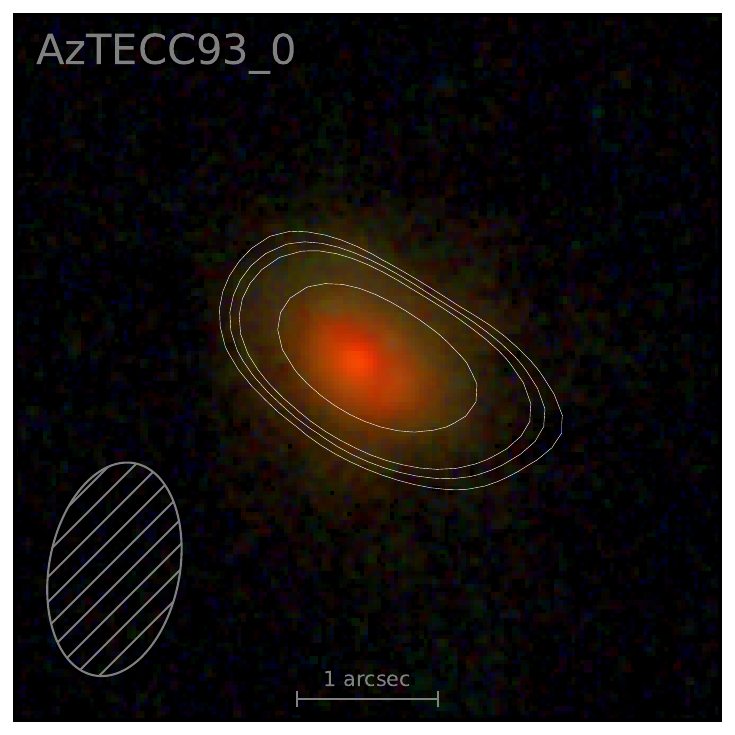}
\includegraphics[width=0.24\textwidth]{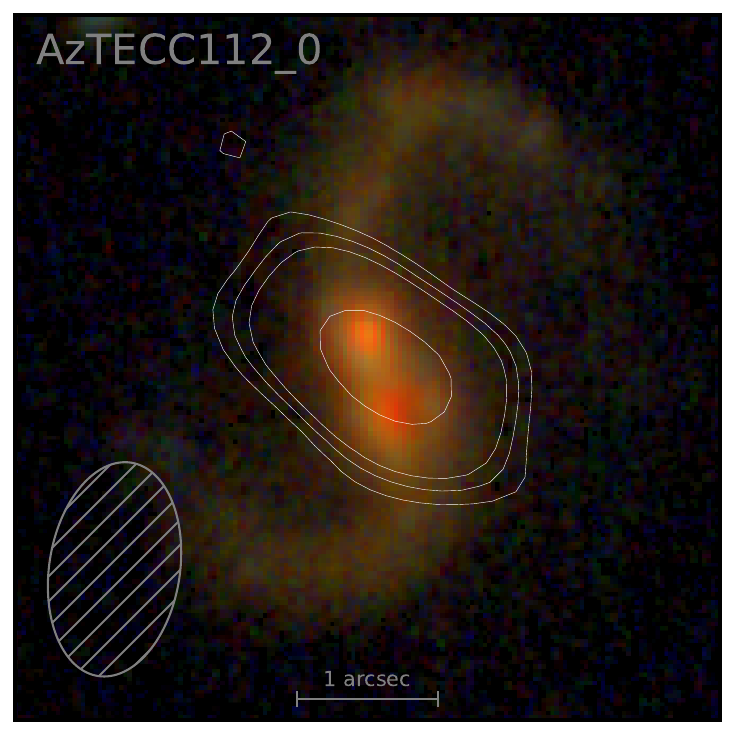}
\includegraphics[width=0.24\textwidth]{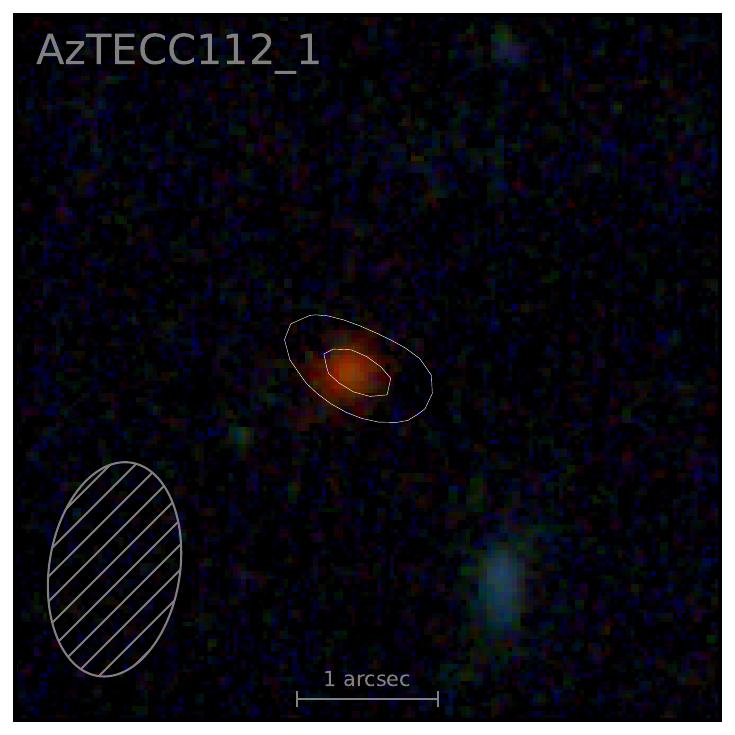}
\includegraphics[width=0.24\textwidth]{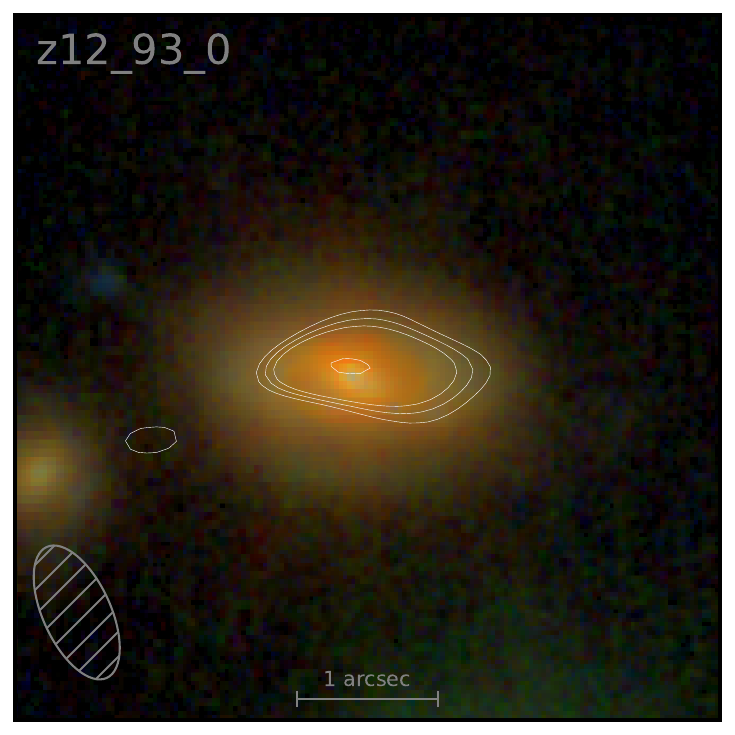}
\includegraphics[width=0.24\textwidth]{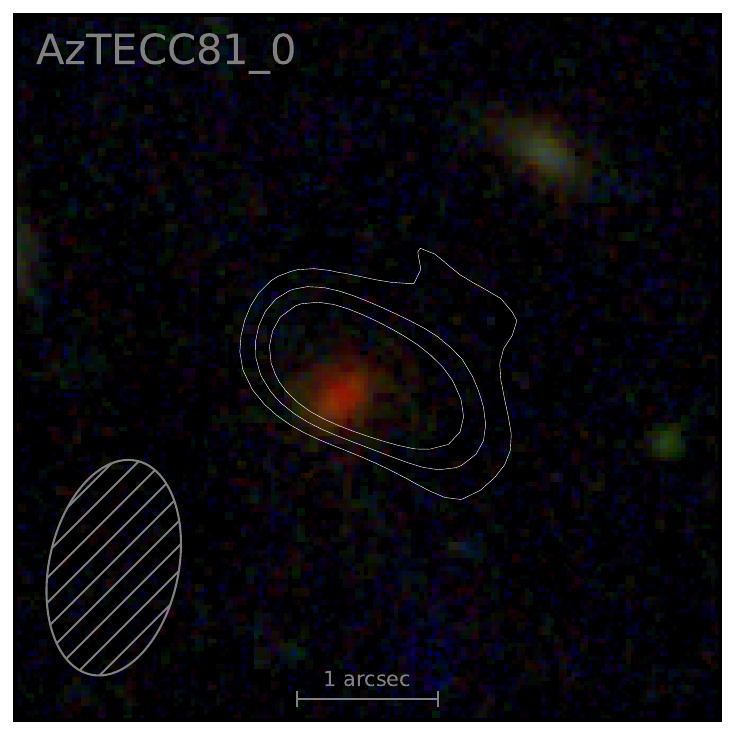}
\includegraphics[width=0.24\textwidth]{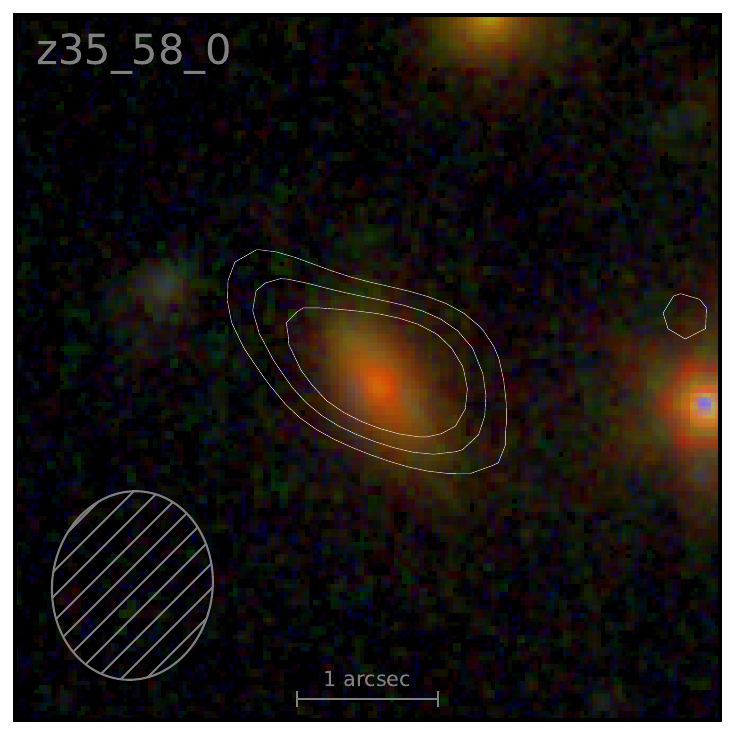}
\includegraphics[width=0.24\textwidth]{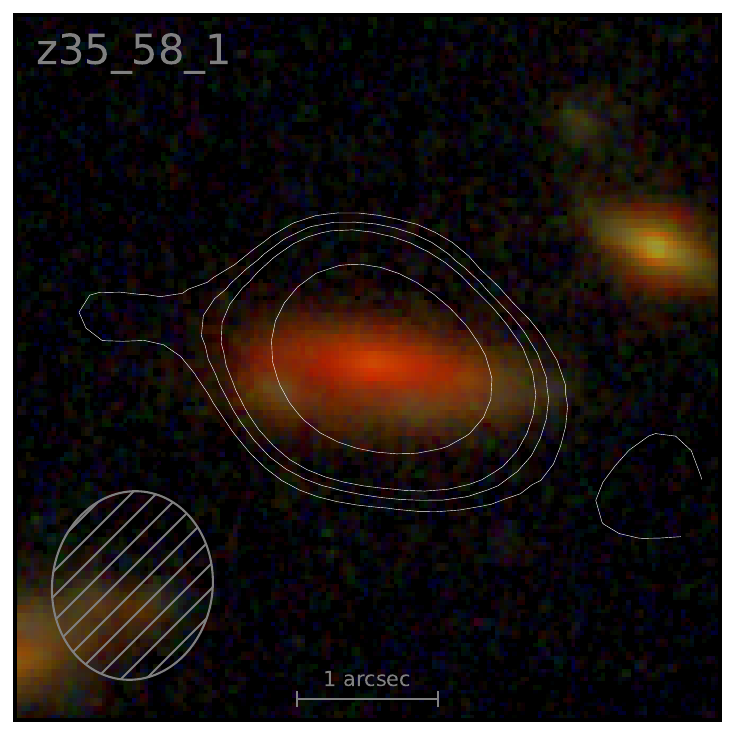}
\includegraphics[width=0.24\textwidth]{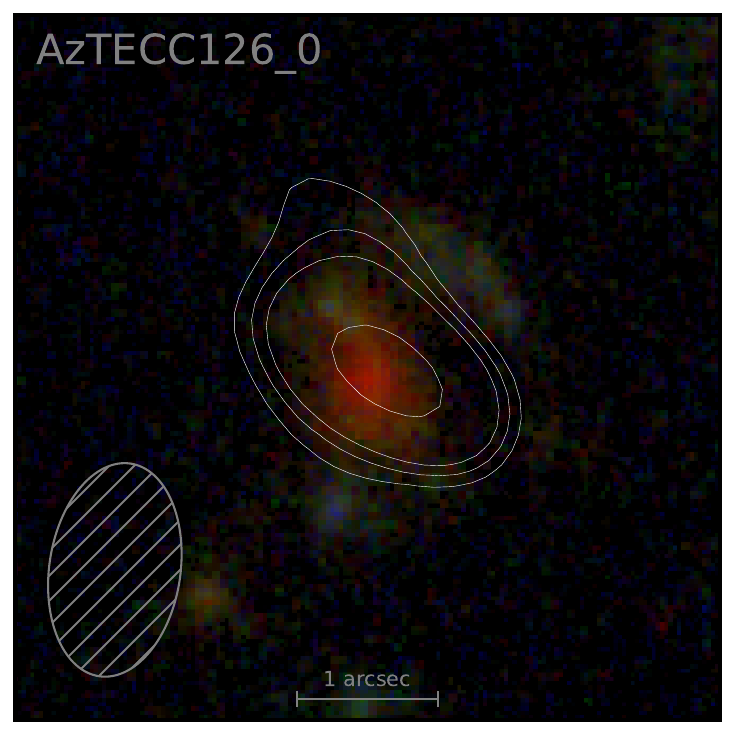}
\includegraphics[width=0.24\textwidth]{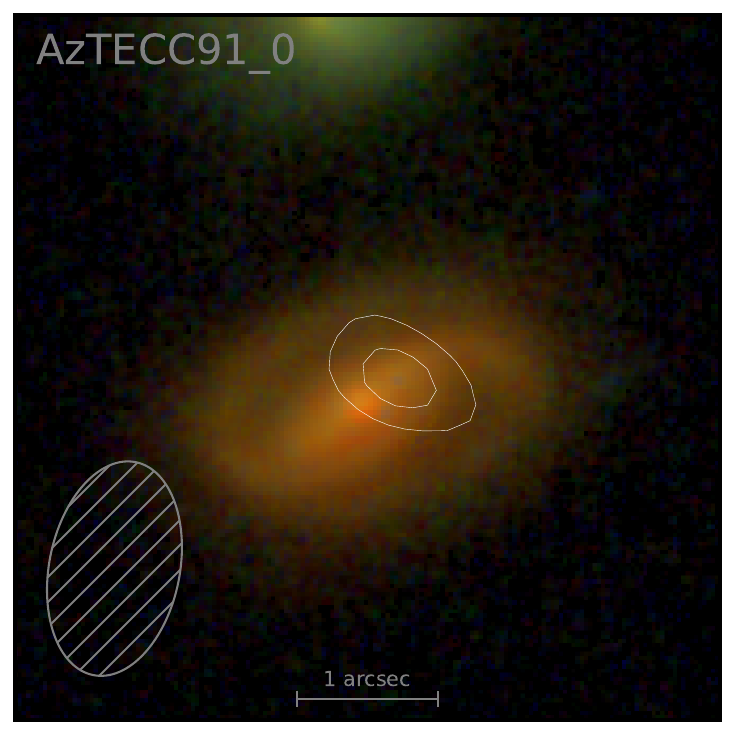}
\includegraphics[width=0.24\textwidth]{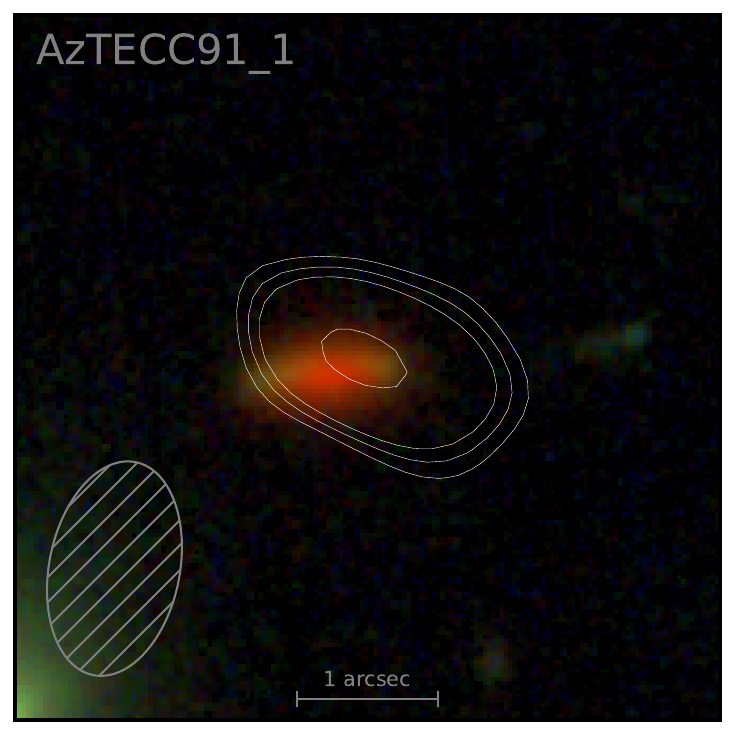}
\includegraphics[width=0.24\textwidth]{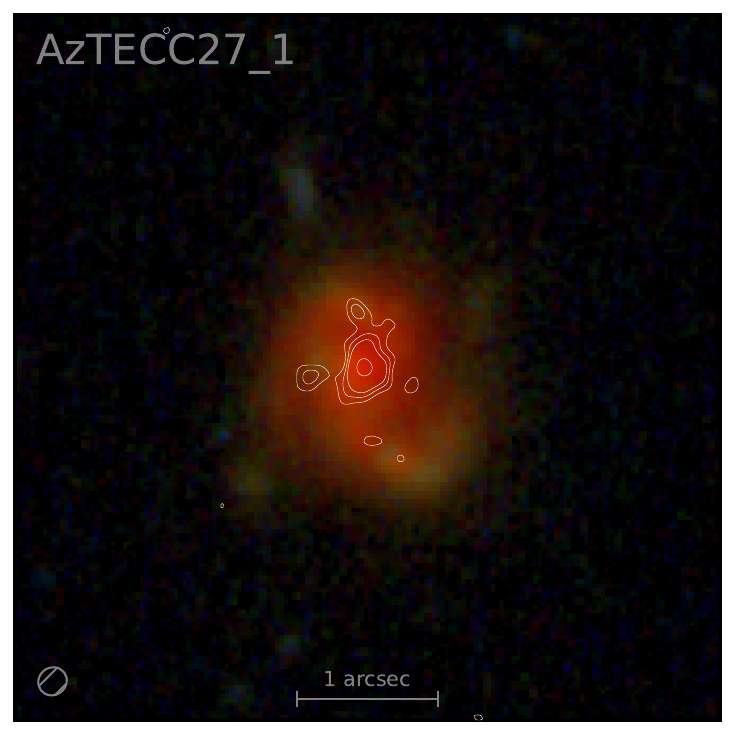}
\includegraphics[width=0.24\textwidth]{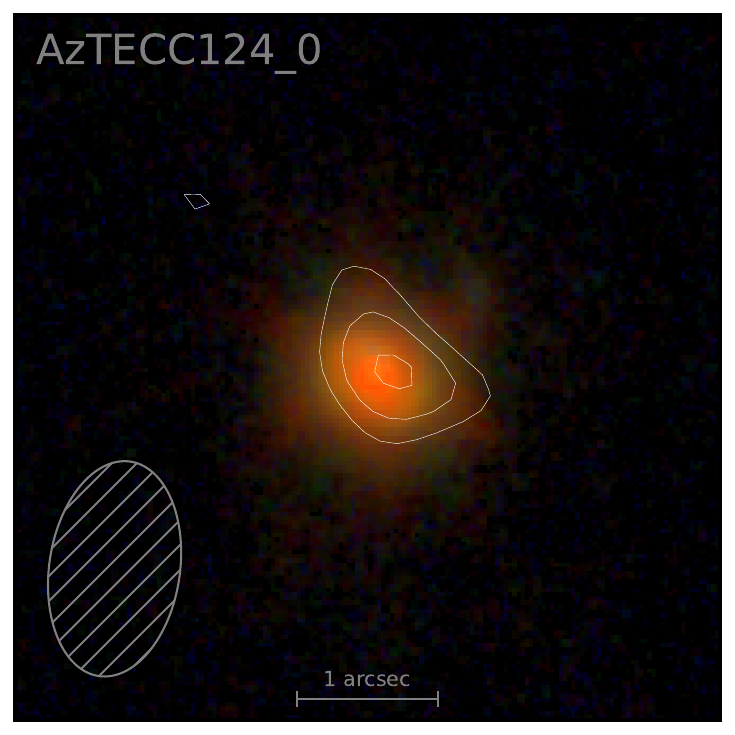}
\includegraphics[width=0.24\textwidth]{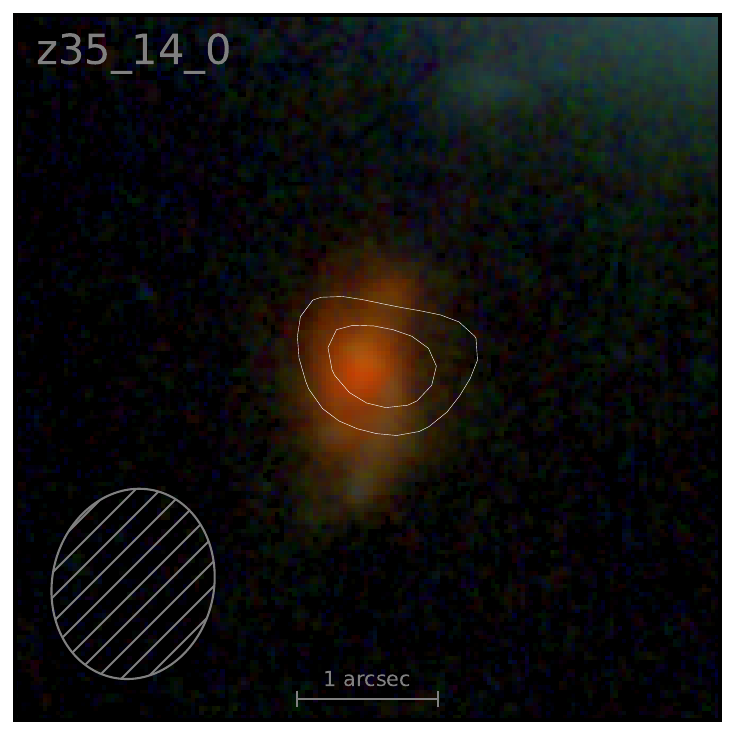}
\includegraphics[width=0.24\textwidth]{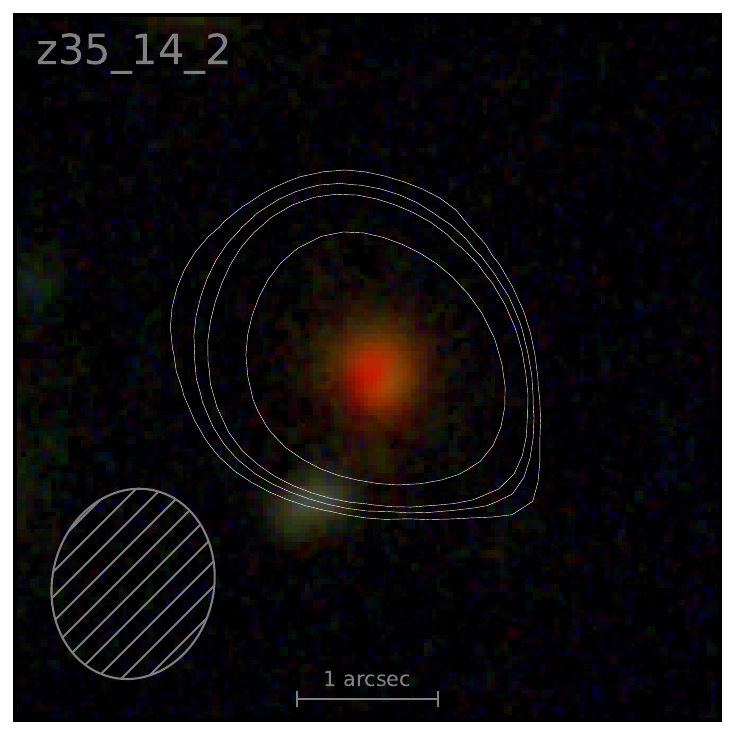}
\includegraphics[width=0.24\textwidth]{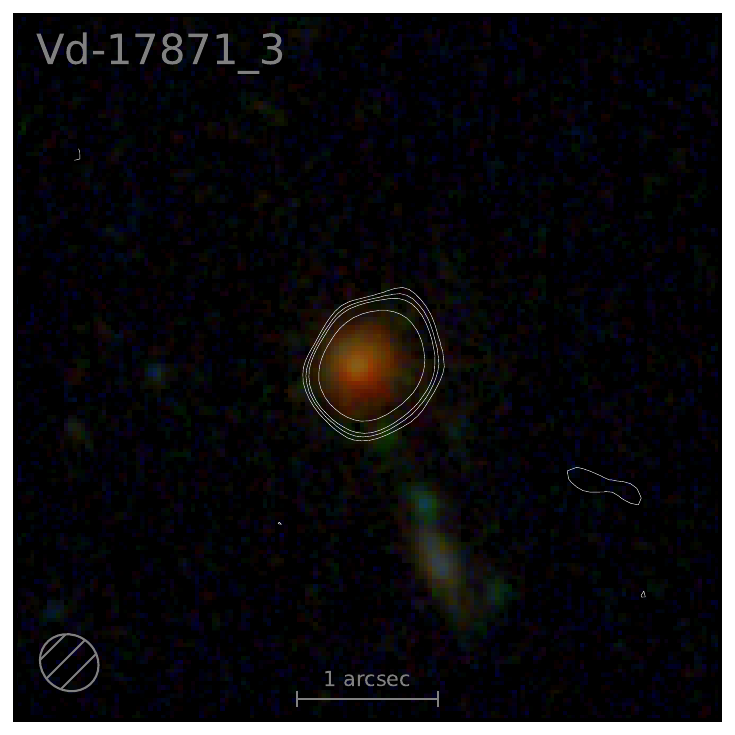}
\includegraphics[width=0.24\textwidth]{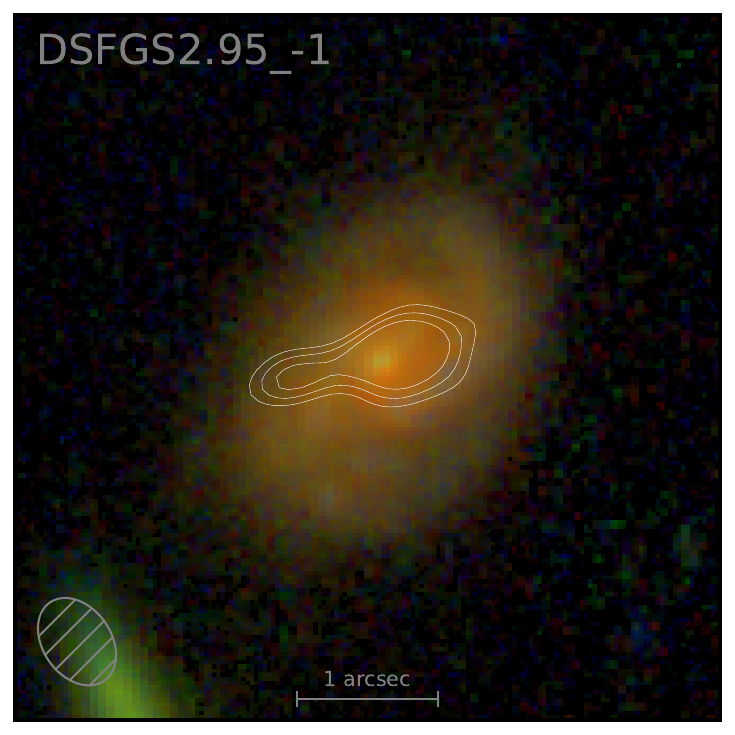}
\end{figure*}
\begin{figure*}
\centering
    \includegraphics[width=0.24\textwidth]{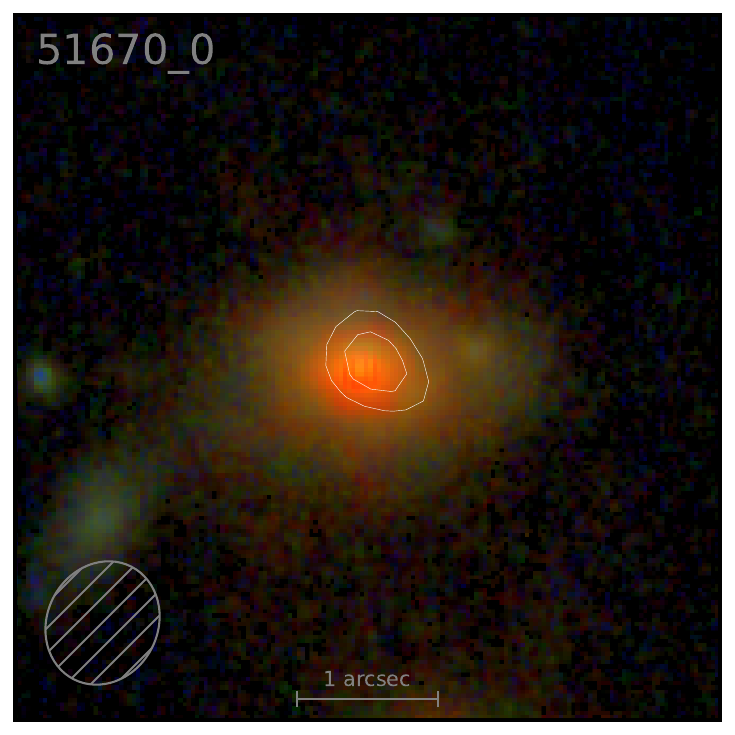}
    \includegraphics[width=0.24\textwidth]{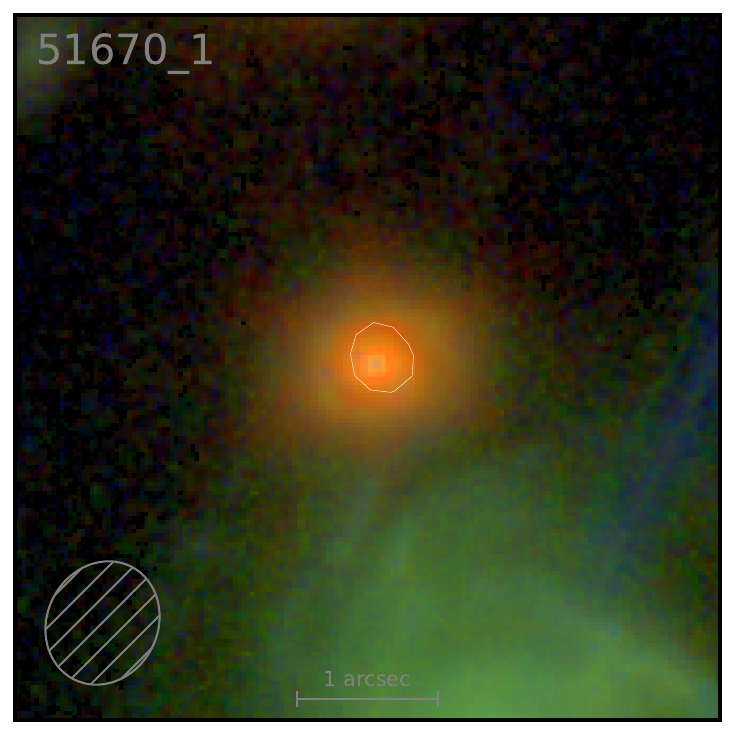}
    \includegraphics[width=0.24\textwidth]{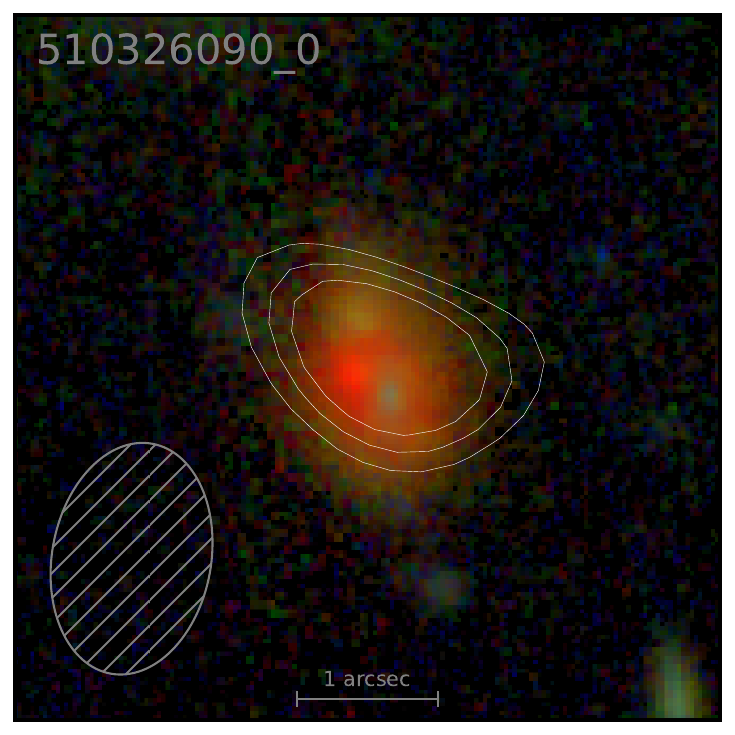}
    \includegraphics[width=0.24\textwidth]{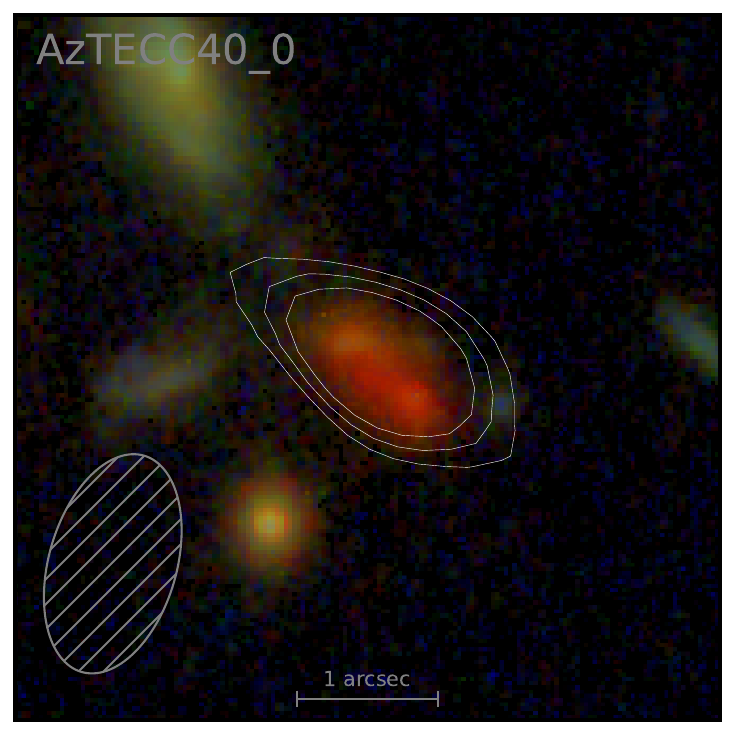}
    \includegraphics[width=0.24\textwidth]{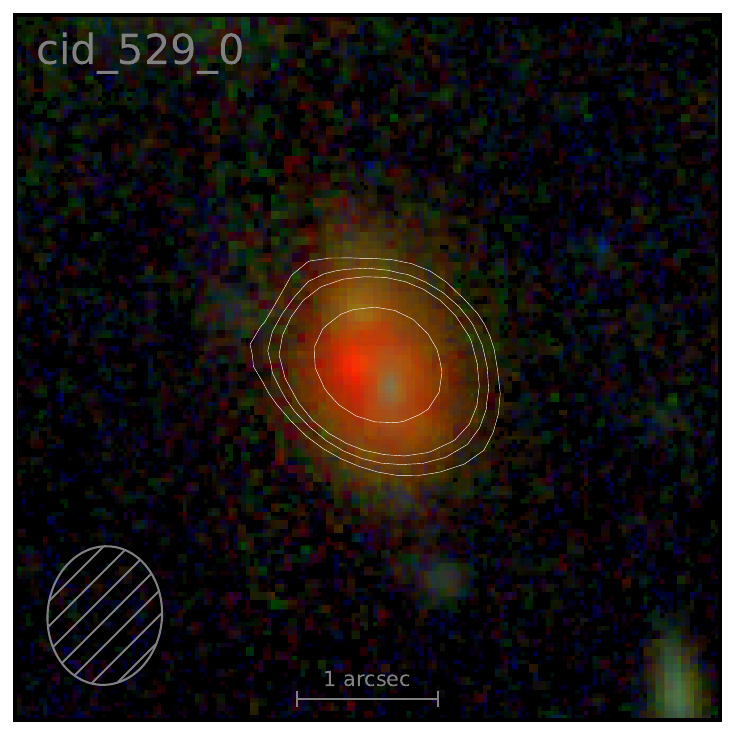}
    \includegraphics[width=0.24\textwidth]{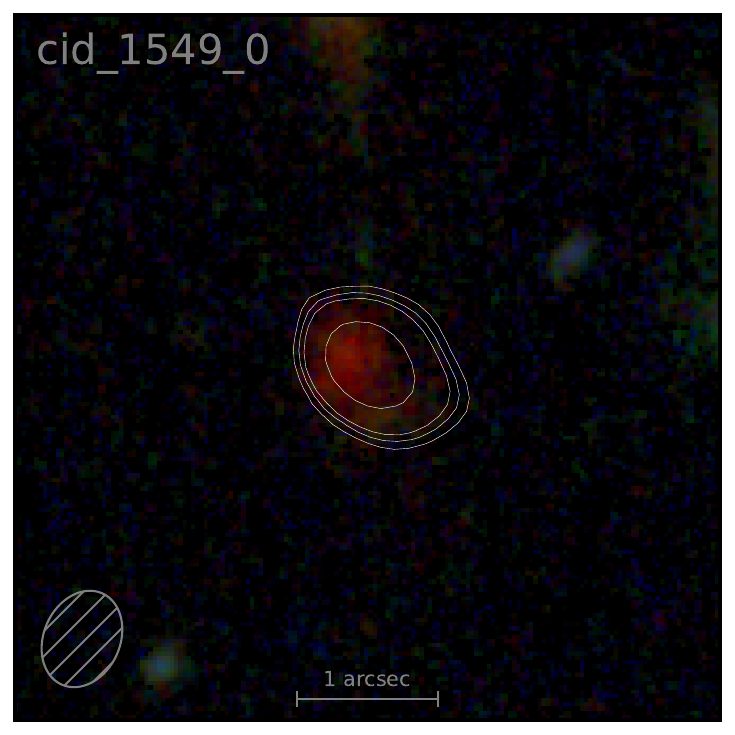}
    \includegraphics[width=0.24\textwidth]{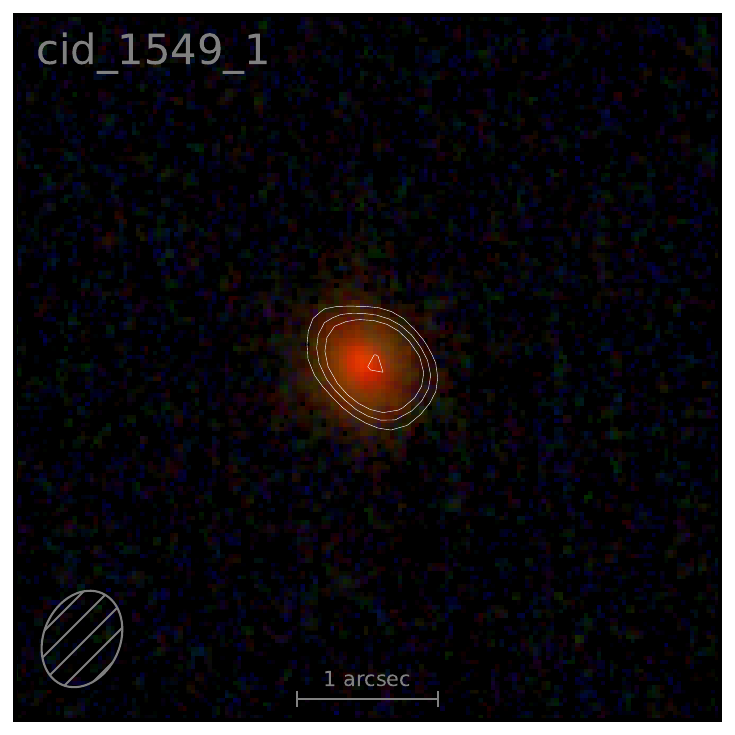}
    \includegraphics[width=0.24\textwidth]{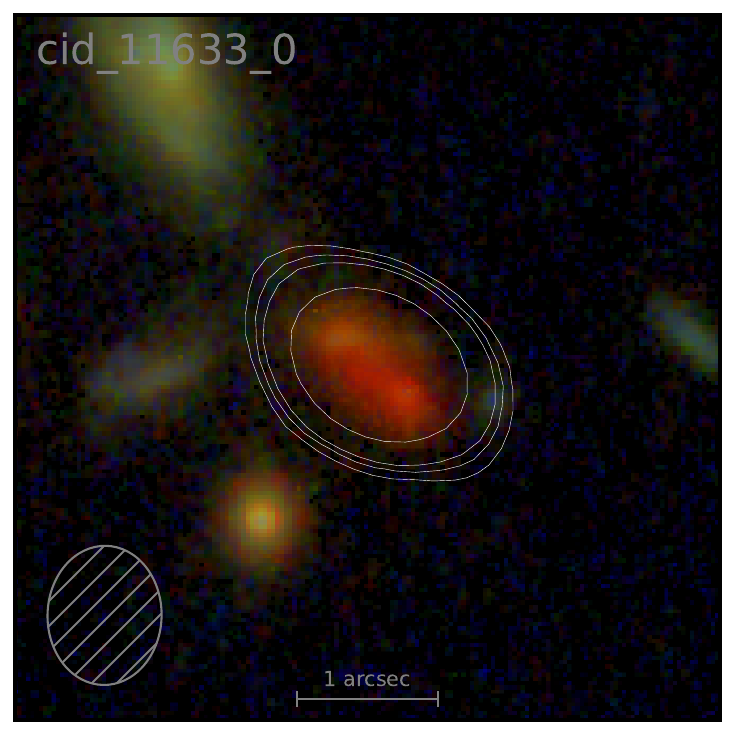}
    \includegraphics[width=0.24\textwidth]{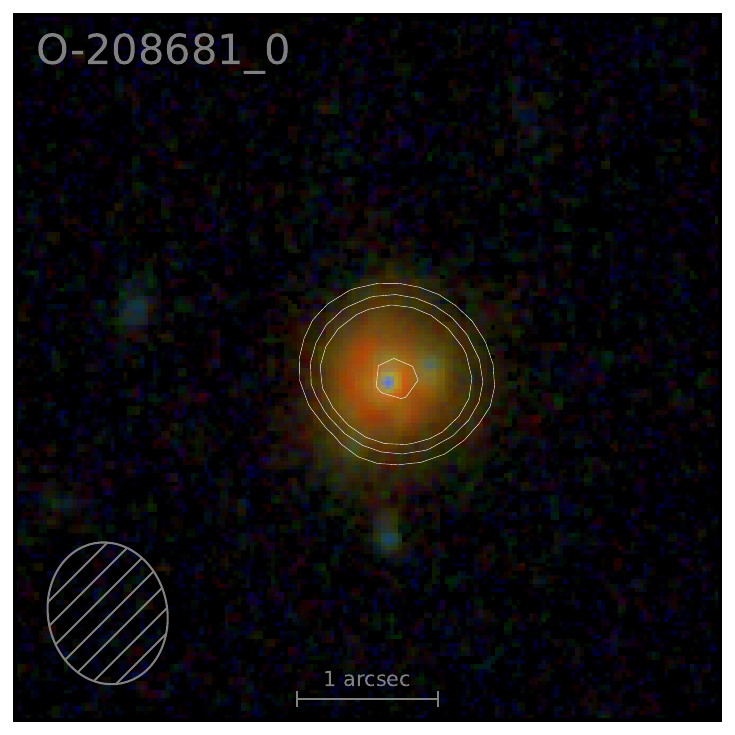}
    \includegraphics[width=0.24\textwidth]{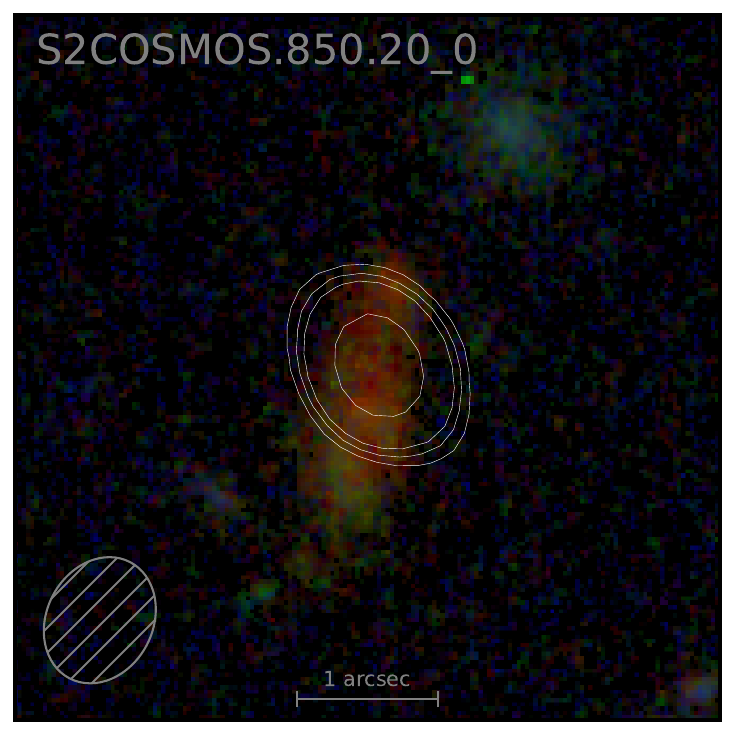}
    \includegraphics[width=0.24\textwidth]{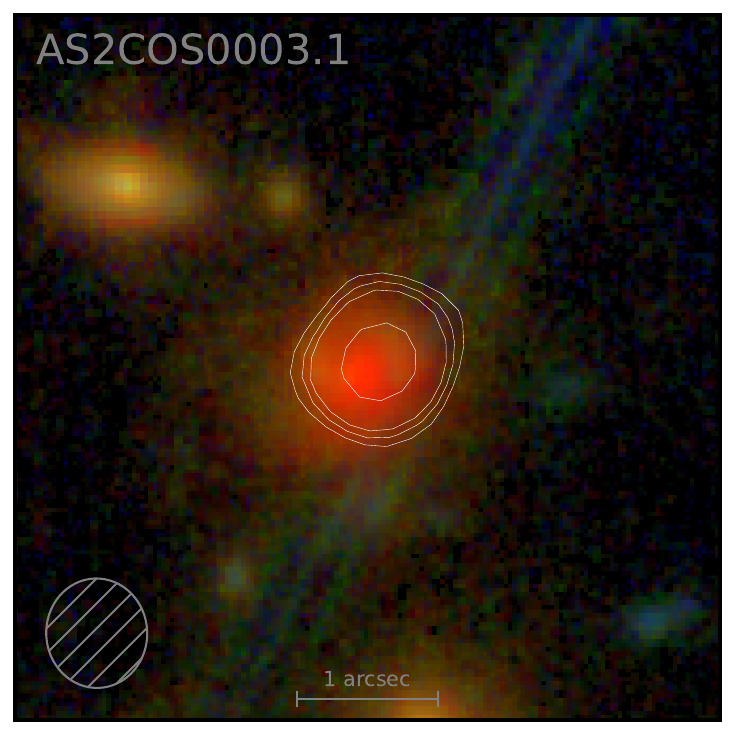}
    \includegraphics[width=0.24\textwidth]{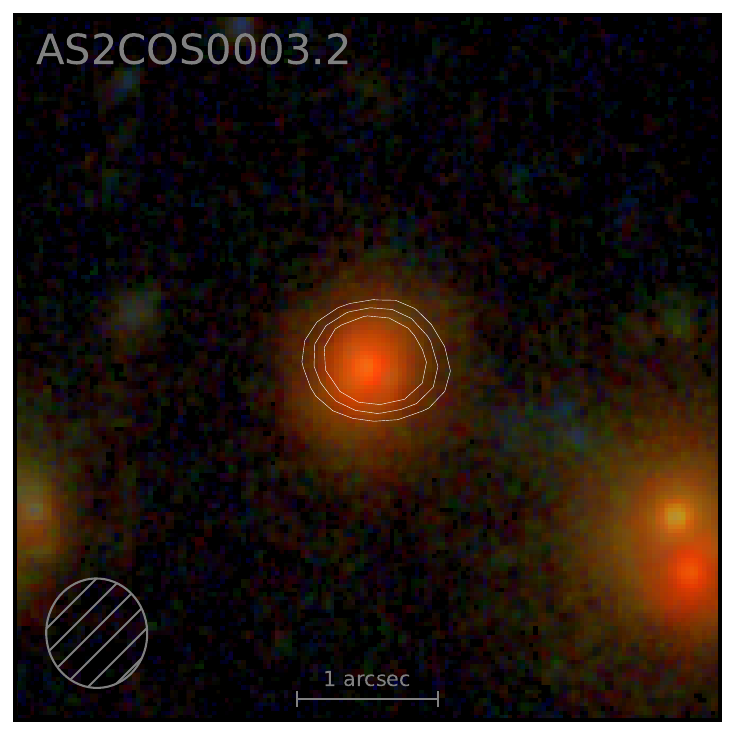}
    \includegraphics[width=0.24\textwidth]{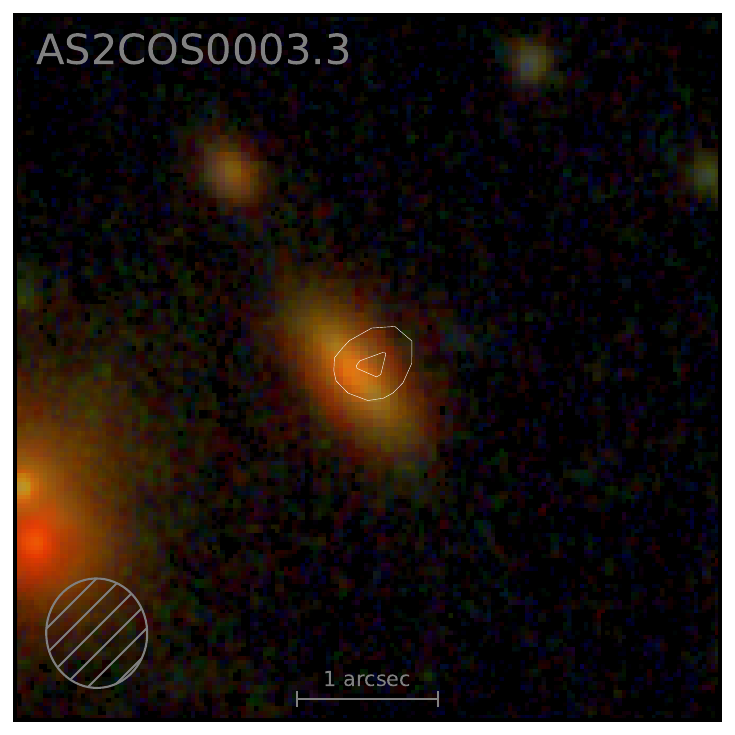}
    \includegraphics[width=0.24\textwidth]{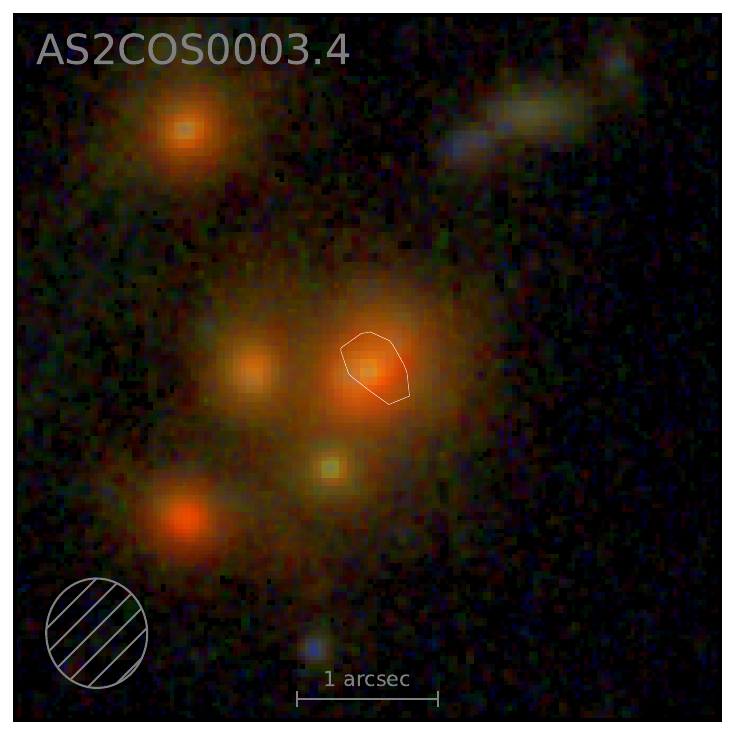}
    \includegraphics[width=0.24\textwidth]{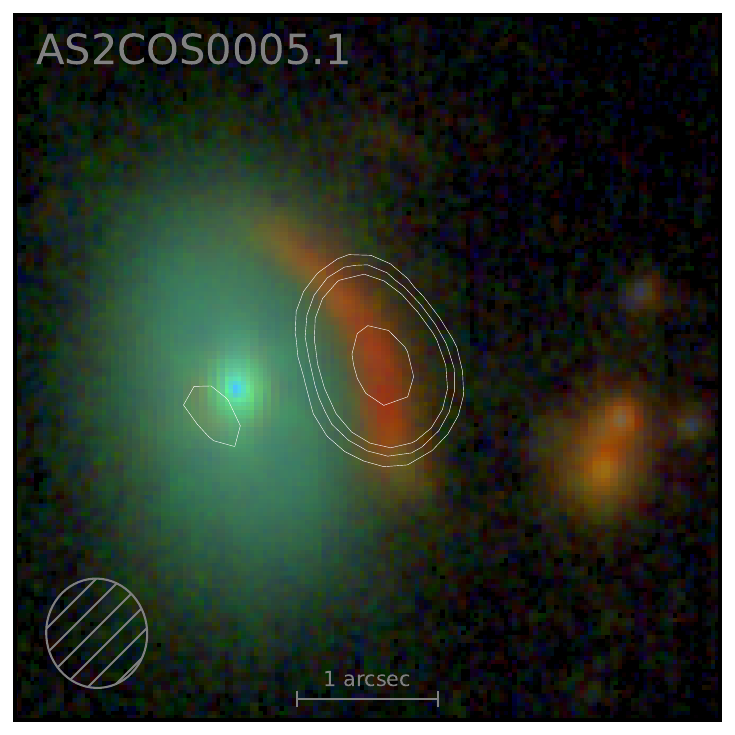}
    \includegraphics[width=0.24\textwidth]{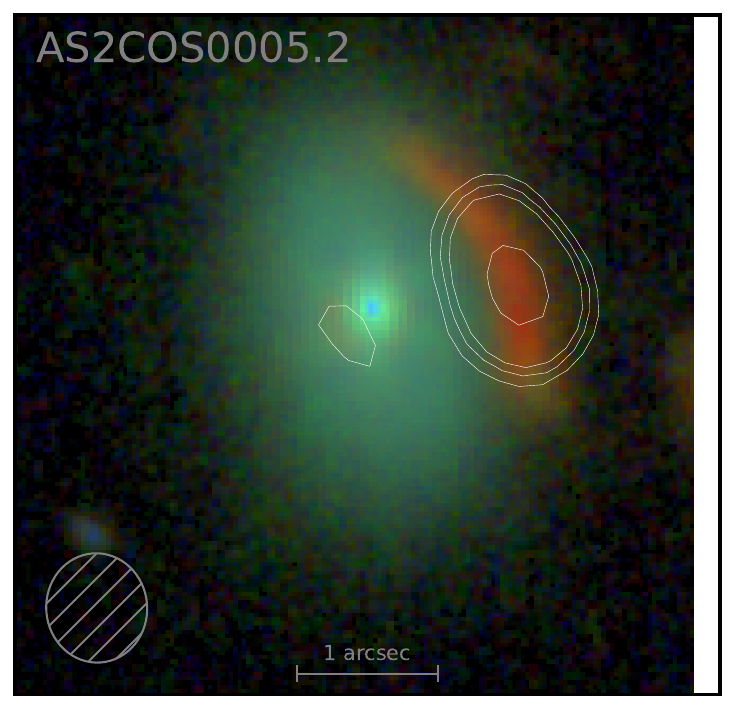}
    \includegraphics[width=0.24\textwidth]{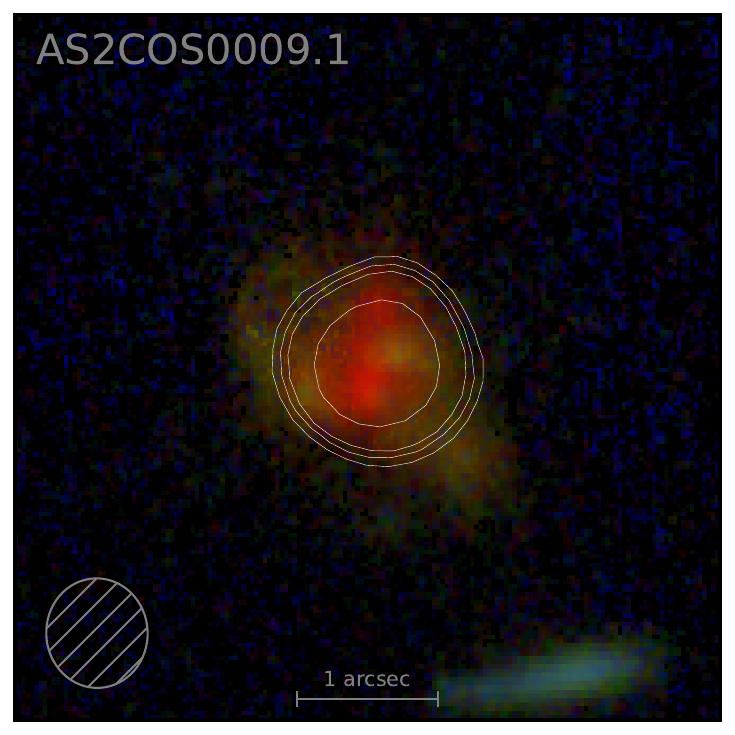}
    \includegraphics[width=0.24\textwidth]{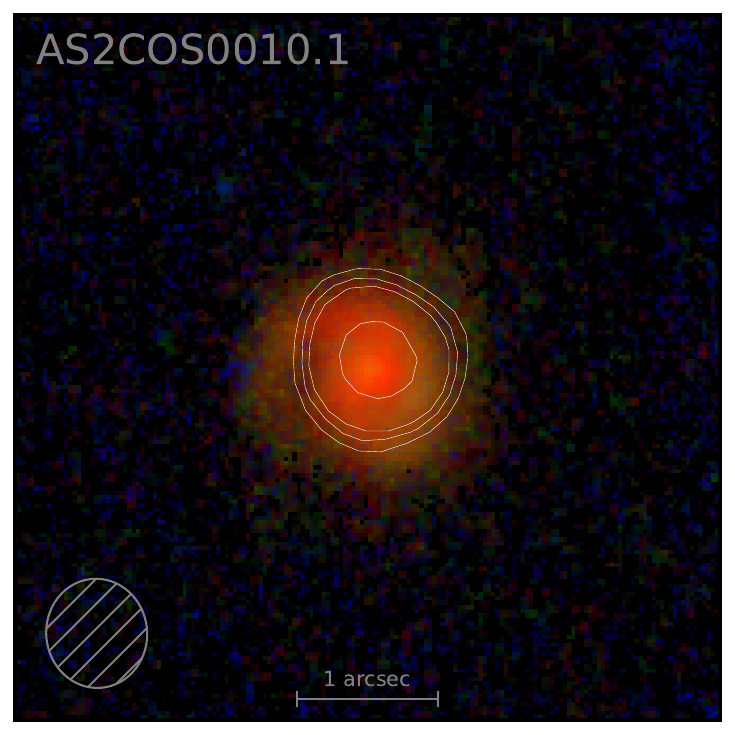}
    \includegraphics[width=0.24\textwidth]{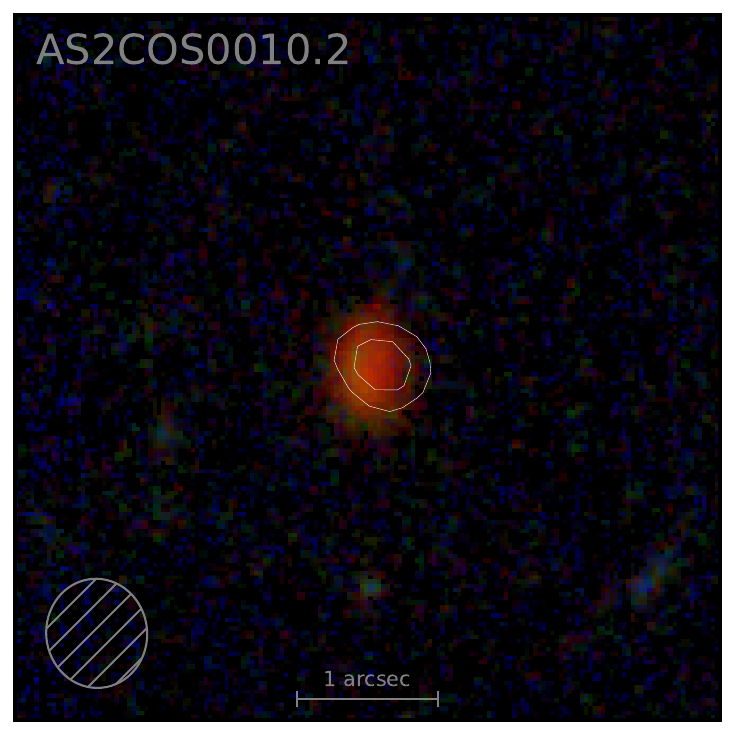}
    \includegraphics[width=0.24\textwidth]{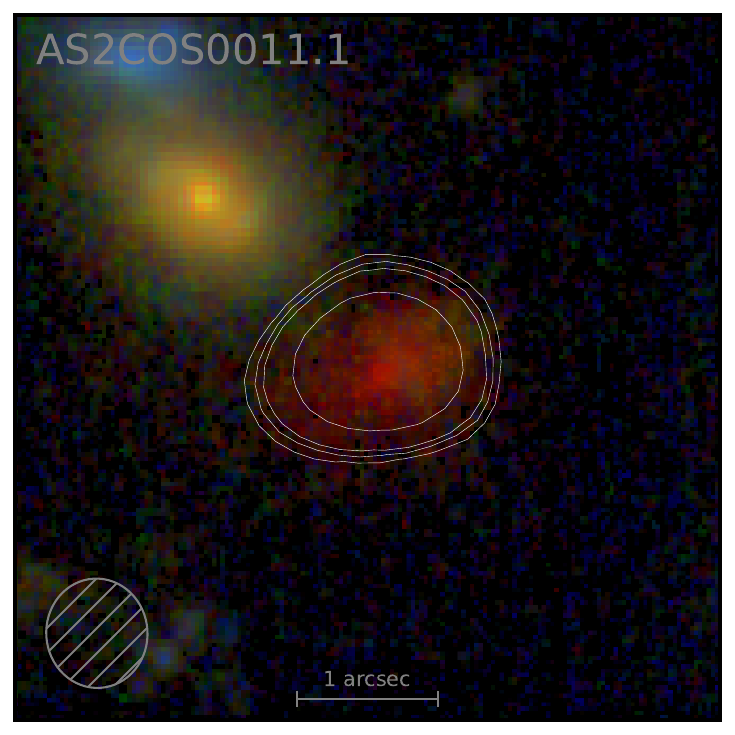}
\end{figure*}
\begin{figure*}
\centering  
\includegraphics[width=0.24\textwidth]{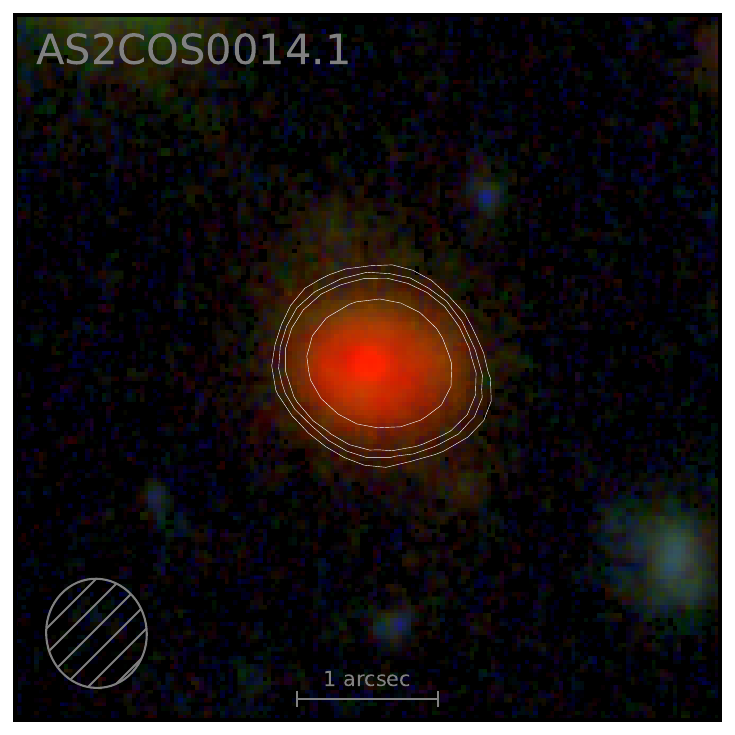}
\includegraphics[width=0.24\textwidth]{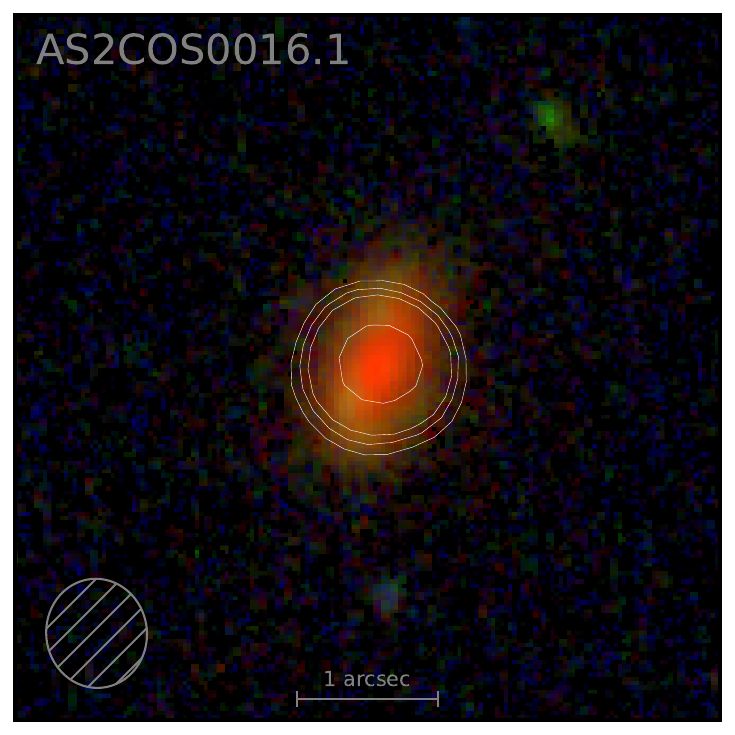}
\includegraphics[width=0.24\textwidth]{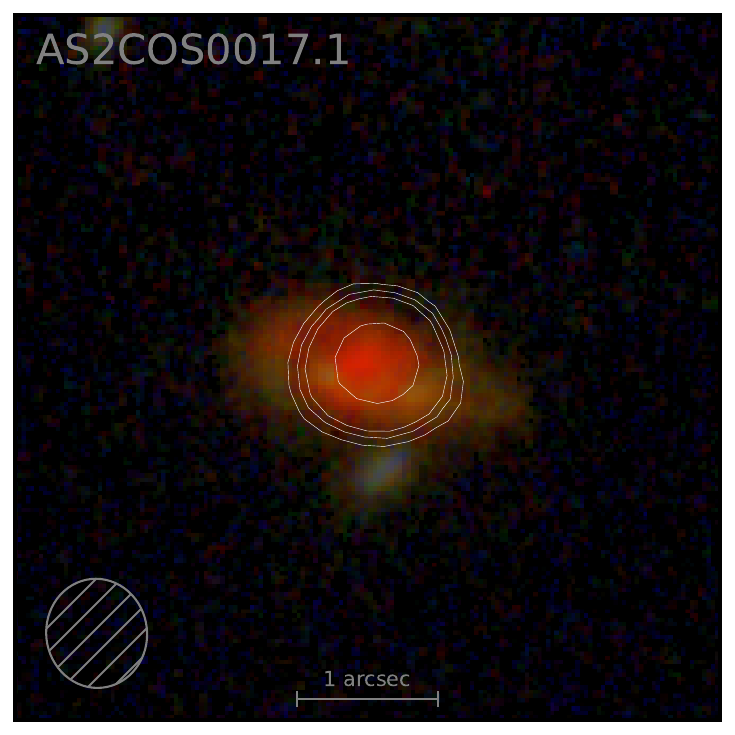}
\includegraphics[width=0.24\textwidth]{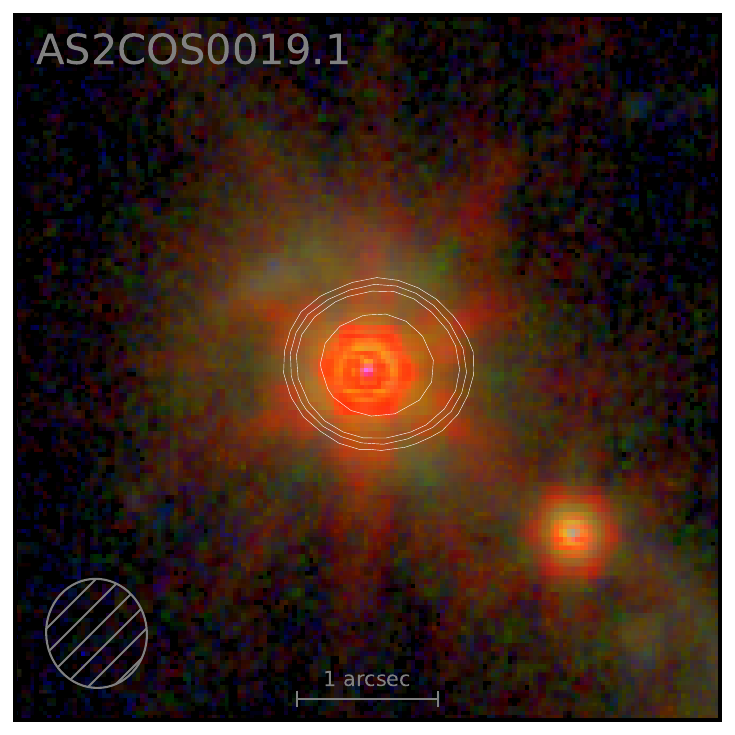}
\includegraphics[width=0.24\textwidth]{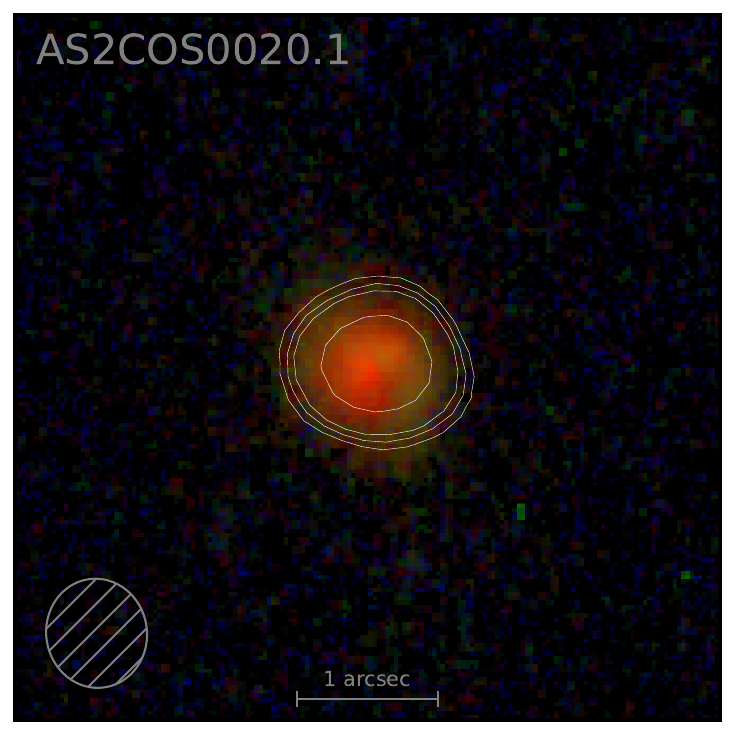}
\includegraphics[width=0.24\textwidth]{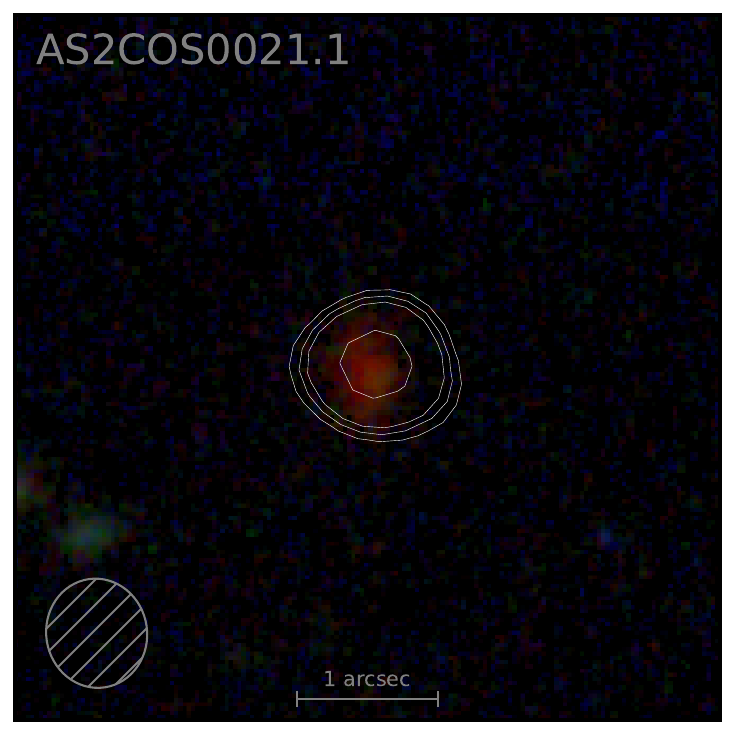}
\includegraphics[width=0.24\textwidth]{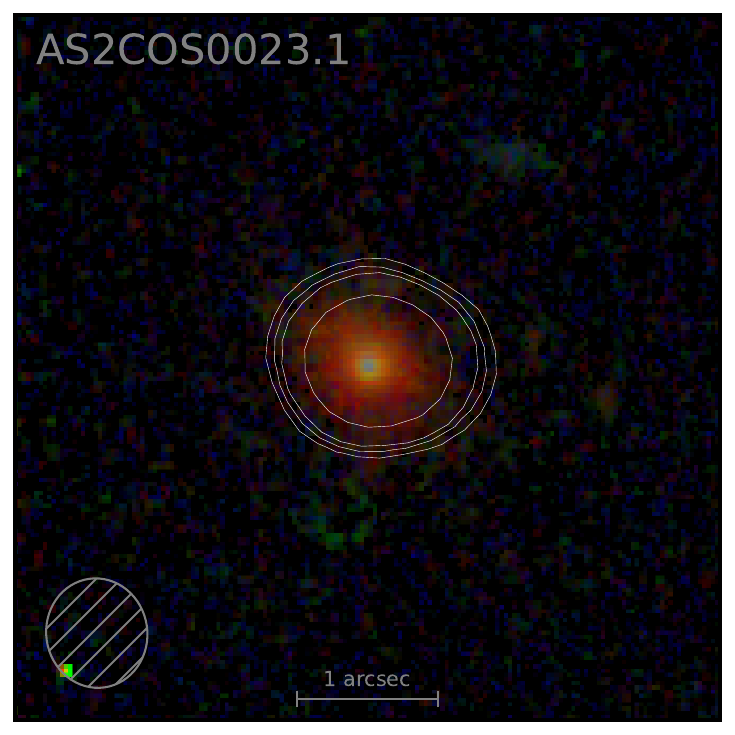}
\includegraphics[width=0.24\textwidth]{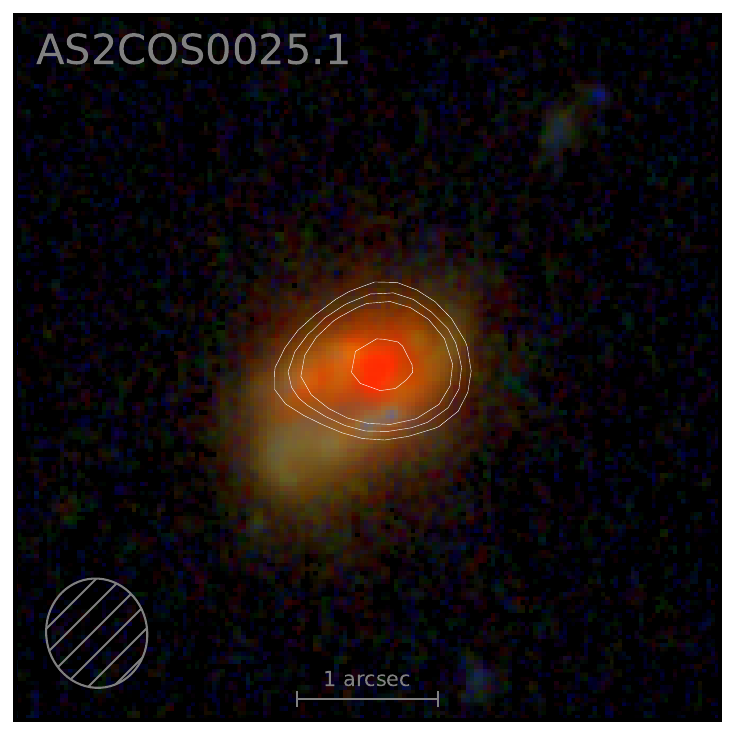}
\includegraphics[width=0.24\textwidth]{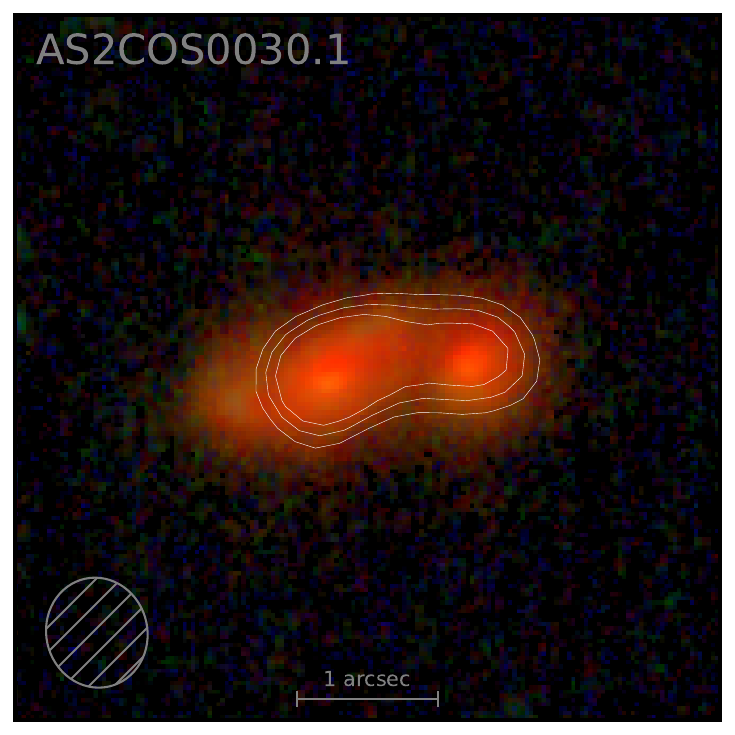}
\includegraphics[width=0.24\textwidth]{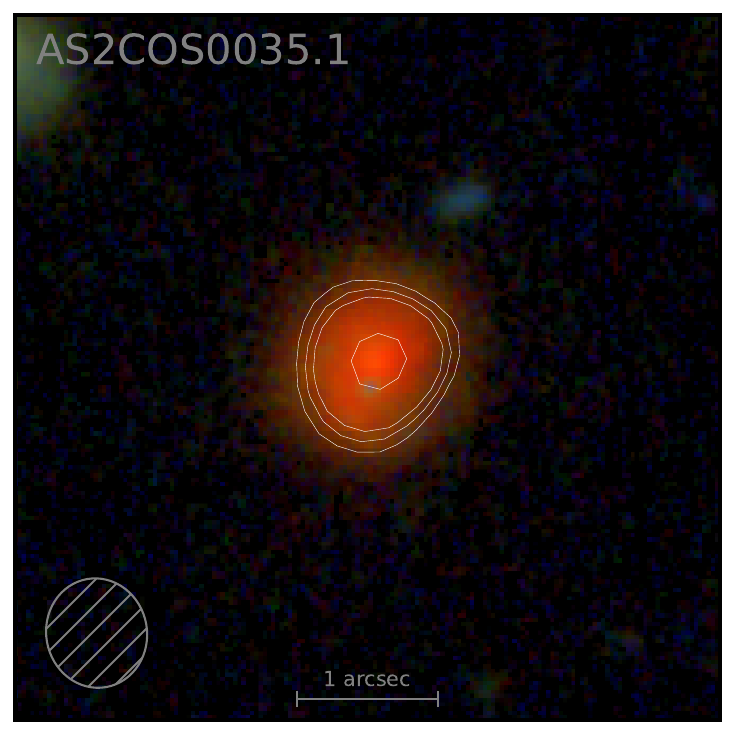}
\includegraphics[width=0.24\textwidth]{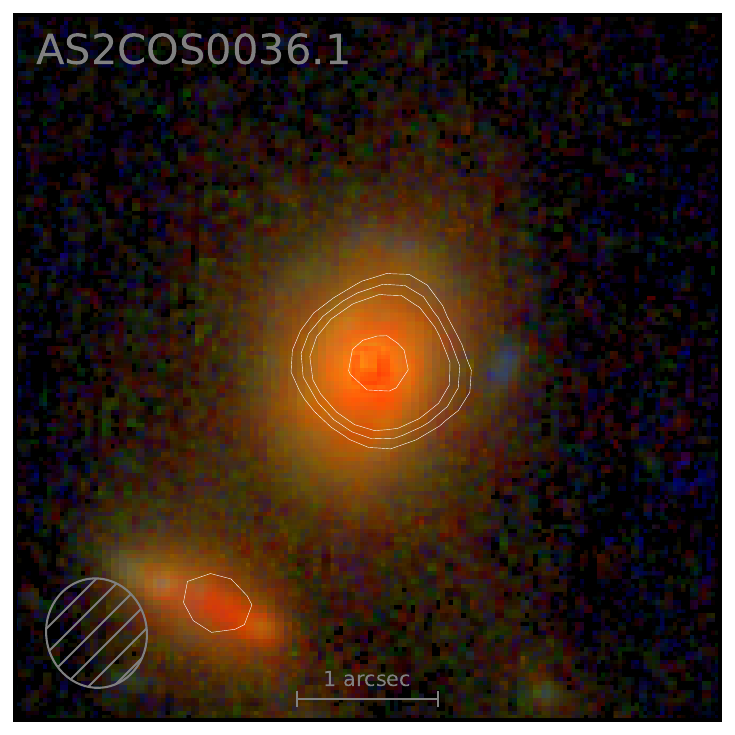}
\includegraphics[width=0.24\textwidth]{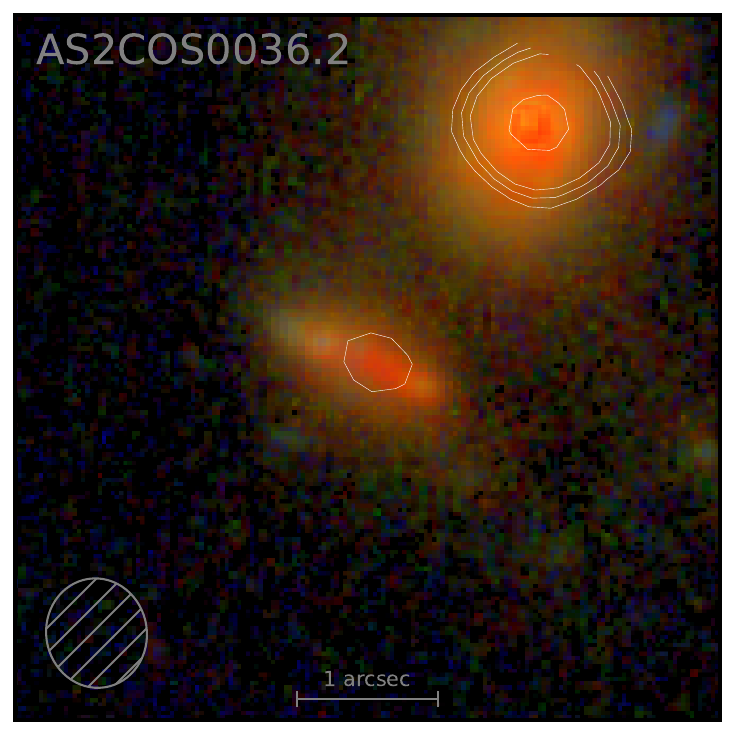}
\includegraphics[width=0.24\textwidth]{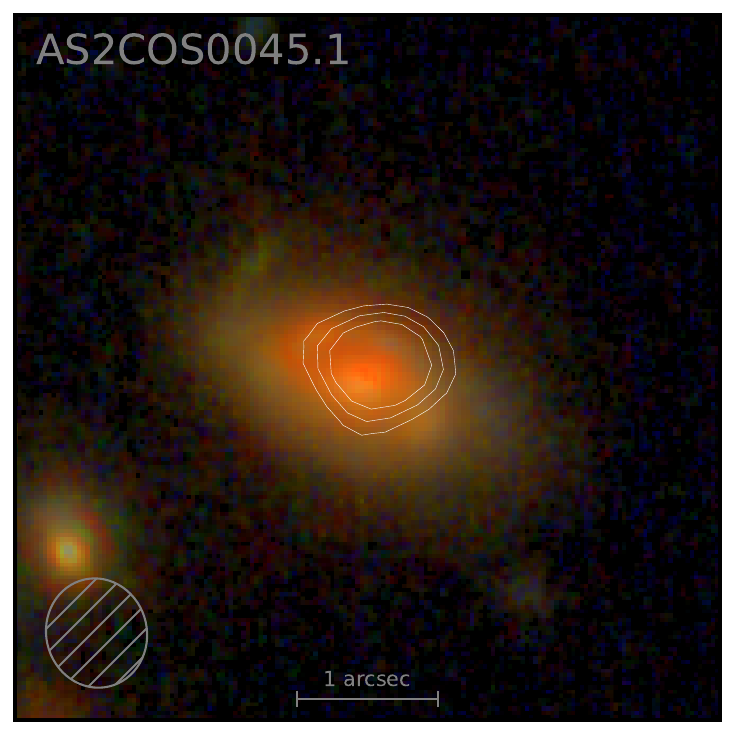}
\includegraphics[width=0.24\textwidth]{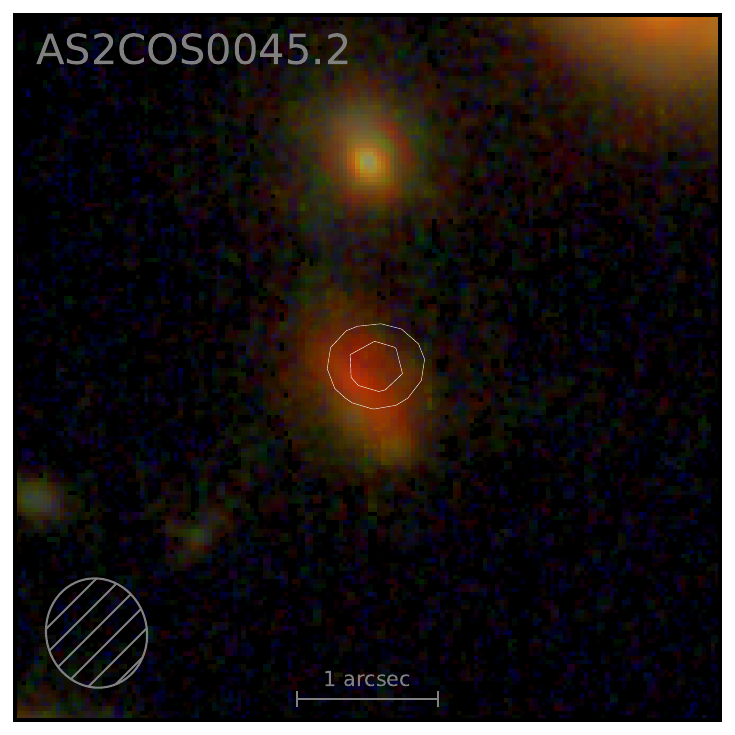}
\includegraphics[width=0.24\textwidth]{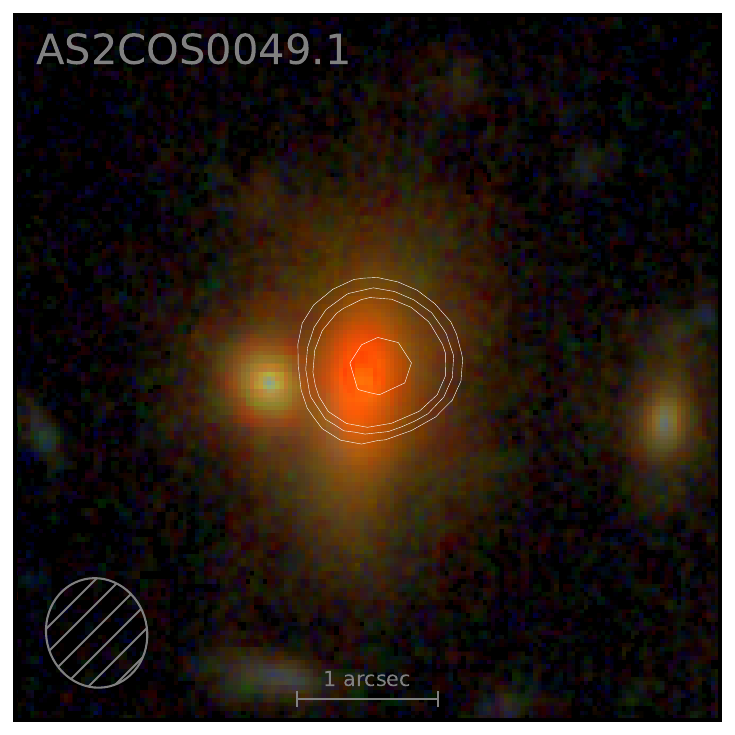}
\includegraphics[width=0.24\textwidth]{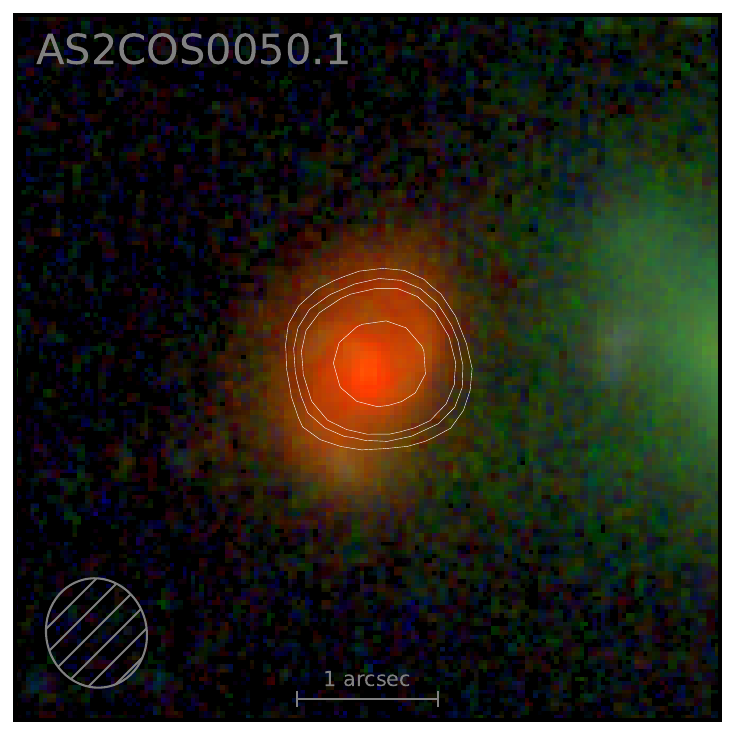}
\includegraphics[width=0.24\textwidth]{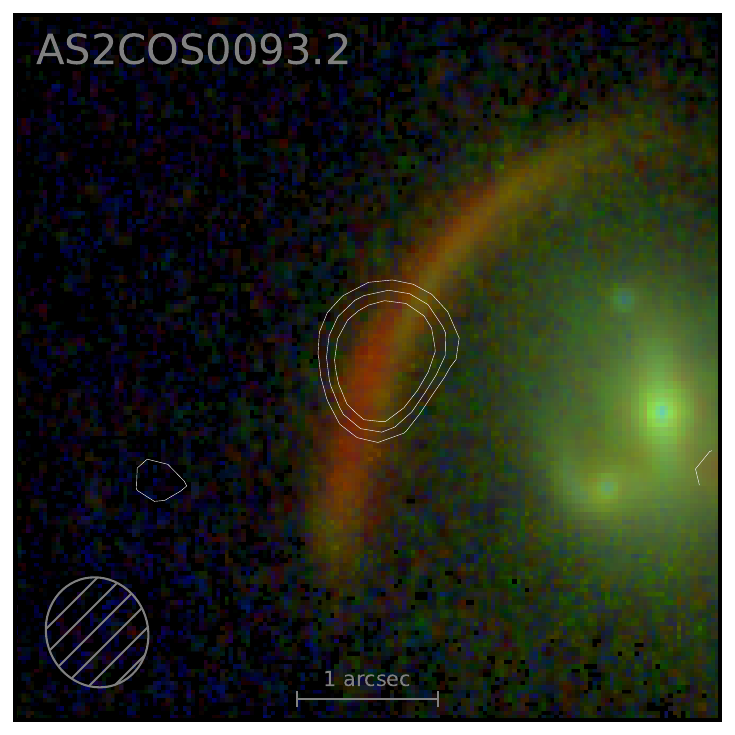}
\includegraphics[width=0.24\textwidth]{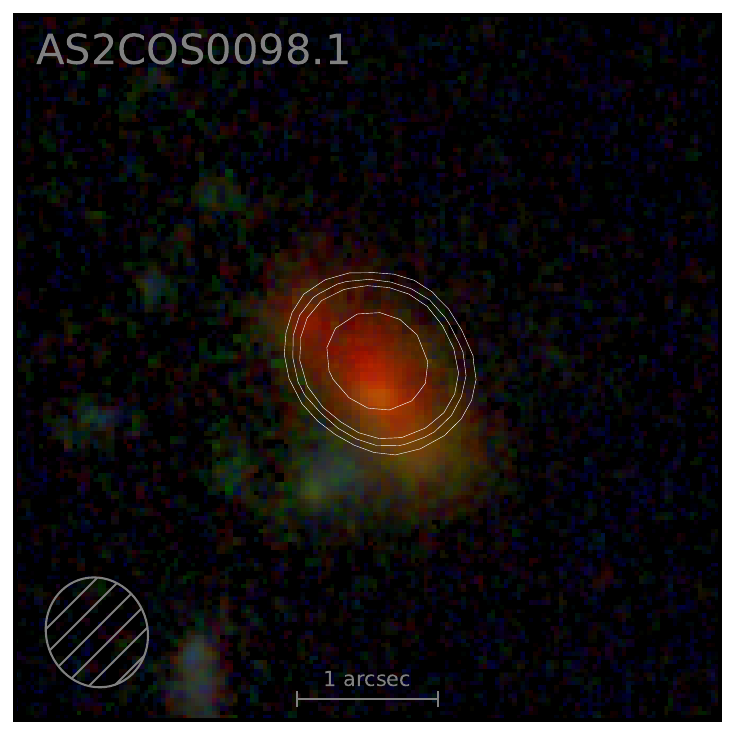}
\includegraphics[width=0.24\textwidth]{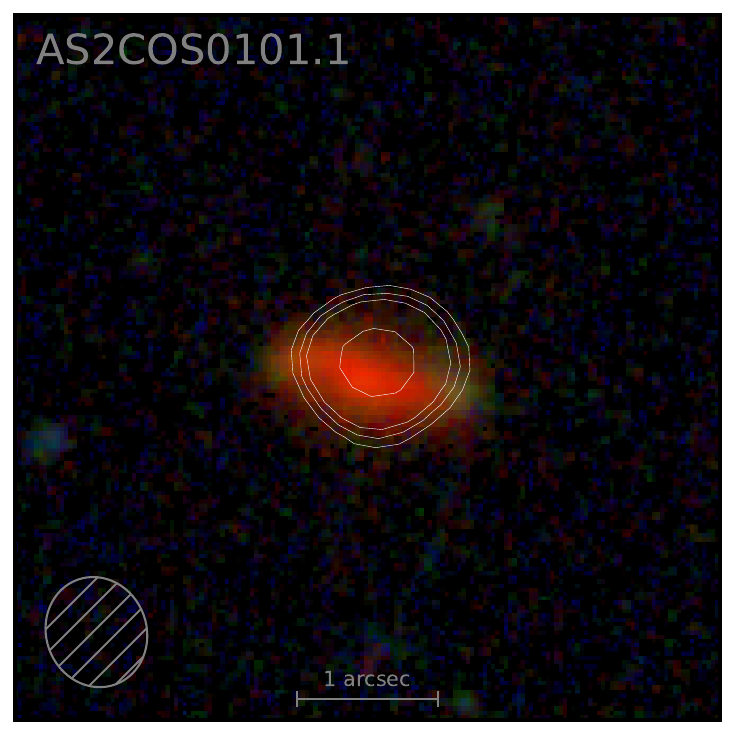}
\includegraphics[width=0.24\textwidth]{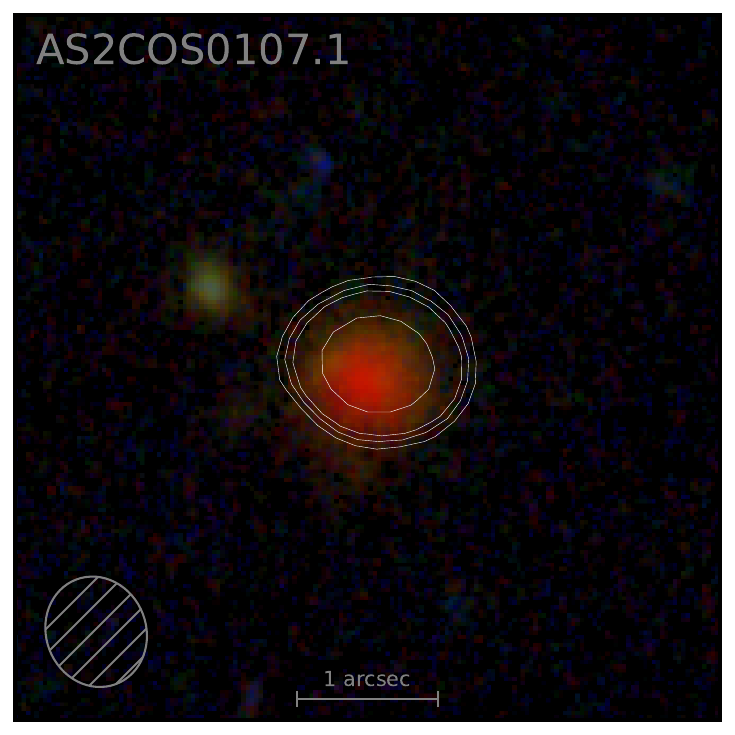}
\end{figure*}
\begin{figure*}
\centering
\includegraphics[width=0.24\textwidth]{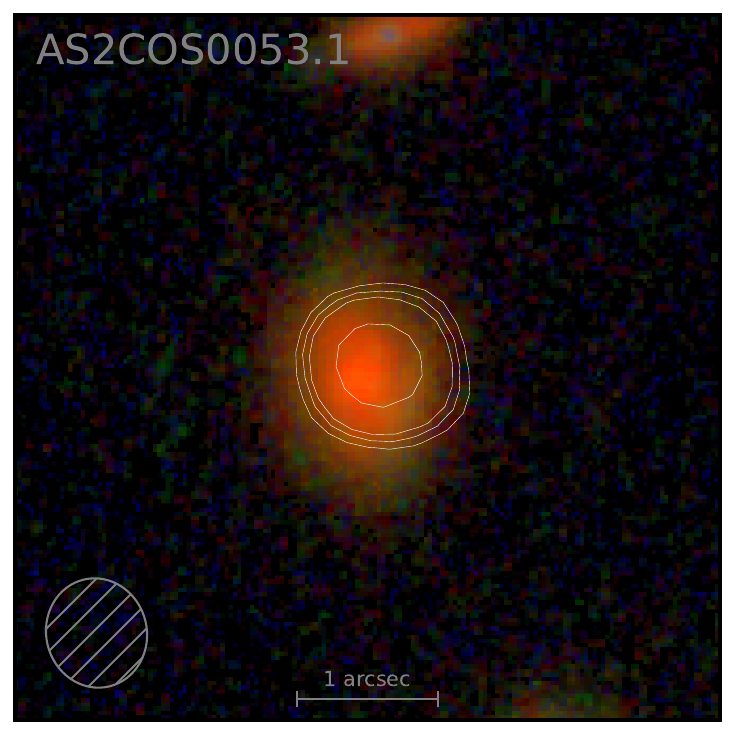}
\includegraphics[width=0.24\textwidth]{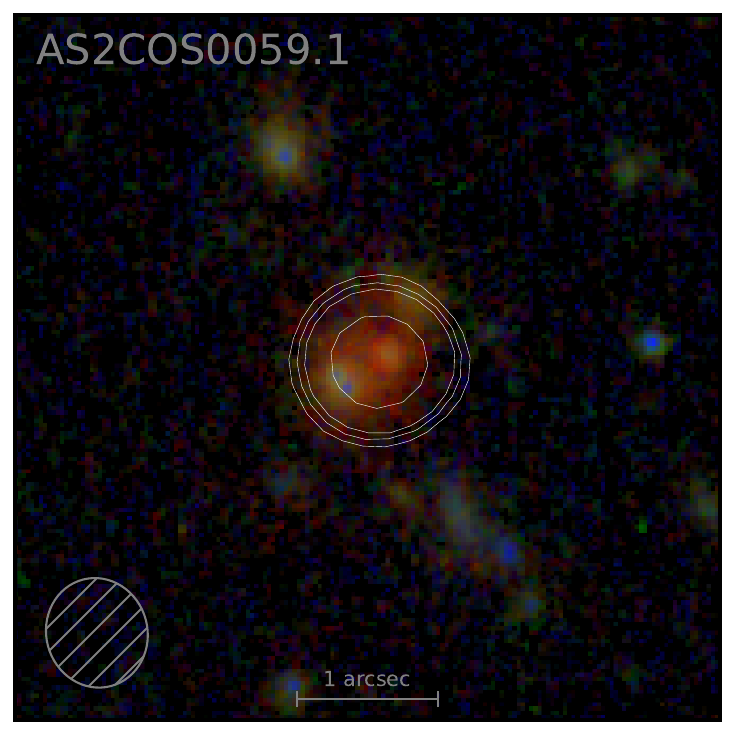}
\includegraphics[width=0.24\textwidth]{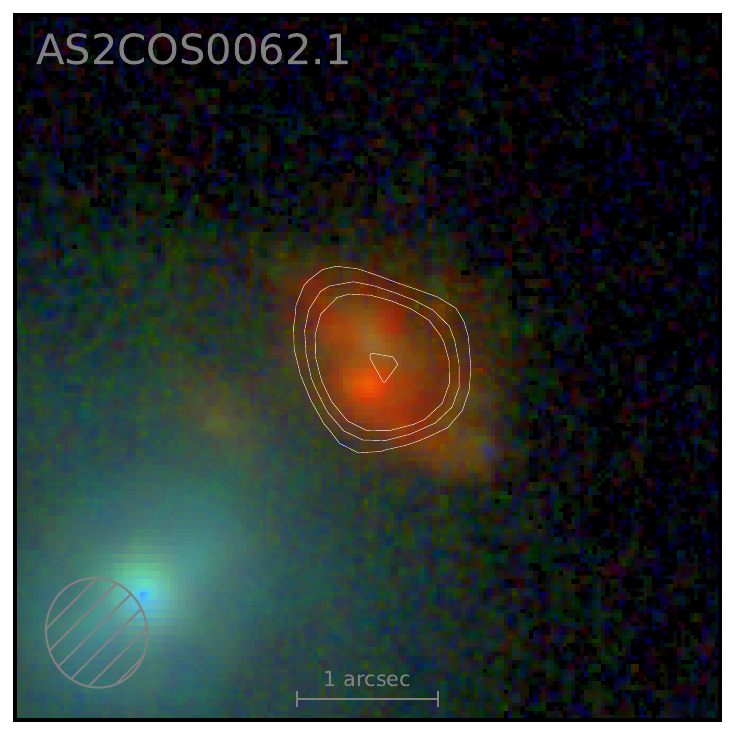}
\includegraphics[width=0.24\textwidth]{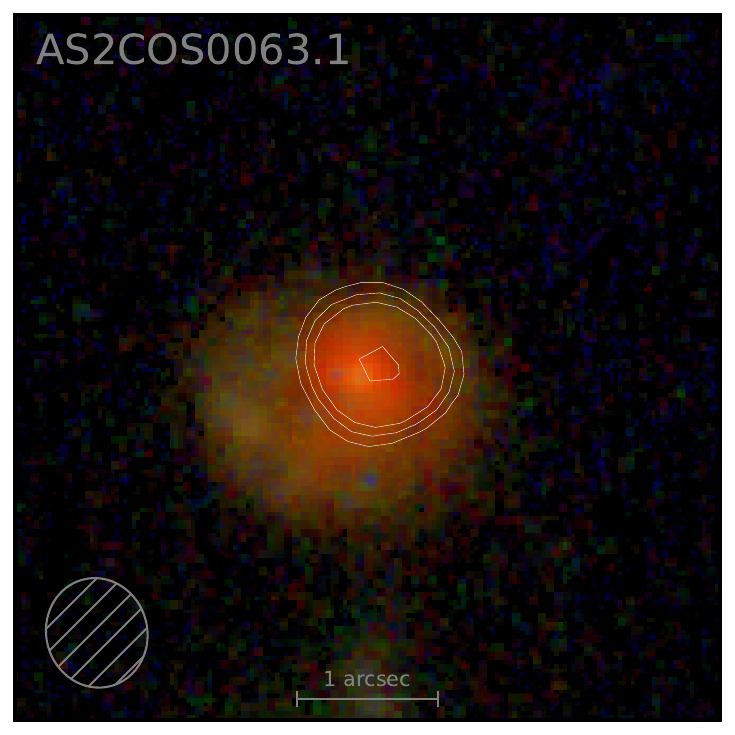}
\includegraphics[width=0.24\textwidth]{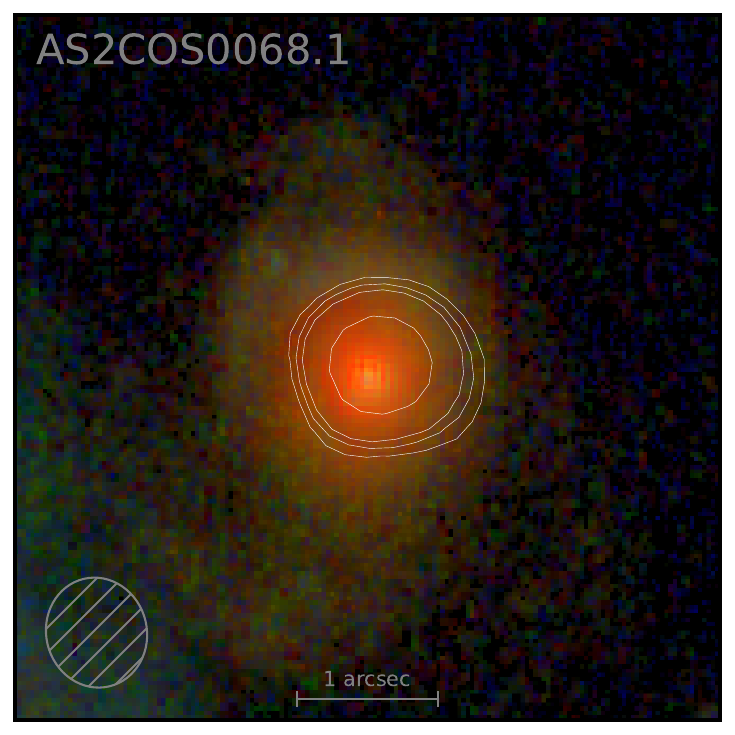}
\includegraphics[width=0.24\textwidth]{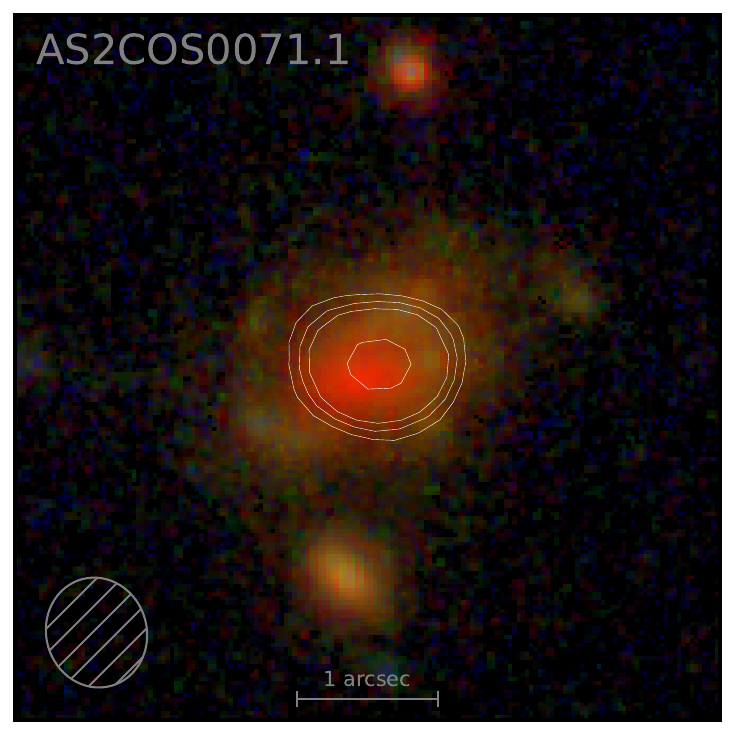}
\includegraphics[width=0.24\textwidth]{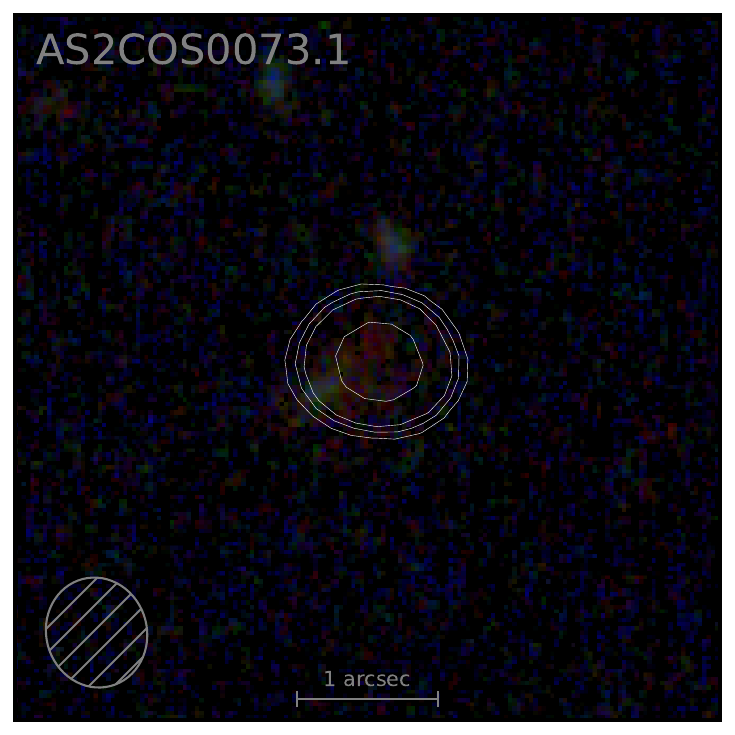}
\includegraphics[width=0.24\textwidth]{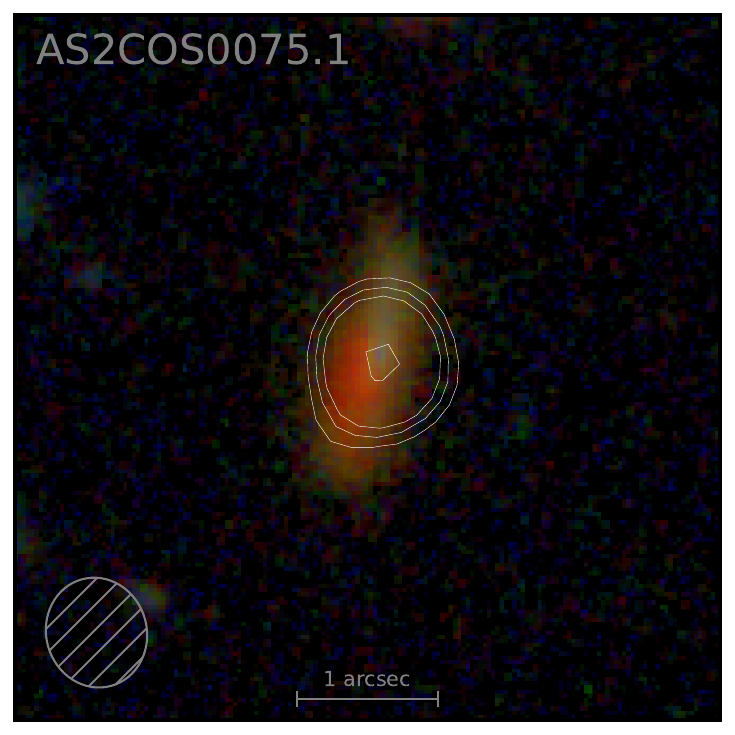}
\includegraphics[width=0.24\textwidth]{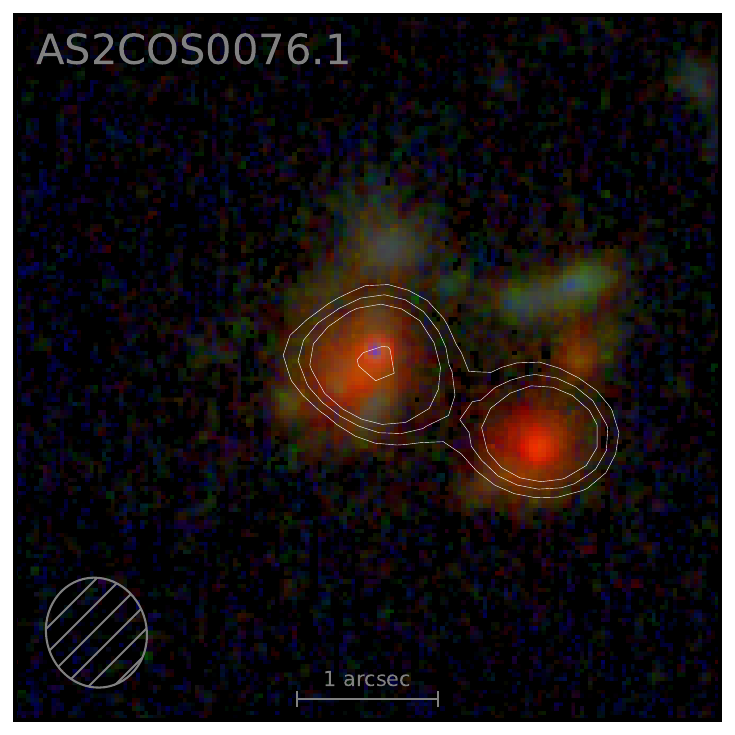}
\includegraphics[width=0.24\textwidth]{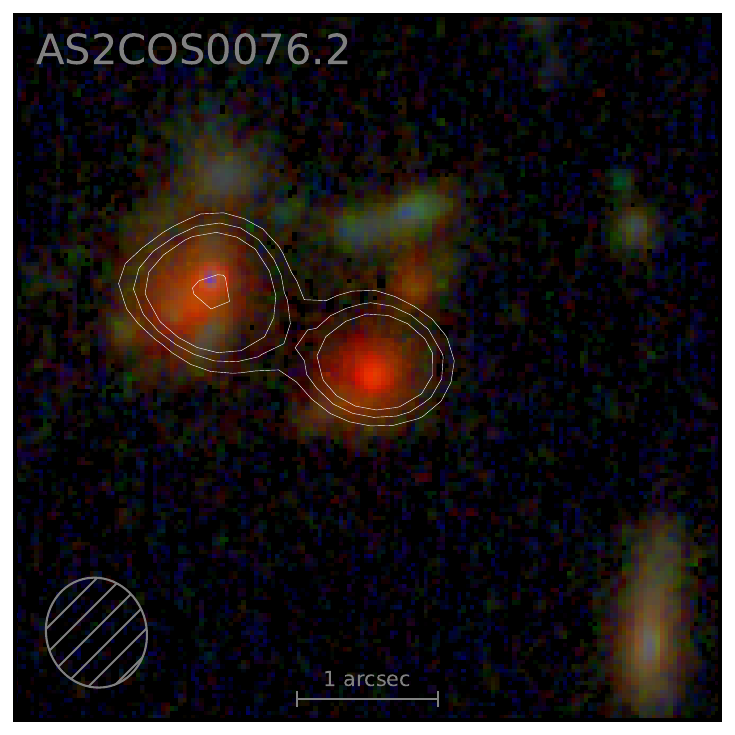}
\includegraphics[width=0.24\textwidth]{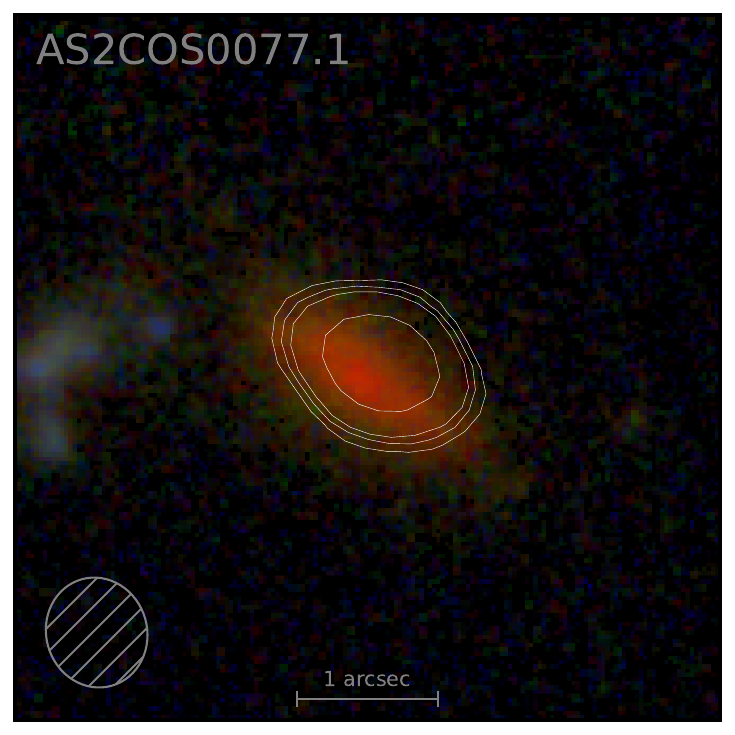}
\includegraphics[width=0.24\textwidth]{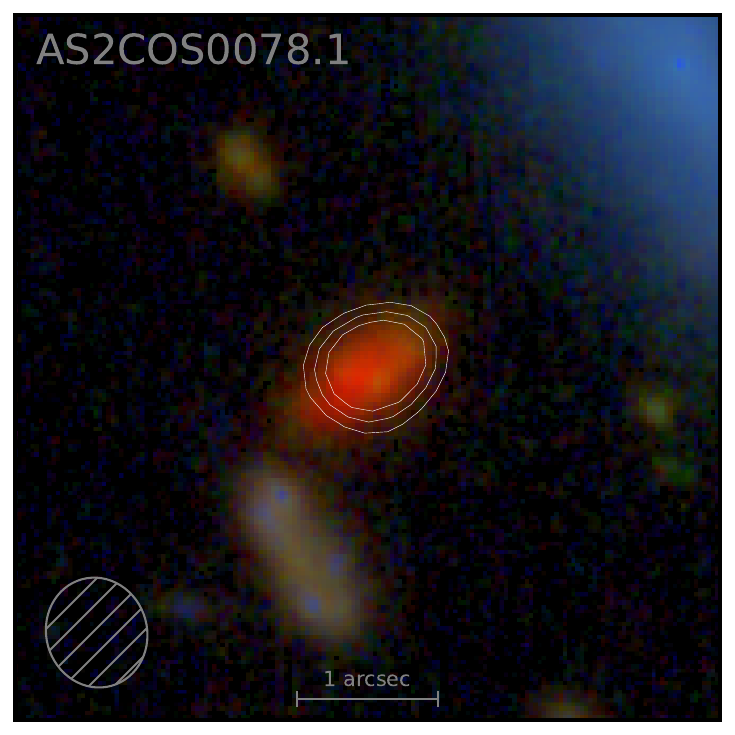}
\includegraphics[width=0.24\textwidth]{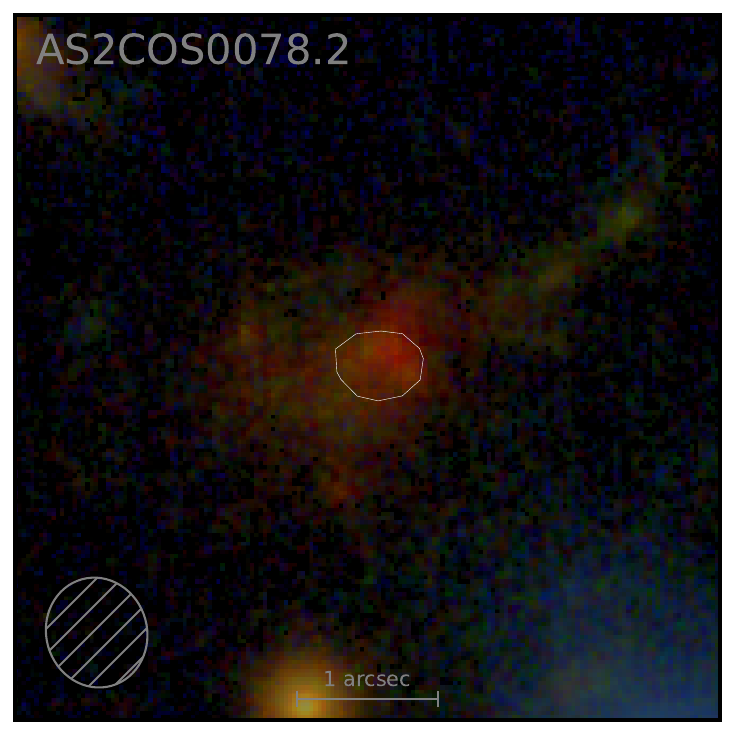}
\includegraphics[width=0.24\textwidth]{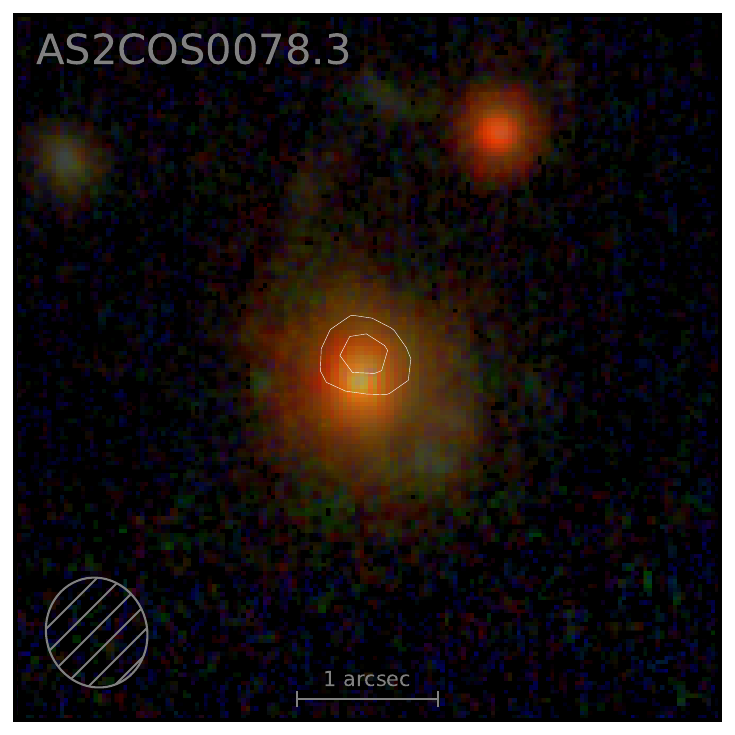}
\includegraphics[width=0.24\textwidth]{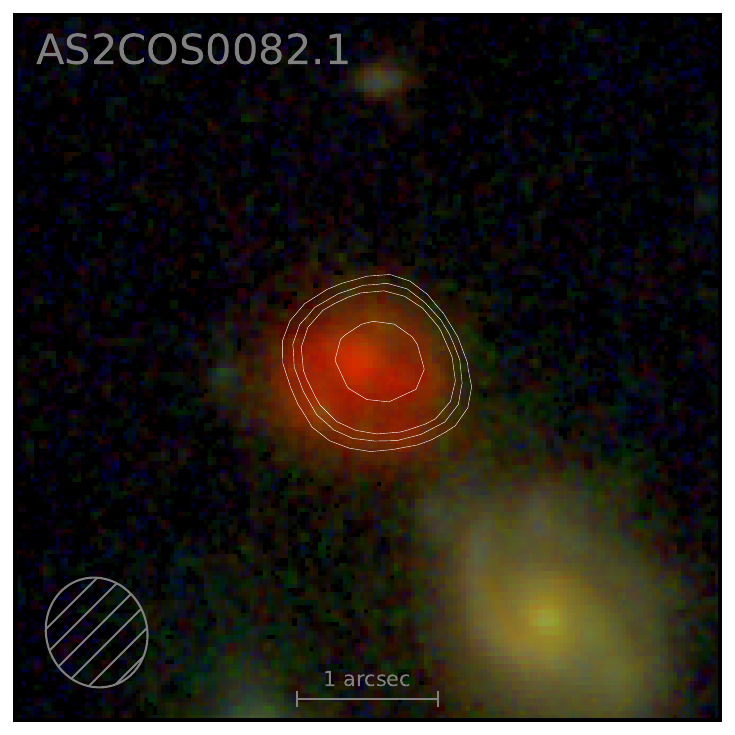}
\includegraphics[width=0.24\textwidth]{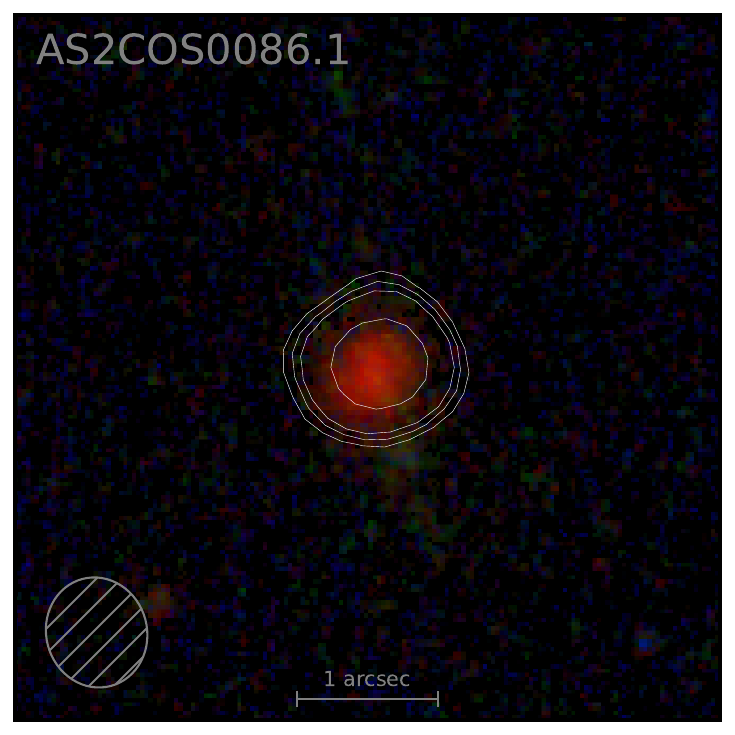}
\includegraphics[width=0.24\textwidth]{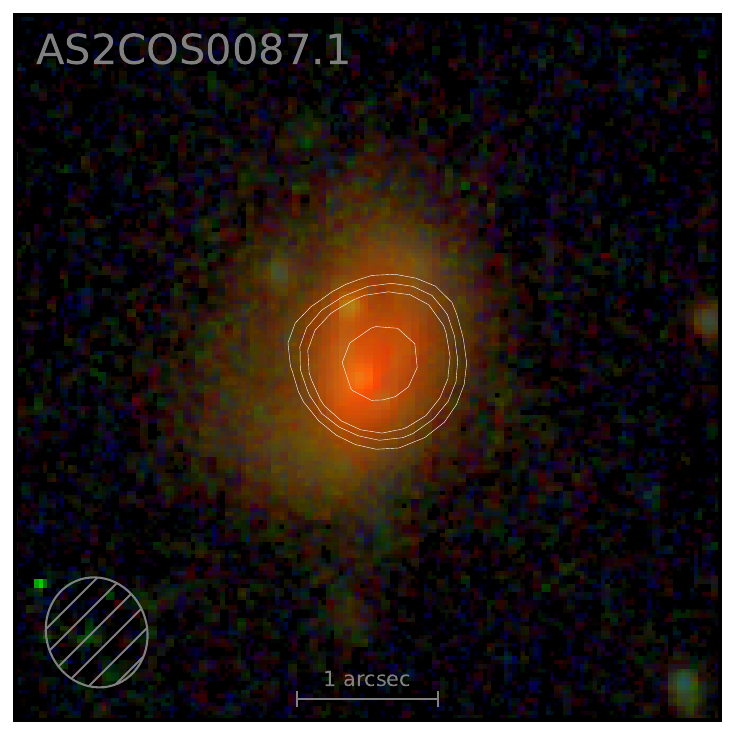}
\includegraphics[width=0.24\textwidth]{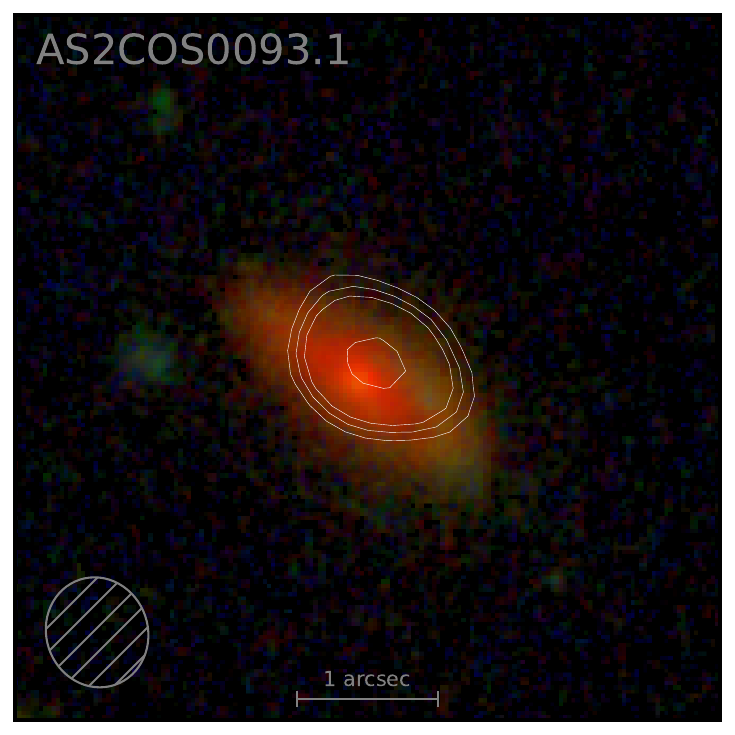}
\includegraphics[width=0.24\textwidth]{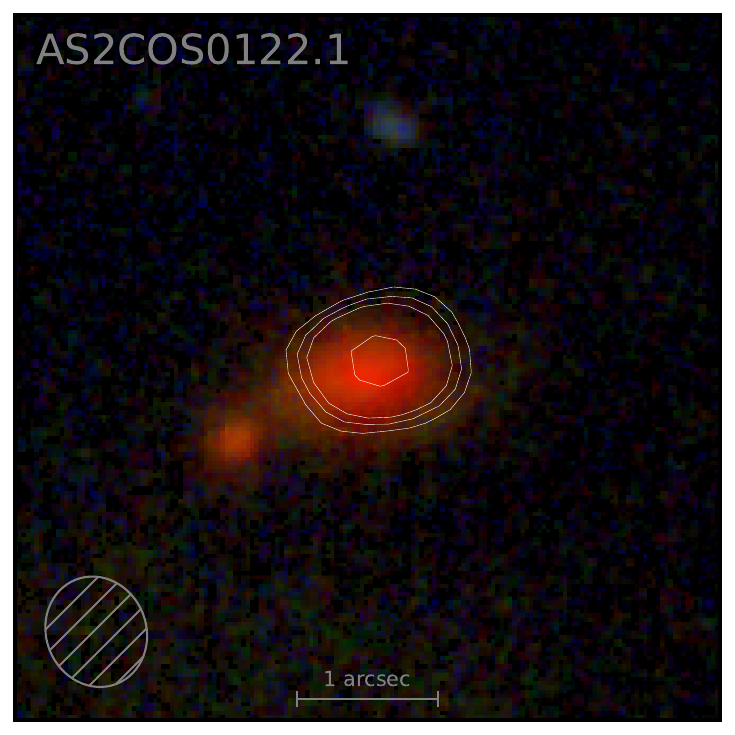}
\includegraphics[width=0.24\textwidth]{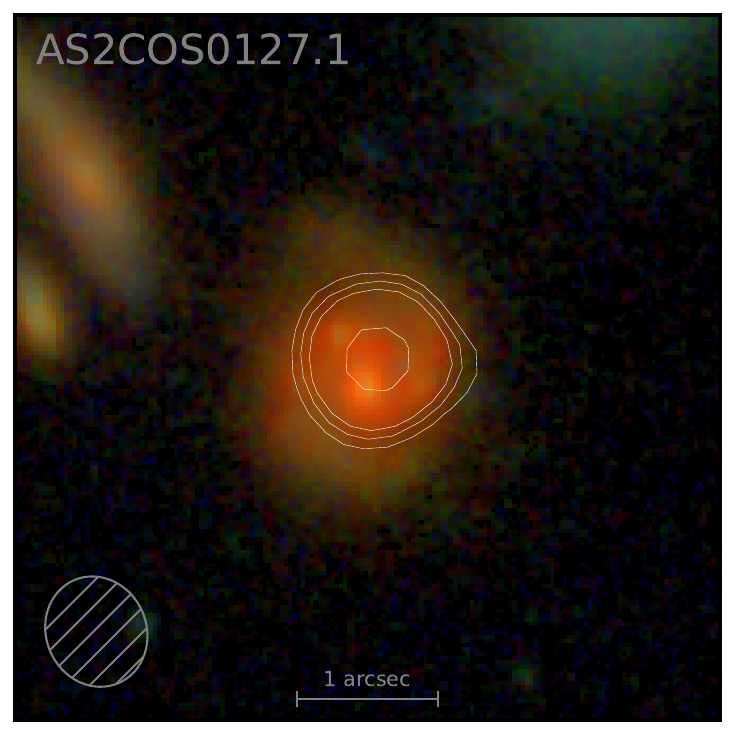}
\end{figure*}


\clearpage
\bibliography{references}{}

\begin{thebibliography}{}
\expandafter\ifx\csname natexlab\endcsname\relax\def\natexlab#1{#1}\fi
\providecommand{\url}[1]{\href{#1}{#1}}
\providecommand{\dodoi}[1]{doi:~\href{http://doi.org/#1}{\nolinkurl{#1}}}
\providecommand{\doeprint}[1]{\href{http://ascl.net/#1}{\nolinkurl{http://ascl.net/#1}}}
\providecommand{\doarXiv}[1]{\href{https://arxiv.org/abs/#1}{\nolinkurl{https://arxiv.org/abs/#1}}}

\bibitem[{{Akins} {et~al.}(2024){Akins}, {Casey}, {Lambrides}, {Allen}, {Andika}, {Brinch}, {Champagne}, {Cooper}, {Ding}, {Drakos}, {Faisst}, {Finkelstein}, {Franco}, {Fujimoto}, {Gentile}, {Gillman}, {Gozaliasl}, {Harish}, {Hayward}, {Hirschmann}, {Ilbert}, {Kartaltepe}, {Kocevski}, {Koekemoer}, {Kokorev}, {Liu}, {Long}, {McCracken}, {McKinney}, {Onoue}, {Paquereau}, {Renzini}, {Rhodes}, {Robertson}, {Shuntov}, {Silverman}, {Tanaka}, {Toft}, {Trakhtenbrot}, {Valentino}, \& {Zavala}}]{Akins2024}
{Akins}, H.~B., {Casey}, C.~M., {Lambrides}, E., {et~al.} 2024, arXiv e-prints, arXiv:2406.10341, \dodoi{10.48550/arXiv.2406.10341}

\bibitem[{{Alberts} \& {Noble}(2022)}]{AlbertsNoble2022}
{Alberts}, S., \& {Noble}, A. 2022, Universe, 8, 554, \dodoi{10.3390/universe8110554}

\bibitem[{{Algera} {et~al.}(2024){Algera}, {Inami}, {Sommovigo}, {Fudamoto}, {Schneider}, {Graziani}, {Dayal}, {Bouwens}, {Aravena}, {da Cunha}, {Ferrara}, {Hygate}, {van Leeuwen}, {De Looze}, {Palla}, {Pallottini}, {Smit}, {Stefanon}, {Topping}, \& {van der Werf}}]{Algera2024}
{Algera}, H. S.~B., {Inami}, H., {Sommovigo}, L., {et~al.} 2024, \mnras, 527, 6867, \dodoi{10.1093/mnras/stad3111}

\bibitem[{{{\'A}lvarez Crespo} {et~al.}(2021){{\'A}lvarez Crespo}, {Smoli{\'c}}, {Finoguenov}, {Barrufet}, \& {Aravena}}]{AlvarezCrespo2021}
{{\'A}lvarez Crespo}, N., {Smoli{\'c}}, V., {Finoguenov}, A., {Barrufet}, L., \& {Aravena}, M. 2021, \aap, 646, A174, \dodoi{10.1051/0004-6361/202039227}

\bibitem[{{Aretxaga} {et~al.}(2011){Aretxaga}, {Wilson}, {Aguilar}, {Alberts}, {Scott}, {Scoville}, {Yun}, {Austermann}, {Downes}, {Ezawa}, {Hatsukade}, {Hughes}, {Kawabe}, {Kohno}, {Oshima}, {Perera}, {Tamura}, \& {Zeballos}}]{Aretxaga2011}
{Aretxaga}, I., {Wilson}, G.~W., {Aguilar}, E., {et~al.} 2011, \mnras, 415, 3831, \dodoi{10.1111/j.1365-2966.2011.18989.x}

\bibitem[{{Barger} {et~al.}(2000){Barger}, {Cowie}, \& {Richards}}]{Barger2000}
{Barger}, A.~J., {Cowie}, L.~L., \& {Richards}, E.~A. 2000, \aj, 119, 2092, \dodoi{10.1086/301341}

\bibitem[{{Barger} {et~al.}(1998){Barger}, {Cowie}, {Sanders}, {Fulton}, {Taniguchi}, {Sato}, {Kawara}, \& {Okuda}}]{Barger1998}
{Barger}, A.~J., {Cowie}, L.~L., {Sanders}, D.~B., {et~al.} 1998, \nat, 394, 248, \dodoi{10.1038/28338}

\bibitem[{{Barger} {et~al.}(2007){Barger}, {Cowie}, \& {Wang}}]{Barger2007}
{Barger}, A.~J., {Cowie}, L.~L., \& {Wang}, W.~H. 2007, \apj, 654, 764, \dodoi{10.1086/509102}

\bibitem[{{Barrufet} {et~al.}(2023){Barrufet}, {Oesch}, {Weibel}, {Brammer}, {Bezanson}, {Bouwens}, {Fudamoto}, {Gonzalez}, {Gottumukkala}, {Illingworth}, {Heintz}, {Holden}, {Labbe}, {Magee}, {Naidu}, {Nelson}, {Stefanon}, {Smit}, {van Dokkum}, {Weaver}, \& {Williams}}]{Barrufet2022}
{Barrufet}, L., {Oesch}, P.~A., {Weibel}, A., {et~al.} 2023, \mnras, 522, 449, \dodoi{10.1093/mnras/stad947}

\bibitem[{{Barrufet} {et~al.}(2024){Barrufet}, {Oesch}, {Marques-Chaves}, {Arellano-Cordova}, {Baggen}, {Carnall}, {Cullen}, {Dunlop}, {Gottumukkala}, {Fudamoto}, {Illingworth}, {Magee}, {McLure}, {McLeod}, {Micha{\l}owski}, {Stefanon}, {van Dokkum}, \& {Weibel}}]{Barrufet2024}
{Barrufet}, L., {Oesch}, P., {Marques-Chaves}, R., {et~al.} 2024, arXiv e-prints, arXiv:2404.08052, \dodoi{10.48550/arXiv.2404.08052}

\bibitem[{{Battisti} {et~al.}(2019){Battisti}, {da Cunha}, {Grasha}, {Salvato}, {Daddi}, {Davies}, {Jin}, {Liu}, {Schinnerer}, {Vaccari}, \& {COSMOS Collaboration}}]{magphys-photoz}
{Battisti}, A.~J., {da Cunha}, E., {Grasha}, K., {et~al.} 2019, \apj, 882, 61, \dodoi{10.3847/1538-4357/ab345d}

\bibitem[{{Behroozi} {et~al.}(2010){Behroozi}, {Conroy}, \& {Wechsler}}]{Behroozi2010}
{Behroozi}, P.~S., {Conroy}, C., \& {Wechsler}, R.~H. 2010, \apj, 717, 379, \dodoi{10.1088/0004-637X/717/1/379}

\bibitem[{{Bertin} \& {Arnouts}(1996)}]{sourceextractor}
{Bertin}, E., \& {Arnouts}, S. 1996, \aaps, 117, 393, \dodoi{10.1051/aas:1996164}

\bibitem[{{B{\'e}thermin} {et~al.}(2017){B{\'e}thermin}, {Wu}, {Lagache}, {Davidzon}, {Ponthieu}, {Cousin}, {Wang}, {Dor{\'e}}, {Daddi}, \& {Lapi}}]{Bethermin2017}
{B{\'e}thermin}, M., {Wu}, H.-Y., {Lagache}, G., {et~al.} 2017, \aap, 607, A89, \dodoi{10.1051/0004-6361/201730866}

\bibitem[{{Boquien} {et~al.}(2019){Boquien}, {Burgarella}, {Roehlly}, {Buat}, {Ciesla}, {Corre}, {Inoue}, \& {Salas}}]{Boquien2019}
{Boquien}, M., {Burgarella}, D., {Roehlly}, Y., {et~al.} 2019, \aap, 622, A103, \dodoi{10.1051/0004-6361/201834156}

\bibitem[{{Borys} {et~al.}(2005){Borys}, {Smail}, {Chapman}, {Blain}, {Alexander}, \& {Ivison}}]{Borys2005}
{Borys}, C., {Smail}, I., {Chapman}, S.~C., {et~al.} 2005, \apj, 635, 853, \dodoi{10.1086/491617}

\bibitem[{{Boucaud} {et~al.}(2016){Boucaud}, {Bocchio}, {Abergel}, {Orieux}, {Dole}, \& {Hadj-Youcef}}]{pypher}
{Boucaud}, A., {Bocchio}, M., {Abergel}, A., {et~al.} 2016, A\&A, 596, A63, \dodoi{10.1051/0004-6361/201629080}

\bibitem[{Bradley {et~al.}(2023)Bradley, Sip{\H o}cz, Robitaille, Tollerud, Vin{\'{\i}}cius, Deil, Barbary, Wilson, Busko, Donath, G{\"u}nther, Cara, Lim, Me{\ss}linger, Conseil, Bostroem, Droettboom, Bray, Bratholm, Barentsen, Craig, Rathi, Pascual, Perren, Georgiev, de~Val-Borro, Kerzendorf, Bach, Quint, \& Souchereau}]{photutils}
Bradley, L., Sip{\H o}cz, B., Robitaille, T., {et~al.} 2023, astropy/photutils: 1.8.0, 1.8.0,  Zenodo, \dodoi{10.5281/zenodo.7946442}

\bibitem[{{Brinch} {et~al.}(2024){Brinch}, {Greve}, {Sanders}, {McPartland}, {Chartab}, {Gillman}, {Vijayan}, {Lee}, {Brammer}, {Casey}, {Ilbert}, {Jin}, {Magdis}, {McCracken}, {Sillassen}, {Toft}, \& {Zavala}}]{Brinch2024}
{Brinch}, M., {Greve}, T.~R., {Sanders}, D.~B., {et~al.} 2024, \mnras, 527, 6591, \dodoi{10.1093/mnras/stad3409}

\bibitem[{{Brisbin} {et~al.}(2017){Brisbin}, {Miettinen}, {Aravena}, {Smol{\v{c}}i{\'c}}, {Delvecchio}, {Jiang}, {Magnelli}, {Albrecht}, {Arancibia}, {Aussel}, {Baran}, {Bertoldi}, {B{\'e}thermin}, {Capak}, {Casey}, {Civano}, {Hayward}, {Ilbert}, {Karim}, {Le Fevre}, {Marchesi}, {McCracken}, {Navarrete}, {Novak}, {Riechers}, {Padilla}, {Salvato}, {Scott}, {Schinnerer}, {Sheth}, \& {Tasca}}]{Brisbin2017}
{Brisbin}, D., {Miettinen}, O., {Aravena}, M., {et~al.} 2017, \aap, 608, A15, \dodoi{10.1051/0004-6361/201730558}

\bibitem[{{Calvi} {et~al.}(2023){Calvi}, {Castignani}, \& {Dannerbauer}}]{Calvi2023}
{Calvi}, R., {Castignani}, G., \& {Dannerbauer}, H. 2023, \aap, 678, A15, \dodoi{10.1051/0004-6361/202346200}

\bibitem[{{Capak} {et~al.}(2007){Capak}, {Aussel}, {Ajiki}, {McCracken}, {Mobasher}, {Scoville}, {Shopbell}, {Taniguchi}, {Thompson}, {Tribiano}, {Sasaki}, {Blain}, {Brusa}, {Carilli}, {Comastri}, {Carollo}, {Cassata}, {Colbert}, {Ellis}, {Elvis}, {Giavalisco}, {Green}, {Guzzo}, {Hasinger}, {Ilbert}, {Impey}, {Jahnke}, {Kartaltepe}, {Kneib}, {Koda}, {Koekemoer}, {Komiyama}, {Leauthaud}, {Le Fevre}, {Lilly}, {Liu}, {Massey}, {Miyazaki}, {Murayama}, {Nagao}, {Peacock}, {Pickles}, {Porciani}, {Renzini}, {Rhodes}, {Rich}, {Salvato}, {Sanders}, {Scarlata}, {Schiminovich}, {Schinnerer}, {Scodeggio}, {Sheth}, {Shioya}, {Tasca}, {Taylor}, {Yan}, \& {Zamorani}}]{Capak2007}
{Capak}, P., {Aussel}, H., {Ajiki}, M., {et~al.} 2007, \apjs, 172, 99, \dodoi{10.1086/519081}

\bibitem[{{Carnall} {et~al.}(2019){Carnall}, {Leja}, {Johnson}, {McLure}, {Dunlop}, \& {Conroy}}]{Carnall2019}
{Carnall}, A.~C., {Leja}, J., {Johnson}, B.~D., {et~al.} 2019, \apj, 873, 44, \dodoi{10.3847/1538-4357/ab04a2}

\bibitem[{{Casey}(2016)}]{Casey2016}
{Casey}, C.~M. 2016, \apj, 824, 36, \dodoi{10.3847/0004-637X/824/1/36}

\bibitem[{{Casey} {et~al.}(2014{\natexlab{a}}){Casey}, {Narayanan}, \& {Cooray}}]{Casey2014}
{Casey}, C.~M., {Narayanan}, D., \& {Cooray}, A. 2014{\natexlab{a}}, \physrep, 541, 45, \dodoi{10.1016/j.physrep.2014.02.009}

\bibitem[{{Casey} {et~al.}(2009){Casey}, {Chapman}, {Beswick}, {Biggs}, {Blain}, {Hainline}, {Ivison}, {Muxlow}, \& {Smail}}]{Casey2009}
{Casey}, C.~M., {Chapman}, S.~C., {Beswick}, R.~J., {et~al.} 2009, \mnras, 399, 121, \dodoi{10.1111/j.1365-2966.2009.15291.x}

\bibitem[{{Casey} {et~al.}(2013){Casey}, {Chen}, {Cowie}, {Barger}, {Capak}, {Ilbert}, {Koss}, {Lee}, {Le Floc'h}, {Sanders}, \& {Williams}}]{Casey2013}
{Casey}, C.~M., {Chen}, C.-C., {Cowie}, L.~L., {et~al.} 2013, \mnras, 436, 1919, \dodoi{10.1093/mnras/stt1673}

\bibitem[{{Casey} {et~al.}(2014{\natexlab{b}}){Casey}, {Scoville}, {Sanders}, {Lee}, {Cooray}, {Finkelstein}, {Capak}, {Conley}, {De Zotti}, {Farrah}, {Fu}, {Le Floc'h}, {Ilbert}, {Ivison}, \& {Takeuchi}}]{Casey2014b}
{Casey}, C.~M., {Scoville}, N.~Z., {Sanders}, D.~B., {et~al.} 2014{\natexlab{b}}, \apj, 796, 95, \dodoi{10.1088/0004-637X/796/2/95}

\bibitem[{{Casey} {et~al.}(2015){Casey}, {Cooray}, {Capak}, {Fu}, {Kovac}, {Lilly}, {Sanders}, {Scoville}, \& {Treister}}]{Casey2015}
{Casey}, C.~M., {Cooray}, A., {Capak}, P., {et~al.} 2015, \apjl, 808, L33, \dodoi{10.1088/2041-8205/808/2/L33}

\bibitem[{{Casey} {et~al.}(2018){Casey}, {Zavala}, {Spilker}, {da Cunha}, {Hodge}, {Hung}, {Staguhn}, {Finkelstein}, \& {Drew}}]{Casey2018}
{Casey}, C.~M., {Zavala}, J.~A., {Spilker}, J., {et~al.} 2018, \apj, 862, 77, \dodoi{10.3847/1538-4357/aac82d}

\bibitem[{{Casey} {et~al.}(2019){Casey}, {Zavala}, {Aravena}, {B{\'e}thermin}, {Caputi}, {Champagne}, {Clements}, {da Cunha}, {Drew}, {Finkelstein}, {Hayward}, {Kartaltepe}, {Knudsen}, {Koekemoer}, {Magdis}, {Man}, {Manning}, {Scoville}, {Sheth}, {Spilker}, {Staguhn}, {Talia}, {Taniguchi}, {Toft}, {Treister}, \& {Yun}}]{Casey2019}
{Casey}, C.~M., {Zavala}, J.~A., {Aravena}, M., {et~al.} 2019, \apj, 887, 55, \dodoi{10.3847/1538-4357/ab52ff}

\bibitem[{{Casey} {et~al.}(2021){Casey}, {Zavala}, {Manning}, {Aravena}, {B{\'e}thermin}, {Caputi}, {Champagne}, {Clements}, {Drew}, {Finkelstein}, {Fujimoto}, {Hayward}, {Dekel}, {Kokorev}, {Lagos}, {Long}, {Magdis}, {Man}, {Mitsuhashi}, {Popping}, {Spilker}, {Staguhn}, {Talia}, {Toft}, {Treister}, {Weaver}, \& {Yun}}]{Casey2021}
{Casey}, C.~M., {Zavala}, J.~A., {Manning}, S.~M., {et~al.} 2021, \apj, 923, 215, \dodoi{10.3847/1538-4357/ac2eb4}

\bibitem[{{Casey} {et~al.}(2023{\natexlab{a}}){Casey}, {Kartaltepe}, {Drakos}, {Franco}, {Harish}, {Paquereau}, {Ilbert}, {Rose}, {Cox}, {Nightingale}, {Robertson}, {Silverman}, {Koekemoer}, {Massey}, {McCracken}, {Rhodes}, {Akins}, {Allen}, {Amvrosiadis}, {Arango-Toro}, {Bagley}, {Bongiorno}, {Capak}, {Champagne}, {Chartab}, {Ch{\'a}vez Ortiz}, {Chworowsky}, {Cooke}, {Cooper}, {Darvish}, {Ding}, {Faisst}, {Finkelstein}, {Fujimoto}, {Gentile}, {Gillman}, {Gould}, {Gozaliasl}, {Hayward}, {He}, {Hemmati}, {Hirschmann}, {Jahnke}, {Jin}, {Khostovan}, {Kokorev}, {Lambrides}, {Laigle}, {Larson}, {Leung}, {Liu}, {Liaudat}, {Long}, {Magdis}, {Mahler}, {Mainieri}, {Manning}, {Maraston}, {Martin}, {McCleary}, {McKinney}, {McPartland}, {Mobasher}, {Pattnaik}, {Renzini}, {Rich}, {Sanders}, {Sattari}, {Scognamiglio}, {Scoville}, {Sheth}, {Shuntov}, {Sparre}, {Suzuki}, {Talia}, {Toft}, {Trakhtenbrot}, {Urry}, {Valentino}, {Vanderhoof}, {Vardoulaki}, {Weaver}, {Whitaker}, {Wilkins}, {Yang}, \& {Zavala}}]{cosmos-web}
{Casey}, C.~M., {Kartaltepe}, J.~S., {Drakos}, N.~E., {et~al.} 2023{\natexlab{a}}, \apj, 954, 31, \dodoi{10.3847/1538-4357/acc2bc}

\bibitem[{{Casey} {et~al.}(2023{\natexlab{b}}){Casey}, {Kartaltepe}, {Drakos}, {Franco}, {Ilbert}, {Rose}, {Cox}, {Nightingale}, {Robertson}, {Silverman}, {Koekemoer}, {Massey}, {McCracken}, {Rhodes}, {Akins}, {Amvrosiadis}, {Arango-Toro}, {Bagley}, {Capak}, {Champagne}, {Chartab}, {Chavez Ortiz}, {Cooke}, {Cooper}, {Darvish}, {Ding}, {Faisst}, {Finkelstein}, {Fujimoto}, {Gentile}, {Gillman}, {Gould}, {Gozaliasl}, {Harish}, {Hayward}, {He}, {Hemmati}, {Hirschmann}, {Jin}, {Khostovan}, {Kokorev}, {Lambrides}, {Laigle}, {Leung}, {Liu}, {Liaudat}, {Long}, {Magdis}, {Mahler}, {Mainieri}, {Manning}, {Maraston}, {Martin}, {McCleary}, {McKinney}, {McPartland}, {Mobasher}, {Pattnaik}, {Renzini}, {Rich}, {Sanders}, {Sattari}, {Scognamiglio}, {Scoville}, {Sheth}, {Shuntov}, {Sparre}, {Suzuki}, {Talia}, {Toft}, {Trakhtenbrot}, {Urry}, {Valentino}, {Vanderhoof}, {Vardoulaki}, {Weaver}, {Whitaker}, {Wilkins}, {Yang}, \& {Zavala}}]{Casey2022}
---. 2023{\natexlab{b}}, arXiv e-prints, arXiv:2211.07865.
\newblock \doarXiv{2211.07865}

\bibitem[{{Chabrier}(2003)}]{Chabrier2003}
{Chabrier}, G. 2003, \pasp, 115, 763, \dodoi{10.1086/376392}

\bibitem[{{Chapman} {et~al.}(2003){Chapman}, {Blain}, {Ivison}, \& {Smail}}]{Chapman2003}
{Chapman}, S.~C., {Blain}, A.~W., {Ivison}, R.~J., \& {Smail}, I.~R. 2003, \nat, 422, 695, \dodoi{10.1038/nature01540}

\bibitem[{{Chapman} {et~al.}(2005){Chapman}, {Blain}, {Smail}, \& {Ivison}}]{Chapman2005}
{Chapman}, S.~C., {Blain}, A.~W., {Smail}, I., \& {Ivison}, R.~J. 2005, \apj, 622, 772, \dodoi{10.1086/428082}

\bibitem[{{Chapman} {et~al.}(2004){Chapman}, {Smail}, {Windhorst}, {Muxlow}, \& {Ivison}}]{Chapman2004}
{Chapman}, S.~C., {Smail}, I., {Windhorst}, R., {Muxlow}, T., \& {Ivison}, R.~J. 2004, \apj, 611, 732, \dodoi{10.1086/422383}

\bibitem[{{Charlot} \& {Fall}(2000)}]{Charlot2000}
{Charlot}, S., \& {Fall}, S.~M. 2000, \apj, 539, 718, \dodoi{10.1086/309250}

\bibitem[{{Chen} {et~al.}(2013){Chen}, {Cowie}, {Barger}, {Casey}, {Lee}, {Sanders}, {Wang}, \& {Williams}}]{Chen2013}
{Chen}, C.-C., {Cowie}, L.~L., {Barger}, A.~J., {et~al.} 2013, \apj, 776, 131, \dodoi{10.1088/0004-637X/776/2/131}

\bibitem[{{Chen} {et~al.}(2022){Chen}, {Gao}, {Hsu}, {Liao}, {Ling}, {Lo}, {Smail}, {Wang}, \& {Wang}}]{Chen2022}
{Chen}, C.-C., {Gao}, Z.-K., {Hsu}, Q.-N., {et~al.} 2022, \apjl, 939, L7, \dodoi{10.3847/2041-8213/ac98c6}

\bibitem[{{Chen} {et~al.}(2021){Chen}, {Fang}, {Lin}, {Zhang}, {Chen}, \& {Kong}}]{Chen2021}
{Chen}, Z., {Fang}, G., {Lin}, Z., {et~al.} 2021, \apj, 906, 71, \dodoi{10.3847/1538-4357/abc9bb}

\bibitem[{{Cheng} {et~al.}(2022){Cheng}, {Yan}, {Huang}, {Willmer}, {Ma}, \& {Orellana-Gonz{\'a}lez}}]{Cheng2022}
{Cheng}, C., {Yan}, H., {Huang}, J.-S., {et~al.} 2022, \apjl, 936, L19, \dodoi{10.3847/2041-8213/ac8d08}

\bibitem[{{Cheng} {et~al.}(2019){Cheng}, {Clements}, {Greenslade}, {Cairns}, {Andreani}, {Bremer}, {Conversi}, {Cooray}, {Dannerbauer}, {De Zotti}, {Eales}, {Gonz{\'a}lez-Nuevo}, {Ibar}, {Leeuw}, {Ma}, {Micha{\l}owski}, {Nayyeri}, {Riechers}, {Scott}, {Temi}, {Vaccari}, {Valtchanov}, {van Kampen}, \& {Wang}}]{Cheng2019}
{Cheng}, T., {Clements}, D.~L., {Greenslade}, J., {et~al.} 2019, \mnras, 490, 3840, \dodoi{10.1093/mnras/stz2640}

\bibitem[{{Civano} {et~al.}(2016){Civano}, {Marchesi}, {Comastri}, {Urry}, {Elvis}, {Cappelluti}, {Puccetti}, {Brusa}, {Zamorani}, {Hasinger}, {Aldcroft}, {Alexander}, {Allevato}, {Brunner}, {Capak}, {Finoguenov}, {Fiore}, {Fruscione}, {Gilli}, {Glotfelty}, {Griffiths}, {Hao}, {Harrison}, {Jahnke}, {Kartaltepe}, {Karim}, {LaMassa}, {Lanzuisi}, {Miyaji}, {Ranalli}, {Salvato}, {Sargent}, {Scoville}, {Schawinski}, {Schinnerer}, {Silverman}, {Smolcic}, {Stern}, {Toft}, {Trakhtenbrot}, {Treister}, \& {Vignali}}]{Civano2016}
{Civano}, F., {Marchesi}, S., {Comastri}, A., {et~al.} 2016, \apj, 819, 62, \dodoi{10.3847/0004-637X/819/1/62}

\bibitem[{{Clements} {et~al.}(2014){Clements}, {Braglia}, {Hyde}, {P{\'e}rez-Fournon}, {Bock}, {Cava}, {Chapman}, {Conley}, {Cooray}, {Farrah}, {Gonz{\'a}lez Solares}, {Marchetti}, {Marsden}, {Oliver}, {Roseboom}, {Schulz}, {Smith}, {Vaccari}, {Vieira}, {Viero}, {Wang}, {Wardlow}, {Zemcov}, \& {de Zotti}}]{Clements2014}
{Clements}, D.~L., {Braglia}, F.~G., {Hyde}, A.~K., {et~al.} 2014, \mnras, 439, 1193, \dodoi{10.1093/mnras/stt2253}

\bibitem[{{Clements} {et~al.}(2016){Clements}, {Braglia}, {Petitpas}, {Greenslade}, {Cooray}, {Valiante}, {De Zotti}, {O'Halloran}, {Holdship}, {Morris}, {P{\'e}rez-Fournon}, {Herranz}, {Riechers}, {Baes}, {Bremer}, {Bourne}, {Dannerbauer}, {Dariush}, {Dunne}, {Eales}, {Fritz}, {Gonzalez-Nuevo}, {Hopwood}, {Ibar}, {Ivison}, {Leeuw}, {Maddox}, {Micha{\l}owski}, {Negrello}, {Omont}, {Oteo}, {Serjeant}, {Valtchanov}, {Vieira}, {Wardlow}, \& {van der Werf}}]{Clements2016}
{Clements}, D.~L., {Braglia}, F., {Petitpas}, G., {et~al.} 2016, \mnras, 461, 1719, \dodoi{10.1093/mnras/stw1224}

\bibitem[{{Cochrane} {et~al.}(2024){Cochrane}, {Angl{\'e}s-Alc{\'a}zar}, {Cullen}, \& {Hayward}}]{Cochrane2024}
{Cochrane}, R.~K., {Angl{\'e}s-Alc{\'a}zar}, D., {Cullen}, F., \& {Hayward}, C.~C. 2024, \apj, 961, 37, \dodoi{10.3847/1538-4357/ad02f8}

\bibitem[{{Cooper} {et~al.}(2022){Cooper}, {Casey}, {Zavala}, {Champagne}, {da Cunha}, {Long}, {Spilker}, \& {Staguhn}}]{Cooper2022}
{Cooper}, O.~R., {Casey}, C.~M., {Zavala}, J.~A., {et~al.} 2022, \apj, 930, 32, \dodoi{10.3847/1538-4357/ac616d}

\bibitem[{{Cornish} {et~al.}(2024){Cornish}, {Wardlow}, {Wade}, {Sobral}, {Brandt}, {Cox}, {Dannerbauer}, {Decarli}, {Gullberg}, {Knudsen}, {Stott}, {Swinbank}, {Walter}, \& {van der Werf}}]{Cornish2024}
{Cornish}, T.~M., {Wardlow}, J., {Wade}, H., {et~al.} 2024, arXiv e-prints, arXiv:2408.00063.
\newblock \doarXiv{2408.00063}

\bibitem[{{Cowie} {et~al.}(2018){Cowie}, {Gonz{\'a}lez-L{\'o}pez}, {Barger}, {Bauer}, {Hsu}, \& {Wang}}]{Cowie2018}
{Cowie}, L.~L., {Gonz{\'a}lez-L{\'o}pez}, J., {Barger}, A.~J., {et~al.} 2018, \apj, 865, 106, \dodoi{10.3847/1538-4357/aadc63}

\bibitem[{{da Cunha} {et~al.}(2008){da Cunha}, {Charlot}, \& {Elbaz}}]{magphys2008}
{da Cunha}, E., {Charlot}, S., \& {Elbaz}, D. 2008, \mnras, 388, 1595, \dodoi{10.1111/j.1365-2966.2008.13535.x}

\bibitem[{{da Cunha} {et~al.}(2015){da Cunha}, {Walter}, {Smail}, {Swinbank}, {Simpson}, {Decarli}, {Hodge}, {Weiss}, {van der Werf}, {Bertoldi}, {Chapman}, {Cox}, {Danielson}, {Dannerbauer}, {Greve}, {Ivison}, {Karim}, \& {Thomson}}]{daCunha2015}
{da Cunha}, E., {Walter}, F., {Smail}, I.~R., {et~al.} 2015, \apj, 806, 110, \dodoi{10.1088/0004-637X/806/1/110}

\bibitem[{{da Cunha} {et~al.}(2021){da Cunha}, {Hodge}, {Casey}, {Algera}, {Kaasinen}, {Smail}, {Walter}, {Brandt}, {Dannerbauer}, {Decarli}, {Groves}, {Knudsen}, {Swinbank}, {Weiss}, {van der Werf}, \& {Zavala}}]{daCunha2021}
{da Cunha}, E., {Hodge}, J.~A., {Casey}, C.~M., {et~al.} 2021, \apj, 919, 30, \dodoi{10.3847/1538-4357/ac0ae0}

\bibitem[{{Damjanov} {et~al.}(2018){Damjanov}, {Zahid}, {Geller}, {Fabricant}, \& {Hwang}}]{Damjanov2018}
{Damjanov}, I., {Zahid}, H.~J., {Geller}, M.~J., {Fabricant}, D.~G., \& {Hwang}, H.~S. 2018, \apjs, 234, 21, \dodoi{10.3847/1538-4365/aaa01c}

\bibitem[{{Dannerbauer} {et~al.}(2014){Dannerbauer}, {Kurk}, {De Breuck}, {Wylezalek}, {Santos}, {Koyama}, {Seymour}, {Tanaka}, {Hatch}, {Altieri}, {Coia}, {Galametz}, {Kodama}, {Miley}, {R{\"o}ttgering}, {Sanchez-Portal}, {Valtchanov}, {Venemans}, \& {Ziegler}}]{Dannerbauer2014}
{Dannerbauer}, H., {Kurk}, J.~D., {De Breuck}, C., {et~al.} 2014, \aap, 570, A55, \dodoi{10.1051/0004-6361/201423771}

\bibitem[{{Delvecchio} {et~al.}(2021){Delvecchio}, {Daddi}, {Sargent}, {Jarvis}, {Elbaz}, {Jin}, {Liu}, {Whittam}, {Algera}, {Carraro}, {D'Eugenio}, {Delhaize}, {Kalita}, {Leslie}, {Moln{\'a}r}, {Novak}, {Prandoni}, {Smol{\v{c}}i{\'c}}, {Ao}, {Aravena}, {Bournaud}, {Collier}, {Randriamampandry}, {Randriamanakoto}, {Rodighiero}, {Schober}, {White}, \& {Zamorani}}]{Delvecchio2021}
{Delvecchio}, I., {Daddi}, E., {Sargent}, M.~T., {et~al.} 2021, \aap, 647, A123, \dodoi{10.1051/0004-6361/202039647}

\bibitem[{{Draine} \& {Li}(2007)}]{DraineLi2007}
{Draine}, B.~T., \& {Li}, A. 2007, \apj, 657, 810, \dodoi{10.1086/511055}

\bibitem[{{Draine} {et~al.}(2014){Draine}, {Aniano}, {Krause}, {Groves}, {Sandstrom}, {Braun}, {Leroy}, {Klaas}, {Linz}, {Rix}, {Schinnerer}, {Schmiedeke}, \& {Walter}}]{Draine2014}
{Draine}, B.~T., {Aniano}, G., {Krause}, O., {et~al.} 2014, \apj, 780, 172, \dodoi{10.1088/0004-637X/780/2/172}

\bibitem[{{Drew} \& {Casey}(2022)}]{Drew2022}
{Drew}, P.~M., \& {Casey}, C.~M. 2022, \apj, 930, 142, \dodoi{10.3847/1538-4357/ac6270}

\bibitem[{{Dudzevi{\v{c}}i{\={u}}t{\.{e}}} {et~al.}(2020){Dudzevi{\v{c}}i{\={u}}t{\.{e}}}, {Smail}, {Swinbank}, {Stach}, {Almaini}, {da Cunha}, {An}, {Arumugam}, {Birkin}, {Blain}, {Chapman}, {Chen}, {Conselice}, {Coppin}, {Dunlop}, {Farrah}, {Geach}, {Gullberg}, {Hartley}, {Hodge}, {Ivison}, {Maltby}, {Scott}, {Simpson}, {Simpson}, {Thomson}, {Walter}, {Wardlow}, {Weiss}, \& {van der Werf}}]{Dudzeviciute2020}
{Dudzevi{\v{c}}i{\={u}}t{\.{e}}}, U., {Smail}, I., {Swinbank}, A.~M., {et~al.} 2020, \mnras, 494, 3828, \dodoi{10.1093/mnras/staa769}

\bibitem[{{Dunlop} {et~al.}(2021){Dunlop}, {Abraham}, {Ashby}, {Bagley}, {Best}, {Bongiorno}, {Bouwens}, {Bowler}, {Brammer}, {Bremer}, {Calabro'}, {Carnall}, {Castellano}, {Cirasuolo}, {Conselice}, {Cullen}, {Dave}, {Dayal}, {Dekel}, {Dickinson}, {Duncan}, {Elbaz}, {Ellis}, {Ferguson}, {Ferrara}, {Finkelstein}, {Fontana}, {Furlanetto}, {Fynbo}, {Gallerani}, {Gardner}, {Giavalisco}, {Grazian}, {Grogin}, {Harikane}, {Hopkins}, {Ilbert}, {Illingworth}, {Juneau}, {Jung}, {Kartaltepe}, {Kassin}, {Kauffmann}, {Khochfar}, {Kirkpatrick}, {Kocevski}, {Koekemoer}, {Labbe}, {Laporte}, {Larson}, {Lucas}, {Magee}, {Mason}, {McCracken}, {McLeod}, {McLure}, {Merlin}, {Mesinger}, {Milvang-Jensen}, {Newman}, {Oesch}, {Ouchi}, {Pacifici}, {Papovich}, {Peacock}, {Peeples}, {Pentericci}, {Perez-Gonzalez}, {Pirzkal}, {Pope}, {Pye}, {Reddy}, {Robertson}, {Salvato}, {Santini}, {Schaerer}, {Shapley}, {Simons}, {Smit}, {Smith}, {Snyder}, {Somerville}, {Stanway}, {Stefanon}, {Tasca}, {Tikkanen}, {Tresse}, {Trump}, {Whitaker},
  {Wilkins}, {Wright}, {Wyithe}, {van Dokkum}, \& {van der Werf}}]{PRIMER}
{Dunlop}, J.~S., {Abraham}, R.~G., {Ashby}, M. L.~N., {et~al.} 2021, {PRIMER: Public Release IMaging for Extragalactic Research}, JWST Proposal. Cycle 1, ID. \#1837

\bibitem[{{Eales} {et~al.}(1999){Eales}, {Lilly}, {Gear}, {Dunne}, {Bond}, {Hammer}, {Le F{\`e}vre}, \& {Crampton}}]{Eales1999}
{Eales}, S., {Lilly}, S., {Gear}, W., {et~al.} 1999, \apj, 515, 518, \dodoi{10.1086/307069}

\bibitem[{{Epinat} {et~al.}(2024){Epinat}, {Contini}, {Mercier}, {Ciesla}, {Lemaux}, {Johnson}, {Richard}, {Brinchmann}, {Boogaard}, {Carton}, {Michel-Dansac}, {Bacon}, {Krajnovi{\'c}}, {Finley}, {Schroetter}, {Ventou}, {Abril-Melgarejo}, {Boselli}, {Bouch{\'e}}, {Kollatschny}, {Kova{\v{c}}}, {Paalvast}, {Soucail}, {Urrutia}, \& {Weilbacher}}]{Epinat2024}
{Epinat}, B., {Contini}, T., {Mercier}, W., {et~al.} 2024, \aap, 683, A205, \dodoi{10.1051/0004-6361/202348038}

\bibitem[{{Euclid Collaboration} {et~al.}(2022){Euclid Collaboration}, {Moneti}, {McCracken}, {Shuntov}, {Kauffmann}, {Capak}, {Davidzon}, {Ilbert}, {Scarlata}, {Toft}, {Weaver}, {Chary}, {Cuby}, {Faisst}, {Masters}, {McPartland}, {Mobasher}, {Sanders}, {Scaramella}, {Stern}, {Szapudi}, {Teplitz}, {Zalesky}, {Amara}, {Auricchio}, {Bodendorf}, {Bonino}, {Branchini}, {Brau-Nogue}, {Brescia}, {Brinchmann}, {Capobianco}, {Carbone}, {Carretero}, {Castander}, {Castellano}, {Cavuoti}, {Cimatti}, {Cledassou}, {Congedo}, {Conselice}, {Conversi}, {Copin}, {Corcione}, {Costille}, {Cropper}, {Da Silva}, {Degaudenzi}, {Douspis}, {Dubath}, {Duncan}, {Dupac}, {Dusini}, {Farrens}, {Ferriol}, {Fosalba}, {Frailis}, {Franceschi}, {Fumana}, {Garilli}, {Gillis}, {Giocoli}, {Granett}, {Grazian}, {Grupp}, {Haugan}, {Hoekstra}, {Holmes}, {Hormuth}, {Hudelot}, {Jahnke}, {Kermiche}, {Kiessling}, {Kilbinger}, {Kitching}, {Kohley}, {K{\"u}mmel}, {Kunz}, {Kurki-Suonio}, {Ligori}, {Lilje}, {Lloro}, {Maiorano}, {Mansutti}, {Marggraf},
  {Markovic}, {Marulli}, {Massey}, {Maurogordato}, {Meneghetti}, {Merlin}, {Meylan}, {Moresco}, {Moscardini}, {Munari}, {Niemi}, {Padilla}, {Paltani}, {Pasian}, {Pedersen}, {Pires}, {Poncet}, {Popa}, {Pozzetti}, {Raison}, {Rebolo}, {Rhodes}, {Rix}, {Roncarelli}, {Rossetti}, {Saglia}, {Schneider}, {Secroun}, {Seidel}, {Serrano}, {Sirignano}, {Sirri}, {Stanco}, {Tallada-Cresp{\'\i}}, {Taylor}, {Tereno}, {Toledo-Moreo}, {Torradeflot}, {Wang}, {Welikala}, {Weller}, {Zamorani}, {Zoubian}, {Andreon}, {Bardelli}, {Camera}, {Graci{\'a}-Carpio}, {Medinaceli}, {Mei}, {Polenta}, {Romelli}, {Sureau}, {Tenti}, {Vassallo}, {Zacchei}, {Zucca}, {Baccigalupi}, {Balaguera-Antol{\'\i}nez}, {Bernardeau}, {Biviano}, {Bolzonella}, {Bozzo}, {Burigana}, {Cabanac}, {Cappi}, {Carvalho}, {Casas}, {Castignani}, {Colodro-Conde}, {Coupon}, {Courtois}, {Di Ferdinando}, {Farina}, {Finelli}, {Flose-Reimberg}, {Fotopoulou}, {Galeotta}, {Ganga}, {Garcia-Bellido}, {Gaztanaga}, {Gozaliasl}, {Hook}, {Joachimi}, {Kansal}, {Keihanen},
  {Kirkpatrick}, {Lindholm}, {Mainetti}, {Maino}, {Maoli}, {Martinelli}, {Martinet}, {Maturi}, {Metcalf}, {Morgante}, {Morisset}, {Nucita}, {Patrizii}, {Potter}, {Renzi}, {Riccio}, {S{\'a}nchez}, {Sapone}, {Schirmer}, {Schultheis}, {Scottez}, {Sefusatti}, {Teyssier}, {Tubio}, {Tutusaus}, {Valiviita}, {Viel}, \& {Hildebrandt}}]{Moneti2022}
{Euclid Collaboration}, {Moneti}, A., {McCracken}, H.~J., {et~al.} 2022, \aap, 658, A126, \dodoi{10.1051/0004-6361/202142361}

\bibitem[{{Flores-Cacho} {et~al.}(2016){Flores-Cacho}, {Pierini}, {Soucail}, {Montier}, {Dole}, {Pointecouteau}, {Pell{\'o}}, {Le Floc'h}, {Nesvadba}, {Lagache}, {Guery}, \& {Ca{\~n}ameras}}]{FloresCacho2016}
{Flores-Cacho}, I., {Pierini}, D., {Soucail}, G., {et~al.} 2016, \aap, 585, A54, \dodoi{10.1051/0004-6361/201425226}

\bibitem[{{Franco} {et~al.}(2018){Franco}, {Elbaz}, {B{\'e}thermin}, {Magnelli}, {Schreiber}, {Ciesla}, {Dickinson}, {Nagar}, {Silverman}, {Daddi}, {Alexander}, {Wang}, {Pannella}, {Le Floc'h}, {Pope}, {Giavalisco}, {Maury}, {Bournaud}, {Chary}, {Demarco}, {Ferguson}, {Finkelstein}, {Inami}, {Iono}, {Juneau}, {Lagache}, {Leiton}, {Lin}, {Magdis}, {Messias}, {Motohara}, {Mullaney}, {Okumura}, {Papovich}, {Pforr}, {Rujopakarn}, {Sargent}, {Shu}, \& {Zhou}}]{Franco2018}
{Franco}, M., {Elbaz}, D., {B{\'e}thermin}, M., {et~al.} 2018, \aap, 620, A152, \dodoi{10.1051/0004-6361/201832928}

\bibitem[{{Gall} {et~al.}(2011){Gall}, {Andersen}, \& {Hjorth}}]{Gall2011}
{Gall}, C., {Andersen}, A.~C., \& {Hjorth}, J. 2011, \aap, 528, A13, \dodoi{10.1051/0004-6361/201015286}

\bibitem[{{Geach} {et~al.}(2017){Geach}, {Dunlop}, {Halpern}, {Smail}, {van der Werf}, {Alexander}, {Almaini}, {Aretxaga}, {Arumugam}, {Asboth}, {Banerji}, {Beanlands}, {Best}, {Blain}, {Birkinshaw}, {Chapin}, {Chapman}, {Chen}, {Chrysostomou}, {Clarke}, {Clements}, {Conselice}, {Coppin}, {Cowley}, {Danielson}, {Eales}, {Edge}, {Farrah}, {Gibb}, {Harrison}, {Hine}, {Hughes}, {Ivison}, {Jarvis}, {Jenness}, {Jones}, {Karim}, {Koprowski}, {Knudsen}, {Lacey}, {Mackenzie}, {Marsden}, {McAlpine}, {McMahon}, {Meijerink}, {Micha{\l}owski}, {Oliver}, {Page}, {Peacock}, {Rigopoulou}, {Robson}, {Roseboom}, {Rotermund}, {Scott}, {Serjeant}, {Simpson}, {Simpson}, {Smith}, {Spaans}, {Stanley}, {Stevens}, {Swinbank}, {Targett}, {Thomson}, {Valiante}, {Wake}, {Webb}, {Willott}, {Zavala}, \& {Zemcov}}]{Geach2017}
{Geach}, J.~E., {Dunlop}, J.~S., {Halpern}, M., {et~al.} 2017, \mnras, 465, 1789, \dodoi{10.1093/mnras/stw2721}

\bibitem[{{Gentile} {et~al.}(2024){Gentile}, {Talia}, {Behiri}, {Zamorani}, {Barchiesi}, {Vignali}, {Pozzi}, {Bethermin}, {Enia}, {Faisst}, {Giulietti}, {Gruppioni}, {Lapi}, {Massardi}, {Smol{\v{c}}i{\'c}}, {Vaccari}, \& {Cimatti}}]{Gentile2024}
{Gentile}, F., {Talia}, M., {Behiri}, M., {et~al.} 2024, \apj, 962, 26, \dodoi{10.3847/1538-4357/ad1519}

\bibitem[{{Gillman} {et~al.}(2023){Gillman}, {Gullberg}, {Brammer}, {Vijayan}, {Lee}, {Bl{\'a}nquez}, {Brinch}, {Greve}, {Jermann}, {Jin}, {Kokorev}, {Liu}, {Magdis}, {Rizzo}, \& {Valentino}}]{Gillman2023}
{Gillman}, S., {Gullberg}, B., {Brammer}, G., {et~al.} 2023, \aap, 676, A26, \dodoi{10.1051/0004-6361/202346531}

\bibitem[{{Gillman} {et~al.}(2024){Gillman}, {Smail}, {Gullberg}, {Swinbank}, {Vijayan}, {Lee}, {Brammer}, {Dudzevi{\v{c}}i{\={u}}t{\.{e}}}, {Greve}, {Almaini}, {Brinch}, {Chapman}, {Chen}, {Ikarashi}, {Matsuda}, {Wang}, {Walter}, \& {van der Werf}}]{Gillman2024}
{Gillman}, S., {Smail}, I., {Gullberg}, B., {et~al.} 2024, arXiv e-prints, arXiv:2406.03544, \dodoi{10.48550/arXiv.2406.03544}

\bibitem[{{G{\'o}mez-Guijarro} {et~al.}(2019){G{\'o}mez-Guijarro}, {Riechers}, {Pavesi}, {Magdis}, {Leung}, {Valentino}, {Toft}, {Aravena}, {Chapman}, {Clements}, {Dannerbauer}, {Oliver}, {P{\'e}rez-Fournon}, \& {Valtchanov}}]{GomezGuijarro2019}
{G{\'o}mez-Guijarro}, C., {Riechers}, D.~A., {Pavesi}, R., {et~al.} 2019, \apj, 872, 117, \dodoi{10.3847/1538-4357/ab002a}

\bibitem[{{Gottumukkala} {et~al.}(2024){Gottumukkala}, {Barrufet}, {Oesch}, {Weibel}, {Allen}, {Alcalde Pampliega}, {Nelson}, {Williams}, {Brammer}, {Fudamoto}, {Gonz{\'a}lez}, {Heintz}, {Illingworth}, {Magee}, {Naidu}, {Shuntov}, {Stefanon}, {Toft}, {Valentino}, \& {Xiao}}]{Gottumukkala2024}
{Gottumukkala}, R., {Barrufet}, L., {Oesch}, P.~A., {et~al.} 2024, \mnras, 530, 966, \dodoi{10.1093/mnras/stae754}

\bibitem[{{Greenslade} {et~al.}(2018){Greenslade}, {Clements}, {Cheng}, {De Zotti}, {Scott}, {Valiante}, {Eales}, {Bremer}, {Dannerbauer}, {Birkinshaw}, {Farrah}, {Harrison}, {Micha{\l}owski}, {Valtchanov}, {Oteo}, {Baes}, {Cooray}, {Negrello}, {Wang}, {van der Werf}, {Dunne}, \& {Dye}}]{Greenslade2018}
{Greenslade}, J., {Clements}, D.~L., {Cheng}, T., {et~al.} 2018, \mnras, 476, 3336, \dodoi{10.1093/mnras/sty023}

\bibitem[{{Gullberg} {et~al.}(2019){Gullberg}, {Smail}, {Swinbank}, {Dudzevi{\v{c}}i{\={u}}t{\.{e}}}, {Stach}, {Thomson}, {Almaini}, {Chen}, {Conselice}, {Cooke}, {Farrah}, {Ivison}, {Maltby}, {Micha{\l}owski}, {Simpson}, {Scott}, {Wardlow}, \& {Weiss}}]{Gullberg2019}
{Gullberg}, B., {Smail}, I., {Swinbank}, A.~M., {et~al.} 2019, \mnras, 490, 4956, \dodoi{10.1093/mnras/stz2835}

\bibitem[{{Guo} {et~al.}(2023){Guo}, {Jogee}, {Finkelstein}, {Chen}, {Wise}, {Bagley}, {Barro}, {Wuyts}, {Kocevski}, {Kartaltepe}, {McGrath}, {Ferguson}, {Mobasher}, {Giavalisco}, {Lucas}, {Zavala}, {Lotz}, {Grogin}, {Huertas-Company}, {Vega-Ferrero}, {Hathi}, {Arrabal Haro}, {Dickinson}, {Koekemoer}, {Papovich}, {Pirzkal}, {Yung}, {Backhaus}, {Bell}, {Calabr{\`o}}, {Cleri}, {Coogan}, {Cooper}, {Costantin}, {Croton}, {Davis}, {Dekel}, {Franco}, {Gardner}, {Holwerda}, {Hutchison}, {Pandya}, {P{\'e}rez-Gonz{\'a}lez}, {Ravindranath}, {Rose}, {Trump}, {de la Vega}, \& {Wang}}]{Guo2023}
{Guo}, Y., {Jogee}, S., {Finkelstein}, S.~L., {et~al.} 2023, \apjl, 945, L10, \dodoi{10.3847/2041-8213/acacfb}

\bibitem[{{Hainline} {et~al.}(2011){Hainline}, {Blain}, {Smail}, {Alexander}, {Armus}, {Chapman}, \& {Ivison}}]{Hainline2011}
{Hainline}, L.~J., {Blain}, A.~W., {Smail}, I., {et~al.} 2011, \apj, 740, 96, \dodoi{10.1088/0004-637X/740/2/96}

\bibitem[{{Hasinger} {et~al.}(2018){Hasinger}, {Capak}, {Salvato}, {Barger}, {Cowie}, {Faisst}, {Hemmati}, {Kakazu}, {Kartaltepe}, {Masters}, {Mobasher}, {Nayyeri}, {Sanders}, {Scoville}, {Suh}, {Steinhardt}, \& {Yang}}]{Hasinger2018}
{Hasinger}, G., {Capak}, P., {Salvato}, M., {et~al.} 2018, \apj, 858, 77, \dodoi{10.3847/1538-4357/aabacf}

\bibitem[{{Hayward} {et~al.}(2018){Hayward}, {Chapman}, {Steidel}, {Golob}, {Casey}, {Smith}, {Zitrin}, {Blain}, {Bremer}, {Chen}, {Coppin}, {Farrah}, {Ibar}, {Micha{\l}owski}, {Sawicki}, {Scott}, {van der Werf}, {Fazio}, {Geach}, {Gurwell}, {Petitpas}, \& {Wilner}}]{Hayward2018}
{Hayward}, C.~C., {Chapman}, S.~C., {Steidel}, C.~C., {et~al.} 2018, \mnras, 476, 2278, \dodoi{10.1093/mnras/sty304}

\bibitem[{{Hezaveh} {et~al.}(2013){Hezaveh}, {Marrone}, {Fassnacht}, {Spilker}, {Vieira}, {Aguirre}, {Aird}, {Aravena}, {Ashby}, {Bayliss}, {Benson}, {Bleem}, {Bothwell}, {Brodwin}, {Carlstrom}, {Chang}, {Chapman}, {Crawford}, {Crites}, {De Breuck}, {de Haan}, {Dobbs}, {Fomalont}, {George}, {Gladders}, {Gonzalez}, {Greve}, {Halverson}, {High}, {Holder}, {Holzapfel}, {Hoover}, {Hrubes}, {Husband}, {Hunter}, {Keisler}, {Lee}, {Leitch}, {Lueker}, {Luong-Van}, {Malkan}, {McIntyre}, {McMahon}, {Mehl}, {Menten}, {Meyer}, {Mocanu}, {Murphy}, {Natoli}, {Padin}, {Plagge}, {Reichardt}, {Rest}, {Ruel}, {Ruhl}, {Sharon}, {Schaffer}, {Shaw}, {Shirokoff}, {Stalder}, {Staniszewski}, {Stark}, {Story}, {Vanderlinde}, {Wei{\ss}}, {Welikala}, \& {Williamson}}]{Hezaveh2013}
{Hezaveh}, Y.~D., {Marrone}, D.~P., {Fassnacht}, C.~D., {et~al.} 2013, \apj, 767, 132, \dodoi{10.1088/0004-637X/767/2/132}

\bibitem[{{Hodge} \& {da Cunha}(2020)}]{Hodge2020}
{Hodge}, J.~A., \& {da Cunha}, E. 2020, Royal Society Open Science, 7, 200556, \dodoi{10.1098/rsos.200556}

\bibitem[{{Hodge} {et~al.}(2013){Hodge}, {Karim}, {Smail}, {Swinbank}, {Walter}, {Biggs}, {Ivison}, {Weiss}, {Alexander}, {Bertoldi}, {Brandt}, {Chapman}, {Coppin}, {Cox}, {Danielson}, {Dannerbauer}, {De Breuck}, {Decarli}, {Edge}, {Greve}, {Knudsen}, {Menten}, {Rix}, {Schinnerer}, {Simpson}, {Wardlow}, \& {van der Werf}}]{Hodge2013}
{Hodge}, J.~A., {Karim}, A., {Smail}, I., {et~al.} 2013, \apj, 768, 91, \dodoi{10.1088/0004-637X/768/1/91}

\bibitem[{{Hodge} {et~al.}(2016){Hodge}, {Swinbank}, {Simpson}, {Smail}, {Walter}, {Alexander}, {Bertoldi}, {Biggs}, {Brandt}, {Chapman}, {Chen}, {Coppin}, {Cox}, {Dannerbauer}, {Edge}, {Greve}, {Ivison}, {Karim}, {Knudsen}, {Menten}, {Rix}, {Schinnerer}, {Wardlow}, {Weiss}, \& {van der Werf}}]{Hodge2016}
{Hodge}, J.~A., {Swinbank}, A.~M., {Simpson}, J.~M., {et~al.} 2016, \apj, 833, 103, \dodoi{10.3847/1538-4357/833/1/103}

\bibitem[{{Hodge} {et~al.}(2024){Hodge}, {da Cunha}, {Kendrew}, {Li}, {Smail}, {Westoby}, {Nayak}, {Swinbank}, {Chen}, {Walter}, {van der Werf}, {Cracraft}, {Battisti}, {Brandt}, {Calistro Rivera}, {Chapman}, {Cox}, {Dannerbauer}, {Decarli}, {Frias Castillo}, {Greve}, {Knudsen}, {Leslie}, {Menten}, {Rybak}, {Schinnerer}, {Wardlow}, \& {Weiss}}]{Hodge2024}
{Hodge}, J.~A., {da Cunha}, E., {Kendrew}, S., {et~al.} 2024, arXiv e-prints, arXiv:2407.15846, \dodoi{10.48550/arXiv.2407.15846}

\bibitem[{{Hopkins} {et~al.}(2008){Hopkins}, {Hernquist}, {Cox}, \& {Kere{\v{s}}}}]{Hopkins2008}
{Hopkins}, P.~F., {Hernquist}, L., {Cox}, T.~J., \& {Kere{\v{s}}}, D. 2008, \apjs, 175, 356, \dodoi{10.1086/524362}

\bibitem[{{Horowitz} {et~al.}(2022){Horowitz}, {Lee}, {Ata}, {M{\"u}ller}, {Krolewski}, {Prochaska}, {Hennawi}, {White}, {Schlegel}, {Rich}, {Nugent}, {Suzuki}, {Kashino}, {Koekemoer}, \& {Lemaux}}]{Horowitz2022}
{Horowitz}, B., {Lee}, K.-G., {Ata}, M., {et~al.} 2022, \apjs, 263, 27, \dodoi{10.3847/1538-4365/ac982d}

\bibitem[{{Howell} {et~al.}(2010){Howell}, {Armus}, {Mazzarella}, {Evans}, {Surace}, {Sanders}, {Petric}, {Appleton}, {Bothun}, {Bridge}, {Chan}, {Charmandaris}, {Frayer}, {Haan}, {Inami}, {Kim}, {Lord}, {Madore}, {Melbourne}, {Schulz}, {U}, {Vavilkin}, {Veilleux}, \& {Xu}}]{Howell2010}
{Howell}, J.~H., {Armus}, L., {Mazzarella}, J.~M., {et~al.} 2010, \apj, 715, 572, \dodoi{10.1088/0004-637X/715/1/572}

\bibitem[{{Hughes} {et~al.}(1998){Hughes}, {Serjeant}, {Dunlop}, {Rowan-Robinson}, {Blain}, {Mann}, {Ivison}, {Peacock}, {Efstathiou}, {Gear}, {Oliver}, {Lawrence}, {Longair}, {Goldschmidt}, \& {Jenness}}]{Hughes1998}
{Hughes}, D.~H., {Serjeant}, S., {Dunlop}, J., {et~al.} 1998, \nat, 394, 241, \dodoi{10.1038/28328}

\bibitem[{{Hung} {et~al.}(2016){Hung}, {Casey}, {Chiang}, {Capak}, {Cowley}, {Darvish}, {Kacprzak}, {Kova{\v{c}}}, {Lilly}, {Nanayakkara}, {Spitler}, {Tran}, \& {Yuan}}]{Hung2016}
{Hung}, C.-L., {Casey}, C.~M., {Chiang}, Y.-K., {et~al.} 2016, \apj, 826, 130, \dodoi{10.3847/0004-637X/826/2/130}

\bibitem[{{Hurley} {et~al.}(2017){Hurley}, {Oliver}, {Betancourt}, {Clarke}, {Cowley}, {Duivenvoorden}, {Farrah}, {Griffin}, {Lacey}, {Le Floc'h}, {Papadopoulos}, {Sargent}, {Scudder}, {Vaccari}, {Valtchanov}, \& {Wang}}]{Hurley2017}
{Hurley}, P.~D., {Oliver}, S., {Betancourt}, M., {et~al.} 2017, \mnras, 464, 885, \dodoi{10.1093/mnras/stw2375}

\bibitem[{{Ivezi{\'c}} {et~al.}(2020){Ivezi{\'c}}, {Connolly}, {VanderPlas}, \& {Gray}}]{astroML-book}
{Ivezi{\'c}}, {\v{Z}}., {Connolly}, A.~J., {VanderPlas}, J.~T., \& {Gray}, A. 2020, {Statistics, Data Mining, and Machine Learning in Astronomy. A Practical Python Guide for the Analysis of Survey Data, Updated Edition}, \dodoi{10.1515/9780691197050}

\bibitem[{{Jim{\'e}nez-Andrade} {et~al.}(2020){Jim{\'e}nez-Andrade}, {Zavala}, {Magnelli}, {Casey}, {Liu}, {Romano-D{\'\i}az}, {Schinnerer}, {Harrington}, {Aretxaga}, {Karim}, {Staguhn}, {Burnham}, {Monta{\~n}a}, {Smol{\v{c}}i{\'c}}, {Yun}, {Bertoldi}, \& {Hughes}}]{JimenezAndrade2020}
{Jim{\'e}nez-Andrade}, E.~F., {Zavala}, J.~A., {Magnelli}, B., {et~al.} 2020, \apj, 890, 171, \dodoi{10.3847/1538-4357/ab6dec}

\bibitem[{{Jin} {et~al.}(2019){Jin}, {Daddi}, {Magdis}, {Liu}, {Schinnerer}, {Papadopoulos}, {Gu}, {Gao}, \& {Calabr{\`o}}}]{Jin2019}
{Jin}, S., {Daddi}, E., {Magdis}, G.~E., {et~al.} 2019, \apj, 887, 144, \dodoi{10.3847/1538-4357/ab55d6}

\bibitem[{{Jin} {et~al.}(2024){Jin}, {Sillassen}, {Hodge}, {Magdis}, {Casey}, {Rizzo}, {Koekemoer}, {Valentino}, {Kokorev}, {Magnelli}, {Gobat}, {Gillman}, {Franco}, {Faisst}, {Kartaltepe}, {Schinnerer}, {Toft}, {Algera}, {Harish}, {Lee}, {Liu}, {Shuntov}, {Talia}, \& {Vijayan}}]{Jin2024}
{Jin}, S., {Sillassen}, N.~B., {Hodge}, J., {et~al.} 2024, arXiv e-prints, arXiv:2407.07585, \dodoi{10.48550/arXiv.2407.07585}

\bibitem[{{Jogee} {et~al.}(2004){Jogee}, {Barazza}, {Rix}, {Shlosman}, {Barden}, {Wolf}, {Davies}, {Heyer}, {Beckwith}, {Bell}, {Borch}, {Caldwell}, {Conselice}, {Dahlen}, {H{\"a}ussler}, {Heymans}, {Jahnke}, {Knapen}, {Laine}, {Lubell}, {Mobasher}, {McIntosh}, {Meisenheimer}, {Peng}, {Ravindranath}, {Sanchez}, {Somerville}, \& {Wisotzki}}]{Jogee2004}
{Jogee}, S., {Barazza}, F.~D., {Rix}, H.-W., {et~al.} 2004, \apjl, 615, L105, \dodoi{10.1086/426138}

\bibitem[{{Karim} {et~al.}(2013){Karim}, {Swinbank}, {Hodge}, {Smail}, {Walter}, {Biggs}, {Simpson}, {Danielson}, {Alexander}, {Bertoldi}, {de Breuck}, {Chapman}, {Coppin}, {Dannerbauer}, {Edge}, {Greve}, {Ivison}, {Knudsen}, {Menten}, {Schinnerer}, {Wardlow}, {Wei{\ss}}, \& {van der Werf}}]{Karim2013}
{Karim}, A., {Swinbank}, A.~M., {Hodge}, J.~A., {et~al.} 2013, \mnras, 432, 2, \dodoi{10.1093/mnras/stt196}

\bibitem[{{Kartaltepe} {et~al.}(2015){Kartaltepe}, {Sanders}, {Silverman}, {Kashino}, {Chu}, {Zahid}, {Hasinger}, {Kewley}, {Matsuoka}, {Nagao}, {Riguccini}, {Salvato}, {Schawinski}, {Taniguchi}, {Treister}, {Capak}, {Daddi}, \& {Ohta}}]{Kartaltepe2015_spec}
{Kartaltepe}, J.~S., {Sanders}, D.~B., {Silverman}, J.~D., {et~al.} 2015, \apjl, 806, L35, \dodoi{10.1088/2041-8205/806/2/L35}

\bibitem[{{Kashino} {et~al.}(2019){Kashino}, {Silverman}, {Sanders}, {Kartaltepe}, {Daddi}, {Renzini}, {Rodighiero}, {Puglisi}, {Valentino}, {Juneau}, {Arimoto}, {Nagao}, {Ilbert}, {Le F{\`e}vre}, \& {Koekemoer}}]{Kashino2019}
{Kashino}, D., {Silverman}, J.~D., {Sanders}, D., {et~al.} 2019, \apjs, 241, 10, \dodoi{10.3847/1538-4365/ab06c4}

\bibitem[{{Kirkpatrick} {et~al.}(2015){Kirkpatrick}, {Pope}, {Sajina}, {Roebuck}, {Yan}, {Armus}, {D{\'{\i}}az-Santos}, \& {Stierwalt}}]{Kirkpatrick2015}
{Kirkpatrick}, A., {Pope}, A., {Sajina}, A., {et~al.} 2015, \apj, 814, 9, \dodoi{10.1088/0004-637X/814/1/9}

\bibitem[{{Koekemoer} {et~al.}(2007){Koekemoer}, {Aussel}, {Calzetti}, {Capak}, {Giavalisco}, {Kneib}, {Leauthaud}, {Le F{\`e}vre}, {McCracken}, {Massey}, {Mobasher}, {Rhodes}, {Scoville}, \& {Shopbell}}]{Koekemoer2007}
{Koekemoer}, A.~M., {Aussel}, H., {Calzetti}, D., {et~al.} 2007, \apjs, 172, 196, \dodoi{10.1086/520086}

\bibitem[{{Kokorev} {et~al.}(2023){Kokorev}, {Jin}, {Magdis}, {Caputi}, {Valentino}, {Dayal}, {Trebitsch}, {Brammer}, {Fujimoto}, {Bauer}, {Iani}, {Kohno}, {Bl{\'a}nquez Ses{\'e}}, {G{\'o}mez-Guijarro}, {Rinaldi}, \& {Navarro-Carrera}}]{Kokorev2023}
{Kokorev}, V., {Jin}, S., {Magdis}, G.~E., {et~al.} 2023, \apjl, 945, L25, \dodoi{10.3847/2041-8213/acbd9d}

\bibitem[{{Kokorev} {et~al.}(2024){Kokorev}, {Caputi}, {Greene}, {Dayal}, {Trebitsch}, {Cutler}, {Fujimoto}, {Labb{\'e}}, {Miller}, {Iani}, {Navarro-Carrera}, \& {Rinaldi}}]{Kokorev2024}
{Kokorev}, V., {Caputi}, K.~I., {Greene}, J.~E., {et~al.} 2024, arXiv e-prints, arXiv:2401.09981, \dodoi{10.48550/arXiv.2401.09981}

\bibitem[{{Kriek} {et~al.}(2015){Kriek}, {Shapley}, {Reddy}, {Siana}, {Coil}, {Mobasher}, {Freeman}, {de Groot}, {Price}, {Sanders}, {Shivaei}, {Brammer}, {Momcheva}, {Skelton}, {van Dokkum}, {Whitaker}, {Aird}, {Azadi}, {Kassis}, {Bullock}, {Conroy}, {Dav{\'e}}, {Kere{\v{s}}}, \& {Krumholz}}]{Kriek2015}
{Kriek}, M., {Shapley}, A.~E., {Reddy}, N.~A., {et~al.} 2015, \apjs, 218, 15, \dodoi{10.1088/0067-0049/218/2/15}

\bibitem[{{Krogager} {et~al.}(2014){Krogager}, {Zirm}, {Toft}, {Man}, \& {Brammer}}]{Krogager2014}
{Krogager}, J.~K., {Zirm}, A.~W., {Toft}, S., {Man}, A., \& {Brammer}, G. 2014, \apj, 797, 17, \dodoi{10.1088/0004-637X/797/1/17}

\bibitem[{{Lagos} {et~al.}(2020){Lagos}, {da Cunha}, {Robotham}, {Obreschkow}, {Valentino}, {Fujimoto}, {Magdis}, \& {Tobar}}]{Lagos2020}
{Lagos}, C. d.~P., {da Cunha}, E., {Robotham}, A. S.~G., {et~al.} 2020, \mnras, 499, 1948, \dodoi{10.1093/mnras/staa2861}

\bibitem[{{Lagos} {et~al.}(2019){Lagos}, {Robotham}, {Trayford}, {Tobar}, {Bravo}, {Bellstedt}, {Davies}, {Driver}, {Elahi}, {Obreschkow}, \& {Power}}]{Lagos2019}
{Lagos}, C. d.~P., {Robotham}, A. S.~G., {Trayford}, J.~W., {et~al.} 2019, \mnras, 489, 4196, \dodoi{10.1093/mnras/stz2427}

\bibitem[{{Le Bail} {et~al.}(2023){Le Bail}, {Daddi}, {Elbaz}, {Dickinson}, {Giavalisco}, {Magnelli}, {Gomez-Guijarro}, {Kalita}, {Koekemoer}, {Holwerda}, {Bournaud}, {de la Vega}, {Calabro}, {Dekel}, {Cheng}, {Bisigello}, {Franco}, {Costantin}, {Lucas}, {Perez-Gonzalez}, {Lu}, {Wilkins}, {Arrabal Haro}, {Bagley}, {Finkelstein}, {Kartaltepe}, {Papovich}, {Pirzkal}, \& {Yung}}]{LeBail2023}
{Le Bail}, A., {Daddi}, E., {Elbaz}, D., {et~al.} 2023, arXiv e-prints, arXiv:2307.07599, \dodoi{10.48550/arXiv.2307.07599}

\bibitem[{{Le Floc'h} {et~al.}(2009){Le Floc'h}, {Aussel}, {Ilbert}, {Riguccini}, {Frayer}, {Salvato}, {Arnouts}, {Surace}, {Feruglio}, {Rodighiero}, {Capak}, {Kartaltepe}, {Heinis}, {Sheth}, {Yan}, {McCracken}, {Thompson}, {Sanders}, {Scoville}, \& {Koekemoer}}]{LeFloch2009}
{Le Floc'h}, E., {Aussel}, H., {Ilbert}, O., {et~al.} 2009, \apj, 703, 222, \dodoi{10.1088/0004-637X/703/1/222}

\bibitem[{{Liang} {et~al.}(2019){Liang}, {Feldmann}, {Kere{\v{s}}}, {Scoville}, {Hayward}, {Faucher-Gigu{\`e}re}, {Schreiber}, {Ma}, {Hopkins}, \& {Quataert}}]{Liang2019}
{Liang}, L., {Feldmann}, R., {Kere{\v{s}}}, D., {et~al.} 2019, \mnras, 489, 1397, \dodoi{10.1093/mnras/stz2134}

\bibitem[{{Lilly} {et~al.}(2007){Lilly}, {Le F{\`e}vre}, {Renzini}, {Zamorani}, {Scodeggio}, {Contini}, {Carollo}, {Hasinger}, {Kneib}, {Iovino}, {Le Brun}, {Maier}, {Mainieri}, {Mignoli}, {Silverman}, {Tasca}, {Bolzonella}, {Bongiorno}, {Bottini}, {Capak}, {Caputi}, {Cimatti}, {Cucciati}, {Daddi}, {Feldmann}, {Franzetti}, {Garilli}, {Guzzo}, {Ilbert}, {Kampczyk}, {Kovac}, {Lamareille}, {Leauthaud}, {Le Borgne}, {McCracken}, {Marinoni}, {Pello}, {Ricciardelli}, {Scarlata}, {Vergani}, {Sanders}, {Schinnerer}, {Scoville}, {Taniguchi}, {Arnouts}, {Aussel}, {Bardelli}, {Brusa}, {Cappi}, {Ciliegi}, {Finoguenov}, {Foucaud}, {Franceschini}, {Halliday}, {Impey}, {Knobel}, {Koekemoer}, {Kurk}, {Maccagni}, {Maddox}, {Marano}, {Marconi}, {Meneux}, {Mobasher}, {Moreau}, {Peacock}, {Porciani}, {Pozzetti}, {Scaramella}, {Schiminovich}, {Shopbell}, {Smail}, {Thompson}, {Tresse}, {Vettolani}, {Zanichelli}, \& {Zucca}}]{Lilly2007}
{Lilly}, S.~J., {Le F{\`e}vre}, O., {Renzini}, A., {et~al.} 2007, \apjs, 172, 70, \dodoi{10.1086/516589}

\bibitem[{{Long} {et~al.}(2023){Long}, {Casey}, {del P. Lagos}, {Lambrides}, {Zavala}, {Champagne}, {Cooper}, \& {Cooray}}]{Long2023}
{Long}, A.~S., {Casey}, C.~M., {del P. Lagos}, C., {et~al.} 2023, \apj, 953, 11, \dodoi{10.3847/1538-4357/acddde}

\bibitem[{{Lutz} {et~al.}(2011){Lutz}, {Poglitsch}, {Altieri}, {Andreani}, {Aussel}, {Berta}, {Bongiovanni}, {Brisbin}, {Cava}, {Cepa}, {Cimatti}, {Daddi}, {Dominguez-Sanchez}, {Elbaz}, {F{\"o}rster Schreiber}, {Genzel}, {Grazian}, {Gruppioni}, {Harwit}, {Le Floc'h}, {Magdis}, {Magnelli}, {Maiolino}, {Nordon}, {P{\'e}rez Garc{\'\i}a}, {Popesso}, {Pozzi}, {Riguccini}, {Rodighiero}, {Saintonge}, {Sanchez Portal}, {Santini}, {Shao}, {Sturm}, {Tacconi}, {Valtchanov}, {Wetzstein}, \& {Wieprecht}}]{Lutz2011}
{Lutz}, D., {Poglitsch}, A., {Altieri}, B., {et~al.} 2011, \aap, 532, A90, \dodoi{10.1051/0004-6361/201117107}

\bibitem[{{Madau} \& {Dickinson}(2014)}]{MadauDickinson2014}
{Madau}, P., \& {Dickinson}, M. 2014, \araa, 52, 415, \dodoi{10.1146/annurev-astro-081811-125615}

\bibitem[{{Manning} {et~al.}(2022){Manning}, {Casey}, {Zavala}, {Magdis}, {Drew}, {Champagne}, {Aravena}, {B{\'e}thermin}, {Clements}, {Finkelstein}, {Fujimoto}, {Hayward}, {Hodge}, {Ilbert}, {Kartaltepe}, {Knudsen}, {Koekemoer}, {Man}, {Sanders}, {Sheth}, {Spilker}, {Staguhn}, {Talia}, {Treister}, \& {Yun}}]{Manning2022}
{Manning}, S.~M., {Casey}, C.~M., {Zavala}, J.~A., {et~al.} 2022, \apj, 925, 23, \dodoi{10.3847/1538-4357/ac366a}

\bibitem[{{Marchesi} {et~al.}(2016){Marchesi}, {Civano}, {Elvis}, {Salvato}, {Brusa}, {Comastri}, {Gilli}, {Hasinger}, {Lanzuisi}, {Miyaji}, {Treister}, {Urry}, {Vignali}, {Zamorani}, {Allevato}, {Cappelluti}, {Cardamone}, {Finoguenov}, {Griffiths}, {Karim}, {Laigle}, {LaMassa}, {Jahnke}, {Ranalli}, {Schawinski}, {Schinnerer}, {Silverman}, {Smolcic}, {Suh}, \& {Trakhtenbrot}}]{Marchesi2016}
{Marchesi}, S., {Civano}, F., {Elvis}, M., {et~al.} 2016, \apj, 817, 34, \dodoi{10.3847/0004-637X/817/1/34}

\bibitem[{{Marinova} \& {Jogee}(2007)}]{Marinova2007}
{Marinova}, I., \& {Jogee}, S. 2007, \apj, 659, 1176, \dodoi{10.1086/512355}

\bibitem[{{McKinney} {et~al.}(2021){McKinney}, {Hayward}, {Rosenthal}, {Mart{\'\i}nez-Galarza}, {Pope}, {Sajina}, \& {Smith}}]{McKinney2021agn}
{McKinney}, J., {Hayward}, C.~C., {Rosenthal}, L.~J., {et~al.} 2021, \apj, 921, 55, \dodoi{10.3847/1538-4357/ac185f}

\bibitem[{{McKinney} {et~al.}(2023{\natexlab{a}}){McKinney}, {Manning}, {Cooper}, {Long}, {Akins}, {Casey}, {Faisst}, {Franco}, {Hayward}, {Lambrides}, {Magdis}, {Whitaker}, {Yun}, {Champagne}, {Drakos}, {Gentile}, {Gillman}, {Gozaliasl}, {Ilbert}, {Jin}, {Koekemoer}, {Kokorev}, {Liu}, {Rich}, {Robertson}, {Valentino}, {Weaver}, {Zavala}, {Allen}, {Kartaltepe}, {McCracken}, {Paquereau}, {Rhodes}, {Shuntov}, \& {Toft}}]{McKinney2023}
{McKinney}, J., {Manning}, S.~M., {Cooper}, O.~R., {et~al.} 2023{\natexlab{a}}, \apj, 956, 72, \dodoi{10.3847/1538-4357/acf614}

\bibitem[{{McKinney} {et~al.}(2023{\natexlab{b}}){McKinney}, {Finnerty}, {Casey}, {Franco}, {Long}, {Fujimoto}, {Zavala}, {Cooper}, {Akins}, {Pope}, {Armus}, {Soifer}, {Larson}, {Matthews}, {Melbourne}, \& {Cushing}}]{McKinney2023agn}
{McKinney}, J., {Finnerty}, L., {Casey}, C.~M., {et~al.} 2023{\natexlab{b}}, \apjl, 946, L39, \dodoi{10.3847/2041-8213/acc322}

\bibitem[{{McMullin} {et~al.}(2007){McMullin}, {Waters}, {Schiebel}, {Young}, \& {Golap}}]{casa}
{McMullin}, J.~P., {Waters}, B., {Schiebel}, D., {Young}, W., \& {Golap}, K. 2007, in Astronomical Society of the Pacific Conference Series, Vol. 376, Astronomical Data Analysis Software and Systems XVI, ed. R.~A. {Shaw}, F.~{Hill}, \& D.~J. {Bell}, 127

\bibitem[{{Micha{\l}owski} {et~al.}(2012){Micha{\l}owski}, {Dunlop}, {Cirasuolo}, {Hjorth}, {Hayward}, \& {Watson}}]{Michalowski2012}
{Micha{\l}owski}, M.~J., {Dunlop}, J.~S., {Cirasuolo}, M., {et~al.} 2012, \aap, 541, A85, \dodoi{10.1051/0004-6361/201016308}

\bibitem[{{Micha{\l}owski} {et~al.}(2014){Micha{\l}owski}, {Hayward}, {Dunlop}, {Bruce}, {Cirasuolo}, {Cullen}, \& {Hernquist}}]{Michalowski2014}
{Micha{\l}owski}, M.~J., {Hayward}, C.~C., {Dunlop}, J.~S., {et~al.} 2014, \aap, 571, A75, \dodoi{10.1051/0004-6361/201424174}

\bibitem[{{Miettinen} {et~al.}(2017){Miettinen}, {Delvecchio}, {Smol{\v{c}}i{\'c}}, {Aravena}, {Brisbin}, {Karim}, {Magnelli}, {Novak}, {Schinnerer}, {Albrecht}, {Aussel}, {Bertoldi}, {Capak}, {Casey}, {Hayward}, {Ilbert}, {Intema}, {Jiang}, {Le F{\`e}vre}, {McCracken}, {Mu{\~n}oz Arancibia}, {Navarrete}, {Padilla}, {Riechers}, {Salvato}, {Scott}, {Sheth}, \& {Tasca}}]{Miettinen2017}
{Miettinen}, O., {Delvecchio}, I., {Smol{\v{c}}i{\'c}}, V., {et~al.} 2017, \aap, 606, A17, \dodoi{10.1051/0004-6361/201730762}

\bibitem[{{Miller} {et~al.}(2015){Miller}, {Hayward}, {Chapman}, \& {Behroozi}}]{Miller2015}
{Miller}, T.~B., {Hayward}, C.~C., {Chapman}, S.~C., \& {Behroozi}, P.~S. 2015, \mnras, 452, 878, \dodoi{10.1093/mnras/stv1267}

\bibitem[{{Mohan} \& {Rafferty}(2015)}]{pybdsf}
{Mohan}, N., \& {Rafferty}, D. 2015, {PyBDSF: Python Blob Detection and Source Finder}, Astrophysics Source Code Library, record ascl:1502.007

\bibitem[{{Moster} {et~al.}(2011){Moster}, {Somerville}, {Newman}, \& {Rix}}]{Moster2011}
{Moster}, B.~P., {Somerville}, R.~S., {Newman}, J.~A., \& {Rix}, H.-W. 2011, \apj, 731, 113, \dodoi{10.1088/0004-637X/731/2/113}

\bibitem[{{Murphy} {et~al.}(2011){Murphy}, {Chary}, {Dickinson}, {Pope}, {Frayer}, \& {Lin}}]{Murphy2011}
{Murphy}, E.~J., {Chary}, R.~R., {Dickinson}, M., {et~al.} 2011, \apj, 732, 126, \dodoi{10.1088/0004-637X/732/2/126}

\bibitem[{{Nanayakkara} {et~al.}(2016){Nanayakkara}, {Glazebrook}, {Kacprzak}, {Yuan}, {Tran}, {Spitler}, {Kewley}, {Straatman}, {Cowley}, {Fisher}, {Labbe}, {Tomczak}, {Allen}, \& {Alcorn}}]{Nanayakkara2016}
{Nanayakkara}, T., {Glazebrook}, K., {Kacprzak}, G.~G., {et~al.} 2016, \apj, 828, 21, \dodoi{10.3847/0004-637X/828/1/21}

\bibitem[{{Oke}(1974)}]{Oke1974}
{Oke}, J.~B. 1974, \apjs, 27, 21, \dodoi{10.1086/190287}

\bibitem[{{Oliver} {et~al.}(2012){Oliver}, {Bock}, {Altieri}, {Amblard}, {Arumugam}, {Aussel}, {Babbedge}, {Beelen}, {B{\'e}thermin}, {Blain}, {Boselli}, {Bridge}, {Brisbin}, {Buat}, {Burgarella}, {Castro-Rodr{\'\i}guez}, {Cava}, {Chanial}, {Cirasuolo}, {Clements}, {Conley}, {Conversi}, {Cooray}, {Dowell}, {Dubois}, {Dwek}, {Dye}, {Eales}, {Elbaz}, {Farrah}, {Feltre}, {Ferrero}, {Fiolet}, {Fox}, {Franceschini}, {Gear}, {Giovannoli}, {Glenn}, {Gong}, {Gonz{\'a}lez Solares}, {Griffin}, {Halpern}, {Harwit}, {Hatziminaoglou}, {Heinis}, {Hurley}, {Hwang}, {Hyde}, {Ibar}, {Ilbert}, {Isaak}, {Ivison}, {Lagache}, {Le Floc'h}, {Levenson}, {Faro}, {Lu}, {Madden}, {Maffei}, {Magdis}, {Mainetti}, {Marchetti}, {Marsden}, {Marshall}, {Mortier}, {Nguyen}, {O'Halloran}, {Omont}, {Page}, {Panuzzo}, {Papageorgiou}, {Patel}, {Pearson}, {P{\'e}rez-Fournon}, {Pohlen}, {Rawlings}, {Raymond}, {Rigopoulou}, {Riguccini}, {Rizzo}, {Rodighiero}, {Roseboom}, {Rowan-Robinson}, {S{\'a}nchez Portal}, {Schulz}, {Scott}, {Seymour}, {Shupe},
  {Smith}, {Stevens}, {Symeonidis}, {Trichas}, {Tugwell}, {Vaccari}, {Valtchanov}, {Vieira}, {Viero}, {Vigroux}, {Wang}, {Ward}, {Wardlow}, {Wright}, {Xu}, \& {Zemcov}}]{Oliver2012}
{Oliver}, S.~J., {Bock}, J., {Altieri}, B., {et~al.} 2012, \mnras, 424, 1614, \dodoi{10.1111/j.1365-2966.2012.20912.x}

\bibitem[{{Onodera} {et~al.}(2015){Onodera}, {Carollo}, {Renzini}, {Cappellari}, {Mancini}, {Arimoto}, {Daddi}, {Gobat}, {Strazzullo}, {Tacchella}, \& {Yamada}}]{Onodera2015}
{Onodera}, M., {Carollo}, C.~M., {Renzini}, A., {et~al.} 2015, \apj, 808, 161, \dodoi{10.1088/0004-637X/808/2/161}

\bibitem[{{Popescu} {et~al.}(2023){Popescu}, {Pope}, {Lee}, {Alberts}, {Chiang}, {Lee}, {Brodwin}, {McKinney}, \& {Ramakrishnan}}]{Popescu2023}
{Popescu}, R., {Pope}, A., {Lee}, K.-S., {et~al.} 2023, \apj, 958, 12, \dodoi{10.3847/1538-4357/acee79}

\bibitem[{{Popping} {et~al.}(2017){Popping}, {Somerville}, \& {Galametz}}]{Popping2017}
{Popping}, G., {Somerville}, R.~S., \& {Galametz}, M. 2017, \mnras, 471, 3152, \dodoi{10.1093/mnras/stx1545}

\bibitem[{{Price} {et~al.}(2023){Price}, {Suess}, {Williams}, {Bezanson}, {Khullar}, {Nelson}, {Wang}, {Weaver}, {Fujimoto}, {Kokorev}, {Greene}, {Brammer}, {Cutler}, {Dayal}, {Furtak}, {Labbe}, {Leja}, {Miller}, {Nanayakkara}, {Pan}, \& {Whitaker}}]{Price2023}
{Price}, S.~H., {Suess}, K.~A., {Williams}, C.~C., {et~al.} 2023, arXiv e-prints, arXiv:2310.02500, \dodoi{10.48550/arXiv.2310.02500}

\bibitem[{{Reddy} {et~al.}(2015){Reddy}, {Kriek}, {Shapley}, {Freeman}, {Siana}, {Coil}, {Mobasher}, {Price}, {Sanders}, \& {Shivaei}}]{Reddy2015}
{Reddy}, N.~A., {Kriek}, M., {Shapley}, A.~E., {et~al.} 2015, \apj, 806, 259, \dodoi{10.1088/0004-637X/806/2/259}

\bibitem[{{Rujopakarn} {et~al.}(2023){Rujopakarn}, {Williams}, {Daddi}, {Schramm}, {Sun}, {Alberts}, {Rieke}, {Tan}, {Tacchella}, {Giavalisco}, \& {Silverman}}]{Rujopakarn2023}
{Rujopakarn}, W., {Williams}, C.~C., {Daddi}, E., {et~al.} 2023, \apjl, 948, L8, \dodoi{10.3847/2041-8213/accc82}

\bibitem[{{Sanders} \& {Mirabel}(1996)}]{Sanders1996}
{Sanders}, D.~B., \& {Mirabel}, I.~F. 1996, \araa, 34, 749, \dodoi{10.1146/annurev.astro.34.1.749}

\bibitem[{{Sanders} {et~al.}(2007{\natexlab{a}}){Sanders}, {Salvato}, {Aussel}, {Ilbert}, {Scoville}, {Surace}, {Frayer}, {Sheth}, {Helou}, {Brooke}, {Bhattacharya}, {Yan}, {Kartaltepe}, {Barnes}, {Blain}, {Calzetti}, {Capak}, {Carilli}, {Carollo}, {Comastri}, {Daddi}, {Ellis}, {Elvis}, {Fall}, {Franceschini}, {Giavalisco}, {Hasinger}, {Impey}, {Koekemoer}, {Le F{\`e}vre}, {Lilly}, {Liu}, {McCracken}, {Mobasher}, {Renzini}, {Rich}, {Schinnerer}, {Shopbell}, {Taniguchi}, {Thompson}, {Urry}, \& {Williams}}]{Sanders2007}
{Sanders}, D.~B., {Salvato}, M., {Aussel}, H., {et~al.} 2007{\natexlab{a}}, \apjs, 172, 86, \dodoi{10.1086/517885}

\bibitem[{{Sanders} {et~al.}(2007{\natexlab{b}}){Sanders}, {Salvato}, {Aussel}, {Ilbert}, {Scoville}, {Surace}, {Frayer}, {Sheth}, {Helou}, {Brooke}, {Bhattacharya}, {Yan}, {Kartaltepe}, {Barnes}, {Blain}, {Calzetti}, {Capak}, {Carilli}, {Carollo}, {Comastri}, {Daddi}, {Ellis}, {Elvis}, {Fall}, {Franceschini}, {Giavalisco}, {Hasinger}, {Impey}, {Koekemoer}, {Le F{\`e}vre}, {Lilly}, {Liu}, {McCracken}, {Mobasher}, {Renzini}, {Rich}, {Schinnerer}, {Shopbell}, {Taniguchi}, {Thompson}, {Urry}, \& {Williams}}]{SCOSMOS}
---. 2007{\natexlab{b}}, \apjs, 172, 86, \dodoi{10.1086/517885}

\bibitem[{{Schinnerer} {et~al.}(2010){Schinnerer}, {Sargent}, {Bondi}, {Smol{\v{c}}i{\'c}}, {Datta}, {Carilli}, {Bertoldi}, {Blain}, {Ciliegi}, {Koekemoer}, \& {Scoville}}]{Schinnerer2010}
{Schinnerer}, E., {Sargent}, M.~T., {Bondi}, M., {et~al.} 2010, \apjs, 188, 384, \dodoi{10.1088/0067-0049/188/2/384}

\bibitem[{{Schreiber} {et~al.}(2018){Schreiber}, {Elbaz}, {Pannella}, {Ciesla}, {Wang}, \& {Franco}}]{Schreiber2018}
{Schreiber}, C., {Elbaz}, D., {Pannella}, M., {et~al.} 2018, \aap, 609, A30, \dodoi{10.1051/0004-6361/201731506}

\bibitem[{{Scoville} {et~al.}(2007){Scoville}, {Aussel}, {Brusa}, {Capak}, {Carollo}, {Elvis}, {Giavalisco}, {Guzzo}, {Hasinger}, {Impey}, {Kneib}, {LeFevre}, {Lilly}, {Mobasher}, {Renzini}, {Rich}, {Sanders}, {Schinnerer}, {Schminovich}, {Shopbell}, {Taniguchi}, \& {Tyson}}]{Scoville2007}
{Scoville}, N., {Aussel}, H., {Brusa}, M., {et~al.} 2007, \apjs, 172, 1, \dodoi{10.1086/516585}

\bibitem[{{Shah} {et~al.}(2020){Shah}, {Kartaltepe}, {Magagnoli}, {Cox}, {Wetherell}, {Vanderhoof}, {Calabro}, {Chartab}, {Conselice}, {Croton}, {Donley}, {de Groot}, {de la Vega}, {Hathi}, {Ilbert}, {Inami}, {Kocevski}, {Koekemoer}, {Lemaux}, {Mantha}, {Marchesi}, {Martig}, {Masters}, {McGrath}, {McIntosh}, {Moreno}, {Nayyeri}, {Pampliega}, {Salvato}, {Snyder}, {Straughn}, {Treister}, \& {Weston}}]{Shah2020}
{Shah}, E.~A., {Kartaltepe}, J.~S., {Magagnoli}, C.~T., {et~al.} 2020, \apj, 904, 107, \dodoi{10.3847/1538-4357/abbf59}

\bibitem[{{Shivaei} {et~al.}(2015){Shivaei}, {Reddy}, {Shapley}, {Kriek}, {Siana}, {Mobasher}, {Coil}, {Freeman}, {Sanders}, {Price}, {de Groot}, \& {Azadi}}]{Shivaei2015}
{Shivaei}, I., {Reddy}, N.~A., {Shapley}, A.~E., {et~al.} 2015, \apj, 815, 98, \dodoi{10.1088/0004-637X/815/2/98}

\bibitem[{{Simpson} {et~al.}(2014){Simpson}, {Swinbank}, {Smail}, {Alexander}, {Brandt}, {Bertoldi}, {de Breuck}, {Chapman}, {Coppin}, {da Cunha}, {Danielson}, {Dannerbauer}, {Greve}, {Hodge}, {Ivison}, {Karim}, {Knudsen}, {Poggianti}, {Schinnerer}, {Thomson}, {Walter}, {Wardlow}, {Wei{\ss}}, \& {van der Werf}}]{Simpson2014}
{Simpson}, J.~M., {Swinbank}, A.~M., {Smail}, I., {et~al.} 2014, \apj, 788, 125, \dodoi{10.1088/0004-637X/788/2/125}

\bibitem[{{Simpson} {et~al.}(2019){Simpson}, {Smail}, {Swinbank}, {Chapman}, {Chen}, {Geach}, {Matsuda}, {Wang}, {Wang}, {Yang}, {Ao}, {Asquith}, {Bourne}, {Coogan}, {Coppin}, {Gullberg}, {Hine}, {Ho}, {Hwang}, {Ivison}, {Kato}, {Lacaille}, {Lewis}, {Liu}, {Micha{\l}owski}, {Oteo}, {Sawicki}, {Scholtz}, {Smith}, {Thomson}, \& {Wardlow}}]{Simpson2019}
{Simpson}, J.~M., {Smail}, I., {Swinbank}, A.~M., {et~al.} 2019, \apj, 880, 43, \dodoi{10.3847/1538-4357/ab23ff}

\bibitem[{{Simpson} {et~al.}(2020){Simpson}, {Smail}, {Dudzevi{\v{c}}i{\={u}}t{\.{e}}}, {Matsuda}, {Hsieh}, {Wang}, {Swinbank}, {Stach}, {An}, {Birkin}, {Ao}, {Bunker}, {Chapman}, {Chen}, {Coppin}, {Ikarashi}, {Ivison}, {Mitsuhashi}, {Saito}, {Umehata}, {Wang}, \& {Zhao}}]{Simpson2020}
{Simpson}, J.~M., {Smail}, I., {Dudzevi{\v{c}}i{\={u}}t{\.{e}}}, U., {et~al.} 2020, \mnras, 495, 3409, \dodoi{10.1093/mnras/staa1345}

\bibitem[{{Smail} {et~al.}(1997){Smail}, {Ivison}, \& {Blain}}]{Smail1997}
{Smail}, I., {Ivison}, R.~J., \& {Blain}, A.~W. 1997, \apjl, 490, L5, \dodoi{10.1086/311017}

\bibitem[{{Smail} {et~al.}(2021){Smail}, {Dudzevi{\v{c}}i{\={u}}t{\.{e}}}, {Stach}, {Almaini}, {Birkin}, {Chapman}, {Chen}, {Geach}, {Gullberg}, {Hodge}, {Ikarashi}, {Ivison}, {Scott}, {Simpson}, {Swinbank}, {Thomson}, {Walter}, {Wardlow}, \& {van der Werf}}]{Smail2021}
{Smail}, I., {Dudzevi{\v{c}}i{\={u}}t{\.{e}}}, U., {Stach}, S.~M., {et~al.} 2021, \mnras, 502, 3426, \dodoi{10.1093/mnras/stab283}

\bibitem[{{Smol{\v{c}}i{\'c}} {et~al.}(2017{\natexlab{a}}){Smol{\v{c}}i{\'c}}, {Novak}, {Bondi}, {Ciliegi}, {Mooley}, {Schinnerer}, {Zamorani}, {Navarrete}, {Bourke}, {Karim}, {Vardoulaki}, {Leslie}, {Delhaize}, {Carilli}, {Myers}, {Baran}, {Delvecchio}, {Miettinen}, {Banfield}, {Balokovi{\'c}}, {Bertoldi}, {Capak}, {Frail}, {Hallinan}, {Hao}, {Herrera Ruiz}, {Horesh}, {Ilbert}, {Intema}, {Jeli{\'c}}, {Kl{\"o}ckner}, {Krpan}, {Kulkarni}, {McCracken}, {Laigle}, {Middleberg}, {Murphy}, {Sargent}, {Scoville}, \& {Sheth}}]{Smolcic2017}
{Smol{\v{c}}i{\'c}}, V., {Novak}, M., {Bondi}, M., {et~al.} 2017{\natexlab{a}}, \aap, 602, A1, \dodoi{10.1051/0004-6361/201628704}

\bibitem[{{Smol{\v{c}}i{\'c}} {et~al.}(2017{\natexlab{b}}){Smol{\v{c}}i{\'c}}, {Miettinen}, {Tomi{\v{c}}i{\'c}}, {Zamorani}, {Finoguenov}, {Lemaux}, {Aravena}, {Capak}, {Chiang}, {Civano}, {Delvecchio}, {Ilbert}, {Jurlin}, {Karim}, {Laigle}, {Le F{\`e}vre}, {Marchesi}, {McCracken}, {Riechers}, {Salvato}, {Schinnerer}, {Tasca}, \& {Toft}}]{Smolcic2017overdense}
{Smol{\v{c}}i{\'c}}, V., {Miettinen}, O., {Tomi{\v{c}}i{\'c}}, N., {et~al.} 2017{\natexlab{b}}, \aap, 597, A4, \dodoi{10.1051/0004-6361/201526989}

\bibitem[{{Solimano} {et~al.}(2024){Solimano}, {Gonz{\'a}lez-L{\'o}pez}, {Aravena}, {Herrera-Camus}, {De Looze}, {F{\"o}rster Schreiber}, {Spilker}, {Tadaki}, {Assef}, {Barcos-Mu{\~n}oz}, {Davies}, {D{\'\i}az-Santos}, {Ferrara}, {Fisher}, {Guaita}, {Ikeda}, {Johnston}, {Lutz}, {Mitsuhashi}, {Moya-Sierralta}, {Rela{\~n}o}, {Naab}, {Posses}, {Telikova}, {{\"U}bler}, {van der Giessen}, \& {Villanueva}}]{Solimano2024}
{Solimano}, M., {Gonz{\'a}lez-L{\'o}pez}, J., {Aravena}, M., {et~al.} 2024, arXiv e-prints, arXiv:2401.04919, \dodoi{10.48550/arXiv.2401.04919}

\bibitem[{{Sommovigo} {et~al.}(2022){Sommovigo}, {Ferrara}, {Pallottini}, {Dayal}, {Bouwens}, {Smit}, {da Cunha}, {De Looze}, {Bowler}, {Hodge}, {Inami}, {Oesch}, {Endsley}, {Gonzalez}, {Schouws}, {Stark}, {Stefanon}, {Aravena}, {Graziani}, {Riechers}, {Schneider}, {van der Werf}, {Algera}, {Barrufet}, {Fudamoto}, {Hygate}, {Labb{\'e}}, {Li}, {Nanayakkara}, \& {Topping}}]{Sommovigo2022}
{Sommovigo}, L., {Ferrara}, A., {Pallottini}, A., {et~al.} 2022, \mnras, 513, 3122, \dodoi{10.1093/mnras/stac302}

\bibitem[{{Storey-Fisher} {et~al.}(2024){Storey-Fisher}, {Hogg}, {Rix}, {Eilers}, {Fabbian}, {Blanton}, \& {Alonso}}]{StoreyFisher2024}
{Storey-Fisher}, K., {Hogg}, D.~W., {Rix}, H.-W., {et~al.} 2024, \apj, 964, 69, \dodoi{10.3847/1538-4357/ad1328}

\bibitem[{{Stott} {et~al.}(2016){Stott}, {Swinbank}, {Johnson}, {Tiley}, {Magdis}, {Bower}, {Bunker}, {Bureau}, {Harrison}, {Jarvis}, {Sharples}, {Smail}, {Sobral}, {Best}, \& {Cirasuolo}}]{Stott2016}
{Stott}, J.~P., {Swinbank}, A.~M., {Johnson}, H.~L., {et~al.} 2016, \mnras, 457, 1888, \dodoi{10.1093/mnras/stw129}

\bibitem[{{Swinbank} {et~al.}(2004){Swinbank}, {Smail}, {Chapman}, {Blain}, {Ivison}, \& {Keel}}]{Swinbank2004}
{Swinbank}, A.~M., {Smail}, I., {Chapman}, S.~C., {et~al.} 2004, \apj, 617, 64, \dodoi{10.1086/425171}

\bibitem[{{Swinbank} {et~al.}(2008){Swinbank}, {Lacey}, {Smail}, {Baugh}, {Frenk}, {Blain}, {Chapman}, {Coppin}, {Ivison}, {Gonzalez}, \& {Hainline}}]{Swinbank2008}
{Swinbank}, A.~M., {Lacey}, C.~G., {Smail}, I., {et~al.} 2008, \mnras, 391, 420, \dodoi{10.1111/j.1365-2966.2008.13911.x}

\bibitem[{{Talia} {et~al.}(2021){Talia}, {Cimatti}, {Giulietti}, {Zamorani}, {Bethermin}, {Faisst}, {Le F{\`e}vre}, \& {Smol{\c{c}}i{\'c}}}]{Talia2021}
{Talia}, M., {Cimatti}, A., {Giulietti}, M., {et~al.} 2021, \apj, 909, 23, \dodoi{10.3847/1538-4357/abd6e3}

\bibitem[{{Toft} {et~al.}(2014){Toft}, {Smol{\v{c}}i{\'c}}, {Magnelli}, {Karim}, {Zirm}, {Michalowski}, {Capak}, {Sheth}, {Schawinski}, {Krogager}, {Wuyts}, {Sanders}, {Man}, {Lutz}, {Staguhn}, {Berta}, {Mccracken}, {Krpan}, \& {Riechers}}]{Toft2014}
{Toft}, S., {Smol{\v{c}}i{\'c}}, V., {Magnelli}, B., {et~al.} 2014, \apj, 782, 68, \dodoi{10.1088/0004-637X/782/2/68}

\bibitem[{{Trenti} \& {Stiavelli}(2008)}]{Trenti2008}
{Trenti}, M., \& {Stiavelli}, M. 2008, \apj, 676, 767, \dodoi{10.1086/528674}

\bibitem[{{Valentino} {et~al.}(2020){Valentino}, {Tanaka}, {Davidzon}, {Toft}, {G{\'o}mez-Guijarro}, {Stockmann}, {Onodera}, {Brammer}, {Ceverino}, {Faisst}, {Gallazzi}, {Hayward}, {Ilbert}, {Kubo}, {Magdis}, {Selsing}, {Shimakawa}, {Sparre}, {Steinhardt}, {Yabe}, \& {Zabl}}]{Valentino2020}
{Valentino}, F., {Tanaka}, M., {Davidzon}, I., {et~al.} 2020, \apj, 889, 93, \dodoi{10.3847/1538-4357/ab64dc}

\bibitem[{{VanderPlas} {et~al.}(2012){VanderPlas}, {Connolly}, {Ivezic}, \& {Gray}}]{astroML}
{VanderPlas}, J., {Connolly}, A.~J., {Ivezic}, Z., \& {Gray}, A. 2012, in Proceedings of Conference on Intelligent Data Understanding (CIDU, 47--54, \dodoi{10.1109/CIDU.2012.6382200}

\bibitem[{{Viero} {et~al.}(2022){Viero}, {Sun}, {Chung}, {Moncelsi}, \& {Condon}}]{Viero2022}
{Viero}, M.~P., {Sun}, G., {Chung}, D.~T., {Moncelsi}, L., \& {Condon}, S.~S. 2022, \mnras, 516, L30, \dodoi{10.1093/mnrasl/slac075}

\bibitem[{{Walter} {et~al.}(2012){Walter}, {Decarli}, {Carilli}, {Bertoldi}, {Cox}, {da Cunha}, {Daddi}, {Dickinson}, {Downes}, {Elbaz}, {Ellis}, {Hodge}, {Neri}, {Riechers}, {Weiss}, {Bell}, {Dannerbauer}, {Krips}, {Krumholz}, {Lentati}, {Maiolino}, {Menten}, {Rix}, {Robertson}, {Spinrad}, {Stark}, \& {Stern}}]{Walter2012}
{Walter}, F., {Decarli}, R., {Carilli}, C., {et~al.} 2012, \nat, 486, 233, \dodoi{10.1038/nature11073}

\bibitem[{{Wang} {et~al.}(2013){Wang}, {Brandt}, {Luo}, {Smail}, {Alexander}, {Danielson}, {Hodge}, {Karim}, {Lehmer}, {Simpson}, {Swinbank}, {Walter}, {Wardlow}, {Xue}, {Chapman}, {Coppin}, {Dannerbauer}, {De Breuck}, {Menten}, \& {van der Werf}}]{Wang2013}
{Wang}, S.~X., {Brandt}, W.~N., {Luo}, B., {et~al.} 2013, \apj, 778, 179, \dodoi{10.1088/0004-637X/778/2/179}

\bibitem[{{Wang} {et~al.}(2016{\natexlab{a}}){Wang}, {Elbaz}, {Schreiber}, {Pannella}, {Shu}, {Willner}, {Ashby}, {Huang}, {Fontana}, {Dekel}, {Daddi}, {Ferguson}, {Dunlop}, {Ciesla}, {Koekemoer}, {Giavalisco}, {Boutsia}, {Finkelstein}, {Juneau}, {Barro}, {Koo}, {Micha{\l}owski}, {Orellana}, {Lu}, {Castellano}, {Bourne}, {Buitrago}, {Santini}, {Faber}, {Hathi}, {Lucas}, \& {P{\'e}rez-Gonz{\'a}lez}}]{Wang2016}
{Wang}, T., {Elbaz}, D., {Schreiber}, C., {et~al.} 2016{\natexlab{a}}, \apj, 816, 84, \dodoi{10.3847/0004-637X/816/2/84}

\bibitem[{{Wang} {et~al.}(2016{\natexlab{b}}){Wang}, {Elbaz}, {Daddi}, {Finoguenov}, {Liu}, {Schreiber}, {Mart{\'\i}n}, {Strazzullo}, {Valentino}, {van der Burg}, {Zanella}, {Ciesla}, {Gobat}, {Le Brun}, {Pannella}, {Sargent}, {Shu}, {Tan}, {Cappelluti}, \& {Li}}]{WangTao2016}
{Wang}, T., {Elbaz}, D., {Daddi}, E., {et~al.} 2016{\natexlab{b}}, \apj, 828, 56, \dodoi{10.3847/0004-637X/828/1/56}

\bibitem[{{Wardlow} {et~al.}(2011){Wardlow}, {Smail}, {Coppin}, {Alexander}, {Brandt}, {Danielson}, {Luo}, {Swinbank}, {Walter}, {Wei{\ss}}, {Xue}, {Zibetti}, {Bertoldi}, {Biggs}, {Chapman}, {Dannerbauer}, {Dunlop}, {Gawiser}, {Ivison}, {Knudsen}, {Kov{\'a}cs}, {Lacey}, {Menten}, {Padilla}, {Rix}, \& {van der Werf}}]{Wardlow2011}
{Wardlow}, J.~L., {Smail}, I., {Coppin}, K.~E.~K., {et~al.} 2011, \mnras, 415, 1479, \dodoi{10.1111/j.1365-2966.2011.18795.x}

\bibitem[{{Weaver} {et~al.}(2022){Weaver}, {Kauffmann}, {Ilbert}, {McCracken}, {Moneti}, {Toft}, {Brammer}, {Shuntov}, {Davidzon}, {Hsieh}, {Laigle}, {Anastasiou}, {Jespersen}, {Vinther}, {Capak}, {Casey}, {McPartland}, {Milvang-Jensen}, {Mobasher}, {Sanders}, {Zalesky}, {Arnouts}, {Aussel}, {Dunlop}, {Faisst}, {Franx}, {Furtak}, {Fynbo}, {Gould}, {Greve}, {Gwyn}, {Kartaltepe}, {Kashino}, {Koekemoer}, {Kokorev}, {Le F{\`e}vre}, {Lilly}, {Masters}, {Magdis}, {Mehta}, {Peng}, {Riechers}, {Salvato}, {Sawicki}, {Scarlata}, {Scoville}, {Shirley}, {Silverman}, {Sneppen}, {Smolc̆i{\'c}}, {Steinhardt}, {Stern}, {Tanaka}, {Taniguchi}, {Teplitz}, {Vaccari}, {Wang}, \& {Zamorani}}]{Weaver2022}
{Weaver}, J.~R., {Kauffmann}, O.~B., {Ilbert}, O., {et~al.} 2022, \apjs, 258, 11, \dodoi{10.3847/1538-4365/ac3078}

\bibitem[{{Wei{\ss}} {et~al.}(2009){Wei{\ss}}, {Ivison}, {Downes}, {Walter}, {Cirasuolo}, \& {Menten}}]{Weiss2009}
{Wei{\ss}}, A., {Ivison}, R.~J., {Downes}, D., {et~al.} 2009, \apjl, 705, L45, \dodoi{10.1088/0004-637X/705/1/L45}

\bibitem[{{Williams} {et~al.}(2019){Williams}, {Labbe}, {Spilker}, {Stefanon}, {Leja}, {Whitaker}, {Bezanson}, {Narayanan}, {Oesch}, \& {Weiner}}]{Williams2019}
{Williams}, C.~C., {Labbe}, I., {Spilker}, J., {et~al.} 2019, \apj, 884, 154, \dodoi{10.3847/1538-4357/ab44aa}

\bibitem[{{Williams} {et~al.}(2023){Williams}, {Alberts}, {Ji}, {Hainline}, {Lyu}, {Rieke}, {Endsley}, {Suess}, {Johnson}, {Florian}, {Shivaei}, {Rujopakarn}, {Baker}, {Bhatawdekar}, {Boyett}, {Bunker}, {Carniani}, {Charlot}, {Curtis-Lake}, {DeCoursey}, {de Graaff}, {Egami}, {Eisenstein}, {Gibson}, {Hausen}, {Helton}, {Maiolino}, {Maseda}, {Nelson}, {Perez-Gonzalez}, {Rieke}, {Robertson}, {Sun}, {Tacchella}, {Willmer}, \& {Willott}}]{Williams2023}
{Williams}, C.~C., {Alberts}, S., {Ji}, Z., {et~al.} 2023, arXiv e-prints, arXiv:2311.07483, \dodoi{10.48550/arXiv.2311.07483}

\bibitem[{{Wu} {et~al.}(2023){Wu}, {Cai}, {Sun}, {Bian}, {Lin}, {Li}, {Li}, {Bauer}, {Egami}, {Fan}, {Gonz{\'a}lez-L{\'o}pez}, {Li}, {Wang}, {Yang}, {Zhang}, \& {Zou}}]{Wu2023}
{Wu}, Y., {Cai}, Z., {Sun}, F., {et~al.} 2023, \apjl, 942, L1, \dodoi{10.3847/2041-8213/aca652}

\bibitem[{{Zavala} {et~al.}(2021){Zavala}, {Casey}, {Manning}, {Aravena}, {Bethermin}, {Caputi}, {Clements}, {Cunha}, {Drew}, {Finkelstein}, {Fujimoto}, {Hayward}, {Hodge}, {Kartaltepe}, {Knudsen}, {Koekemoer}, {Long}, {Magdis}, {Man}, {Popping}, {Sanders}, {Scoville}, {Sheth}, {Staguhn}, {Toft}, {Treister}, {Vieira}, \& {Yun}}]{Zavala2021}
{Zavala}, J.~A., {Casey}, C.~M., {Manning}, S.~M., {et~al.} 2021, \apj, 909, 165, \dodoi{10.3847/1538-4357/abdb27}

\bibitem[{{Zavala} {et~al.}(2023){Zavala}, {Buat}, {Casey}, {Finkelstein}, {Burgarella}, {Bagley}, {Ciesla}, {Daddi}, {Dickinson}, {Ferguson}, {Franco}, {Jim{\'e}nez-Andrade}, {Kartaltepe}, {Koekemoer}, {Bail}, {Murphy}, {Papovich}, {Tacchella}, {Wilkins}, {Aretxaga}, {Behroozi}, {Champagne}, {Fontana}, {Giavalisco}, {Grazian}, {Grogin}, {Kewley}, {Kocevski}, {Kirkpatrick}, {Lotz}, {Pentericci}, {P{\'e}rez-Gonz{\'a}lez}, {Pirzkal}, {Ravindranath}, {Somerville}, {Trump}, {Yang}, {Yung}, {Almaini}, {Amor{\'\i}n}, {Annunziatella}, {Haro}, {Backhaus}, {Barro}, {Bell}, {Bhatawdekar}, {Bisigello}, {Buitrago}, {Calabr{\`o}}, {Castellano}, {Ch{\'a}vez Ortiz}, {Chworowsky}, {Cleri}, {Cohen}, {Cole}, {Cooke}, {Cooper}, {Cooray}, {Costantin}, {Cox}, {Croton}, {Dav{\'e}}, {de La Vega}, {Dekel}, {Elbaz}, {Estrada-Carpenter}, {Fern{\'a}ndez}, {Finkelstein}, {Freundlich}, {Fujimoto}, {Garc{\'\i}a-Argum{\'a}nez}, {Gardner}, {Gawiser}, {G{\'o}mez-Guijarro}, {Guo}, {Hamilton}, {Hathi}, {Holwerda}, {Hirschmann},
  {Huertas-Company}, {Hutchison}, {Iyer}, {Jaskot}, {Jha}, {Jogee}, {Juneau}, {Jung}, {Kassin}, {Kurczynski}, {Larson}, {Leung}, {Long}, {Lucas}, {Magnelli}, {Mantha}, {Matharu}, {McGrath}, {McIntosh}, {Medrano}, {Merlin}, {Mobasher}, {Morales}, {Newman}, {Nicholls}, {Pandya}, {Rafelski}, {Ronayne}, {Rose}, {Ryan}, {Santini}, {Seill{\'e}}, {Shah}, {Shen}, {Simons}, {Snyder}, {Stanway}, {Straughn}, {Teplitz}, {Vanderhoof}, {Vega-Ferrero}, {Wang}, {Weiner}, {Willmer}, {Wuyts}, \& {(The Ceers Team)}}]{Zavala2022}
{Zavala}, J.~A., {Buat}, V., {Casey}, C.~M., {et~al.} 2023, \apjl, 943, L9, \dodoi{10.3847/2041-8213/acacfe}

\bibitem[{{Zimmerman} {et~al.}(2024){Zimmerman}, {Narayanan}, {Whitaker}, \& {Dav{\`e}}}]{Zimmerman2024}
{Zimmerman}, D.~T., {Narayanan}, D., {Whitaker}, K.~E., \& {Dav{\`e}}, R. 2024, arXiv e-prints, arXiv:2401.06719, \dodoi{10.48550/arXiv.2401.06719}

\end{thebibliography}
\bibliographystyle{aasjournal}

\end{document}